\documentstyle[twoside,12pt,fleqn,epsfig]{report}
\def\photonatomright{\begin{picture}(3,1.5)(0,0)
                                \put(0,-0.75){\tencircw \symbol{2}}
                                \put(1.5,-0.75){\tencircw \symbol{1}}
                                \put(1.5,0.75){\tencircw \symbol{3}}
                                \put(3,0.75){\tencircw \symbol{0}}
                      \end{picture}}

\def\photonatomup{\begin{picture}(1.5,3)(0,0)
                             \put(-0.75,3){\tencircw \symbol{3}}
                             \put(-0.75,1.5){\tencircw \symbol{2}}
                             \put(0.75,1.5){\tencircw \symbol{0}}
                             \put(0.75,0){\tencircw \symbol{1}}
                   \end{picture}}
\def\gluonatomright{\begin{picture}(3.5,3.25)(0,0)
                                \put(2.5,0){\tencircw \symbol{9}}
                                \put(1,0){\tencircw \symbol{10}}
                                \put(1,3){\tencircw \symbol{3}}
                                \put(2.5,3){\tencircw \symbol{0}}
                     \end{picture}}

\def\photonright{\begin{picture}(30,1.5)(0,0)
                     \multiput(0,0)(3,0){10}{\photonatomright}
                  \end{picture}}

\def\photonrighthalf{\begin{picture}(30,1.5)(0,0)
                     \multiput(0,0)(3,0){5}{\photonatomright}
                  \end{picture}}

\def\photonup{\begin{picture}(1.5,30)(0,0)
                  \multiput(0,0)(0,3){10}{\photonatomup}
              \end{picture}}
\def\photonuphalf{\begin{picture}(1.5,15)(0,0)
                      \multiput(0,0)(0,3){5}{\photonatomup}
                  \end{picture}}
\def\gluonrighthalf{\begin{picture}(15,3.25)(0,0)
                         \multiput(0,0)(3,0){5}{\gluonatomright}
                    \end{picture}}
\def\fermionup{\begin{picture}(1,30)(0,0)
                     \put(0,0){\vector(0,1){15}}
                     \put(0,15){\line(0,1){15}}
               \end{picture}}
\def\fermionuphalf{\begin{picture}(1,15)(0,0)
                         \put(0,0){\vector(0,1){7.5}}
                         \put(0,7.5){\line(0,1){7.5}}
                   \end{picture}}

\def\fermionleft{\begin{picture}(30,1)(0,0)
                       \put(30,0){\vector(-1,0){16}}
                       \put(15,0){\line(-1,0){15}}
                 \end{picture}}

\def\fermionright{\begin{picture}(30,1)(0,0)
                        \put(0,0){\vector(1,0){16}}
                        \put(15,0){\line(1,0){15}}
                  \end{picture}}
\def\fermionrighthalf{\begin{picture}(15,1)(0,0)
                            \put(0,0){\vector(1,0){7.5}}
                            \put(7.5,0){\line(1,0){7.5}}
                      \end{picture}}

\def\fermionulhalf{\begin{picture}(7.5,7.5)(0,0)
                        \put(0,0){\vector(-1,1){4.5}}
                        \put(-3.75,3.75){\line(-1,1){3.75}}
                  \end{picture}}

\def\fermionurhalf{\begin{picture}(7.5,7.5)(0,0)
                        \put(-7.5,-7.5){\vector(1,1){4.5}}
                        \put(-3.75,-3.75){\line(1,1){3.75}}
                  \end{picture}}

\def\fermionull{\begin{picture}(30,15)(0,0)
                        \put(0,0){\vector(-2,1){15}}
                        \put(-15,7.5){\line(-2,1){15}}
                  \end{picture}}
\def\fermionullhalf{\begin{picture}(15,7.5)(0,0)
                        \put(0,0){\vector(-2,1){7.5}}
                        \put(-7.5,3.75){\line(-2,1){7.5}}
                  \end{picture}}
\def\fermionurr{\begin{picture}(30,15)(0,0)
                        \put(-30,-15){\vector(2,1){15}}
                        \put(-15,-7.5){\line(2,1){15}}
                  \end{picture}}
\def\fermionurrhalf{\begin{picture}(15,7.5)(0,0)
                        \put(-15,-7.5){\vector(2,1){7.5}}
                        \put(-7.5,-3.75){\line(2,1){7.5}}
                  \end{picture}}
\def\fermiondrr{\begin{picture}(30,15)(0,0)
                        \put(0,0){\vector(2,-1){15}}
                        \put(15,-7.5){\line(2,-1){15}}
                \end{picture}}

\def\fermiondll{\begin{picture}(30,15)(0,0)
                        \put(30,15){\vector(-2,-1){15}}
                        \put(15,7.5){\line(-2,-1){15}}
                  \end{picture}}

\def\gaugebosonright{\begin{picture}(30,1)(0,0)
                            \put(0,0){\line(1,0){0.75}}
                            \multiput(2.25,0)(3,0){9}{\line(1,0){1.5}}
                            \put(29.25,0){\line(1,0){0.75}}
                     \end{picture}}

\newenvironment{Feynman}[3]{\begin{center}
                            \setlength{\unitlength}{#3 mm}
                            \begin{picture}(#1)(#2)
                            \thicklines
                           }{\end{picture} \end{center}}
\begin{document}
\renewcommand{\topfraction}{0.8}
\renewcommand{\bottomfraction}{0.8}
\renewcommand{\thefootnote}{\fnsymbol{footnote}}
\renewcommand{\medskip}{\vspace{.4cm} \\}
\renewcommand{\bigskip}{\vspace{.5cm} \\}
%
\def\"#1{{\accent127#1}}
\catcode`\"=\active
\catcode`\@=11
\def"#1{{\accent127#1}\penalty\@M \hskip\z@skip}
\catcode`\@=12
\let\3\ss
\lccode25=25
\hyphenation{brems-strah-lung}
\hyphenation{mathe-ma-tics}
\hyphenation{pa-ra-me-ter}
\hyphenation{pa-ra-me-ters}
\hyphenation{coun-ter-term}
\hyphenation{coun-ter-terms}
\hyphenation{ma-ni-festa-tion}
\hyphenation{be-hav-ior}
\hyphenation{pre-dic-ta-bi-li-ty}
\hyphenation{ref-er-ence}
\hyphenation{ex-per-i-men-tal}
\hyphenation{ex-per-i-ment}
\hyphenation{re-nor-mal-i-za-bi-li-ty}
\hyphenation{re-nor-mal-i-za-tion}
\hyphenation{re-nor-mal-ize}
\hyphenation{be-tween}
\hyphenation{sum-ma-tion}
\hyphenation{Feyn-man}
\hyphenation{ref-er-ence}
\hyphenation{e-qua-tion}
\newcommand{\para}{\vspace{.2cm} \\}
\newcommand{\bpara}{\vspace{.5cm} \\}
\newcommand{\scal}{\vrule width 0pt height 7mm depth 4mm}
\newcommand{\scalfrac}{\vrule width 0pt height 5mm depth 3mm}
\newcommand{\scaltab}{\vrule width 0pt height 5mm depth 2mm}
\newcommand{\scaleqn}{\vrule width 0pt height 5mm depth 3mm}
\newcommand{\ds}{\displaystyle}
\newcommand{\mycos}{\cos\!}
\newcommand{\mysin}{\sin\!}
\newcommand{\myln}{\ln\!}
\newcommand{\mysp}{{\rm Li_2\!}}
\newcommand{\mytri}{{\rm Li_3\!}}
\newcommand{\ba}{\begin{eqnarray}}
\newcommand{\ea}{\end{eqnarray}}
\newcommand{\bas}{\begin{eqnarray*}}
\newcommand{\eas}{\end{eqnarray*}}
\newcommand{\beq}{\begin{equation}}
\newcommand{\eeq}{\end{equation}}
\newcommand{\seteps}{\stackrel{\ds \epsilon}{\rule[-0cm]{0cm}{-0cm}}}
%
\newcommand{\sm}{Standard Model}
\newcommand{\inter}{in\-ter\-ac\-tion~}
\newcommand{\inters}{in\-ter\-ac\-tions~}
\newcommand{\xsec}{cross-sec\-tion~}
\newcommand{\xsecs}{cross-sec\-tions~}
\newcommand{\radcor}{ra\-dia\-tive cor\-rec\-tions~}
\newcommand{\lumi}{lu\-mi\-no\-si\-ty~}
\newcommand{\ene}{center of mass ener\-gy~}
\newcommand{\enes}{center of mass ener\-gies~}
\newcommand{\mc}{Monte-Carlo~}
\newcommand{\nl}{\nonumber \\}
\newcommand{\miniskip}{\vspace{.15cm} \\}
\newcommand{\smskip}{\vspace{.3cm} \\}
\newcommand{\hugeskip}{\vspace{.8cm} \\}
\newcommand{\numreal}{I\!\!R}
\newcommand{\numcomp}{{\rm C} \put(-5.5,.3){\line(0,1){7.4}}
                              \put(-5.2,.2){\line(0,1){7.6}} }
\newcommand{\ieps}{{\rm i}\varepsilon}
\newcommand{\bc}{\bigcirc}
\newcommand{\iospi}{\frac{{\rm i}}{16\,\pi^2}}
\newcommand{\lz}{l_0}
\newcommand{\ri}{{\rm i}}
\newcommand{\cc}{{\cal C}}
\newcommand{\cd}{{\cal D}}
\newcommand{\cf}{{\cal F}}
\newcommand{\cl}{{\cal L}}
\newcommand{\cm}{{\cal M}}
\newcommand{\cp}{{\cal P}}
\newcommand{\cs}{{\cal S}}
\newcommand{\ct}{{\cal T}}
\newcommand{\varth}{\vartheta}
\newcommand{\mz}{$M_Z$}
\newcommand{\mzs}{$M_Z^2$}
\newcommand{\mzp}{$M_{Z'}$}
\newcommand{\mzps}{$M_{Z'}^2$}
\newcommand{\mw}{$M_W$}
\newcommand{\mws}{$M_W^2$}
\newcommand{\mmu}{$m_{\mu}$}
\newcommand{\me}{$m_e$}
\newcommand{\mb}{$m_b$}
\newcommand{\zz}{$Z^0$}
\newcommand{\zzp}{$Z'$}
\newcommand{\wpl}{$W^+$}
\newcommand{\wmi}{$W^-$}
\newcommand{\fone}{$f_1$}
\newcommand{\ftwo}{$f_2$}
\newcommand{\bfone}{$\bar f_1$}
\newcommand{\bftwo}{$\bar f_2$}
\newcommand{\fpone}{$\fone\bfone$}
\newcommand{\fptwo}{$\ftwo\bftwo$}
\newcommand{\mupl}{$\mu^+$}
\newcommand{\mumi}{$\mu^-$}
\newcommand{\mm}{\mupl\mumi}
\newcommand{\epl}{$e^+$}
\newcommand{\emi}{$e^-$}
\newcommand{\ee}{\epl\emi}
\newcommand{\bbar}{$\bar b$}
\newcommand{\bb}{$b$\bbar}
\newcommand{\rs}{$\sqrt{s}$}
\newcommand{\tata}{$\tau^+ \tau^-$}
\newcommand{\mmbb}{\protect{\mm\bb}}
\newcommand{\etoff}{\protect{\epl\emi$\to$\fone\bfone\ftwo\bftwo}}
\newcommand{\etomb}{\protect{\epl\emi$\to$\mmbb}}
\newcommand{\oal}{${\cal O}(\alpha)$}
\newcommand{\oals}{${\cal O}(\alpha^2)$}
\newcommand{\permille}{$^0 \!\!\!\: / \! _{00}$}
\newcommand{\gamz}{$\Gamma_Z$}
\newcommand{\gamzp}{$\Gamma_{Z'}$}
\newcommand{\gamhad}{$\Gamma_{Had}$}
\newcommand{\gamtot}{$\Gamma_{Tot}$}
\newcommand{\gamlep}{$\Gamma_{Lept}$}
\newcommand{\gammm}{$\Gamma_{\mu^+\mu^-}$}
\newcommand{\egam}{$E_{\gamma}$}
\newcommand{\ebeam}{$E_{beam}$}
\newcommand{\siw}{$\sin{\!\theta_w}$}
\newcommand{\cow}{$\cos{\!\theta_w}$}
\newcommand{\sws}{$\sin^2{\!\theta_w}$}
\newcommand{\cws}{$\cos^2{\!\theta_w}$}
\newcommand{\swsq}{$s^2_w$}
\newcommand{\cwsq}{$c^2_w$}
\newcommand{\spr}{$s'$}
\newcommand{\gmu}{$G_{\mu}$}
\newcommand{\al}{$\alpha$}
\newcommand{\alpi}{$\alpha/\pi$}
\newcommand{\dmu}{$\partial_{\mu}$}
\newcommand{\dmuup}{$\partial^{\mu}$}
\newcommand{\dmukov}{${\cal D}_{\mu}$}
\newcommand{\dmukon}{${\cal D}^{\mu}$}
\newcommand{\MZ}{M_Z}
\newcommand{\MZS}{M_Z^2}
\newcommand{\MZP}{M_{Z'}}
\newcommand{\MZPS}{M_{Z'}^2}
\newcommand{\MW}{M_W}
\newcommand{\MWS}{M_W^2}
\newcommand{\MMU}{m_{\mu}}
\newcommand{\ME}{m_e}
\newcommand{\MES}{m_e^2}
\newcommand{\MB}{m_b}
\newcommand{\ZZ}{Z^0}
\newcommand{\ZZP}{Z'}
\newcommand{\WPL}{W^+}
\newcommand{\WMI}{W^-}
\newcommand{\FONE}{f_1}
\newcommand{\FTWO}{f_2}
\newcommand{\BFONE}{\bar f_1}
\newcommand{\BFTWO}{\bar f_2}
\newcommand{\FPONE}{\FONE\BFONE}
\newcommand{\FPTWO}{\FTWO\BFTWO}
\newcommand{\MUPl}{\mu^+}
\newcommand{\MUMI}{\mu^-}
\newcommand{\MM}{\MUPL\MUMI}
\newcommand{\EPL}{e^+}
\newcommand{\EMI}{e^-}
\newcommand{\EE}{\EPL\EMI}
\newcommand{\BBAR}{\bar b}
\newcommand{\BB}{b\BBAR}
\newcommand{\RS}{\sqrt{s}}
\newcommand{\TATA}{\tau^+ \tau^-}
\newcommand{\MMBB}{\protect{\MM\BB}}
\newcommand{\ETOFF}{\protect{\EPL\EMI\to\FONE\BFONE\FTWO\BFTWO}}
\newcommand{\ETOMB}{\protect{\EPL\EMI\to\MMBB}}
\newcommand{\OAL}{{\cal O}(\alpha)}
\newcommand{\OALS}{{\cal O}(\alpha^2)}
\newcommand{\SLAM}{\sqrt{\lambda}}
\newcommand{\SLAMP}{\sqrt{\lambda^{'}}}
\newcommand{\LAMP}{\lambda^{'}}
\newcommand{\SLAMB}{\sqrt{\bar{\lambda}}}
\newcommand{\LAMB}{\bar{\lambda}}
\newcommand{\sprm}{s'_{\!-}}
\newcommand{\sprp}{s'_{\!+}}
\newcommand{\SONE}{s_{12}}
\newcommand{\STWO}{s_{34}}
\newcommand{\GAMZ}{\Gamma_Z}
\newcommand{\GAMZP}{\Gamma_{Z'}}
\newcommand{\GAMHAD}{\Gamma_{Had}}
\newcommand{\GAMTOT}{\Gamma_{Tot}}
\newcommand{\GAMLEP}{\Gamma_{Lept}}
\newcommand{\GAMMM}{\Gamma_{\mu^+\mu^-}}
\newcommand{\EGAM}{E_{\gamma}}
\newcommand{\EBEAM}{E_{beam}}
\newcommand{\SIW}{\sin{\!\theta_w}}
\newcommand{\COW}{\cos{\!\theta_w}}
\newcommand{\SWS}{\sin^2{\!\theta_w}}
\newcommand{\CWS}{\cos^2{\!\theta_w}}
\newcommand{\SWSQ}{s^2_w}
\newcommand{\CWSQ}{c^2_w}
\newcommand{\SPR}{s'}
\newcommand{\GNU}{G_{\mu}}
\newcommand{\AL}{\alpha}
\newcommand{\ALPI}{\alpha/\pi}
\newcommand{\DMU}{\partial_{\mu}}
\newcommand{\DMUUP}{\partial^{\mu}}
\newcommand{\DMUKOV}{{\cal D}_{\mu}}
\newcommand{\DMUKON}{{\cal D}^{\mu}}
\newcommand{\DNU}{\partial_{\nu}}
\newcommand{\DNUUP}{\partial^{\nu}}
\newcommand{\DNUKOV}{{\cal D}_{\nu}}
\newcommand{\DNUKON}{{\cal D}^{\nu}}
\newcommand{\WMU}{\mbox{\boldmath$W_\mu$}}
\newcommand{\WMUUP}{\mbox{\boldmath$W^\mu$}}
\newcommand{\RE}{\Re\!e}
\newcommand{\IM}{\Im\!m}
\newcommand{\dza}{\delta Z_{\!A}}
\newcommand{\dze}{\delta Z_e}
\newcommand{\dzp}{\delta Z_\psi}
\newcommand{\dzm}{\delta Z_m}
\newcommand{\dmf}{\delta m}
\newcommand{\mten}{g_{\mu\nu}}
\newcommand{\lhs}{{\sc l.h.s.}}
\newcommand{\rhs}{{\sc r.h.s.}}
\newcommand{\dagg}[1]{#1 \hspace{-.18cm} / \hspace{.05cm}}
\newtheorem{lemma}{Lemma}
%
%
\evensidemargin=0.2cm
\begin{titlepage}
    \begin{center}
       {\LARGE \bf
         Initial State Radiative Corrections to \\
         \zz~Pair Production in \ee~Annihilation\vspace{.35cm} \\
         ~-- The Semi-Analytical Approach --
       }\vspace{1.5cm} \\
       {\Large D i s s e r t a t i o n} \vspace{.5cm} \\
       \begin{large}
         zur Erlangung des akademischen Grades \vspace{.5cm} \\
         {\bf doctor rerum naturalium} \\
         {\bf (Dr. rer. nat.)} \vspace{1cm} \\
         eingereicht an der \\
         Mathematisch-Naturwissenschaftlichen Fakult"at~I \\
         der Humboldt-Universit"at zu Berlin \vspace{.5cm} \\
         von \\
         {\bf Diplom-Physiker Dietrich Lehner} \\
         geb. am 2. November 1967 in M"unchen \vspace{1cm} \\
         Pr"asidentin der Humboldt-Universit"at zu Berlin: \\
         Prof. Dr. M. D"urkop \vspace{.5cm} \\
         Dekan der Mathematisch-Naturwissenschaftlichen Fakult"at~I: \\
         Prof. Dr. M. v. Ortenberg\vspace{1cm} \\
         \begin{tabular}{lll}
           Gutachter: & 1. & Prof. Dr. F. Jegerlehner \\
                      & 2. & Prof. Dr. M. M"uller-Preu{\ss}ker\\
                      & 3. & Prof. Dr. P. S"oding
         \end{tabular}
         \vspace{1cm} \\
         \hspace*{-.4cm} Tag der m"undlichen Pr"ufung: 7. Dezember 1995
       \end{large}
     \end{center}
\newpage
\pagestyle{empty}
\vspace*{20cm}
\noindent
Vollst"andiger Abdruck der von der
Mathematisch-Naturwissenschaftlichen Fakult"at~I der
Humboldt-Universit"at zu Berlin genehmigten Dissertation unter
Verbesserung zweier orthographischer Fehler.
\end{titlepage}
%
%
\evensidemargin=-0.35cm
\renewcommand{\thepage}{\roman{page}}
\setcounter{page}{1}
\evensidemargin=0.2cm
\begin{abstract}
\noindent
A total and differential cross-section calculation for the Electroweak
Standard Model reaction $\EE \rightarrow (\ZZ\ZZ) \rightarrow
f_1\bar{f_1}f_2\bar{f_2}(\gamma)$ including the effects of the finite
\zz~width and, for the first time, all QED initial state radiative
corrections (ISR) is presented. A semi-analytical technique is used:
All angular phase space integrations, five for the tree level process,
seven if ISR is included, are carried out analytically. The remaining
phase space variables are the invariant masses of the two decaying
\zz~bosons for tree level plus the reduced center of mass energy
squared in the ISR case. Invariant masses are submitted to high
precision numerical integration. Semi-analytical and numerical results
for \ee~center of mass energies between $\RS\!=\!130\:$GeV and
$1\:$TeV are reported. It is shown that the radiatively corrected
\xsec splits into a universal, leading logarithm part with the tree
level \xsec factorizing on one hand and a non-universal contribution
on the other hand. A generalization to the case of Standard Model
neutral boson pair production, $\EE \rightarrow (\ZZ\ZZ, \ZZ\gamma,
\gamma\gamma) \rightarrow f_1\bar{f_1}f_2\bar{f_2}(\gamma)$, is worked
out in detail. The relevance of the presented calculation for the
phenomenology of \ee~collider physics is discussed, especially for $W$
and Higgs physics both of which will be important topics in the near
future. All significant technical details of the calculation are
explained.
\vspace{1.7cm}
\begin{center}
  {\bf Zusammenfassung}
\end{center}
Im Rahmen des Standardmodells elektroschwacher Wechselwirkungen wird
eine Berechnung
totaler und differentieller Wirkungsquerschnitte f"ur die Reaktion
$\EE \rightarrow (\ZZ\ZZ) \rightarrow f_1\bar{f_1}f_2\bar{f_2}(\gamma)$
vorgestellt. Dabei werden die Effekte der endlichen Bosonlebensdauer
und erstmalig die voll\-st"an\-di\-gen
Eingangszustands-QED-Strahlungskorrekturen (ISR) ber"ucksichtigt. Es
wird eine semianalytische Technik angewandt: S"amtliche
Winkelvariablen des Phasenraums, f"unf f"ur die Born-Approximation,
sieben im ISR-Fall, werden analytisch integriert. Die verbleibenden
Phasenraumvariablen sind invariante Mas\-sen, n"amlich die invarianten
Mas\-sen der zerfallenden \zz-Bosonen und im ISR-Fall zus"atzlich die
reduzierte Schwer\-punkts\-ener\-gie. "Uber die invarianten Massen
wird mit einem sehr pr"azisen numerischen Algorithmus integriert.
Es wird "uber semianalytische und
numerische Er\-geb\-nis\-se f"ur \ee-Schwer\-punkts\-ener\-gien zwischen
$\RS\!=\!130\:$GeV und $1\:$TeV berichtet. Dabei wird gezeigt, da{\ss} der
Wirkungsquerschnitt mit ISR in einen universellen
Anteil f"uhrender Logarithmen mit dem Born-Wirkungsquerschnitt als
Faktor einerseits und einen nichtuniversellen Beitrag andererseits
zerlegt werden kann. Eine Verallgemeinerung der
Resultate auf die Standardmodell-Paarproduktion neutraler Bosonen,
$\EE \rightarrow (\ZZ\ZZ, \ZZ\gamma, \gamma\gamma) \rightarrow
f_1\bar{f_1}f_2\bar{f_2}(\gamma)$ wird detailliert ausgearbeitet. Die
ph"anomenologische Bedeutung der vor\-ge\-stel\-lten Rechnungen f"ur die
\ee~Beschleunigerphysik wird diskutiert, insbesondere die Bedeutung
f"ur die in naher Zukunft wichtigen Themen der $W$- ~und der Higgs-Physik.
Alle wesentlichen technischen Einzelheiten der Rechnung sind erl"autert.
\end{abstract}
\evensidemargin=-0.35cm
\tableofcontents
\listoffigures
\listoftables
\clearpage
\renewcommand{\thepage}{\arabic{page}}
\setcounter{page}{1}
%
\chapter{Introduction}
\label{intro}
%
It is the goal of modern physics to achieve reduction of nature's
complexity to as few as possible fundamental principles. The method
is increasing abstraction, the tool mathematics, the result
unification of, up to that date, distinct phenomena. An important
success pointing in that direction was Maxwell's formulation of the
theory of electromagnetism in 1864~\cite{maxwell}. The electromagnetic
theory was not only the origin of the theory of
relativity~\cite{einstein05}, but
it could also be cast into a Lorentz-covariant form. In the late
twenties, originating from Dirac's work~\cite{dirac27},
electromagnetism continued its career as a quantum field
theory~\cite{pauli29} which was shown to be U(1) gauge invariant by
Weyl~\cite{weyl29}. Quantum Electrodynamics (QED) was born, when
Schwinger reformulated the electromagnetic quantum field theory in a
Lorentz-covariant way~\cite{schwinger48}. Shortly thereafter the
well-known Feynman diagrams were introduced~\cite{feynman49}, and it
only took another year to establish the mathematical foundations of
Feynman's theory~\cite{feynman50}. Today, QED is one of the
best-tested theories we know. A manifestation of this success is the
prediction of the anomalous magnetic moment $a_e$ of the electron up
to $\delta a_e/a_e = 2.4 \times 10^{-8}$ from perturbative
QED~\cite{kinoshita90} in agreement with experiment~\cite{pdb94}.
Other QED predictions matching high precision experiments are the
anomalous magnetic moment $a_\mu$ of the muon~\cite{marciano90} and
the Lamb shift of the hydrogen atom~\cite{kinoshita90,pipkin90}. This
list could be continued. The successes of QED are clear evidence for
the quantum corrections as obtained from relativistic quantum field
theory. Perturbative QED is a very important
ingredient of the Standard Model of Electroweak Interactions (SM).
\para
The second main ingredient of the Standard Model are the weak
interactions which entered history in 1896 when Becquerel discovered
radioactivity in the form of nuclear $\beta$-decay~\cite{becquerel96}.
In 1934 Fermi developed a ``theory of weak interactions'' for
$\beta$-decay which was constructed along the lines of
QED~\cite{fermi33}. The structure of weak interactions was further
clarified with the theoretical consideration of parity violation by Lee
and Yang~\cite{lee56}. Soon thereafter, parity violation was
experimentally confirmed in the famous Wu experiment~\cite{wu57}.
This was when Fermi's original pure vector currents were replaced by a
(V--A) structure which is maintained until today~\cite{salam57}.
Although very successful, Fermi's theory of weak interactions is not
renormalizable and runs into a unitarity problem at center of mass
energies around 600~GeV. This problem was cured by the introduction
of gauge bosons that mediate the weak interaction. The short range of
weak interactions requires massive gauge
bosons via the Yukawa theory~\cite{yukawa35}. However, for the sake
of gauge symmetry, bosons are not allowed to have explicit mass terms
within the framework of a Yang-Mills field theory~\cite{yang54}.
A remedy to this mass prohibition was found with the Higgs(-Kibble)
mechanism of spontaneous symmetry breaking
which ``generates'' gauge boson masses that are proportional to the
non-vanishing vacuum expectation value of a newly introduced scalar
field~\cite{higgs}. In addition, the Higgs mechanism enables the
gauge invariant introduction of chiral fermion masses into a
Yang-Mills theory. In 1971, it could be proven that Yang-Mills
theories with gauge boson masses generated by spontaneous symmetry
breaking are renormalizable~\cite{thooft71}.
\para
It has been a milestone of modern physics when, in parallel to the
development of the Higgs mechanism in the sixties, a renormalizable,
non-abelian $SU(2)_L \times U(1)_Y$ gauge field theory for the unified
description of weak interactions and QED could be found~\cite{gws}
which is today called the Standard Model of Electroweak
Interactions. The groups $SU(2)_L$ and $U(1)_Y$ are associated with
the weak isospin and the weak hypercharge symmetries.
Via the Higgs mechanism three of the four gauge bosons of the
model, namely the $W^\pm$ and the \zz~are endowed with masses, whereas
the photon as the mediator of the electromagnetic force stays massless
as required by the infinite range of QED forces. In the framework of
the Standard Model, the Higgs mechanism entails the introduction of one
scalar boson, the so-called Higgs boson. The photon has a pure
vector coupling to electromagnetically charged particles, $W^\pm$
bosons have maximally parity violating (V--A) coupling structure, and
the \zz~bosons comprise both (V--A) and (V+A) couplings. In the 1950s
and 1960s many weak decays of hadrons were experimentally studied.
With respect to these observations it is a crucial feature of the
Standard Model that it is able to naturally incorporate the hadronic
constituents, the quarks~\cite{gellmann64}, via the
Glashow-Iliopoulos-Maiani mechanism~\cite{glashow70}. This
incorporation was based on the idea of a unitary transformation of the
quark flavor mass eigenstates into weak eigenstates which is today
known as the Cabibbo-Kobayashi-Maskawa matrix~\cite{cabibbo63}. The
present version of the Standard Model is made anomaly-free by the
introduction of color~\cite{holliklect}, a concept originating from
Quantum Chromodynamics~\cite{QCD}.
\para
The experimental confirmation of the Standard Model started in 1973
with the discovery of the predicted neutral current interactions in
the Gargamelle bubble chamber at the CERN proton synchrotron, where
$\stackrel{_{(-)}}{\nu_\mu}$ scattering on electrons and nucleons was
investigated~\cite{NC73}. Soon the Standard Model was further
supported by the
discovery of the weak isospin partner of the strange quark, which was
called ``charm''~\cite{charm74}. The breakthrough came with the
direct observation of the gauge bosons $W^\pm$ and \zz~by the UA1 and
UA2 experiments at the CERN $Sp{\bar p}S$~\cite{UA83}. With the advent
of the SLC and LEP \ee~colliders for resonance production of
\zz~bosons in \ee~annihilation, the Standard Model entered its
precision test era. From
LEP and SLC, the mass \mz~and the width \gamz~of the \zz~as well as
the Weinberg mixing angle $\sin\!^2 \theta^{lept}_{eff}$~are very
accurately known~\cite{LEPcomb95}:
\ba
  \hspace{4cm}
  \MZ & = & (91.1887 \pm 0.0022)~{\rm GeV} \nl
  \GAMZ & = & (~2.4971 \pm 0.0032)~{\rm GeV} \nl
  \sin\!^2 \theta^{lept}_{eff} & = & 0.2315 \pm 0.0004~~. \nonumber
\ea
These impressive numbers, however, are only the peak of an iceberg of
electroweak precision measurements all of which agree
with the Standard Model. The measurements at LEP have reached a level
of precision
which enables to test radiative corrections not only in the QED sector
of the theory. Also the truly electroweak radiative corrections and
thus the quantum corrections of the Standard Model are now
established~\cite{schildknecht94}. The renormalizability of the
Standard Model ensures the sensibility of the perturbation expansion
of these quantum effects.
The latest success in the confirmation of the Standard Model is the
discovery of the last missing fermion of the Model, the top quark, by
the CDF and D0 collaborations at the Fermilab ``Tevatron''
$p {\bar p}$~collider~\cite{top}\footnote[2]{Since the determination
  of the  number of light neutrino generations as $N_\nu = 2.987 \pm
  0.016$~\cite{LEPcomb95}, there is reason to believe that only three
  generations of quarks and leptons exist. Thus, in the framework of
  the Standard Model, the observation of the top quark together with
  the indirect evidence for the $\tau$ neutrino justifies the
  statement that the last missing fermion has been discovered.}.
It is very noteworthy that the direct
measurements of the top quark mass are in good agreement with
indirect mass determinations obtained from precision experiments at
the \zz~pole~\cite{LEPcomb95,top}. Precision measurements can also
severely constrain new physics beyond the Standard Model, and it is
discussed, whether they can put limits on the Higgs mass, if the
precision of top quark mass measurements is further improved.
\para
An upgraded version of LEP, commonly referred to as LEP2, will pass
the threshold for the pair production of heavy gauge bosons in the
very near future and operate
at center of mass energies between 175 and 206 GeV~\cite{lep2en}. This
opens the opportunity to study the properties of the $W$ boson in
$W$~pair and single $W$~production processes. Furthermore,
the non-abelian coupling structure of the electroweak theory will be
investigated directly for the first time through the observation of
trilinear $\gamma\WPL\WMI$ and $\ZZ\WPL\WMI$ interactions. Recently,
the different non-standard trilinear boson couplings have been subject
of several studies~\cite{trilin}\footnote[3]{It should be mentioned
  that non-standard trilinear boson couplings are heavily
  constrained by unitarity requirements~\cite{baur88}.}.
If LEP2 does not yield the discovery of the Higgs boson, it will at
least increase the lower Higgs mass limit to approximately the machine
energy minus 100 GeV, i.e.
$\sqrt{s}-100~{\rm GeV}$~\cite{felcini94,sopczak95}.
The above-mentioned physics tasks at LEP2 involve the analysis of
four-fermion final states. At present, the latter are heavily studied,
and several experimental analyses were already carried out at
LEP1~\cite{DLdipl,4fexp}. This thesis is a contribution to the
clarification of the physics of four-fermion final states in
\ee~annihilation. A study of the process
\beq
  \EE \rightarrow (\ZZ\ZZ) \rightarrow
  f_1\bar{f_1}f_2\bar{f_2}(\gamma)~,
  ~~~~~~~~~~~~~~~~~~~~~ f_1\!\neq\!f_2~,~~~f_i\!\neq\!e^\pm,\nu_e
  \label{eezz4f}
\eeq
will be presented, including the first complete treatment of initial
state radiation in first order perturbation theory~\cite{dl94}.
Process~(\ref{eezz4f}) is not only observable at LEP2, but also
interesting at planned higher energy \ee~colliders.
Process~(\ref{eezz4f}) yields a higher order test of the
Standard Model and can easily be extended to nonstandard neutral
current physics with additional neutral bosons replacing the
\zz's. Another strong motivation for this study is the importance of
process~(\ref{eezz4f}) for $W$~pair physics and for Higgs searches at
LEP2. The relevance to $W$~pair physics is twofold.
Process~(\ref{eezz4f}) does not only constitute a double resonance
background to some decay channels in $W$~pair production, but it was
also important to confirm the generality of properties found for
W pair production in the framework of the ``current splitting
technique''~\cite{dimaww,teupitz94}. As far as Higgs searches at LEP2
are concerned the \zz~pair double resonance background to four-fermion
final states with a b quark pair {\it must} be taken into account. This
thesis presents sufficiently precise results for studies of
the \zz~pair double resonance background to Higgs searches. Further
importance of the presented results on process~(\ref{eezz4f}) lies in
their straightforward generalizability to include new heavy neutral
bosons. It should be investigated in how far \zz$Z'$ pair
production in \ee~annihilation can put competitive limits on the mass
and/or couplings of a new boson $Z'$.
%
%
On-shell \zz~pair production has been discussed long
ago~\cite{Brown78}. Numerical calculations including all
\oal~electroweak corrections\footnote[2]{That is the electroweak
  corrections of first order in the fine structure constant
  $\alpha=e^2/(4\pi)$.}
except hard photon brems\-strah\-lung were reported in~\cite{Denner88}
for on-shell and in~\cite{Denner90} for off-shell \zz~bosons. In this
thesis, initial state radiation (ISR) to \zz~pair production will be
treated completely, including hard bremsstrahlung corrections.
Contrary to the more common approaches to process~(\ref{eezz4f}) and
other neutral current processes with Monte Carlo phase space
integration~\cite{excal,NCMC,eebb}, this thesis
follows a ``semi-analytical'' approach where all angular phase space
variables are integrated analytically and the remaining invariant
masses are subjected to a high precision numerical integration. Thus
the precision of the cross-section calculation is
physics dominated, meaning that the error on the computed \xsec
essentially originates from theoretical systematic errors such as the
disregard of other radiative corrections or the small final state mass
approximation. Another advantage of semi-analytical calculations is
their transparency, i.e. the fact that, at tree level, they yield very
compact expressions for cross-sections and invariant mass
distributions. Even with initial state radiation included,
semi-analytical \xsec formulae allow to identify physically
interesting properties. Having in mind that, using the
semi-analytical method, it is difficult to implement experimental cuts
on all relevant angular variables it is clear that semi-analytical and
Monte Carlo calculations of four-fermion final states complement each
other. However, it should be born in mind that experimental cuts on
invariant masses and the boson scattering angle can be easily
implemented in semi-analytical calculations, and it is also possible
to obtain semi-analytical distributions in many interesting
variables~\cite{aldist95}.
\para
Despite its stunning successes, the Standard Model has conceptual
drawbacks, and it is expected that there will be new physics at
higher energies. Firstly, and more philosophically, the Standard Model
relies on ``too many'' arbitrary assumptions and parameters. This
fostered the hope to find an underlying theory that can e.g. explain
the origin of mass, of electric charge, or color. Secondly, one would
like to find a quantum field theory that unifies the electroweak with
the other known interactions, namely the strong interaction described
by Quantum Chromodynamics (QCD) and gravitation. Thirdly, the fact
that the Standard Model is not asymptotically free suggests that it is
a kind of ``low energy'' effective theory of a more fundamental
one~\cite{haber85}. In addition it is known that, within the framework
of the Standard Model, the scattering of longitudinally polarized
$W$~bosons violates the Froissart limit for the proper high energy
behavior of cross-sections. Finally, there is the well-known
fine-tuning or naturalness problem of the Standard Model which is
related to a quadratic divergence in the radiative corrections to the
Higgs mass. This problem necessitates new phenomena at a scale of
${\cal O}$(1 TeV)~\cite{haber85}. A candidate to solve the fine-tuning
problem is Supersymmetry which introduces a fundamental symmetry
between bosons and fermions and thus, via the relative minus sign in
fermion loops compared to boson loops and via supersymmetric relations
between masses and couplings, cancels the problematic quadratic
divergence. Supersymmetric theories are free of quadratic
divergences~\cite{haber85,nilles84}. A further interesting class of
theories beyond the Standard Model are String Theories. In spite of
their mathematical difficulties, String Theories are attractive,
because they are finite to all orders of perturbation theory and
ultraviolet divergence free, because they rely on few free parameters,
because they may be able to explain mass hierarchies, and because they
yield coupling constant unification. In addition, String Theories
comprise gravitation, possibly solve the problem of the non-vanishing
cosmological constant~\cite{thomas}, and they may also be on their way
to reach phenomenological predictability~\cite{lopez94}. As a final
remark it must, however, be stressed that so far no experimental
evidence for physics beyond the Standard Model exists.
\para
This thesis is organized as follows. In Chapter~\ref{born}, the tree
level results for process~(\ref{eezz4f}) will be discussed, followed
by the presentation of QED initial state radiation (ISR) results in
chapter~\ref{isr}. In
chapter~\ref{xsZZGG}, results from preceding chapters are generalized
to include s-channel exchange photons. Following this body of the
thesis a set of appendices contains notations, definitions, and
technical details of the presented calculations. Appendix~\ref{metric}
gives the metric for this thesis and appendix~\ref{SM} introduces the
Standard Model Lagrangian, the Feynman rules, and the renormalization
that were used. Appendix~\ref{rellogs} contains the definitions of
polylogarithms and in appendix~\ref{phasespa} the 2$\to$4 and
2$\to$5 particle phase space parametrizations needed for the
evaluation of \xsecs at tree level and with ISR are given. In
appendix~\ref{xscalc}, all important details of the calculation of the
tree level and ISR \xsecs are collected. Appendix~\ref{loops}
presents techniques for the evaluation of loop integrals as encountered
in the virtual initial state QED corrections to process~(\ref{eezz4f})
and appendix~\ref{tabofint} contains all analytical integrals required
for the angular phase space integrations to be carried out in the \xsec
calculation for process~(\ref{eezz4f}).
\vspace{1.5cm}
%
\chapter{The \zz~Pair Born Cross-Section}
\label{born}
\begin{figure}[t]
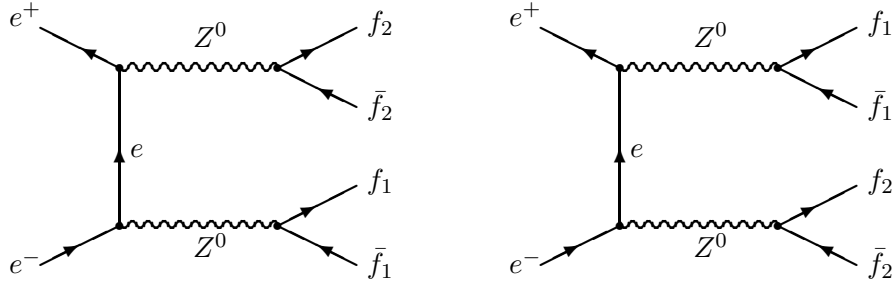

\vspace*{.2cm}
\begin{minipage}[tbh]{7.8cm}{
\begin{center}
\begin{Feynman}{75,60}{-28,0}{0.7}
%
\put(5,15){\fermionurrhalf}
\put(5,45){\fermionullhalf}
\put(5,15){\fermionup}
\put(5,45){\photonright}
\put(5,15){\photonright}
\put(5,45){\circle*{1.5}}
\put(5,15){\circle*{1.5}}
\put(35,15){\circle*{1.5}}
\put(35,45){\circle*{1.5}}
\put(50,7.5){\fermionullhalf}
\put(50,22.5){\fermionurrhalf}
\put(50,37.5){\fermionullhalf}
\put(50,52.5){\fermionurrhalf}
\small
\put(7,28){$e$}
\put(-16,6){\emi}
\put(-16,53){\epl}
\put(19,49){\zz}  
\put(19,09){\zz}
\put(52,06){${\bar f}_1$} 
\put(52,22){$ f_1$}
\put(52,36){${\bar f}_2$}
\put(52,52){$ f_2$}
\normalsize
\put(100,15){\fermionurrhalf}
\put(100,45){\fermionullhalf}
\put(100,15){\fermionup}
\put(100,45){\photonright}
\put(100,15){\photonright}
\put(100,45){\circle*{1.5}}
\put(100,15){\circle*{1.5}}
\put(130,15){\circle*{1.5}}
\put(130,45){\circle*{1.5}}
\put(145,7.5){\fermionullhalf}
\put(145,22.5){\fermionurrhalf}
\put(145,37.5){\fermionullhalf}
\put(145,52.5){\fermionurrhalf}
\small
\put(102,28){$e$}
\put(79,6){\emi}
\put(79,53){\epl}
\put(114,49){\zz}  
\put(114,09){\zz}
\put(147,06){${\bar f}_2$} 
\put(147,22){$ f_2$}
\put(147,36){${\bar f}_1$}
\put(147,52){$ f_1$}
\normalsize
\end{Feynman}
\end{center}
}\end{minipage}
\vspace{-.5cm}
\caption[Born level \zz~pair production Feynman diagrams]
{\it The Born level Feynman diagrams for off-shell \zz~pair
  production, the so-called `{\it Conversion}' (or `{\it Crab}')
  diagrams. Left: t-channel. Right: u-channel. }
\label{zzfeyn}
\end{figure}

%
At Born or tree level and for a center of mass energy $\sqrt{s}$,
process~(\ref{eezz4f}) is described by the two Feynman diagrams
depicted in figure~\ref{zzfeyn}.
%
%
The corresponding \xsec is derived in appendix~\ref{bornres}. It can
be written as a simple double convolution formula
\ba
  \sigma^B(s) & = &
    \int\limits_{4m_1^2}^{(\sqrt{s} - 2m_2)^2} d\SONE \, \rho_Z(\SONE)
    \int\limits_{4m_2^2}^{(\sqrt{s} - \sqrt{\SONE})^2} d\STWO \,
    \rho_Z(\STWO) \; \times 2 \cdot B\!R(1) \cdot B\!R(2) \times \nl
    & & \hspace{7.8cm} \times \, \sigma^B_4(s;\SONE,\STWO)
  \label{sigzz1}
\ea
with branching ratios $B\!R(1)$~and $B\!R(2)$\footnote[2]{The
  factor 2 conventionally introduced with the branching ratios,
  simplifies comparisons with on-shell \zz~pair production. This is
  seen from equation~(\ref{OScomp}) and was explained in some detail
  in reference~\cite{dl94}.}
of the \zz~boson into
the fermion-antifermion pairs \fone\bfone~and \ftwo\bftwo~
respectively, with invariant fermion pair masses $\SONE$ and $\STWO$,
and with the \zz~boson Breit-Wigner densities
\ba
  \rho_Z(s_{ij}) & = & \frac{1}{\pi}
    \frac {\sqrt{s_{ij}} \; \Gamma_Z (s_{ij})}
          {|s_{ij} - M_Z^2 + i \sqrt{s_{ij}} \, \Gamma_Z (s_{ij}) |^2}~~~~
     -\!\!\!-\!\!\!-\!\!\!\!
      \longrightarrow_{\hspace{-1cm}_{\Gamma_Z \rightarrow 0}}
      ~~~~\delta (s_{ij} - M_Z^2)~~.
  \label{rhos}
\ea
In the limit of on-shell \zz~bosons their widths vanish and the
Breit-Wigner densities are replaced by Dirac $\delta$ distributions.
The \zz~width is given by
\beq
\Gamma_Z (s_{ij}) =
  \frac{G_{\mu}\, M_Z^2} {24\pi \sqrt{2}} \sqrt{s_{ij}} \cdot
  \sum_f \, \left({g^V_{f\!{\bar f}\!Z}}^2+{g^A_{f\!{\bar f}\!Z}}^2
  \right) \!\cdot\! N_c(f)~~,
\label{gzoff1}
\eeq
where $M_Z$ represents the \zz~mass, $G_{\mu}$ is the Fermi coupling
constant, and $g^V_{f\!{\bar f}\!Z}$~and $g^A_{f\!{\bar f}\!Z}$~are
the vector and axial vector couplings of a fermion-antifermion pair
$f{\bar f}$ to the \zz~boson (see appendix~\ref{feynrules}). The
color factor $N_c(f)$ is unity for leptons and three for quarks. Using
left- and right-handed couplings as introduced in
equation~(\ref{lrcoup}),
\ba
  L_{\EPL\!\EMI Z} & = & -\,\frac{e}{\,4\,\SIW\,\COW} \cdot
    \left(1 \,-\, 2\,\SWS\right) \;\; \equiv \;\;
    \left[ \rule[-.1cm]{0cm}{.8cm} \frac{G_{\mu} M_Z^2}{2 \sqrt{2}}
    \right]^{\frac{1}{2}} \!\cdot\! L_e \nl \nl
  R_{\EPL\!\EMI Z} & = & \hspace{.395cm} \frac{e}{\,4\,\SIW\,\COW} \cdot
    2\,\SWS \hspace{1.52cm} \equiv \;\;
    \left[ \rule[-.1cm]{0cm}{.8cm} \frac{G_{\mu} M_Z^2}{2 \sqrt{2}}
    \right]^{\frac{1}{2}} \!\cdot\! R_e~~, \nonumber
\ea
$\sigma^B_4(s;\SONE,\STWO)$ is obtained after fivefold analytical
integration over the angular degrees of freedom of the $2\!\to\!4$
particle phase space (see appendix~\ref{ps2to4}),
\ba
  \sigma^B_4(s;\SONE,\STWO) & = & \frac{\sqrt{\lambda}}{\,\pi\,s^2}\:
    {\ds \frac{\left(G_{\mu} M_Z^2 \right)^2}{8}}
    \left(L_e^4+R_e^4\right) \cdot {\cal G}_4^{t+u}(s;\SONE,\STWO)~~,
\label{sigzz41}
\ea
with the sub-index 4 indicating that the underlying matrix element
contains four resonant \zz~propagators. In $\sigma^B_4$, final state
fermion masses are
neglected. The factor in front of ${\cal G}_4^{t+u}$ comprises phase
space factors and couplings to the initial state current. The
kinematical function ${\cal G}_4^{t+u}(s;\SONE,\STWO)$ for the summed
contributions from the t-channel, the u-channel, and the t-u
interference is given by
\ba
  {\cal G}_4^{t+u}(s;\SONE,\STWO) & = &
    \frac{\,s^2+(\SONE+\STWO)^2\,}{s-\SONE-\STWO}\,\cl_B - 2~~,
  \label{ZZmuta}
\ea
with
\ba
  {\cal L}_B & = & {\cal L}(s;\SONE,\STWO) \; = \;
    \frac{1}{\sqrt{\lambda}} \,
    \ln \frac{s-\SONE-\STWO+\sqrt{\lambda}}
             {s-\SONE-\STWO-\sqrt{\lambda}}~~, \nl
  \lambda~ & = & \lambda(s,\SONE,\STWO)~~,
  \nonumber
\ea
\vspace{-.8cm}
\ba
  \lambda(a,b,c) \; \equiv \; a^2 + b^2 + c^2 - 2ab - 2ac -
  2bc~~. \nonumber
\ea
${\cal G}_4^{t+u}$ was rederived and fully agrees with earlier
results~\cite{Brown78,baier66}. As is shown in
reference~\cite{dl94}, the on-shell limit of equation~(\ref{ZZmuta})
recovers the result given in~\cite{Brown78}. ${\cal G}_4^{t+u}$ is
also known in the context of ${\cal O}(\alpha^2)$ corrections to
\ee~annihilation into fermion pairs~\cite{kniehl88}.
${\cal G}_4^{t+u}$ separartes into contributions from the t- and
u-channels and a t-u interference contribution (compare
equation~(\ref{borncon})):
\ba
  {\cal G}_4^{t+u}(s;\SONE,\STWO) & = & {\cal G}_4^t + {\cal G}_4^u +
                                        {\cal G}_4^{tu} \nonumber
\ea
\ba
  {\cal G}_4^t & = & {\cal G}_4^u \;\; = \;\;
    \frac{1}{8\,\SONE\STWO} \left[ \rule[0cm]{0cm}{.6cm}
      \frac{\lambda}{6} + 2s(\SONE\!+\!\STWO) - 8\SONE\STWO +
      4\SONE\STWO(s\!-\!\SONE\!-\!\STWO){\cal L}_B \right] \nl
  {\cal G}_4^{tu} & = & \hspace{1.09cm} - \, \frac{1}{8\,\SONE\STWO}
    \left[ \rule[0cm]{0cm}{.7cm} \frac{\lambda}{3} +
      4s(\SONE\!+\!\STWO) +
      16\,s\,\SONE\STWO \left( \rule[0cm]{0cm}{.6cm}
      1 - \frac{s}{s\!-\!\SONE\!-\!\STWO} \right)
      {\cal L}_B \right]~~. \nl
  \label{borncont}
\ea
If considered separately, the contributions ${\cal G}_4^{t}$~from the
t-channel, ${\cal G}_4^{u}$~from the u-channel, and
${\cal G}_4^{tu}$~from the t-u interference violate unitarity. As can
be seen from equations~(\ref{sigzz41}) and~(\ref{borncont}),
every contribution yields a total \xsec behavior proportional to $s$ as
$s\!\to\!\infty$. Thus each single contribution violates the
Froissart limit~\cite{itzykson}, and only summation of the
contributions ${\cal G}_4^{t}$, ${\cal G}_4^{u}$, and ${\cal G}_4^{tu}$,
corresponding to the gauge invariant set of
Feynman diagrams shown in figure~\ref{zzfeyn}, delivers a reasonable
result with proper high energy behavior. This is a standard result of
Yang-Mills theories~\cite{alles77}. It is noteworthy that, adding the
three contributions from t-channel, u-channel, and t-u-interference,
the factors $1/(\SONE\STWO)$ contained in the individual kinematical
functions given in equation~(\ref{borncont}) cancel. Consequently, the presence
of the overall damping factor $(\SONE\STWO)/s^2$ in
$d^2 \sigma^B/(d\SONE d\STWO)$ should be considered a result of the
gauge cancellations ensuring the unitarity of $\sigma^B$.
\para
A graphical representation of $\SLAM/s \!\cdot\! {\cal G}_4^{t+u}$ is
given in plots a) and b) of figure~\ref{kinfunZZ} for two different
ranges of the variables $1-\SONE/s$~and $1-\STWO/s$. As $\SLAM/s
\!\cdot\! {\cal G}_4^{t+u}$ is a dimensionless quantity, it only
depends on $\SONE/s$~and $\STWO/s$. From plot c) of
figure~\ref{kinfunZZ} one can see how the double-resonance behavior
entering the \xsec via the product $\rho_Z(\SONE) \, \rho_Z(\STWO)$ of
Breit-Wigner densities dominates the character of the
twofold differential \xsec $d^2 \sigma^B/(d\SONE\,d\STWO)$: Two
intersecting resonances are seen in plot c) whereas the behavior of
${\cal G}_4^{t+u}$ in dependence on $\SONE$~and $\STWO$~is rather
suppressed. Final \xsec results are obtained from
equation~(\ref{sigzz1}) by twofold numerical
integration\footnote[2]{All numerical programs for this thesis were
  written in {\sc Fortran}. The author is grateful to D. Bardin for
  providing the integration routines {\tt SIMP} and {\tt FDSIMP}
  together with routines for the evaluation of Dilogarithm and
  Trilogarithm functions.}.
%
\begin{figure}[p]
  \vspace{19.5cm}
  \caption[The kinematical function ${\cal G}_4^{t+u}$]
    {\it \underline{Plots a) \& b):} Behavior of the dimensionless
      kinematical function $\SLAM/s \cdot{\cal G}_4^{t+u}$ for the
      two different ranges $\,0 \!\leq\! 1\!-\!s_{ij}/s \!<\!
      1\!-\!s_{ij,min}/s\,$ and $\,0.15 \!\leq\! 1\!-\!s_{ij}/s
      \!\leq\! 0.995$. ~The var\-ia\-bles
      $1\!-\!\SONE/s$~and $1\!-\!\STWO/s$~were chosen for the sake
      of clearer graphical presentation. Note that the
      used graphics package PAW~\cite{paw} has problems at the
      lower edges of phase space because of the very steep gradient of
      the plotted function. The flat region in the foreground
      represents zero level and lies outside the phase space. The edge
      of phase space is clearly seen from plot b).~~\underline{Plot
      c):} Multiplication of $\SLAM/s \!\cdot\!{\cal G}_4^{t+u}$ with
      $s\!\cdot\! \rho_Z(\SONE) \rho_Z(\STWO)$ at $\sqrt{s} = 200$~GeV
      emphasizes the dominance of the double-resonance character of
      process~(\ref{eezz4f}).}
  \label{kinfunZZ}
\end{figure}
%
\para
To be able to compare the on-shell with the off-shell case, a
``total'' \xsec is introduced via
\beq
  \sigma^B_{tot}(s) \; \equiv \;
    \frac{\sigma^B(s)}{\,2 \cdot B\!R(1) \cdot B\!R(2)\,}~~,
  \label{OScomp}
\eeq
which, in the narrow width limit, corresponds to the on-shell \zz~pair
cross-section. The effect of the finite \zz~width on the total
\xsec(\ref{OScomp}) is illustrated in figure~\ref{plboZZ}.
%
\begin{figure}[tb]
  \vspace{17.5cm}
  \caption[The \zz~pair Born \xsec]
    {\it The total \zz~pair production \xsec $\sigma^B_{tot}(s)$ for
      process~(\ref{eezz4f}) and the effect of the finite \zz~width.
      The numerical input for the \zz~pair production figures is:
      $G_{\mu} \!=\! 1.16639\times10^{-5}~GeV^{-2},
      ~M_Z\!=\!91.173~GeV,~\Gamma_Z(M_Z)\!=\!2.487~GeV,
      ~\SWS\!=\!0.2325.$ Final state fermion masses are neglected. The
      numerical precision is estimated to be better than
      0.01$\,$\permille.}
  \label{plboZZ}
  \vspace{-.2cm}
\end{figure}
%
The off-shellness of the
\zz~bosons causes the characteristic smearing of the excitation curve
with a dumped peak and stretched tails. The change in the \xsec
induced by the finite width is of the order $\Gamma_Z/M_Z$ and
ranges from $-7\%$~at the \xsec peak to $+5\%$~at 1 TeV. The
peak is shifted by approximately $+4.5$ GeV. Using a constant
width, i.e. $\sqrt{s} \, \Gamma_Z(s) \rightarrow M_Z \Gamma_Z$,
corresponds to a redefinition of the \zz~boson mass~\cite{massdef}:
$M_Z \rightarrow {\bar M}_Z\!=\!M_Z + \frac{1}{2} \Gamma_Z^2/M_Z$,
resulting in ${\bar M}_Z \approx M_Z + 34$ MeV.
\para
In the case of $W$ pair production, gauge
violating imaginary parts from the introduction of the finite width
are only properly cancelled by imaginary parts from the trilinear
vertex correction~\cite{baur95,passarino95}. Such gauge invariance problems
due to the naive introduction of the finite boson width into the
propagators are not present in \zz~pair production. Possible gauge
violations from the introduction of the \zz~width as an $s$-dependent
or $s$-independent quantity are phenomenologically irrelevant and may
be trivially resolved by adjusting the definition of the width. The
absence of the more serious gauge violations in off-shell \zz~pair
production is understood from the fact that gauge cancellations
already happen at the level of $\sigma^B_4$ because of the identical
coupling structure of the t- and the u-channel contributions.
%
\chapter{${\cal O}(\alpha)$ Initial State Radiation}
\label{isr}
\begin{figure}[t]
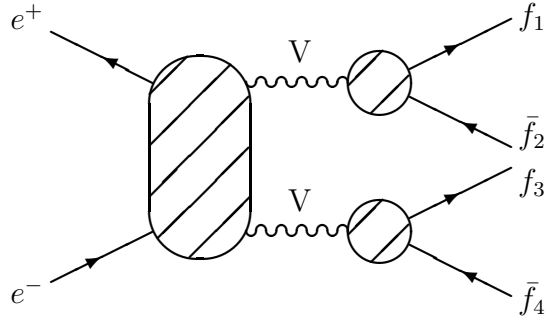

  \vspace*{.2cm}
  \begin{Feynman}{75,60}{-14.5,0}{0.9}
%
    \put(5,19){\fermionurrhalf}
    \put(5,41){\fermionullhalf}
    \put(10.5,15.2){\line(1,1){8.9}}
    \put(10.5,22.7){\line(1,1){8.7}}
    \put(10.5,22.7){\line(-1,-1){4.7}}
    \put(10.5,30.2){\line(1,1){8.7}}
    \put(10.5,30.2){\line(-1,-1){5.85}}
    \put(10.5,37.7){\line(1,1){6}}
    \put(10.5,37.7){\line(-1,-1){5.85}}
    \put(10.1,44.8){\line(-1,-1){5.5}}
    \put(12,30){\oval(15,30)}
    \put(18.65,19){\photonrighthalf}
    \put(18.65,41){\photonrighthalf}
    \put(34,40.7){\line(1,1){5}}
    \put(37.3,36.5){\line(1,1){5.5}}
    \put(34,18.7){\line(1,1){5}}
    \put(37.3,14.5){\line(1,1){5.5}}
    \put(38.5,19){\circle{10}}
    \put(38.5,41){\circle{10}}
    \put(58,50.5){\fermionurrhalf}
    \put(58,28.5){\fermionurrhalf}
    \put(58,31.5){\fermionullhalf}
    \put(58,9.5){\fermionullhalf}
    \put(-16,8.5){\emi}
    \put(-16,49){\epl}
    \put(25,44){V}
    \put(25,22){V}
    \put(59,7.5){${\bar f}_4$}
    \put(59,25){$ f_3$}
    \put(59,32){${\bar f}_2$}
    \put(59,49.5){$ f_1$}
  \end{Feynman}
  \vspace{-1cm}
  \caption[Generic two-boson production Feynman diagram for
  \ee~annihilation]
    {\it The generic two-boson production and decay Feynman diagram
      for \ee~annihilation. $V$~stands for any vector boson, the
      $f_i,~i=1,2,3,4$~represent the decay fermions.}
  \label{twores}
\end{figure}

In \ee~annihilation, initial state radiation (ISR) is known to
represent the bulk of the radiative corrections. The
${\cal O}(\alpha)$ `amputated' Feynman diagrams for initial state
bremsstrahlung to process~(\ref{eezz4f}) are shown in
figure~\ref{zzbrem}, the corresponding virtual initial state
correction diagrams are given in figure~\ref{zzvirt}. External leg
self energies are absorbed into on-shell renormalization.
For both the t- and the u-channel, \zz~bosons have to be attached
to the two unconnected vertices of each `amputated' Feynman diagram.
%
%
\begin{figure}[h]
\vspace{.6cm}
\begin{minipage}[tbh]{15.cm} {
\begin{center}
\begin{Feynman}{150,60}{-43,0}{0.9}
%
\small
\put(-26,52.5){$e^+$}  
\put(-26,06.5){$e^-$}
\put(5,49){$\gamma(p)$}
\normalsize
\put(-5,45){\line(-2,1){15.00}}
\put(-16.25,50.625){\vector(-2,1){1}}  
\put(-8.75,46.875){\vector(-2,1){1}}
\put(-5,15){\fermionurrhalf}
\put(-5,15){\fermionup}
\put(-12.5,48.75){\photonrighthalf}
\put(-5,45){\circle*{1.5}}
\put(-5,15){\circle*{1.5}}
\put(-12.5,48.75){\circle*{1.5}}
%
\put(85,45){\fermionullhalf}
\put(85,15){\fermionup}
\put(76.5,10.75){\photonrighthalf}
\put(85,45){\circle*{1.5}}
\put(85,15){\circle*{1.5}}
\put(76.5,10.75){\circle*{1.5}}
\put(70,7.5){\line(2,1){15.00}}
\put(73.75,9.375){\vector(2,1){1}}  
\put(83,14){\vector(2,1){1}}
%
\put(40,15){\fermionurrhalf}
\put(40,45){\fermionullhalf}
\put(40,30){\fermionuphalf}
\put(40,15){\fermionuphalf}
\put(40,30){\photonrighthalf}
\put(40,45){\circle*{1.5}}
\put(40,15){\circle*{1.5}}
\put(40,30){\circle*{1.5}}
\end{Feynman}
\end{center}
}\end{minipage}
\vspace{-.5cm}
\caption[Feynman diagrams for initial state bremsstrahlung]
{\it The amputated initial state bremsstrahlung diagrams for \zz~pair
  production.}
\label{zzbrem}
\vspace{.3cm}
\begin{minipage}[tbh]{15cm} {
\begin{center}
\begin{Feynman}{150,60}{-44,0}{0.9}
%
\small
\put(-42,52){$e^+$}  
\put(-42,6){$e^-$}
\put(-21,15){\fermionurrhalf}
\put(-21,45){\fermionullhalf}
\put(-21,15){\fermionuphalf}
\put(-21,30){\photonuphalf}
\put(-32,37){$\gamma(p)$}
\put(-21,37.5){\oval(15,15)[r]}
\put(-15.6,42.8){\vector(-1,1){1}}  
\put(-15.55,32.2){\vector(1,1){1}}  
\put(-21,45){\circle*{1.5}}
\put(-21,15){\circle*{1.5}}
\put(-21,30){\circle*{1.5}}
\put(-13.5,37.5){\circle*{1.5}}
\put(16,15){\fermionurrhalf}
\put(16,45){\fermionullhalf}
\put(16,30){\fermionuphalf}
\put(16,15){\photonuphalf}
\put(16,22.5){\oval(15,15)[r]}
\put(21.35,27.8){\vector(-1,1){1}}  
\put(21.35,17.2){\vector(1,1){1}}  
\put(16,45){\circle*{1.5}}
\put(16,15){\circle*{1.5}}
\put(16,30){\circle*{1.5}}
\put(23.50,22.5){\circle*{1.5}}
\put(53,15){\fermionurrhalf}
\put(53,45){\fermionullhalf}
\put(53,15){\line(0,1){7.5}}
\put(53,18.75){\vector(0,1){1}}  
\put(53,41.25){\vector(0,1){1}}  
\put(53,37.5){\line(0,1){7.5}}
\put(53,22.5){\photonuphalf}
\put(53,30){\oval(15,15)[r]}
\put(60.48,30){\vector(0,1){1}}  
\put(53,15){\circle*{1.5}}
\put(53,22.5){\circle*{1.5}}
\put(53,37.5){\circle*{1.5}}
\put(53,45){\circle*{1.5}}
\put(90,15){\fermionurrhalf}
\put(105,22.5){\fermionurrhalf}
\put(90,45){\fermionullhalf}
\put(105,37.5){\fermionullhalf}
\put(105,22.5){\fermionuphalf}
\put(90,15){\photonup}
\put(90,45){\circle*{1.5}}
\put(90,15){\circle*{1.5}}
\put(105,22.5){\circle*{1.5}}
\put(105,37.5){\circle*{1.5}}
\end{Feynman}
\end{center}
}\end{minipage}
\vspace{-.6cm}
\caption[Feynman diagrams for virtual initial state corrections]
{\it The amputated virtual initial state correction diagrams for
  \zz~pair production.}
\vspace{.3cm}
\label{zzvirt}
\end{figure}
\clearpage
\noindent
The computation of the virtual \oal~\xsec starts from the
$2\!\to\!4$~particle phase space given in equation~(\ref{gamma4}),
appendix~\ref{ps2to4}. A \xsec differential in $\SONE$~and $\STWO$ is
obtained by fivefold analytical angular integration of the
interference between
the Born and the virtual matrix elements over the boson scattering
angle and over the fermion decay azimuthal and polar angles in the
corresponding boson rest frames. The computation of \oal~initial state
bremsstrahlung starts from the $2\!\to\!5$~particle phase space given
in equation~(\ref{ph25el}), appendix~\ref{ps2to5}. Then, a sevenfold
analytical angular integration of the bremsstrahlung matrix element
squared is carried out over the photon scattering angle, the boson
scattering azimuthal and polar angles in the two boson rest frame, and
over the fermion decay azimuthal and polar angles in the corresponding
boson rest frames.
After these five- and sevenfold angular integrations, the differential
\xsec for the \zz~pair production process~(\ref{eezz4f}) including
\oal~ISR with soft photon exponentiation is given by
\ba
  \frac{d^2 \sigma^{ISR}}{\,d\SONE \, d\STWO\,} & = &
  \int\limits_{(\sqrt{\SONE}+\sqrt{\STWO})^2}^s \!\!\!\!
  \frac{ds'}{s} \; \rho_Z(\SONE) \; \rho_Z(\STWO) \!\cdot\!
  2 B\!R(1) B\!R(2) \cdot
  \left[ \beta_e v^{\beta_e - 1} {\cal S} + {\cal H} \right]
  \label{compqed}
\ea
with $\beta_e \!=\!2 \alpha/\pi \, [ \ln (s/m_e^2) - 1 ]$~and
$v=1-s'/s$. This is exactly the result derived in appendix~\ref{expo}
and finally given in equation~(\ref{expooal}), except that the
sequence of integration has been changed to match the 2$\to$5 phase
space parametrization described in equation~(\ref{par25b}),
appendix~\ref{ps2to5}. The soft+virtual and hard photonic parts
${\cal S}$ and ${\cal H}$ in equation~(\ref{compqed}) were calculated
analytically. Both ${\cal S}$ and ${\cal H}$
separate into a universal part with the Born \xsec at the reduced
center of mass energy squared $s'$ factorizing and a non-universal
part:
\ba
  {\cal S}(s,s';\SONE,\STWO) & = &
  \left[1 + {\bar S_1}(s) \right] \!\cdot\! \sigma^B_4(s';\SONE,\STWO)
  \; + \; \sigma_{\hat S}(s';\SONE,\STWO) ~~~~~~,
  \nl
  \hspace{-1cm}
  {\cal H}(s,s';\SONE,\STWO) & = &
  \underbrace{{\bar H_1}(s,s') \!\cdot\! \sigma^B_4(s';\SONE,\STWO)~~~}_{\rm
    Universal~Part} \;\: + \;
  \underbrace{\sigma_{\hat H}(s,s';\SONE,\STWO)}_{\rm
    Non-universal~Part}~~.
  \label{xsfull}
\ea
The \oal~radiators ${\bar S_1}$~and ${\bar H_1}$~were explicitly
derived in the full calculation. They emerged from the virtual and
bremsstrahlung matrix elements after five- and sevenfold angular
integration. Their derivation is outlined in appendix~\ref{expo}:
\ba
  {\bar S_1} & = & \frac{\,\alpha\,}{\pi}
    \left[ \rule[0cm]{-.1cm}{.35cm} \frac{\pi^2}{3} - \frac{1}{2} \right]
    \; + \; \frac{3}{4} \, \beta_e~~, \nl
  {\bar H_1} & = & - \frac{1}{2} \left[ \rule[0cm]{-.1cm}{.65cm}
    \; 1+\frac{s'}{s} \right] \!\cdot\! \beta_e~~.
  \label{radiators}
\ea
${\cal O}(\alpha^2)$~corrections may be taken into account in the
universal corrections by adding the radiators ${\bar S_2}$~and
${\bar H_2}$~to ${\bar S_1}$~and ${\bar H_1}$ respectively.
${\bar S_2}$~and ${\bar H_2}$~are e.g. found in
reference~\cite{berends88}. The factorizing parts of the \xsecs
${\cal S}$~and ${\cal H}$~are called ``universal'', because they are
independent of the process topology, and especially of the initial
state topology of the radiation process\footnote[2]{In this statement
it is implicitly assumed that the considered initial state topologies
have unit charge initial state fermion lines connecting the initial
state \epl~and \emi.}. In fact, the universal
radiators ${\bar S_1}$~and ${\bar H_1}$ are also found in the
literature on fermion pair production in \ee~annihilation,
e.g. in reference~\cite{berends88}. It must be stressed that, although
it was expected to find a separation of the \xsec into universal and
non-universal contributions, the working out of the separation was a
technically difficult task, especially for the t-u interference
contribution. The virtual+soft and hard, non-universal,
non-factorizable contributions in equation~(\ref{xsfull}) are given by
\ba
  \sigma_{\hat S}(s';\SONE,\STWO) & = & \frac{\alpha}{\pi} \cdot
    \frac{\,\left(G_{\mu} M_Z^2 \right)^2\,}{8\pi s'} \:
    \left(L_e^4 \!+\! R_e^4\right) \cdot
    \sigma^V_{4,nonuni}(s';\SONE,\STWO)
    \nl\nl
  \sigma_{\hat H}(s,s';\SONE,\STWO) & = & \frac{\alpha}{\pi} \cdot
    \frac{\,\left(G_{\mu} M_Z^2 \right)^2\,}{8\pi s} \:
    \left(L_e^4 \!+\! R_e^4\right) \cdot
    \sigma^R_{4,nonuni}(s,s';\SONE,\STWO)
  \label{nunixs}
\ea
with $\sigma^V_{4,nonuni}$~and $\sigma^R_{4,nonuni}$~as derived in
appendix~\ref{xscalc}. The full expressions are given in
equation~(\ref{nunivxs}), appendix~\ref{virtres} and in
equation~(\ref{brnunitu}), appendix~\ref{bremres}.
Equations~(\ref{nunivxs}) and~(\ref{brnunitu}) represent a main result
of this thesis, but will not be rewritten here because of their
considerable length. The result~(\ref{nunixs}) is new even in its
limit for the case of on-shell \zz~bosons.
It is seen from equations~(\ref{nunivxs}) and~(\ref{brnunitu})
that the non-universal contributions contain many dilogarithm
and trilogarithm functions and are rather involved. One should
note that, in equation~(\ref{nunixs}), the virtual+soft non-universal
contribution is evaluated at $s'$ instead of $s$. The reason for this
treatment is to avoid an unphysical, $\delta$~distribution-like
concentration of non-universal virtual+soft ISR corrections at zero
radiative energy loss. One of the reasons for the appearance of
non-universal ISR contributions is the angular dependence of t- and
u-channel propagators. For both virtual and bremsstrahlung corrections
the angular behavior of t- and u-channel propagators is much more
complicated than the angular behavior of s-channel propagators
(see appendices~\ref{ps2to5}, \ref{virtres}, and \ref{bremres} for
details). Total \xsecs are computed from equation~(\ref{compqed}) by
threefold numerical integration over $s',~\STWO,$~and $\SONE$.
\para
Several other remarkable properties of equation~(\ref{xsfull}) are
worth mention. The leading ISR corrections to the \xsec with the
mass singularities $\beta_e$ factorize and are contained in the
universal \xsec contributions. This is true not only for the overall
cross-section, but also for the individual contributions from the t-
and the u-channel and from their interference. As one would expect,
non-universal contributions do not contain the mass singularity
$\beta_e$. As a consequence, non-universal ISR corrections are
suppressed with respect to the universal ones. The numerical influence
of universal and complete ISR corrections on the \xsec for
process~(\ref{eezz4f}) is presented in figure~\ref{plISRZZ}.
Some discussion of the results presented in figure~\ref{plISRZZ} is in
order. Around threshold, in the energy range of LEP2, universal ISR
peaks at large negative contributions with a relative correction of
almost --30\% at the \zz~pair threshold itself.
Below $\sqrt{s}$=280~GeV, ISR radiatively reduces the \zz~pair
\xsec \hspace{-.16cm}. At higher energies ISR corrections are positive
and develop a radiative tail increasing the Born off-shell \xsec by
+14\% at $\sqrt{s}$=1~TeV. These patterns are familiar from the
\zz~pole, which is not surprising, because single \zz~production is a
resonant process too, and because the radiators
${\bar S_1}$~and ${\bar H_1}$~describe the complete \oal~ISR at the
\zz~pole~\cite{lep1qed}. The radiative tail, already present in
universal ISR, is further pronounced by non-universal ISR, whereas
non-universal corrections are hardly noticeable around threshold and
therefore irrelevant in the LEP2 energy regime. The explanation for
the radiative tail in the universal corrections is that radiative
reduction of $s$ to $s'$ increases the relevant Born \xsec
$\sigma^B_4(s',\SONE\STWO)$ (see equations~(\ref{sigzz41})
and~(\ref{xsfull})) in the region of negative slope of
$\sigma^B_4$. The additional corrections due to non-universal
contributions range from 0.3\% below threshold up
to 2.5\% at 1~TeV. In the energy range of LEP2, non-universal
corrections do not exceed 0.4\%. Comparison of figures~\ref{plboZZ}
and~\ref{plISRZZ} shows that the high energy tail due to ISR is much
more pronounced than the one due to the off-shellness of the
\zz~bosons. It should also be realized that ISR shifts the \xsec
maximum by 9~GeV from $\sqrt{s}$=211~GeV to $\sqrt{s}$=220~GeV.
%
\begin{figure}[t]
  \vspace{14cm}
  \caption[The \zz~pair \xsec with initial state radiative corrections]
    {\it The total off-shell \zz~pair production \xsec
      $\sigma_{tot}(s)$. The solid line represents the Born \xsec
      $\sigma^B_{tot}(s)$~as in figure~\ref{plboZZ}. The
      dash-dotted line gives the \xsec with universal ISR corrections
      only, i.e. $\sigma_{\hat S}$ and $\sigma_{\hat H}$ are set to
      zero. The dotted line gives the \xsec with complete ISR
      from eq.~(\ref{compqed}). In the inset, the
      relative deviations of the universally and completely ISR
      corrected \xsecs from the off-shell Born \xsec
      $\sigma^B_{tot}(s)$~are given. The numerical input for
      the ISR corrections is: $\alpha\!=\!1/137.0359895,~m_e \!=\!
      0.51099906~MeV$. The numerical precision is around
      0.1$\,$\permille.}
  \vspace*{-.5cm}
  \label{plISRZZ}
\end{figure}
%
\clearpage
\noindent
Investigating the unitarity for the ISR corrected cross-section, one
observes from equations~(\ref{compqed}) to~(\ref{radiators}) that the
$1/s$ behavior observed in the Born
\xsec is altered by factors $\ln(s/m_e^2)$ originating from the mass
singularity $\beta_e$. Therefore the unitary behavior due to the
gauge cancellations discussed in chapter~\ref{born} is carried through
to the universal part of the ISR corrections. For $s\!\to\!\!\infty\,$,
the universally ISR corrected \xsec behaves better than $(\ln^3 s)/s$
and thus approaches zero. As
is seen from table~\ref{taxsZZ}, the absolute value of the
non-universal corrections reaches a maximum around $\sqrt{s}$=450~GeV
and then decreases with increasing $\sqrt{s}$. This already hints at
the unitary behavior of the non-universal ISR corrections which is
ensured by the overall damping factor already discussed in
chapter~\ref{born}. However, an important difference between
the appearance of the factor $(\SONE\STWO)/s^2$ in the tree level and
in the non-universally ISR corrected \xsec exists: All ISR
\xsec contributions, namely the ISR t-channel, the ISR u-channel, and
the ISR t-u interference contributions, contain this {\it screening}
factor, whereas at tree level its appearance is only a result of the
gauge cancellations.
%
\begin{table}[t]
  \vspace*{.5cm}
  \begin{center}
  \begin{tabular}{|c||c|c|c|} \hline
  & & & \vspace{-.3cm} \\
    $\sqrt{s}$ [GeV]
  & $\sigma^B_{tot}$ [pb]
  & $\sigma^{ISR,uni}_{tot}$ [pb]
  & $\sigma^{ISR,compl.}_{tot}$ [pb] \\
  & & & \vspace{-.3cm} \\
  \hline
  130.0 & 0.0015 & 0.0013 & 0.0013 \\
  150.0 & 0.0067 & 0.0055 & 0.0056 \\
  165.0 & 0.0215 & 0.0175 & 0.0176 \\
  180.0 & 0.1657 & 0.1227 & 0.1231 \\
  190.0 & 0.8971 & 0.6995 & 0.7021 \\
  200.0 & 1.1397 & 0.9641 & 0.9678 \\
  212.0 & 1.1898 & 1.0629 & 1.0671 \\
  230.0 & 1.1316 & 1.0611 & 1.0657 \\
  260.0 & 0.9799 & 0.9613 & 0.9663 \\
  300.0 & 0.8099 & 0.8211 & 0.8264 \\
  400.0 & 0.5503 & 0.5811 & 0.5866 \\
  500.0 & 0.4074 & 0.4399 & 0.4454 \\
  600.0 & 0.3171 & 0.3481 & 0.3534 \\
  800.0 & 0.2111 & 0.2374 & 0.2422 \\
 1000.0 & 0.1524 & 0.1744 & 0.1787 \\
  \hline
  \end{tabular}
  \end{center}
  \vspace{-.4cm}
  \caption[Off-shell \zz~pair production \xsec with and without
  initial state radiative corrections]
    {\it Numerical values for the off-shell \zz~pair production \xsec
      at tree level, with universal ISR, and with complete ISR as
      obtained from equations~(\ref{sigzz1}) and~(\ref{compqed}) by
      numerical integration.}
  \label{taxsZZ}
\end{table}
Thus the screening factor appears separately in all the following
non-universal \xsec parts,
\beq
  \rho(\SONE) \, \rho(\STWO) \cdot \sigma^{t,u,tu}_{{\hat S},{\hat H}}
    \;\sim\; \frac{\,\SONE\,\STWO\,}{s^2}~~,
\eeq
and it ensures the unitary behavior, i.e. the smallness of
non-universal ISR
corrections, at the level of individual contributions. This is more
than one could expect, because \xsec unitarity only requires a
unitarity-conserving mechanism for the sum of all contributions.
\para
To summarize this chapter, the previously unknown full initial state
QED corrections to \zz~pair production were calculated and separated
into a universal leading logarithm and a non-universal part. It was
illustrated how high energy unitarity is conserved for the
radiatively corrected cross-section. Complementary results for
\wpl\wmi~pair production were reported in~\cite{dimaww} where, instead
of a u-channel amplitude two s-channel amplitudes containing the
\wpl\wmi\zz~and \wpl\wmi$\gamma$ vertices are present. Taking these
results into account, one can claim knowledge of all initial state QED
corrections for the whole class of four-fermion production via
two-boson production processes in \ee~annihilation. The
corres\-ponding generic two-boson production and decay Feynman diagram
is shown in figure~\ref{twores}.
%
%
%
\chapter{Generalization to Neutral Current Pair Production}
\label{xsZZGG}
%
Four-fermion production via massive boson pairs is easily accessible
through experimental cuts on fermion pair invariant masses. But still
one would like to know not only the four-fermion \xsec proceeding
via \zz~pair production, but the full four-fermion generalization
\beq
  \EE \rightarrow f_1\bar{f_1}f_2\bar{f_2}~,
  ~~~~~~~~~~~~ f_1\!\neq\!f_2~,
  ~~~f_i\!\neq\!e^\pm,\stackrel{_{(-)}}{\nu_e}
  \label{ee4ffull}
\eeq
of process~(\ref{eezz4f}). Process~(\ref{ee4ffull}) is described by 24
electroweak Feynman diagrams~\cite{alNC}. If $f_1$~and $f_2$ are
quarks, QCD contributions from 8 gluon exchange diagrams must be taken
into account~\cite{pittau94}, and if massive fermions are produced,
associated Higgs production must not be forgotten. In the electroweak
case, if $f_1$~and $f_2$~belong to the same isospin doublet, both
charged current (CC) and neutral current (NC) processes contribute to
reaction~(\ref{ee4ffull}).
The various aspects of the CC reactions have been studied by many
authors~\cite{dimaww,teupitz94,muta86,wim91,durhamww,coulomb,wim94}.
Interferences between CC and NC contributions should be small due
to different invariant masses favored by the production of fermion
pairs via boson decays. Born level semi-analytical calculations for
the NC case are available, and the inclusion of initial state
radiation (ISR) in a leading logarithm approximation is
unproblematic~\cite{excal,alNC}.
%
\begin{figure}[b]
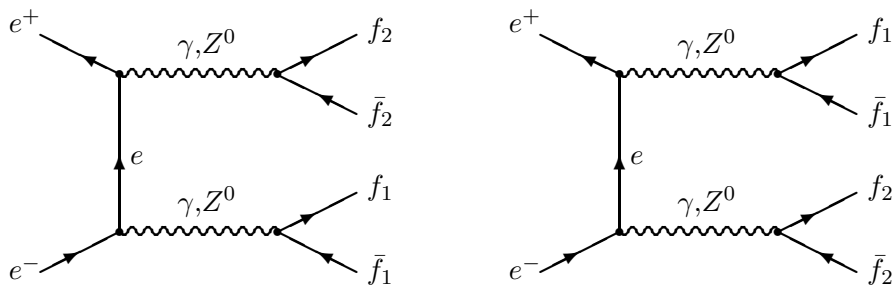

\vspace*{.2cm}
\begin{minipage}[tbh]{7.8cm}{
\begin{center}
\begin{Feynman}{75,60}{-20,0}{0.7}
%
\put(5,15){\fermionurrhalf}
\put(5,45){\fermionullhalf}
\put(5,15){\fermionup}
\put(5,45){\photonright}
\put(5,15){\photonright}
\put(5,45){\circle*{1.5}}
\put(5,15){\circle*{1.5}}
\put(35,15){\circle*{1.5}}
\put(35,45){\circle*{1.5}}
\put(50,7.5){\fermionullhalf}
\put(50,22.5){\fermionurrhalf}
\put(50,37.5){\fermionullhalf}
\put(50,52.5){\fermionurrhalf}
\small
\put(7,28){$e$}
\put(-16,6){\emi}
\put(-16,53){\epl}
\put(16,49){$\gamma$,\zz}
\put(16,19){$\gamma$,\zz}
\put(52,06){${\bar f}_1$}
\put(52,22){$ f_1$}
\put(52,36){${\bar f}_2$}
\put(52,52){$ f_2$}
\normalsize
\put(100,15){\fermionurrhalf}
\put(100,45){\fermionullhalf}
\put(100,15){\fermionup}
\put(100,45){\photonright}
\put(100,15){\photonright}
\put(100,45){\circle*{1.5}}
\put(100,15){\circle*{1.5}}
\put(130,15){\circle*{1.5}}
\put(130,45){\circle*{1.5}}
\put(145,7.5){\fermionullhalf}
\put(145,22.5){\fermionurrhalf}
\put(145,37.5){\fermionullhalf}
\put(145,52.5){\fermionurrhalf}
\small
\put(102,28){$e$}
\put(79,6){\emi}
\put(79,53){\epl}
\put(111,49){$\gamma$,\zz}  
\put(111,19){$\gamma$,\zz}
\put(147,06){${\bar f}_2$} 
\put(147,22){$ f_2$}
\put(147,36){${\bar f}_1$}
\put(147,52){$ f_1$}
\normalsize
\end{Feynman}
\end{center}
}\end{minipage}
\vspace{-.4cm}
\caption[Born level neutral current pair production Feynman diagrams]
{\it The Born level Feynman diagrams for neutral current pair
  production. Left: t-channel. Right: u-channel. }
\label{NCdiag}
\end{figure}
%
%
\begin{table}[t]
  \begin{center}
  \begin{tabular}{|c||c|c|} \hline
  & &  \vspace{-.3cm} \\
  & $\sigma_{NC}^{\mu^+\!\mu^- b \bar b}$ [fb]
  & $\sigma_{Conv}^{\mu^+\!\mu^- b \bar b}$ [fb] \\
  $\sqrt{s}$ [GeV] & All NC diagrams & NC Conversion diagrams \\
  & &  \vspace{-.3cm} \\ \hline
  91.0  & 613.832 & 278.080 \\
  141.0 &  94.469 &  94.477 \\
  191.0 &  53.252 &  53.231 \\
  241.0 &  38.610 &  38.613 \\
  291.0 &  28.396 &  28.422 \\
  341.0 &  22.128 &  22.161 \\
  391.0 &  17.930 &  17.970 \\
  441.0 &  14.946 &  14.982 \\
  491.0 &  12.713 &  12.742 \\
  591.0 &   9.620 &   9.570 \\
  691.0 &   7.551 &   7.603 \\
  791.0 &   5.935 &   6.187 \\
  891.0 &   5.166 &   5.165 \\
  991.0 &   4.237 &   4.382 \\
  \hline
  \end{tabular}
  \end{center}
  \vspace{-.3cm}
  \caption[Neutral current four-fermion production \xsecs]
    {\it Neutral current (NC) $\mu^+\!\mu^- b \bar b$~production
      cross-sections. By ``NC'' it is emphasized that only amplitudes
      containing intermediate \zz~bosons or photons are accounted for.
      Feynman diagrams containing the Higgs boson are excluded. No
      cuts are applied. The author is grateful to A.~Leike for
      supplying the {\sc Fortran} code {\tt 4fAN}~\cite{alNC} for the
      production of the results in this table.}
  \label{NCxsbo}
\end{table}
$\!\!$In figure~\ref{NCdiag}, all NC Feynman diagrams of the so-called
{\it Conversion} type are depicted.
As is seen from table~\ref{NCxsbo}, the {\it Conversion} Feynman
diagrams strongly dominate the \xsec at energies above the
\zz~mass. Interferences between {\it Conversion} type and other
NC diagrams are small and negative. Therefore it is interesting to
investigate the process
\beq
  \EE \rightarrow (\ZZ\ZZ,\ZZ\gamma,\gamma\gamma) \rightarrow
  f_1\bar{f_1}f_2\bar{f_2}(\gamma)~,
  ~~~~~~~~~~~~~~~ f_1\!\neq\!f_2~,
  ~~~f_i\!\neq\!e^\pm, \stackrel{_{(-)}}{\nu_e}
  \label{eeNC4f}
\eeq
with full ISR, proceeding via the tree level diagrams shown in
figure~\ref{NCdiag}.
The exchange photons to be taken into account have no resonance
behavior. Consequently, contrary to the double-resonance case treated
in chapter~\ref{born}, it is not advantageous to express initial state
couplings and boson propagators in terms of Breit-Wigner densities as
given in equation~(\ref{rhos}). Since process~(\ref{eeNC4f}) can be
described by several sets of t-channel, u-channel and t-u interference
contributions, it is appropriate to replace equation~(\ref{sigzz1}) by
\ba
  \sigma^B_{Conv}(s) & = &
    \int\limits_{4m_1^2}^{(\sqrt{s} - 2m_2)^2} \hspace{-.5cm} d\SONE \,
    \int\limits_{4m_2^2}^{(\sqrt{s} - \sqrt{\SONE})^2} \hspace{-.5cm}
    d\STWO \hspace{.1cm}
    \frac{\,d^2\sigma^B_{Conv}(s;\SONE,\STWO)\,}{d\SONE\,d\STWO}
  \label{sigmaNC}
\ea
with the twofold differential \xsec factorizing into a phase space
factor, a function $\cc$ of couplings and propagators, and the
kinematical function ${\cal G}_4^{t+u}$ already known from
equation~(\ref{ZZmuta}),
\ba
  \frac{\,d^2\sigma^B_{Conv}\,(s;\SONE,\STWO)}
       {d\SONE\,d\STWO} & = &
  \frac{\SLAM}{\,\pi\,s^2\,} \cdot \cc(\FONE,\SONE;\FTWO,\STWO) \,
  {\cal G}_4^{t+u}(s;\SONE,\STWO)~~.
  \label{sigmaNC2}
\ea
As ${\cal G}_4^{t+u}$ is, in the language of Feynman diagrams, related
to a pair of t- and u-channel diagrams, the function $\cc$~is obtained
by summing over all 16 such pairs that can be formed from the eight
diagrams of figure~\ref{NCdiag}. This results in
\ba
  \lefteqn{\cc(\FONE,\SONE;\FTWO,\STWO) \;\; = \;\;
    \frac{\,2\,\SONE\,\STWO\,}{(6\pi^2)^2} \times} \nl
    & & \RE \left\{ \rule[-.2cm]{0cm}{1cm}
      \sum_{B_i,B_j,B_k,B_l=\gamma,Z}
        \frac{1}{D_{B_i}(\SONE)}   \frac{1}{D_{B_j}(\STWO)}
        \frac{1}{D_{B_k}^*(\SONE)} \frac{1}{D_{B_l}^*(\STWO)} \times
      \right. \nl
    & & \hspace{1.1cm} \left[ \rule[0cm]{0cm}{.4cm}
      L_{\EPL\!\EMI B_i} \, L_{\EPL\!\EMI B_k} \,
      L_{\EPL\!\EMI B_j} \, L_{\EPL\!\EMI B_l} \; + \;
      R_{\EPL\!\EMI B_i} \, R_{\EPL\!\EMI B_k} \,
      R_{\EPL\!\EMI B_j} \, R_{\EPL\!\EMI B_l} \, \right] \times \nl
    & & \hspace{1.1cm} \left[ \rule[0cm]{0cm}{.4cm}
      L_{\FONE\!\BFONE B_i} \, L_{\FONE\!\BFONE B_k} \; + \;
      R_{\FONE\!\BFONE B_i} \, R_{\FONE\!\BFONE B_k} \, \right]
      \!\cdot\! N_c(f_1) \, N_p(B_i,B_k,m_1,\SONE) \times \nl
    & &
      \hspace{1.1cm}
      \left[ \rule[0cm]{0cm}{.4cm}
      L_{\FTWO\!\BFTWO B_j} \, L_{\FTWO\!\BFTWO B_l} \;\, + \;
      R_{\FTWO\!\BFTWO B_j} \, R_{\FTWO\!\BFTWO B_l} \, \right]
      \!\cdot\! N_c(f_2) \, N_p(B_j,B_l,m_2,\STWO)
  \label{couplings}
\ea
\vspace{-2.03cm}
\ba
  \left. \rule[-.2cm]{0cm}{1cm} \hspace{13.5cm} \right\}
  \nonumber
\ea
where left- and right-handed couplings are used as defined in
equations~(\ref{LRdef}) and (\ref{LRcoup}). The propagator
denominators are given by
\beq
  D_B(s_{ij}) \;\; = \;\; s_{ij} - M_B^2 + i\sqrt{s_{ij}} \,
    \Gamma_B(s_{ij})
\eeq
with their complex conjugates $D_B^*$. $M_{\gamma}=\Gamma_{\gamma}=0$
for the photon and $\Gamma_Z(s_{ij})\!=\!\sqrt{s_{ij}} \, \Gamma_Z / M_Z$
which is a good approximation around the \zz~peak and above. The color
factor $N_c(f)$ is unity for leptons and three for quarks. The phase
space factor
\ba
  N_p(B_i,B_k,m,s) & = & \!\!\! \left\{ \rule[-.3cm]{0cm}{1.2cm} \!\!\!
    \begin{array}{ccl}
      1 & \;\;\; {\rm for} \;\;\; & B_i\neq\gamma \;\; {\rm or} \;\;
        B_k\neq\gamma \vspace{.2cm} \\
      \sqrt{1 - 4m^2/s} \cdot (1 + 2 m^2/s) & \;\;\;
        {\rm for} \;\;\; & B_i = B_k = \gamma
    \end{array} \right.
  \label{psfact}
\ea
correctly takes into account that fermion masses may be neglected if
the corresponding fermion pair couples to a \zz~resonance but may not
if it only couples to a photon. In the latter case, finite \xsec
contributions arise from close to the lower boundary of the invariant
masses $s_{ij}$. It is readily verified that the above generalization
of the \zz~pair results from chapter~\ref{born} is formally carried out
through the replacement
\beq
  \rho_Z(\SONE) \, \rho_Z(\STWO) \!\cdot\! 2 B\!R(1) B\!R(2) \cdot
  \frac{\left(G_{\mu} M_Z^2 \right)^2}{8} \left(L_e^4+R_e^4\right)
  \;\; \longrightarrow \;\; \cc(\FONE,\SONE;\FTWO,\STWO)~.
  \label{ZZtoNC}
  \\ \nonumber
\eeq
The input for numerical calculations was chosen as $G_{\mu}\!=\!1.16639
\!\cdot\! 10^{-5}$~GeV$^{-2}$,~~$M_Z\!=\!91.173$~GeV,
{}~~$\Gamma_Z\!=\!2.4971$~GeV, ~$M_W\!=\!80.22~$GeV,
{}~and $m_e\!=\!0.51099906$~MeV.  For the electroweak mixing angle, the
effective value ~$\SWS\!=\!0.2325$ was chosen. The input for the left-
and right-handed couplings was $Q_e\!=\!-1,~~\ct^{(3)}_{W,e}\!=\!-
\frac{1}{2},$~~and $e\!=\!\sqrt{4\pi\alpha}$~~with $\alpha\!=\!\sqrt{2}
M_W^2 \SWS \, G_{\mu}/\pi$. For the QED corrections, because they are
mainly due to soft photons, the Thomson limit
$\alpha\!=\!1/137.0359895$~\cite{pdb94}~was used. These parameter
choices enabled straightforward comparisons with the results in
reference~\cite{alNC}.
For several center of mass energies in the range of LEP2 and above and
for different final state fermion pair combinations, tree level
results for the full set of {\it Conversion} diagrams were compared to
the appropriate contribution as obtained from the {\sc Fortran} code
{\tt 4fAN} used in reference~\cite{alNC}. Agreement was found to be at
the level of the numerical precision of the integration routine.
\para
Taking ISR into account, means to implement the
replacement~(\ref{ZZtoNC}) into formula~(\ref{compqed}), resulting in
\ba
  \frac{d^2 \sigma^{ISR}_{Conv}}
       {\,d\SONE\,d\STWO\,} & = &
    \int\limits_{(\sqrt{\SONE}+\sqrt{\STWO})^2}^s \!\!\!\!
    \frac{ds'}{s} \; \cc(\FONE,\SONE;\FTWO,\STWO) \cdot \left[\,
    \beta_e v^{\beta_e - 1} \, {\cal S}_{Conv} \,+\, {\cal H}_{Conv}
    \, \right]
  \label{NCISR}
\ea
for the twofold differential \oal~ISR corrected \xsec with soft photon
exponentiation for the NC pair production process~(\ref{eeNC4f}).
Recalling that the factor ~$\left(G_{\mu} M_Z^2 \right)^2\!\!/8
\!\cdot\! \left(L_e^4+R_e^4\right)$ is now absorbed in $\cc$ and
defining
\beq
  \sigma^0_{Conv}(s;\SONE,\STWO) \;\; \equiv \;\;
    \frac{\SLAM}{\,\pi\,s^2} \, {\cal G}_4^{t+u}(s;\SONE,\STWO)
\eeq
one finds from equation~(\ref{sigzz41}) and in analogy to
equations~(\ref{xsfull}) and~(\ref{nunixs})
\ba
    {\cal S}_{Conv}(s,s';\SONE,\STWO) & = &
  \left[1 + {\bar S_1}(s) \right] \!\cdot\!
  \sigma^0_{Conv}(s';\SONE,\STWO) \hspace{0.07cm}
  \; + \; \sigma_{{\hat S},Conv}(s';\SONE,\STWO) ~,
  \nl
  {\cal H}_{Conv}(s,s';\SONE,\STWO) & = &
  \underbrace{{\bar H_1}(s,s') \!\cdot\!
    \sigma^0_{Conv}(s';\SONE,\STWO)~~~\,}_{\rm Universal~Part} \;\: + \;
  \underbrace{\sigma_{{\hat H},Conv}(s,s';\SONE,\STWO)}_{\rm
    Non-universal~Part} \nl
  \label{NCfull}
\ea
with the radiators ${\bar S_1}$~and ${\bar H_1}$ given in
equation~(\ref{radiators}) and
\ba
  \sigma_{{\hat S},Conv}(s';\SONE,\STWO) & = & \frac{\alpha}{\pi}
    \cdot \frac{\,\sigma^V_{4,nonuni}(s';\SONE,\STWO)\,}{\pi s'}~~~~,
  \nl
  \sigma_{{\hat H},Conv}(s,s';\SONE,\STWO) & = & \frac{\alpha}{\pi}
    \cdot \frac{\,\sigma^R_{4,nonuni}(s,s';\SONE,\STWO)\,}{\pi s}~~.
  \label{nunixsNC}
\ea
$\sigma^V_{4,nonuni}$~and $\sigma^R_{4,nonuni}$~are found in
equations~(\ref{nunivxs}) and~(\ref{brnunitu}) respectively.
\para
As it yields a final state of experimental importance for Higgs
searches and as it is also representative for semi-leptonic final
states, the following presentation of results will focus on the
reaction
\ba
  \EPL\EMI \;\to\; (\ZZ\ZZ,\ZZ\gamma,\gamma\gamma) \;\to\;
  \mu^+ \!\mu^- \, b {\bar b}
  \label{mb}
\ea
which has already received considerable attention in the
literature~\cite{eebb,alteup94,alNC}.
In figure~\ref{NCnocut}, \xsecs for reaction~(\ref{mb}) as
obtained from equations~(\ref{sigmaNC}) and~(\ref{NCISR}) are
presented. The \xsec correction due to universal ISR is approximately
12\% at $\sqrt{s}$=130 GeV, increases to 18\% below the \zz~pair
threshold, decreases to 13\% in the region of the \zz~pair-induced
\xsec ``bump'' and then steadily increases to 21\% at 600 GeV. The
additional relative correction from non-universal ISR corrections
increases from
9$\,$\permille~at 130 GeV to 4.2\% at 600 GeV. These numbers can be
extracted from table~\ref{taxsNC} where numerical values for the
\xsecs presented in figure~\ref{NCnocut} are given. The numerical
precision of the values in table~\ref{taxsNC} is around
0.1$\,$\permille\footnote[2]{In the following, numerical precisions
  will not be quoted any more. They are around 0.1$\,$\permille~or
  better throughout this thesis.}.
%
\begin{figure}[p]
  \vspace{5.5cm}
  \caption[The neutral current pair \xsec with ISR corrections]
    {\it The NC~pair production \xsec for process~(\ref{mb}).
      The solid line represents the Born \xsec$\!\!$, the dash-dotted
      line includes universal, and the dotted line includes all ISR
      corrections. In the inset the full NC~pair \xsec is compared
      to the contributions from \zz~and photon pair production. The
      muon and b quark masses were taken as $m_\mu\!=\!0.105658389\,$
      GeV and $m_b\!=\!4.3\,$ GeV~\cite{pdb94}.}
  \label{NCnocut}
\end{figure}
%
%
\begin{figure}[p]
  \vspace{5cm}
  \caption[Effect of cuts on the neutral current pair \xsec]
    {\it The effect of cuts of ~$2\!\cdot\!\Gamma_Z$~and
      ~$5\!\cdot\!\Gamma_Z$ around the \zz~mass $M_Z$~on the NC pair
      (`All Graphs') and \zz~pair (`ZZ Graphs') cross-sections.
      ~$\Gamma_Z=2.4971$~GeV. ~The cuts were applied to both fermion
      pair invariant masses. All curves show \xsecs for
      reaction~(\ref{mb}) with universal ISR corrections.}
  \vspace*{-3cm}
  \label{NCZZcuts}
\end{figure}
\begin{table}[thb]
  \begin{center}
  \begin{tabular}{|c||c|c|c|c|c|} \hline
  & & & & & \vspace{-.3cm} \\
    $\sqrt{s}$ [GeV]
  & $\sigma^B_{\mu^+\!\mu^- b {\bar b}}$ [fb]
  & $\sigma^{ISR,uni}_{\mu^+\!\mu^- b {\bar b}}$ [fb]
  & $\sigma^{ISR,compl.}_{\mu^+\!\mu^- b {\bar b}}$ [fb]
  & $\sigma^{ISR,uni}_{\mu^+\!\mu^- b {\bar b}}$ [fb]
  & $\sigma^{ISR,compl.}_{\mu^+\!\mu^- b {\bar b}}$ [fb] \\
  & & & & & \vspace{-.3cm} \\
  & & & & $s'/s \!\geq\! 0.9$ & $s'/s \!\geq\! 0.9$ \\
  & & & & & \vspace{-.3cm} \\ \hline \hline
  130.0 & 131.4968 & 147.4554 & 148.7304 & 111.5411 & 112.2462 \\
  150.0 & ~84.3863 & ~97.9860 & ~99.2177 & ~71.1558 & ~71.7510 \\
  165.0 & ~65.8298 & ~77.5651 & ~78.7483 & ~55.3251 & ~55.8606 \\
  180.0 & ~55.0144 & ~64.8390 & ~65.9749 & ~45.9486 & ~46.4371 \\
  190.0 & ~56.4100 & ~63.8071 & ~64.9309 & ~46.6443 & ~47.1260 \\
  200.0 & ~54.0441 & ~60.9149 & ~62.0168 & ~44.9442 & ~45.4106 \\
  212.0 & ~49.8441 & ~56.4425 & ~57.5120 & ~41.5066 & ~41.9482 \\
  230.0 & ~43.7651 & ~49.9910 & ~51.0124 & ~36.4290 & ~36.8391 \\
  260.0 & ~35.7550 & ~41.3205 & ~42.2653 & ~29.7064 & ~30.0696 \\
  300.0 & ~28.3717 & ~33.1358 & ~33.9928 & ~23.5131 & ~23.8300 \\
  400.0 & ~18.1597 & ~21.5427 &    --    & ~14.9760 &    --    \\
  500.0 & ~12.9578 & ~15.5310 & ~16.1086 & ~10.6475 & ~10.8406 \\
  600.0 & ~~9.8428 & ~11.8958 & ~12.3895 & ~~8.0646 & ~~8.2251 \\
  800.0 & ~~6.3587 & ~~7.7901 &    --    & ~~5.1868 &    --    \\
 1000.0 & ~~4.5111 & ~~5.5883 &    --    & ~~3.6671 &    --    \\
  \hline
  \end{tabular}
  \end{center}
  \caption[Neutral current boson pair production \xsec with initial
    state radiation and $s'$ cuts]
    {\it Numerical values for the \xsec of process~(\ref{mb}) at tree
      level, with universal ISR, and with complete ISR as obtained
      from equations~(\ref{sigmaNC}) to~(\ref{nunixsNC}). The last two
      columns give ISR corrected \xsec values after cutting on $s'$~as
      indicated.}
  \label{taxsNC}
\end{table}
$\!\!$With respect to \zz~pair production the relative corrections of
both universal and non-universal ISR are enhanced. Contrary to
\zz~pair production, ISR corrections are always positive which is
easily explained by the negative slope of the \xsec curve for all
center of mass energies $\sqrt{s}$. Thus a radiative reduction of
$s$~to $s'$~enhances the \xsec in a
way very similar to the radiative tail in \zz~pair production.
It must be stressed that the absolute value of the non-universal
corrections to reaction~(\ref{mb}) steadily decreases with energy,
which is expected as an effect of the screening property. It is
seen from figure~\ref{NCnocut} that, as expected, ISR smears out the
\zz~pair threshold behavior. In the inset of figure~\ref{NCnocut}, the
full NC pair production is compared to the contributions due to
\zz~and photon pair production for the universally ISR corrected
case. Clearly, the main part of the \xsec is due to contributions
which contain both \zz~bosons and photons.
\para
Figure~\ref{NCZZcuts} shows how the \zz~pair production \xsec is
restored from the full set of {\it Conversion} diagrams through
invariant mass cuts on $\SONE$~and $\STWO$. Also seen is the influence
of the applied cuts on the \zz~pair cross-section.
\para
In figure~\ref{NCevolve} it is shown how the \xsec for
reaction~(\ref{mb}) evolves when invariant mass cuts on the
\mupl\mumi~and the $b {\bar b}$~pairs are tightened. Already cuts of
50 GeV on both invariant pair masses yield an $s$-dependence of the
\xsec very much resembling the double-resonant excitation curve for
\zz~pair production. The cuts
{}~$\left|M_{\mu^+\mu^-} - M_Z\right| \leq 5 \Gamma_Z$~and
{}~$M_{b {\bar b}}\geq 60$~GeV represented by the dash-dotted line in
the figure are of interest for Higgs searches, as the current Higgs
mass limit lies around 60 GeV~\cite{sopczak95}. In these searches,
associated Higgs production is used with the $b {\bar b}$ pair
coupling to the Higgs boson.
%
\begin{figure}[t]
  \vspace{12cm}
  \caption[Evolution of the neutral current pair \xsec with cuts]
    {\it Evolution of the \xsec for reaction~(\ref{mb}) for different
      cuts on the \mupl\mumi~and $b {\bar b}$~invariant pair masses.
      All curves show \xsecs with universal ISR corrections.}
  \label{NCevolve}
\end{figure}
%
\begin{figure}[p]
  \vspace{21cm}
  \caption[Behavior of the \zz~pair \xsec with cuts on $s'$]
    {\it Effect of $s'$~cuts on the $\mu^+\!\mu^- b {\bar b} \,
      (\gamma)$~\xsec for final state production via
      process~(\ref{eezz4f}) (upper plot) and process~(\ref{mb})
      (lower plot). The main windows present universally ISR corrected
      cross-sections, the insets show ratios of completely over
      universally corrected cross-sections. No cuts are applied to
      $\SONE$~and $\STWO$.}
  \label{sprcut}
\end{figure}
\bpara
{}From figure~\ref{sprcut} one can see how \xsecs behave if cuts are
applied to the reduced center of mass energy squared $s'$, i. e. to
the invariant mass squared of the four-fermion system. In
experimental analyses such cuts are very commonly used to reduce
backgrounds. Numerical values after cuts on $s'$~are given in
table~\ref{taxsNC}. Both plots of figure~\ref{sprcut} show \xsecs for
the final state $\mu^+\!\mu^- b {\bar b} (\gamma)$. The upper plot is
for the \zz~pair reaction~(\ref{eezz4f}), the lower one for
process~(\ref{mb}). The main windows present the development of
universally ISR corrected \xsecs when the $s'$~cut is tightened. The
insets, where the ratio of the \xsec with complete over the \xsec with
universal ISR is plotted, give the additional relative \xsec
contribution from non-universal ISR correction. From the upper plot
for $\EPL\!\EMI\!\to\!(\ZZ\ZZ)\!\to\!\mu^+\!\mu^- b {\bar b}$~one
can make three observations. First of all, one notices that the \xsec
with ISR is dominated by rather soft radiation. Secondly, hard photon
ISR\footnote[2]{By hard photon radiation the author denotes the
  radiation of photons strongly reducing $s$. This ``strong
  reduction'' may be several per cent or more than 10\%, depending on
  the context.}
is strongly
suppressed below the double-resonance peak. Only above the peak, such
hard photon events represent a significant \xsec contribution. Both
these results are analogous to similar ones at the single \zz~peak at
LEP1~\cite{lep1qed}. This is, because both processes, single and
double \zz~production, are resonant processes with Breit-Wigner
factors
\ba
  \frac{\sqrt{s} \; \Gamma_Z(s)}
       {|s - M_Z^2 + i \sqrt{s} \, \Gamma_Z (s)|^2} \nl \nonumber
\ea
limiting hard radiation in the peak region. In addition, both
calculations use the same \oal~radiators ${\bar S}_1$~and
${\bar H}_1$. The third observation, made from the
inset of the upper plot, is that non-universal ISR tends to be harder
than universal ISR so that the ratio of the \xsec corrected with
complete ISR over the universally corrected \xsec decreases with a cut
on $s'$. In other words, the non-universal ISR corrections in per cent
decrease with a cut on $s'$. This may be understood as follows:
\oal~non-universal corrections are infrared finite, whereas the
infrared divergent universal corrections underwent a resummation,
namely the soft photon exponentiation. Thus the universal corrections
contain important soft resummed parts, and the non-universal
corrections don't. The opening angle between the two curves in the
inset reflects the increasing importance of hard radiation at higher
energies. Around the double \zz~peak and below however, the ratio of
the completely over the universally corrected \xsec is unaffected by
the $s'$ cut. This means that, below $\sqrt{s}\!\approx\!200$ GeV, no
considerable excess of hard photon radiation due to non-universal
corrections exists. As for universal ISR, where deviations of the
\xsec with the cut $s'/s \geq 0.9$ from the \xsec without cut are only
seen above $\sqrt{s}\!\approx\!200$ GeV too, harder radiation is
disfavored by the steep rise of the tree level \xsec around threshold.
\para
Turning to the lower plot of figure~\ref{sprcut} which is for
reaction~(\ref{mb}), patterns similar to the upper plot of the figure
are encountered. There is, however, no disincentive for hard radiation
due to a steep rise of the cross-section. Therefore, the ISR corrected
\xsecs in the main window are lowered throughout the presented energy
region by all cuts on $s'$. The approach between the curves for no cut
and for $s'/s \geq 0.5$ around 150 GeV must be attributed to the
inhibition of a reduction
of $s'$ below the threshold for single-resonant \zz$\gamma\to\mu^+\!\mu^-
b {\bar b}$ production. From the inset of the lower plot of
figure~\ref{sprcut} one may draw the same conclusions as for the inset of
the upper plot, except that there is no disturbance of hard radiation
due to the rise of the \xsec in the double \zz~resonance region.
\para
To conclude the discussion of figure~\ref{sprcut}, it is mentioned
that the integration of the involved non-universal \xsec contributions
is very CPU time consuming. This means that, contrary to what would be
desirable, it is not possible to mass-produce results with
non-universal ISR and all kinds of cuts. Still, a more thorough
analysis including an investigation of the spectrum of radiated
photons is under way~\cite{spect}.
%
%
\begin{figure}[p]
  \vspace{22.5cm}
  \caption[Cross-sections for different four-fermion final states]
    {\it Universally ISR corrected \xsecs from process~(\ref{eeNC4f})
      for different four-fermion final states containing $b {\bar
        b}$~pairs (upper plot) or $\mu^+\!\mu^-$~pairs (lower plot).}
  \label{xsdiverse}
\end{figure}
\para
Figure~\ref{xsdiverse} shows \xsecs for process~(\ref{eeNC4f})
with different final states. The upper plot of the figure shows final
states containing $b {\bar b}$~pairs, the lower plot the smaller
\xsecs for final states with $\mu^+\!\mu^-$~pairs. Cuts were applied
to require $\sqrt{\SONE}\!\geq\!5$~GeV and
$\sqrt{\STWO}\!\geq\!20$~GeV. Comparing
the final states $u {\bar u} b {\bar b}$~and $d {\bar d} b {\bar b}$,
it is interesting to observe how the different u and d quark electric
charges affect the \xsec below threshold: There, the
$d {\bar d} b {\bar b}$ \xsec is much smaller than the
$u {\bar u} b {\bar b}$~\xsec. Similarly interesting is the
comparison between $\mu^+\!\mu^- b {\bar b}$~and $\nu_\mu {\bar
\nu_\mu} b {\bar b}$. The $\nu_\mu {\bar \nu}_\mu b{\bar b}$~\xsec has
a shape very much like the \zz~pair excitation curve due to the cuts
and the weak coupling of b quarks to photons. The $\mu^+\!\mu^- b
{\bar b}$~\xsec on the other hand receives significant contributions
below the \zz~pair threshold due to the muon's electric charge.
Also rather interesting,
but straightforwardly explained by the cuts and the photonic and weak
couplings, are comparisons between the final states $\mu^+\!\mu^- b
{\bar b}$~versus $\mu^+\!\mu^- c {\bar c}$~and $\mu^+\!\mu^- \nu_\mu
{\bar \nu}_\mu$~versus $\mu^+\!\mu^- \tau^+\!\tau^-$. The fermion
masses for all numerical computations were taken from
reference~\cite{pdb94} and are listed in table~\ref{fermass}.
\para
To conclude the chapter, figure~\ref{mass} presents the influence of
final state fermion masses as taken into account in the
semi-analytical approach. It is recalled that final state fermion
masses are only taken into account through
%
\begin{table}[t]
  \begin{center}
  \begin{tabular}{|c||c|c|c|c|c|c|c|} \hline
  & & & & & & & \vspace{-.3cm} \\
  Fermion & $\nu_\mu$ & $\mu$ & $\tau$ & $u$ & $d$ & $c$ & $b$ \\
  & & & & & & & \vspace{-.3cm} \\ \hline
  & & & & & & & \vspace{-.3cm} \\
  Mass [GeV] & 0 & 0.105658389 & 1.7771 & 0.005 & 0.01 & 1.3 & 4.3 \\
  \hline
  \end{tabular}
  \end{center}
  \vspace{-.3cm}
  \caption[Fermion masses used for numerical computations]
    {\it The fermion masses for the numerical computations presented
      in the figures.}
  \label{fermass}
\end{table}
the factors $N_p(B_i,B_k,m,s)$. Figure~\ref{mass} shows, in the main
window, the two \xsecs of reaction~(\ref{eeNC4f}) into the final
states $\mu^+\!\mu^- d {\bar d}$~and $\mu^+\!\mu^- b {\bar b}$ which
do not differ in couplings, but only in final state quark masses (see
table~\ref{fermass}). The $\mu^+\!\mu^- d {\bar d}$~\xsec is larger
because of its larger phase space due to $m_d<m_b$. If now the
invariant $d {\bar d}$~mass is cut at exactly two times the b quark
mass, the phase spaces for the two final states have equal volume.
However, a difference remains due to the factors $N_p$ for the
quarks. While for the final state $\mu^+\!\mu^- d {\bar d}$
\ba
  N_p(\gamma,\gamma,m_d,\STWO) & = & \sqrt{1 - 4m_d^2/\STWO} \!\cdot\!
  (1 + 2 m_d^2/\STWO) \;\; \approx \;\; 1 \nonumber
\ea
throughout the phase space cut at $\sqrt{\STWO}=2\,m_b$, because of
the, compared to the cut, small $d$~quark mass, this is different for
the final state $\mu^+\!\mu^- b {\bar b}$.
$N_p(\gamma,\gamma,m_b,\STWO)$~is zero at the phase space boundary
$\sqrt{\STWO}=2\,m_b$ and approaches its limiting value 1 to a
precision of 1$\,$\permille~only around $\sqrt{\STWO}=38$~GeV. Thus
the \xsec for the final state $\mu^+\!\mu^- b {\bar b}$~is slightly
smaller than the one for $\mu^+\!\mu^- d {\bar d}$ cut at
$\sqrt{\STWO}=2\,m_b$. This is seen from the inset of
figure~\ref{mass} where the ratio of $\sigma_{\mu^+\!\mu^- d {\bar d}}$
with the cut $\sqrt{\STWO}\!\geq\!2\,m_b$~~over
$\sigma_{\mu^+\!\mu^- b {\bar b}}$~is given.
%
\begin{figure}[t]
  \vspace{16cm}
  \caption[Mass effect in the semi-analytical computations]
    {\it The effects of masses in the \xsecs for $\mu^+\!\mu^- d {\bar
        d}$~and $\mu^+\!\mu^- b {\bar b}$~production according to
      reaction~(\ref{eeNC4f}). The main window shows the \xsec
      enhancement due to the larger phase space for the production of
      lighter fermion pairs. The inset shows the small effect of
      the factors $N_p(\gamma,\gamma,m,s)$, if phase spaces are made
      equally large by applying a cut of $\sqrt{\STWO}=2\,m_b$~to the
      invariant $d {\bar d}$ pair mass. A small bump is seen around
      the \zz~pair threshold. All curves are with universal ISR.}
  \label{mass}
\end{figure}
%
\chapter{Summary, Conclusions, and Outlook}
\label{summary}
\begin{figure}[t]
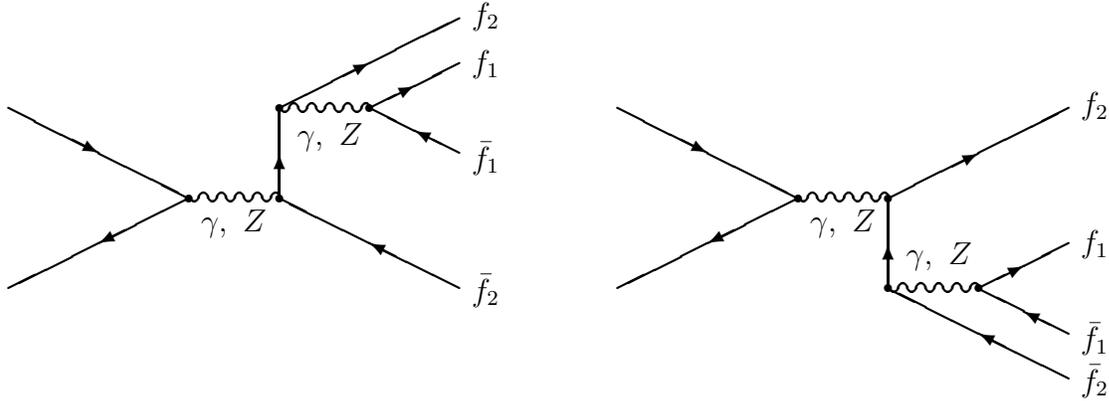

\vspace{1.2cm}
\begin{minipage}[tbh]{7.8cm}{
\begin{center}
\begin{Feynman}{75,60}{-8,0}{0.8}
%
\put(-10,45){\fermiondrr}
\put(-10,15){\fermiondll}
\put(20,30){\photonrighthalf}
\put(65,15){\fermionull}
\put(20,30){\circle*{1.5}}
\put(35,30){\circle*{1.5}}
\put(35,45){\circle*{1.5}}
\put(50,45){\circle*{1.5}}
\put(35,30){\fermionuphalf}
\put(65,60){\fermionurr}
\put(35,45){\photonrighthalf}
\put(65,52.5){\fermionurrhalf}
\put(65,37.5){\fermionullhalf}
\put(22,24.5){$\gamma,\ Z$}
\put(38,39){$\gamma,\ Z$}
\put(67,59){$ f_2$}
\put(67,51){$ f_1$}
\put(67,35){${\bar f_1}$}
\put(67,13){${\bar f_2}$}
%
\end{Feynman}
\end{center}
}\end{minipage}
\begin{minipage}[tbh]{7.8cm}{
\begin{center}
\begin{Feynman}{75,60}{0,0}{0.8}
%
\put(0,45){\fermiondrr}
\put(0,15){\fermiondll}
\put(30,30){\photonrighthalf}
\put(30,30){\circle*{1.5}}
\put(45,30){\circle*{1.5}}
\put(45,15){\circle*{1.5}}
\put(60,15){\circle*{1.5}}
\put(45,15){\fermionuphalf}
\put(75,00){\fermionull}
\put(75,45){\fermionurr}
\put(45,15){\photonrighthalf}
\put(75,22.5){\fermionurrhalf}
\put(75,07.5){\fermionullhalf}
\put(32,24.5){$\gamma,\ Z$}
\put(48,19){$\gamma,\ Z$}
\put(77,44){$ f_2$}
\put(77,21){$ f_1$}
\put(77, 5){${\bar f_1}$}
\put(77,-2){${\bar f_2}$}
%
\end{Feynman}
\end{center}
}\end{minipage}
\vspace*{.2cm}
\caption[Examples of electroweak background Feynman diagrams]
{\it Two examples of annihilation type Feynman diagrams for singly
  resonant electroweak background to reaction~(\ref{eeNC4f}).}
\label{bgdiag}
\end{figure}

\begin{figure}[t]
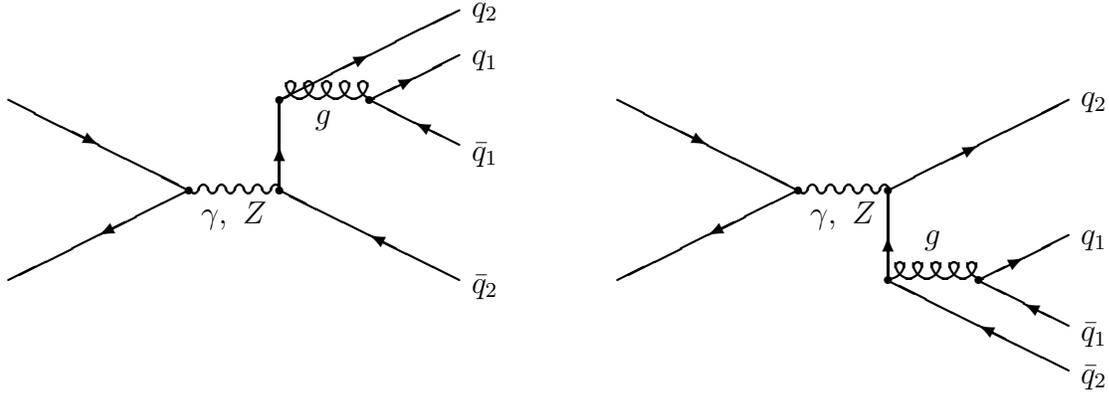

\vspace{1.2cm}
\begin{minipage}[tbh]{7.8cm}{
\begin{center}
\begin{Feynman}{75,60}{-8,0}{0.8}
%
\put(-10,45){\fermiondrr}
\put(-10,15){\fermiondll}
\put(20,30){\photonrighthalf}
\put(65,15){\fermionull}
\put(20,30){\circle*{1.5}}
\put(35,30){\circle*{1.5}}
\put(35,45){\circle*{1.5}}
\put(50,45){\circle*{1.5}}
\put(35,30){\fermionuphalf}
\put(65,60){\fermionurr}
\put(35,45){\gluonrighthalf}
\put(65,52.5){\fermionurrhalf}
\put(65,37.5){\fermionullhalf}
\put(22,24.5){$\gamma,\ Z$}
\put(41,41){$g$}
\put(67,59){$ q_2$}
\put(67,51){$ q_1$}
\put(67,35){${\bar q_1}$}
\put(67,13){${\bar q_2}$}
\end{Feynman}
\end{center}
}\end{minipage}
\begin{minipage}[tbh]{7.8cm}{
\begin{center}
\begin{Feynman}{75,60}{0,0}{0.8}
%
\put(0,45){\fermiondrr}
\put(0,15){\fermiondll}
\put(30,30){\photonrighthalf}
\put(30,30){\circle*{1.5}}
\put(45,30){\circle*{1.5}}
\put(45,15){\circle*{1.5}}
\put(60,15){\circle*{1.5}}
\put(45,15){\fermionuphalf}
\put(75,00){\fermionull}
\put(75,45){\fermionurr}
\put(45,15){\gluonrighthalf}
\put(75,22.5){\fermionurrhalf}
\put(75,07.5){\fermionullhalf}
%
\put(32,24.5){$\gamma,\ Z$}
\put(51,21){$g$}
\put(77,44){$ q_2$}
\put(77,21){$ q_1$}
\put(77, 5){${\bar q_1}$}
\put(77,-2){${\bar q_2}$}
%
\end{Feynman}
\end{center}
}\end{minipage}
\vspace*{.1cm}
\caption[Examples of QCD background Feynman diagrams]
{\it Two examples of annihilation type Feynman diagrams for singly
  resonant QCD background to reaction~(\ref{eeNC4f}). The curled lines
  represent gluons.}
\label{QCDdiag}
\end{figure}

%
Even around the \zz~pole, four-fermion production is an interesting
topic, relevant not only for Higgs boson searches~\cite{sopczak95} but
also for higher order corrections to the \zz~line
shape~\cite{kniehl88,berends88,glover90}.
Above the production thresholds for resonant gauge boson pair
production, \ee~annihilation into four fermions is one of the most
significant Standard Model processes. Thus, double-resonance
production plays a very important r\^{o}le at LEP2 and other future
\ee~colliders, and the study of four-fermion final states will provide
further understanding of gauge boson properties and may be the key to
the mechanism of electroweak symmetry breaking via investigations of
associated Higgs boson production.
\para
In the preceding chapters, the first complete calculation of initial
state QED corrections (ISR) for neutral current double-resonance
production in \ee~annihilation was presented. A semi-analytical
method was used to compute \xsecs and invariant mass distributions for
\zz~and neutral current (NC) pair production processes~(\ref{eezz4f})
and~(\ref{eeNC4f})\footnote[2]{A similar calculation for \wpl\wmi~pair
  production had to face ambiguities of the definition of initial
  state radiation due to the charge transfer from the initial to the
  final state in the t-channel \wpl\wmi~production
  amplitude~\cite{dimaww}.}.
These processes are interesting by themselves and also as background
to \wpl\wmi~physics and Higgs searches. All angular degrees of freedom
were integrated analytically. After the angular integrations,
only two fermion pair invariant masses were left for numerical
integration with a straightforward self-adaptive Simpson algorithm.
In the case of ISR, also the four-fermion invariant mass had to be
integrated numerically. It was shown that boson off-shellness and ISR
both yield considerable corrections to the cross-section, each of
${\cal O}(10\%)$~and up to ${\cal O}(20\%)$~for neutral current pair
production. ISR corrections separate into universal parts with the
Born \xsec factorizing and non-universal parts. Contrary to
non-universal contributions, universal ISR parts
contain the large logarithm $\ln(s/\MES)$~as a result of collinear
photon radiation from the initial state electrons. Thus universal ISR
is enhanced with respect to non-universal ISR. Invariant mass cuts can
be easily implemented to isolate the \zz~pair signal from the full NC
process. Such cuts do not only suppress NC pair production, but also
irreducible backgrounds of the annihilation type. Two examples of
annihilation type background Feynman diagrams  are given in
figure~\ref{bgdiag} (compare reference~\cite{alNC}).
%
%
%
\para
The systematic errors of the presented calculation are
physics dominated, because of the numerical integration's excellent
precision. Errors arise via
\begin{itemize}
  \item The use of the ultrarelativistic approximation (URA) with
    neglect of terms of ${\cal O}(m_e^2/s)$~and
    ${\cal O}(m_e^2/s_{ij})$. Errors due to the URA are very
    small. This is evidently true for \zz~pair production. As all used
    fermion masses are at least one order of magnitude larger than the
    electron mass, and as lower invariant mass cuts of several GeV are
    quite commonly applied, the URA is a valid approximation also for
    the NC pair production process~(\ref{eeNC4f}).
  \item Neglect of final state fermion masses in the matrix elements.
    For light fermions, the effect on the total \xsec is small.
    In the NC pair case with exchange photons, all important mass
    effects are properly taken into account by the factors $N_p$~given
    in equation~(\ref{psfact}). Mass effects are removed by
    even moderate invariant mass cuts. For some cases, such cuts are
    also desirable with respect to non-perturbative threshold effects.
  \item Neglect of final state QED corrections. Compared to ISR, final
    state QED (FSR) is known to be suppressed, but may become
    important at very high energies beyond the TeV level. Interference
    effects between ISR and FSR were found to be suppressed by
    ${\cal O}(\Gamma_B/M_B)$~in resonant pair production of bosons
    B~\cite{IFI}.
  \item Neglect of true electroweak corrections. For
    on-shell \zz~pair production true electroweak effects were found
    to yield corrections around 1\% to 2\%~\cite{Denner88}. This
    estimate may have to be increased to accommodate off-shell boson
    pair production, but no full calculation of electroweak one loop
    corrections for process~(\ref{eezz4f}) is available at present.
    In reference~\cite{Denner88}, a strong dependence of electroweak
    corrections on the \zz~scattering angle was found for center of
    mass energies above 500 GeV.
  \item Theoretical and experimental uncertainties of the parameters
    entered into the numerical calculation.
\end{itemize}
\clearpage
\noindent
In the following, it will be shown how the investigations presented in
this thesis are embedded in the environment of four-fermion final
state physics and in the framework of semi-analytical calculations
dealing with four-fermion production in charged current and neutral
current processes~\cite{dl94,aldist95,dimaww,teupitz94,alteup94,alNC,spect}.
The semi-analytical method yields compact formulae for total and
differential \xsecs at tree level and with leading initial state QED
corrections. At present, semi-analytical \xsecs are available which
are differential in the two boson invariant mass and the boson
scattering angle~\cite{aldist95}. These are easily numerically
integrated to yield singly differential or total cross-sections. In
this thesis, the semi-analytical method was used to present
non-leading initial state QED corrections~\cite{dl94,teupitz94}. The
screening property was found in these non-leading corrections. A
complete initial state QED calculation for a double-resonant
four-fermion production is documented in detail for the first time.
\para
In the presented calculation, some non-resonant annihilation type
background contributions, see figure~\ref{bgdiag}, to
process~(\ref{ee4ffull}),
\ba
  \EE \rightarrow f_1\bar{f_1}f_2\bar{f_2}~,
  ~~~~~~~~~~~~ f_1\!\neq\!f_2~,~~~f_i\!\neq\!e^\pm,
  \stackrel{_{(-)}}{\nu_e}~~,
  \nonumber
\ea
are not taken into account. But already for moderate cuts, the
results given in this thesis represent the dominant contribution
to process~(\ref{ee4ffull}). Process~(\ref{ee4ffull}) was completely
and semi-analytically calculated in
references~\cite{aldist95,alteup94,alNC}, but only at tree
level. However, as background contributions
are small at LEP2 energies and above and as the leading universal ISR
corrections may be straightforwardly implemented by multiplication of
the tree level \xsec with the radiators ${\bar S_1}$~and ${\bar H_1}$
(see equation~(\ref{radiators})), sufficiently precise results are
available for process~(\ref{ee4ffull}). This is still true if
associated Higgs production is considered, because interferences of
the Higgs Feynman diagram are negligible and because ${\bar S_1}$~and
${\bar H_1}$ describe the Higgs diagram's complete ISR
corrections\footnote[2]{Generally, for Feynman diagrams of the {\it
    Annihilation} type with only one vertex in the initial state
  electron current (like the diagrams in figure~\ref{bgdiag}) the
  radiators ${\bar S_1}$~and ${\bar H_1}$~describe the complete
  \oal~ISR. As soon as {\it Conversion} type Feynman diagrams (like
  the diagrams in figure~\ref{NCdiag}) are involved, non-universal ISR
  contributions are present.}.
Thus, for the leptonic\footnote[3]{It should be noted that final
  states with $\mu$- and $\tau$-neutrinos can be calculated
  semi-analytically, but are not experimentally observable
  individually, because the signature from final states with electron
  neutrinos is identical.}
and the important semi-leptonic channels of process~(\ref{ee4ffull}),
the description obtained from this thesis and
references~\cite{aldist95,alteup94,alNC} is sufficient in view of
future \ee~collider data. The treatment of four-quark final states
requires the consideration of eight additional Feynman diagrams of the
annihilation type with decaying gluons radiated from the final state
quarks. Two examples of such diagrams are given in figure~\ref{QCDdiag}.
These diagrams were taken into account in semi-analytical~\cite{alNC}
and in Monte Carlo calculations~\cite{pittau94,boos95}.
%
%
%
The annihilation type QCD background diagrams are described by the
same semi-analytical kinematical functions as the electroweak
annihilation type background diagrams. The differences lie in coupling
constants and color factors. Going a little further one realizes that
hadron jets stemming from light quarks or gluons are experimentally
not really distinguishable so that one must incoherently add several
contributions to obtain a prediction for an experimental signature.
This means that new contributions with the virtual gluon from
figure~\ref{QCDdiag} decaying into two gluons instead of two quarks
must be taken into account. Recently, the reaction ~$\EE \to b{\bar b} +
2\:J\!ets$~~was evaluated~\cite{boos95}. Another problem of the
four-quark final states is related to so-called color reconnection
effects which are mainly of interest for \wpl\wmi~pair production and
the $W$~mass measurement~\cite{reconn}.
\para
It is, at this point, interesting to compare the results for \zz~pair
production with those of
\wpl\wmi~pair production. Both calculations are fundamentally
different in two regards. Firstly, the contributing amplitudes differ:
t- and u-channel amplitudes for \zz~pairs, but t- and s-channel
amplitudes for \wpl\wmi~pairs (see e.g. reference~\cite{teupitz94}).
Secondly, trilinear boson couplings contribute to Standard Model
\wpl\wmi~pair production, but not to Standard Model \zz~pair
production. On the other hand there are, of course, patterns that are
common to \wpl\wmi~and \zz~pair production, such as the
double-resonance behavior and the screening property.
For four-fermion final states containing
fermion-antifermion pairs from the same weak isospin doublet,
double-resonant contributions arise from \wpl\wmi~as well as \zz~pair
production, not to mention the many background contributions to
both. The individual charged current and neutral current contributions
to such final states have already been semi-analytically calculated.
The interferences between CC and NC contributions, though small,
require special treatment. This is due to different particle
pairings from the decays of charged and neutral bosons which
invalidates the unique assignment of the invariant masses $\SONE$~and
$\STWO$~to fermion pairs. This problem is illustrated in
equation~(\ref{WWZZint}) for the double-resonant signals yielding a
$U {\bar U} D {\bar D}$
final state, where U and D denote generic particles with third weak
isospin component $+\frac{1}{2}$ and $-\frac{1}{2}$ respectively:
\vspace{.9cm}
\ba
  {\rm CC:} & & \EE \to \WPL\WMI
  \raisebox{.5cm}[0cm][0cm]
    {$\hspace{-.55cm} \lceil \hspace{-.2cm}$
     \raisebox{.204cm}[0cm][0cm]{$\rightarrow {\bar U} D  ~~~~~
       ({\rm inv.~mass}~~ \SONE)$}}
  \raisebox{-.4cm}[0cm][0cm]
    {$\hspace{-5.6cm} \lfloor \hspace{-.2cm}$
     \raisebox{-.204cm}[0cm][0cm]{$\rightarrow U {\bar D}  ~~~~~
       ({\rm inv.~mass}~~ \STWO)$}}
  \nonumber
\ea
\vspace{.5cm}
\ba
  {\rm NC:} & & \EE \to \ZZ\ZZ
  \raisebox{.5cm}[0cm][0cm]
    {$\hspace{-.35cm} \lceil \hspace{-.2cm}$
     \raisebox{.204cm}[0cm][0cm]{$\rightarrow {\bar U} U  ~~~~~
       ({\rm inv.~mass}~~ \SONE)$}}
  \raisebox{-.4cm}[0cm][0cm]
    {$\hspace{-5.3cm} \lfloor \hspace{-.2cm}$
     \raisebox{-.204cm}[0cm][0cm]{$\rightarrow D {\bar D}  ~~~~~
       ({\rm inv.~mass}~~ \STWO)$~~~~~~.}}
  \label{WWZZint}
 \\ \nonumber
\ea
The complete process $\EE \to U {\bar U} D {\bar D}$ including the
small CC-NC interferences is, at present, only calculated by Monte
Carlo programs (see e.g.~\cite{excal}). Similarly four-fermion final
states containing electrons or electron neutrinos in the final state
have not yet been calculated semi-analytically, but constitute no
problem to Monte Carlo programs~\cite{minami,boosMC}. The additional
complexity of final states with electrons or electron neutrinos is due
to their additional amplitudes with t-channel boson propagators. In
table~\ref{NCclass}, the number of Feynman diagrams contributing to
four-fermion production via neutral boson exchange, i.e. the final
states this thesis is concerned with, are listed.
%
%
\begin{table}[t]
  \vspace{.7cm}
  \begin{center}
    \begin{tabular}{|c|c|c|c|c|c|c|} \hline
      & \raisebox{0.pt}[2.5ex][0.0ex]{${\bar d} d$}
      & ${\bar u} u$ & ${\bar e} e$ & ${\bar \mu} \mu$
      & ${\bar \nu}_{e} \nu_{e}$ & ${\bar \nu}_{\mu} \nu_{\mu}$
      \\ \hline
      \raisebox{0.pt}[2.5ex][0.0ex]{${\bar d} d$}
      & {\tt 4$\cdot $16} & {\it 43} & {48} & {\bf 24} & 21 & {\bf 10}
        \\ \hline
      ${\bar s} s$ & {\bf 32} & {\it 43} & {48}
        & {\bf 24} & {21} & {\bf 10} \\ \hline
      ${\bar u} u$ & {\it 43} & {\tt 4$\cdot$16} & {48}
        & {\bf 24} & {21} & {\bf 10} \\ \hline
      ${\bar e} e$ &{48} &{48} & {\tt 4$\cdot$36} &{48}
        & {\it 56} & {20} \\ \hline
      ${\bar \mu} \mu$  & {\bf 24} & {\bf 24} & {48}
        & {\tt 4$\cdot$12} & {19} & {\it 19} \\ \hline
      ${\bar \tau} \tau$& {\bf 24} & {\bf 24} & {48}
        & {\bf 24} & {19} & {\bf 10} \\ \hline
      ${\bar \nu}_e \nu_{e}$  & {21} & {21} & {\it 56} & {19}
        & {\tt 4$\cdot$9} & {12} \\ \hline
      ${\bar \nu}_{\mu} \nu_{\mu}$ & {\bf 10} & {\bf 10} & {20}
        & {\it 19} & {12} & {\tt 4$\cdot$3} \\ \hline
      ${\bar \nu}_{\tau} \nu_{\tau}$ & {\bf 10} & {\bf 10} & {20}
        & {\bf 10} & {12} & {\bf 6} \\ \hline
    \end{tabular}
    \vspace{-.2cm}
  \end{center}
  \caption[Number of neutral current four-fermion production Feynman
    diagrams]
    {\it Number of tree level Feynman diagrams contributing to the
      production of four-fermion final states via neutral bosons.}
  \vspace{-.4cm}
  \label{NCclass}
\end{table}
%
%
The simplest case (numbers in {\bf boldface}) does not contain
identical particles, electrons, electron neutrinos or particles from
the same isospin doublet. Those final states were calculated
semi-analytically and the calculations presented in this thesis
describe the corresponding dominant contributions for energies above
the \zz~pole. For pairs of identical fermions, antisymmetrization
of the final state as needed to satisfy the Pauli principle increases
the number of Feynman diagrams (given in {\tt typewriter}). Numbers in
{\rm roman} are for final states with electrons or electron neutrinos,
and numbers in {\it italics} correspond to final states with
particles from the same isospin doublet.
Numerous Monte Carlo
approaches~\cite{excal,NCMC,eebb,pittau94,glover90,minami,boosMC,WWMC}
treat four-fermion production in \ee~annihilation. The scopes of the
two independent approaches represented by the semi-analytical and the
Monte Carlo method are different. The semi-analytical method provides
formulae to strongly promote the understanding of four-fermion
production processes. Monte Carlo programs on the other hand aim at
complete sets of Feynman diagrams and are needed by experimentalists
to produce input for their detector simulations and to apply all
experimental cuts. Semi-analytical programs can only be run with
quasi-experimental cuts. Unfortunately, Monte Carlo programs are
often ``black boxes'' to the experimentalist. It becomes clear that
semi-analytical and Monte Carlo calculations are nicely complementary.
Keeping in mind that four-fermion final states
are much richer in structure, much more interesting and much more
complicated than fermion pair final states, the ultimate goal of
semi-analytical calculations is rather obvious, namely to describe
four-fermionic complexity by as much simplicity as possible. The
semi-analytical method has been very successful in this respect,
because a description of many Feynman diagrams and their interferences
by few kinematical functions (such as ${\cal G}_4^{t+u}$, see
equation~(\ref{sigzz41})) was achieved for several interesting four-fermion
final states. Table~\ref{funnum} gives some examples.
%
%
\begin{table}[t]
  \vspace{.25cm}
  \begin{center} \hspace*{-.06cm}
    \begin{tabular}{|c||c|c|c|} \hline
      & \raisebox{0.pt}[.5cm][.2cm]{} $\EE \to \mu^+\mu^- b {\bar b}$
      & $\EE \to u {\bar d} \mu^- {\bar \nu_\mu}$
      & $\EE \to (\ZZ\ZZ,\ZZ\gamma,\gamma\gamma) \to
         f_1\bar{f_1}f_2\bar{f_2}(\gamma)$
      \\ \hline\hline
      \raisebox{0.pt}[.5cm][.2cm]{}
      N$_{diag}$ & 25 & 10 & 8 (+56) \\ \hline
      \raisebox{0.pt}[.5cm][.2cm]{}
      N$_{inter}$ & 325 & 55 & 36 (+532) \\ \hline
      \raisebox{0.pt}[.5cm][.2cm]{}
      N$_{\cal G}$ & 4 & 8 & 1 (+2)~  \\ \hline
      \raisebox{0.pt}[.5cm][.2cm]{}
      Refs. & \cite{alteup94,alNC} & \cite{dimaww,wwback}
      & This thesis \\ \hline
    \end{tabular}
  \end{center}
  \vspace{-.3cm}
  \caption[Number of Feynman diagrams, interferences and kinematical
    functions for example four-fermion final states]
    {\it Examples for the number {\rm N}$_{diag}$ of Feynman diagrams
      contributing to the production of four-fermion final
      states in \ee~annihilation. Also given are the numbers of
      interferences between Feynman diagrams {\rm (N$_{inter}$)} and
      the numbers of kinematical functions {\rm N$_{\cal G}$} needed
      to semi-analytically describe the process. The numbers in
      parentheses in the last column give the additional diagrams,
      interferences, and kinematical functions due to non-universal
      ISR as described in this thesis.}
  \vspace{-.4cm}
  \label{funnum}
\end{table}
\hspace*{1cm}
%
\para
It is of fun\-da\-men\-tal theo\-re\-ti\-cal in\-te\-rest
to con\-ti\-nue the syste\-ma\-tic se\-mi-ana\-ly\-ti\-cal treatment
of \ee~annihilation into four fermions. For the future it is planned
to extend the work of this thesis to include a non-standard, heavy new
neutral boson $Z'$ and to evaluate the corresponding discovery limits
at future \ee~colliders. Further a full calculation for
process~(\ref{ee4ffull}) with all ISR corrections and semi-analytical
calculations of four-fermion final states with electrons and electron
neutrinos are very desirable. It could also be interesting to extend
the presented results to accommodate anomalous couplings between three
neutral gauge bosons. Such couplings are not present in the Standard
Model. Their inclusion would require to take into account an
additional Feynman diagram with s-channel \ee~annihilation into
{\it one} neutral gauge boson which then decays into two neutral
bosons that finally decay into the four final state
fermions\footnote[2]{Depending on cuts, one may also have to account for
annihilation type background diagrams.}. It is worth mentioning that
all initial state QED corrections for such an inclusion are known from
references~\cite{dl94,dimaww} and from this thesis. Anomalous
trilinear neutral gauge bosons couplings are discussed in
reference~\cite{hagiwara87}, but no deviations from the Standard Model
were experimentally found so far~\cite{ntgv}. Finally, it should be
realized that the results presented in this thesis can be validated
for proton-proton collisions at the Large Hadron Collider (LHC) by
folding in parton distributions, changing coupling constants and
introducing color factors.
%
%
\appendix
%
\chapter{Metric and Conventions}
\label{metric}
%
Every calculation in four-dimensional space-time requires the choice
of a metric which is uniquely defined by a metric tensor $\mten$. For
this thesis,
\beq
  \hspace{2.9cm}
  g_{\mu\nu} \;\; = \;\; \left( \rule[-1cm]{0cm}{2cm}
    \begin{array}{cccc}
      \!\! -1 & 0 & 0 & 0 \!\! \\
      \!\!  0 & 1 & 0 & 0 \!\! \\
      \!\!  0 & 0 & 1 & 0 \!\! \\
      \!\!  0 & 0 & 0 & 1 \!\!
    \end{array}
  \right)~,~~~~~~ \mu,\nu = 0,1,2,3
\eeq
is chosen as metric tensor. This is just the negative of the
Bjorken-Drell metric tensor~\cite{bd}. A contravariant 4-vector is
defined by
\beq
  \hspace{2.9cm}
  v^\mu \; \equiv \; \left( v^0,{\vec v} \right) \; = \;
                     \left( v^0, v^1, v^2, v^3 \right)
\eeq
with time-like component $v^0$ and space-like components ${\vec v}$.
The corresponding covariant vector is given by
\beq
  \hspace{2.9cm}
  v_\mu \; = \; g_{\mu\nu} \, v^\nu = \left( -v^0,{\vec v} \right)~,
\eeq
where the convention to sum over repeated indices is adopted. The
summation convention is used throughout this thesis unless stated
otherwise. The scalar product of two 4-vectors is defined as the
contraction with the metric tensor:
\beq
  \hspace{2.9cm}
  v\!\cdot\!w \; \equiv \; v_{\!\mu} w^\mu \; = \; v^\mu w_{\!\mu}
    \; = \;v^\mu \, g_{\mu\nu} \, w^\nu ~.
\eeq
It should be noticed that, with the above convention, time-like
four-vectors have negative and space-like four-vectors have positive
metric. The choice of metric also entails that the invariant mass
squared $M^2_{\cal K}$ of a system ${\cal K}$ with 4-vector $k$ is given
by
\beq
  \hspace{2.9cm}
  M^2_{\cal K} \; = \; (k^0)^2 - {\vec k}^2 \; = \; - k\!\cdot\!k~~.
\eeq
The Dirac algebra is defined by
\ba
  \hspace{2.9cm}
  \left[ \rule[0cm]{0cm}{.35cm}
    \gamma_\mu, \gamma_\nu \right]_+ & = & 2 \, \mten \nl
  \gamma_5 & = & -\,\ri\,\gamma_0 \gamma_1 \gamma_2 \gamma_3~~.
\ea
Finally, it should be mentioned that natural units $c=\hbar=1$ are
used throughout this thesis.
%
\chapter{The Standard Model of Electroweak Interactions}
\label{SM}
%
\section{The Standard Model Lagrangian}
\label{SMfund}
Phenomenologically, the \sm~is based on the empirical fermion content
of nature given in table~\ref{fermions}, on the existence of the four
vector bosons $\gamma, \WPL, \WMI, {\rm and}~\ZZ$ that mediate the
electroweak force, and on the Gell-Mann-Nishijima
relation~\cite{gellmann53}
\beq
  \hspace{3.5cm}
  Q_{\!f} \; = \;  \ct^{(3)}_W + \frac{Y_W}{2}~~.
  \label{gell_mann_nish}
\eeq
These empirical facts can be embedded into a theoretical description
which is characterized by a Lagrangian that is invariant under
$SU(2)_L \times U(1)_{Y_W}$ gauge transformations,
\beq
  \hspace{3.5cm}
  \cl_{SM} \;\; = \;\; \cl_G + \cl_H + \cl_F + \cl_{Yukawa}~~.
  \label{lagrangian}
\eeq
%
%
\begin{table}[t]
  \vspace{.3cm}
  \begin{center}
    \begin{tabular}{|c||c|c|c|c|c|c|} \hline
      \multicolumn{1}{|c} {}
      &\multicolumn{3}{||c|}{\scalfrac Families}
      &\multicolumn{1}{|c|}{$Q_{\!f}$}
      &\multicolumn{1}{|c|}{$\ct^{(3)}_W$}
      &\multicolumn{1}{|c|}{$Y_W$} \\
       \multicolumn{1}{|c|} {}
      &\multicolumn{1}{||c} {1}
      &\multicolumn{1}{c} {2}
      &\multicolumn{1}{c|} {3}
      &\multicolumn{1}{|c|} {}
      &\multicolumn{1}{|c|} {}
      &\multicolumn{1}{|c|} {} \\ \hline \hline
       \multicolumn{1}{|c|} {\vrule width 0pt height 0.5em depth 0mm}
      &\multicolumn{3}{||r|} {}
      &\multicolumn{1}{|r|} {}
      &\multicolumn{1}{|r|} {}
      &\multicolumn{1}{|r|} {} \\
      \multicolumn{1}{|c|}
          {$\begin{array}{l} {\rm Leptons} \\ \scalfrac \end{array}$}
      &\multicolumn{1}{||c}
          {${\left(\begin{array}{c} \scalfrac \nu_e    \\ e
            \end{array}\right)}_L$}
      &\multicolumn{1}{c}
          {${\left(\begin{array}{c} \scalfrac \nu_\mu  \\ \mu
            \end{array}\right)}_L$}
      &\multicolumn{1}{c|}
          {${\left(\begin{array}{c} \scalfrac \nu_\tau \\ \tau
            \end{array}\right)}_L$}
      &\multicolumn{1}{|r|}
          {$\begin{array}{r} \scalfrac  0 \\ -1 \end{array}$}
      &\multicolumn{1}{|r|}
          {$\begin{array}{r} \scalfrac  +\frac{1}{2} \\ -\frac{1}{2}
          \end{array}$}
      &\multicolumn{1}{|r|}
          {$\begin{array}{r} \scalfrac  -1 \\ -1 \end{array}$} \\
      \multicolumn{1}{|c|} {}
      &\multicolumn{1}{||c} {\scalfrac $e_R$}
      &\multicolumn{1}{c} {$\mu_R$}
      &\multicolumn{1}{c|} {$\tau_R$}
      &\multicolumn{1}{|r|}
          {$\begin{array}{r} -1 \end{array}$}
      &\multicolumn{1}{|r|}
          {$\begin{array}{r}  0 \end{array}$}
      &\multicolumn{1}{|r|}
          {$\begin{array}{r} -2 \end{array}$} \\ \hline
       \multicolumn{1}{|c|} {\vrule width 0pt height 0.5em depth 0mm}
      &\multicolumn{3}{||r|} {}
      &\multicolumn{1}{|r|} {}
      &\multicolumn{1}{|r|} {}
      &\multicolumn{1}{|r|} {} \\
      \multicolumn{1}{|c|}
          {$\begin{array}{l} {\rm Quarks} \\ \scalfrac \end{array}$}
      &\multicolumn{1}{||c}
          {${\left(\begin{array}{c} \scalfrac u \\ d'
            \end{array}\right)}_L$}
      &\multicolumn{1}{c}
          {${\left(\begin{array}{c} \scalfrac c \\ s'
            \end{array}\right)}_L$}
      &\multicolumn{1}{c|}
          {${\left(\begin{array}{c} \scalfrac t \\ b'
            \end{array}\right)}_L$}
      &\multicolumn{1}{|r|}
          {$\begin{array}{r} \scalfrac  +\frac{2}{3} \\ -\frac{1}{3}
          \end{array}$}
      &\multicolumn{1}{|r|}
          {$\begin{array}{r} \scalfrac  +\frac{1}{2} \\ -\frac{1}{2}
          \end{array}$}
      &\multicolumn{1}{|r|}
          {$\begin{array}{r} \scalfrac  +\frac{1}{3} \\ +\frac{1}{3}
          \end{array}$} \\
      \multicolumn{1}{|c|} {}
      &\multicolumn{1}{||c} {\scalfrac $u_R$}
      &\multicolumn{1}{c} {$c_R$}
      &\multicolumn{1}{c|} {$t_R$}
      &\multicolumn{1}{|r|}
          {$\begin{array}{r} +\frac{2}{3} \end{array}$}
      &\multicolumn{1}{|r|}
          {$\begin{array}{r}  0 \end{array}$}
      &\multicolumn{1}{|r|}
          {$\begin{array}{r} +\frac{4}{3} \end{array}$} \\
      \multicolumn{1}{|c|} {}
      &\multicolumn{1}{||c} {\scalfrac $d'_R$}
      &\multicolumn{1}{c} {$s'_R$}
      &\multicolumn{1}{c|} {$b'_R$}
      &\multicolumn{1}{|r|}
          {$\begin{array}{r} -\frac{1}{3} \end{array}$}
      &\multicolumn{1}{|r|}
          {$\begin{array}{r}  0 \end{array}$}
      &\multicolumn{1}{|r|}
          {$\begin{array}{r} -\frac{2}{3} \end{array}$} \\ \hline
    \end{tabular}
  \vspace{0.2cm}
  \caption[Fermions and their quantum numbers]
  {\it The classification of fermions into three families of weak
    isospin multiplets. The quantum numbers are the fermion charge
    $Q_{\!f}$, the third component of the weak isospin $\ct^{(3)}_W$,
    and the weak hypercharge
    $Y_W$. The weak eigenstates $(d',s',b')$ are obtained from the
    mass eigenstates $(d,s,b)$ by Cabibbo-Kobayashi-Maskawa rotation.
    For each particle state, an antiparticle state which is obtained
    by charge conjugation exists.}
  \label{fermions}
  \end{center}
  \vspace{-0.6cm}
\end{table}
%
\vspace{.5cm}
{\bf The Gauge Field Lagrangian} \\
The pure gauge field part $\cl_G$ is given in terms of the massless
gauge fields $W^a_\mu~(a=1,2,3)$~and $B_\mu$. The isotriplet
$W^a_\mu$~corresponds to the three generators $\ct_a$ of the isospin
group $SU(2)_L$, belongs to the adjoint representation of $SU(2)$, and
couples to the left-handed doublets only. $B_\mu$ corresponds to the generator
$-Y_W/2$ of the $U(1)$ subgroup of gauge transformations and
couples to hypercharge. Adopting the conventions from
appendix~\ref{metric} one finds
\beq
  \hspace{3cm}
    \cl_G = -\frac{1}{4}\,W^a_{\mu\nu}W^{a,\mu\nu}
            -\frac{1}{4}\,B_{\mu\nu}B^{\mu\nu}~~.
  \label{gaugepart}
\eeq
Summation is also assumed over $a$, and the field strength tensors are
given by
\ba
  \hspace{3cm}
    W^{a,\mu\nu} & = & \DMUUP W^{a,\nu} - \DNUUP W^{a,\mu} +
                       g_2 \epsilon_{abc} W^{b,\mu} W^{c,\nu}~~, \nl
    B^{\mu\nu} & = & \DMUUP B^{\nu} - \DNUUP B^{\mu}~~,
  \label{gaugefields}
\ea
where $\epsilon_{abc}$ are the $SU(2)$ structure constants and
$g_1,g_2$ denote the $U(1)$~and $SU(2)$~coupling constants. The
non-abelian structure of the gauge group manifests itself in the
trilinear and quartic gauge boson self couplings obtained from
$\cl_G$.
\bpara
{\bf The Higgs Field and Higgs - Gauge Field Lagrangian} \\
It is a well-known fact that mass terms of the kind
$M^2 B_{\mu} B^{\mu}$/2 violate gauge invariance (see
e.g.~\cite{aitchison94}) and are therefore forbidden.
To have the carriers \wpl, \wmi, and \zz~massive as is required by
experimental observations, the $SU(2)_L \times U(1)_{Y_W}$ symmetry is
spontaneously broken via the Higgs mechanism~\cite{higgs}, leaving
behind an unbroken $U(1)$ to be interpreted as the electromagnetic
gauge subgroup $U(1)_{em}$. The minimal Higgs mechanism introduces a
complex scalar isodoublet field with $Y_W=1$
\beq
  \hspace{3cm}
    \Phi(x) = \left(
    \begin{array}{c} \phi^+(x) \\ \phi^0(x) \end{array}
    \right)
\eeq
at the space-time point $x$.
Its self-interaction and coupling to gauge bosons is described by
\beq
  \hspace{3cm}
  \cl_H \;\; = \;\; - \left( \DMUKOV \Phi \right)^\dagger
                      \left( \DMUKON \Phi \right) - V(\Phi)
  \label{higgslagr}
\eeq
with the covariant derivative
\beq
  \hspace{3cm}
  \DMUKON \;\; = \;\; \DMUUP \,-\, \ri g_2 \ct_a W^{\mu,a} \,+\,
                                   \ri g_1 \frac{Y_W}{2} B^{\mu}
  \label{covder}
\eeq
and the Higgs potential
\beq
  \hspace{3cm}
  V(\Phi) \;\; = \;\; - \mu^2 \left( \Phi^\dagger \Phi \right) +
          \frac{\lambda}{4} \left( \Phi^\dagger \Phi \right)^2~~.
\eeq
The ground state of the quantum system with Higgs field is obtained by
minimizing the potential $V(\Phi)$. In the ground state, the Higgs
field has a non-vanishing vacuum expectation value which can be chosen
as
\beq
  \hspace{3cm}
  \langle \Phi \rangle_0 \;\; \equiv \;\; \left(
    \begin{array}{c} 0 \\ \frac{v}{\sqrt{2}} \end{array} \right)
    \;\; = \;\; \left(
    \begin{array}{c} 0 \\ \frac{2\mu}{\,\sqrt{2\lambda}\,} \end{array}
    \right)~~.
\eeq
This reduces the manifest $SU(2)$ symmetry to $U(1)_{em}$. The Higgs
field can now be written as
\beq
  \hspace{1.5cm}
    \Phi(x) = \left( \rule[0cm]{0cm}{.8cm}
    \begin{array}{c} \phi^+(x) \\
      \left( \rule[0cm]{0cm}{.35cm}
      v+H(x)+\ri\chi(x) \right)/\sqrt{2} \end{array}
    \right)~~.
\eeq
The fields $\phi^+$~and $\chi$, however, can be gauged away and
therefore do not have any physical meaning. The remaining real part
$H(x)$~of $\phi^0$ can thus be interpreted as a new scalar particle
with its mass acquired via the quartic self coupling of the Higgs
field:
\beq
  \hspace{1.5cm}
  M_H \;\; = \;\; \sqrt{2} \mu~~.
\eeq
The kinetic part of the Higgs Lagrangian, eq.~(\ref{higgslagr}), gives
rise to mass terms for the gauge fields. Since $Y_W=1$ for the Higgs
field, the gauge boson mass terms can be written as
\beq
  \hspace{1.5cm}
  {\ds - \frac{v^2}{8}} \left( W^1_\mu, W^2_\mu, W^3_\mu, B_\mu \right)
  \left( \rule[0cm]{0cm}{1.23cm}
  \begin{array}{cccc}
    \!\! g_2^2 & 0 & 0 & 0 \!\! \\
    \!\! 0 & g_2^2 & 0 & 0 \!\! \\
    \!\! 0 & 0 & g_2^2 & g_1 g_2 \!\! \\
    \!\! 0 & 0 & g_1 g_2 & g_1^2 \!\!
  \end{array}
  \right) \left( \rule[0cm]{0cm}{1.23cm}
  \begin{array}{c}
    \!\! W^{\mu,1} \!\! \\
    \!\! W^{\mu,2} \!\! \\
    \!\! W^{\mu,3} \!\! \\
    \!\! B^\mu \!\!
  \end{array}
  \right)~~.
  \label{physmass}
\eeq
The physical parameters and fields are obtained from determination of
the mass eigenvalues and eigenvectors of
equation~(\ref{physmass}). For the gauge boson mass part of $\cl_H$
one thus finds
\beq
  \hspace{1.5cm}
  - M_W^2 \, W^+_\mu W^{\mu,-} \; - \; {\ds \frac{1}{2} }
  \left( \rule[0cm]{0cm}{.35cm} A_\mu, Z_\mu \right)
  \left( \rule[0cm]{0cm}{.6cm}
  \begin{array}{cc}
    \!\! 0 & 0 \!\! \\
    \!\! 0 & M_Z^2 \!\!
  \end{array}
  \right) \left( \rule[0cm]{0cm}{.65cm}
  \begin{array}{c}
    \!\! A^\mu  \!\! \\
    \!\! Z^\mu  \!\!
  \end{array}
  \right)
\eeq
with the physical fields
\ba
  \hspace{1.5cm}
  W^{\mu \pm} & = &  \frac{1}{\sqrt{2}} \left( \rule[0cm]{0cm}{.35cm}
                  W^{\mu,1} \, \mp \,\ri \, W^{\mu,2}  \right) \nl
  Z^{\mu}\;\: & = & \hspace{.39cm} \COW \: W^{\mu,3} \: + \; \SIW \:
                    B^\mu \nl
  A^{\mu}\;\, & = & - \SIW \: W^{\mu,3} \; + \; \COW \: B^\mu~~,
  \label{physfields}
\ea
the masses
\ba
  \hspace{1.5cm}
   M_W & = & \frac{v}{2} \, g_2 \nl
   M_Z\, & = & \frac{v}{2} \, \sqrt{g_1^2 + g_2^2} \nl
   M_\gamma\; & = & 0~~,
\ea
and the electroweak mixing angle defined by
\beq
  \hspace{1.5cm}
  \COW \;\; = \;\; \frac{g_2}{\sqrt{g_1^2 + g_2^2}} \;\; = \;\;
                   \frac{M_W}{M_Z}~~.
\eeq
Identification of $A_\mu$ with the photon field coupling to the
electric charge $e$~yields an expression for $e$ (compare
equation~(\ref{physfields})):
\beq
  \hspace{1.5cm}
  e \;\; = \;\; \frac{g_1 g_2}{\sqrt{g_1^2 + g_2^2}}
      \; = \;   g_2 \SIW \; = \; g_1 \COW~~.
  \label{positcharge}
\eeq
\vspace{.5cm}
{\bf The Fermion - Gauge Field Lagrangian} \\
As is seen from table~\ref{fermions}, fermions are grouped in
left-handed doublets with field operators
\beq
  \hspace{1cm}
  \psi^L_j \;\; = \;\;
  \left(
  \begin{array}{c}
    \!\! \psi^L_{j+} \!\! \\ \!\! \psi^L_{j-} \!\!
  \end{array}
  \right)
\eeq
and right-handed singlets with field operators
\beq
  \hspace{1cm}
  \psi^R_j \;\; = \;\; \psi^R_{j,\sigma}
\eeq
with the family index $j=1,2,3$ and the isospin index $\sigma = \pm$.
For the left-handed $SU(2)$ doublets $\sigma$ is the sign of the
third isospin component, and for the right-handed $U(1)$ singlets it
denotes the corresponding particle type. The color index of quarks is
suppressed. The Lagrangian for the fermion - gauge field
interaction can be written in terms of the covariant derivative (see
equation~(\ref{covder}))
\beq
  \hspace{1cm}
  \cl_F \;\; = \;\; - \; \sum_j \,
        {\bar \psi}^L_j \gamma^\mu \DMUKOV \psi^L_j \; - \;
        \sum_{j,\sigma} \, {\bar \psi}^R_{j,\sigma} \gamma^\mu
        \DMUKOV \psi^R_{j,\sigma}
\eeq
with
\ba
  \hspace{1cm}
  \psi^{({^L_R})}_{j,\sigma} & \equiv & \frac{1}{2}
       \left( 1 \pm \gamma_5 \right) \psi_{j,\sigma} \nl
  {\bar \psi} & \equiv & \psi^\dagger \,\ri
  \gamma^0~~.
  \nonumber
\ea
The latter definition ensures that ${\bar \psi}$ is identical to the
commonly used one.
\bpara
{\bf The Yukawa Interaction} \\
Because of the different gauge transformation behavior of left- and
right-handed fermion fields, mass terms of the kind $m{\bar \psi}\psi$
are forbidden. Fermion masses can, however, be dynamically generated via
the gauge invariant Yukawa Interaction
\beq
  \hspace{1cm}
  \cl_{Yukawa} \;\, = \;\, - \; \sum_j \; \left\{ \rule[0cm]{0cm}{.5cm}
    \left[ \rule[0cm]{0cm}{.4cm} \,
    g_{j-} \; {\bar \psi}^L_j \, \Phi \, \psi^R_{j-} \; + \;
    g_{j+} \; {\bar \psi}^R_{j+}  \, \Phi^\dagger \, \tau_1 \psi^L_j \,
    \right] \hspace{.2cm} + \hspace{.2cm} h.c. \; \right\}~.
\eeq
Here, we have used the Pauli matrix
\beq
  \hspace{1cm}
  \tau_1 = \left(
    \begin{array}{cc} \!\! 0 & 1 \!\! \\ \!\! 1 & 0 \!\! \end{array}
  \right)~~.
\eeq
The fermion masses are derived from the Yukawa coupling constants
$g_{j\sigma}$ and the non-vanishing vacuum expectation value of the
Higgs field operator. The Yukawa Interaction takes on a very simple
form in the unitary gauge where the unphysical degrees of freedom of
the Higgs field are gauged away~\cite{holliklect},
\ba
  \hspace{1cm}
  \cl_{Yukawa} & = & - \sum_f m_f {\bar \psi}_f \psi_f \;-\;
                       \sum_f \frac{m_f}{v} {\bar \psi}_f \psi_f H~~,
\ea
where the sum runs over all fermions. Note, however, that this is a
description in a specific gauge and not a gauge-invariant Lagrangian
density.
\bpara
{\bf Quantization of the Lagrangian, Gauge Fixing and Ghosts} \\
In general, to be able to quantize the Lagrangian described above, it
is necessary to apply a gauge fixing procedure. Simultaneously, a
compensation of the unphysical content of the gauge fixing Lagrangian
$\cl_{GF}$ is needed. This compensation is provided by the
introduction of a Faddeev-Popov ghost term $\cl_{FP}$. Details of the
procedure can be found in references~\cite{holliklect,wimdiss} or in
standard textbooks~\cite{itzykson,ebert89}. As this thesis is worked
out in the unitary gauge, no additional gauge fixing and ghost terms
are needed, no unphysical particles appear in the actual calculation.
As only QED corrections are treated, the subtleties connected with
unitarity and renormalizability of the unitary gauge do not play
any r\^ole.
\vspace{.8cm}
%
%
\section{Feynman Rules}
\label{feynrules}
One can express the Lagrangian as exposed in appendix~\ref{SMfund}
in terms of physical parameters and derive the Feynman rules for the
Standard Model. Below, Feynman rules relevant for this
thesis are presented. Complete lists of Feynman rules are found in the
literature~\cite{mandl84,boehm86,bd} and in the unconventional, but
excellent textbook by Veltman~\cite{veltman94}.
\vspace{.3cm} \\
{\underline {\bf Rule 1: External Lines}} \\
External fermions with momentum $p$~and helicity index $r$, external
gauge bosons with momentum $k$~and polarization index $\lambda$.
\vspace{.3cm}
\begin{Feynman}{120,60}{0,0}{1.}
  \put(-10,60){\fermionright}
  \put(20,60){\circle*{1.5}}
  \put(50,59){\rm incoming fermion \hspace{.75cm}:~~~$u^r(p,m)$}
  \put(2,56){$p,r$}
  \put(-10,50){\fermionright}
  \put(-10,50){\circle*{1.5}}
  \put(50,49){\rm outgoing fermion \hspace{.82cm}:~~~${\bar u}^r(p,m)$}
  \put(2,46){$p,r$}
  \put(-10,40){\fermionleft}
  \put(20,40){\circle*{1.5}}
  \put(50,39){\rm incoming antifermion\hspace{.19cm}:~~~${\bar u}^r(-p,m)$}
  \put(2,36){$p,r$}
  \put(-10,30){\fermionleft}
  \put(-10,30){\circle*{1.5}}
  \put(50,29){\rm outgoing antifermion\hspace{.26cm}:~~~$u^r(-p,m)$}
  \put(2,26){$p,r$}
  \put(-10,15){\photonrighthalf}
  \put(5,15){\photonrighthalf}
  \put(-10,5){\photonrighthalf}
  \put(5,5){\photonrighthalf}
  \put(20,15){\circle*{1.5}}
  \put(-10,5){\circle*{1.5}}
  \put(19,17){$\alpha$}
  \put(-12,7){$\alpha$}
  \put(2,10){$k,\lambda$}
  \put(2,0){$k,\lambda$}
  \put(50,14){\rm incoming}
  \put(50,4){\rm outgoing}
  \put(70,9){\rm gauge boson:~~~$\varepsilon^{\lambda,\alpha}(k)$}
\end{Feynman}
The fermion spin and vector boson polarization summation conventions
are given by
\ba
  \sum_r u^r(p,m) \, {\bar u}^r(p,m) & = & -\ri
    \left( \rule[0cm]{0cm}{.35cm} \dagg{p} + \ri m \right)~~, \nl
  \sum_\lambda \varepsilon^{\lambda,\alpha}(k) \,
     \varepsilon^{\lambda,\beta}(k) \hspace{.53cm}
     & = & g^{\alpha\beta} + \frac{k^\alpha\,k^\beta}{M_V^2}~~~~.
\ea
{\underline {\bf Rule 2: Propagators}} \\
Propagators with momenta $p$~for fermions, $k$~for photons, and
$q$~for bosons. The imaginary parts $-\ieps$ ensure proper causal
behavior. Define $\dagg{p} \equiv \gamma^\mu p_\mu$.
\vspace{-.3cm}
\begin{Feynman}{120,50}{0,0}{1.}
  \put(-10,40){\fermionright}
  \put(-10,40){\circle*{1.5}}
  \put(20,40){\circle*{1.5}}
  \put(3,36){$p$}
  \put(50,39){\rm fermion \hspace{2.68cm}: \hspace{.5cm}
              ${\ds - \frac{\dagg{p} + \ri m}{\,p^2 + m^2 -\ieps\,}}$}
  \put(-10,25){\photonrighthalf}
  \put(5,25){\photonrighthalf}
  \put(-10,25){\circle*{1.5}}
  \put(20,25){\circle*{1.5}}
  \put(-12,27.5){$\alpha$}
  \put(19,27.5){$\beta$}
  \put(3,21){$k$}
  \put(50,24){\rm photon \hspace{2.78cm}: \hspace{.5cm}
              ${\ds - \frac{\ri\;g_{\alpha\beta}}{\,k^2 - \ieps\,}}$}
  \put(-10,10){\gaugebosonright}
  \put(-10,10){\circle*{1.5}}
  \put(20,10){\circle*{1.5}}
  \put(-12,12.5){$\alpha$}
  \put(19,12.5){$\beta$}
  \put(5,10){\vector(1,0){1.}}
  \put(3,6){$q$}
  \put(50,9){\rm massive gauge boson V: \hspace{.5cm}
     ${\ds - \frac{\,\ri\,(g_{\alpha\beta} + q_\alpha q_\beta/M_V^2)\,}
                  {q^2 + M_V^2 - \ieps} }$ }
\end{Feynman}
{\underline {\bf Rule 3: Vertices}} \\
In this thesis, only neutral current vertices are needed. They can be
cast into a simple generic form: \vspace{.3cm}
\begin{Feynman}{120,15}{0,0}{1.}
  \put(-2,8){\fermionurhalf}
  \put(-2,8){\fermionulhalf}
  \put(-2,8){\photonrighthalf}
  \put(-2,8){\circle*{1.5}}
  \put(-2,10){$\alpha$}
  \put(-13,15){${\bar f}$}
  \put(-13,-1){$f$}
  \put(15,6.5){$B = \gamma,\,\ZZ$}
  \put(70,6.5){$- \: g_B \, g_2 \cdot \gamma^\alpha \left(
                g^V_{f\!{\bar f}\!B} + g^A_{f\!{\bar f}\!B} \,
                \gamma_5 \right) $~~.}
\end{Feynman}
For the couplings $g_B,~g^V_{f\!{\bar f}\!B},$~and $g^A_{f\!{\bar f}\!B}$ of
the fermion $f$~to the photon and the \zz~the relations
\ba
  g_2 \; = \; {\ds \frac{e}{\,\SIW\,}} \hspace{3.5cm} & & \nl \nl
  g_\gamma \; = \; 1 \hspace{4.44cm} & &
  g_Z \; = \; {\ds \frac{1}{\,4\cdot\COW\,}} \nl
  g^V_{f\!{\bar f}\!\gamma} \; = \; \SIW\, Q_{\!f} \hspace{2.92cm} & &
  g^V_{f\!{\bar f}\!Z}  \; = \; 2 \ct^{(3)}_{W,f} - 4\cdot\SWS\,Q_{\!f} \nl
  g^A_{f\!{\bar f}\!\gamma} \; = \; 0 \hspace{4.19cm} & &
  g^A_{f\!{\bar f}\!Z} \; = \; 2 \ct^{(3)}_{W,f}
  \label{GZcoup}
\ea
hold, with the fermion's electric charge $Q_{\!f}$ in units of the
positron charge $e$ and the fermion's weak isospin third component
$\ct^{(3)}_{W,f}\,$. It is often desirable to write
the vertices in terms of left- and right-handed couplings,
\beq
  - \: g_B \, g_2 \cdot \gamma^\alpha \left(
    g^V_{f\!{\bar f}\!B} + g^A_{f\!{\bar f}\!B} \gamma_5 \right)
  \;\; = \;\;
  - \: \gamma^\alpha \left( \rule[0cm]{0cm}{.35cm}
          L_{f\!{\bar f}\!B}\!\cdot\!\left( 1+\gamma_5 \right)
        + R_{f\!{\bar f}\!B}\!\cdot\!\left( 1-\gamma_5 \right) \right)
\eeq
with
\beq
  L_{f\!{\bar f}\!B} \; \equiv \; \frac{\,g_B \, g_2\,}{2}
    \left( g^V_{f\!{\bar f}\!B} + g^A_{f\!{\bar f}\!B} \right)
  \hspace{1.35cm}
  R_{f\!{\bar f}\!B} \; \equiv \; \frac{\,g_B \, g_2\,}{2}
    \left( g^V_{f\!{\bar f}\!B} - g^A_{f\!{\bar f}\!B} \right)~~.
  \label{LRdef}
\eeq
For the above neutral current vertices one obtains
\ba
  L_{f\!{\bar f}\!\gamma} \; = \; \frac{\,e\,Q_{\!f}\,}{2}
  \hspace{2.24cm} & &
  L_{f\!{\bar f}\!Z} \; = \; \frac{e}{\,2\,\SIW\,\COW} \,
    \left(\ct^{(3)}_{W,f} - \SWS\,Q_{\!f}\right)
  \nl \nl
  R_{f\!{\bar f}\!\gamma} \; = \; \frac{\,e\,Q_{\!f}\,}{2}
  \hspace{2.21cm} & &
  R_{f\!{\bar f}\!Z} \; = \; - \frac{e}{\,2\,\SIW\,\COW}
                          \, \SWS\,Q_{\!f}~~.
  \label{LRcoup}
  \\ \nl \nonumber
\ea
{\underline {\bf Rule 4: Ordering}} \\
The spinor factors ($\gamma$-matrices, propagators, 4-spinors)
for each fermion line are ordered from right to left while following
the line in its arrow sense. Multiply the expression by a phase factor
+1 (--1), if an even (odd) permutation is required to write the
fermion operators in the correct normal order.
\vspace{1cm} \\
{\underline {\bf Rule 5: Four-Momentum Conservation}} \\
The four-momenta of lines meeting at a vertex satisfy four-momentum
conservation. For each four-momentum $q$ which is not fixed by
four-momentum conservation, carry out the integration
$\frac{1}{(2\pi)^4} \int d^4 q$.
\vspace{1cm} \\
{\underline {\bf Rule 6: Fermion Loops}} \\
For each closed fermion loop take the trace and multiply by --1.
\vspace{1cm} \\
{\underline {\bf Rule 7: Symmetrization}} \\
Introduce a relative sign +1 (--1) for diagrams that are obtained by an
even (odd) number of permutations of identical final state fermions.
Obtain the matrix element by adding all diagrams.
\vspace{1cm} \\
{\underline {\bf Rule 8: Differential 2$\,\to$n Cross-Section}} \\
${\cal M}_{if}$ be the matrix element for the scattering of a
two-particle initial state $i=\{1,2\}$ to an n-particle final state
$f=\{3,4,...,n+2\}$. Then obtain the differential cross-section by
\ba
  d\sigma & = & \frac{1}{\,4\,\sqrt{(p_1\!\cdot\!p_2)^2 - m_1^2 m_2^2 \,}} \,
               \frac{|{\cal M}_{if}|^2}{S} \:
               (2\pi)^4 \; \delta^{(4)}\! \left( P_i - P_f \right)\,
               d{\tilde p}_3 \cdot \cdot \cdot d{\tilde p}_{n+2}\,
  \nl
  P_i & = & p_1 + p_2 \hspace{3cm}  P_f \; = \; \sum_{k=3}^{n+2}
  p_k~~.
  \label{xsform}
\ea
$S$~yields the symmetry factor, if there are $l$ sets with $k_l$
identical particles,
\ba
  S & = & \prod_l k_l !~~. \nonumber
\ea
With $p^0$ representing the particle energy,
$p^0 = \sqrt{{\vec p}{\,^2}+m^2}$, the Lorentz invariant phase space
element is given by
\ba
  d{\tilde p} & = & \frac{d^3p}{(2\pi)^3\,2 p^0}~~. \nonumber
\ea
In case of unobserved polarization degrees of freedom, the average
over initial state and the sum over final state polarizations has to
be taken. The factors $2m$ left out in equation~(\ref{xsform})
compared to common textbooks like~\cite{itzykson} is cancelled by the
normalization chosen for the fermion spinors $u(p,m)$.
%
%
\section{Renormalization}
\label{renorm}
\begin{figure}[t]
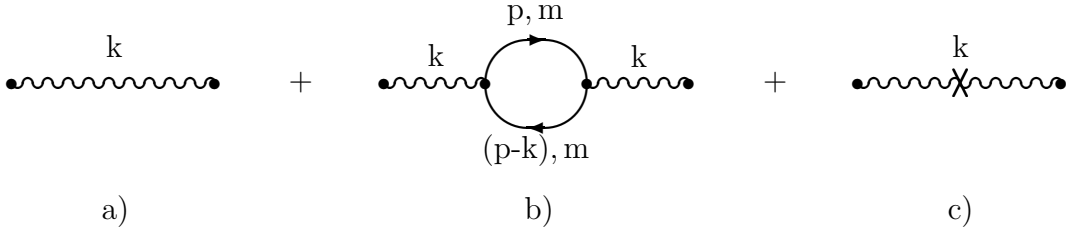

\vspace{1cm}
  \begin{Feynman}{150,20}{-37,-14.5}{0.9}
    \put(-40,0){\photonright}
    \put(-26,4){k}
    \put(-26.7,-20){a)}
    \put(1,-1){+}
    \put(15,0){\photonrighthalf}
    \put(21.5,2.5){k}
    \put(37.5,0){\oval(15,13)[t]}
    \put(37.5,0){\oval(15,13)[b]}
    \put(37.9,6.5){\vector(1,0){1}}
    \put(37.1,-6.5){\vector(-1,0){1}}
    \put(45,0){\photonrighthalf}
    \put(51.5,2.5){k}
    \put(33.2,9.5){p,\,m}
    \put(29.5,-11){(p-k),\,m}
    \put(35.8,-20){b)}
    \put(71,-1){+}
    \put(85,0){\photonright}
    \put(99.15,2.1){\line(1,-2){2}}
    \put(99.15,-1.9){\line(1,2){2}}
    \put(99,4){k}
    \put(98.3,-20){c)}
    \put(-40,0){\circle*{1.5}}
    \put(-10,0){\circle*{1.5}}
    \put(15,0){\circle*{1.5}}
    \put(30,0){\circle*{1.5}}
    \put(45,0){\circle*{1.5}}
    \put(60,0){\circle*{1.5}}
    \put(85,0){\circle*{1.5}}
    \put(115,0){\circle*{1.5}}
  \end{Feynman}
  \caption[Feynman diagrams for the photon propagator renormalization]
    {\it Feynman diagrams for the photon propagator renormalization.
      a) Renormalized photon propagator, ~b) Vacuum polarization, ~and
      c) Photon propagator counterterm.}
  \label{vacpol}
\end{figure}

\begin{figure}[t]
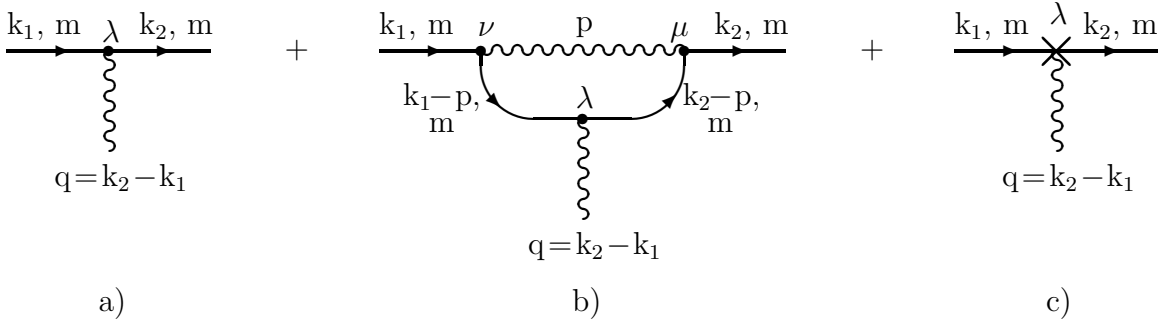

\vspace{1cm}
  \begin{Feynman}{150,40}{-40,-34}{0.9}
    \put(-45,0){\line(1,0){30}}
    \put(-36.5,0){\vector(1,0){1}}
    \put(-21.5,0){\vector(1,0){1}}
    \put(-30,-15){\photonuphalf}
    \put(-31.5,-38.5){a)}
    \put(-30,0){\circle*{1.5}}
    \put(-45,2.5){${\rm k_1,\,m}$}
    \put(-25.5,2.5){${\rm k_2,\,m}$}
    \put(-31,1.3){$\lambda$}
    \put(-38,-20){${\rm q\!=\!k_2\!-\!k_1}$}
    \put(-4,-1){+}
    \put(10,0){\line(1,0){15}}
    \put(18.5,0){\vector(1,0){1}}
    \put(10,2.5){${\rm k_1,\,m}$}
    \put(40,0){\oval(30,20)[b]}
    \put(27,-7.4){\vector(1,-1){1}}
    \put(53,-7.3){\vector(1,1){1}}
    \put(13.5,-8){${\rm k_1\!\!\!\,-\!p,}$}
    \put(17.3,-12){m}
    \put(54.8,-8){${\rm k_2\!\!\!\,-\!p,}$}
    \put(58.5,-12){m}
    \put(25,0){\photonright}
    \put(39,3){p}
    \put(55,0){\line(1,0){15}}
    \put(63.5,0){\vector(1,0){1}}
    \put(59.5,2.5){${\rm k_2,\,m}$}
    \put(40,-25){\photonuphalf}
    \put(38.5,-38.5){b)}
    \put(32,-30){${\rm q\!=\!k_2\!-\!k_1}$}
    \put(39,-8.5){$\lambda$}
    \put(53,2.2){$\mu$}
    \put(24.5,2.2){$\nu$}
    \put(25,0){\circle*{1.5}}
    \put(55,0){\circle*{1.5}}
    \put(40,-10){\circle*{1.5}}
    \put(81,-1){+}
    \put(95,0){\line(1,0){30}}
    \put(103.5,0){\vector(1,0){1}}
    \put(118.5,0){\vector(1,0){1}}
    \put(110,-15){\photonuphalf}
    \put(108.5,-38.5){c)}
    \put(95,2.5){${\rm k_1,\,m}$}
    \put(114,2.5){${\rm k_2,\,m}$}
    \put(102,-20){${\rm q\!=\!k_2\!-\!k_1}$}
    \put(109,4){$\lambda$}
    \put(108,2){\line(1,-1){4}}
    \put(108,-2){\line(1,1){4}}
  \end{Feynman}
  \caption[Feynman diagrams for the electromagnetic vertex
  renormalization]
    {\it Feynman diagrams for the electromagnetic vertex
      renormalization. a) Vertex, ~b) Vertex correction, ~and
      c) Vertex counterterm.}
  \label{renvert}
\end{figure}

In higher order perturbation theory, the relations between the
parameters of the Standard Model Lagrangian~(\ref{lagrangian}) and the
physical observables differ from tree level.  At loop level, the
``bare'' parameters of the tree level Lagrangian have no more physical
meaning, and a redefinition of parameters and fields is required to
recover physical meaning. The procedure of redefinition is called
renormalization and properly removes divergences appearing in loop
diagrams. For a mathematically consistent treatment of divergences,
the theory must be regularized which is most commonly done with
dimensional regularization.
The ``renormalizability'' of the Standard Model assures that
a finite number of so-called renormalization constants suffices to
redefine physical quantities in all orders of perturbation
theory~\cite{thooft71}\footnote[2]{In ``unrenormalizable'' theories,
  additional renormalization constants are needed for each new
  perturbation order.}.
Renormalization schemes differ in the choice of input
parameters and renormalization conditions. As physical results
must be scheme independent, the observables calculated in different
renormalization schemes are identical in infinite order perturbation
theory. Results from $n^{th}$~order perturbation theory obtained in
different schemes, deviate from each other in higher order
contributions~\cite{gmuscheme}. The accuracy of the finite
order approximation is influenced by the choice of input
parameters. Among the various renormalization schemes found in the
literature, the most prominent ones are
\begin{itemize}
  \item The ``modified minimal subtraction'' ($\overline{\rm MS}$)
    scheme~\cite{MSbar,passarino90}, where the r\^{o}le of the
    electroweak mixing angle is emphasized. Renormalization constants
    are fixed by a simple subtraction of the singular parts
    \ba
      \frac{1}{n-4} + \frac{\gamma_E}{2} - \ln(2\sqrt{\pi}) \nonumber
    \ea
    from the two- and three-point functions calculated in dimensional
    regularization~(see appendix~\ref{loops}). The common input
    parameters for the $\overline{\rm MS}$~scheme are
    \ba
      \alpha,~M_W,~M_Z,~M_H,~{\rm~and~the~fermion~masses}~m_f.
      \nonumber
    \ea
    An advantage of the $\overline{\rm MS}$~scheme is that theoretical
    uncertainties due to quark induced self-energy effects are much
    reduced~\cite{passarino90}.
  \item The *~scheme~\cite{kennedy89} which emphasizes vector boson
    propagator effects and running parameters~\cite{jegerlect}.
    The *~scheme is not very commonly used today~\cite{bardinYB95}.
  \item The \gmu~scheme~\cite{gmuscheme,jegerlect} uses the input
    parameters
    \ba
      G_\mu,~M_W,~M_Z,~M_H,~{\rm~and~the~fermion~masses}~m_f.
      \nonumber
    \ea
    It is important that \gmu~does not run from low energies up to the
    vector boson mass scale. The fine structure constant $\alpha$ and
    the Fermi coupling constant \gmu~are related via $\Delta r$, the
    non-QED correction to $\mu$-decay.
  \item The on-shell scheme~\cite{holliklect,OS} which was used for
    this thesis and will now be described in some detail.
\end{itemize}
In the on-shell renormalization scheme, to recover physical meaning at
loop level, so-called multiplicative renormalization of bare fields
$\phi_0$~and bare couplings $g_0$~is applied,
\beq
  \phi_0 \;\; = \;\; Z_\phi^{1/2} \phi \hspace{2cm}
  g_0 \;\; = \;\; Z_g g~~.
\eeq
The renormalized parameters $g$~are finite and fixed by
renormalization conditions. Field renormalization ensures finite Green
functions~\cite{holliklect}. Using the expansions $Z_i = 1\!+\!\delta
Z_i$, the bare Lagrangian is split into a renormalized part and a
counterterm part,
\ba
  {\cal L}(\phi_0,g_0) =  {\cal L}(Z_\phi^{1/2} \phi,Z_g g) =
  {\cal L}(\phi,g) + \delta{\cal L}(\phi,g,\delta Z_\phi,\delta g)~~.
\ea
The renormalized Lagrangian can be expressed in terms of the chosen
set of physical parameters, and from $\delta{\cal L}$~one derives
counterterm Feynman rules in accordance with the on-shell
renormalization conditions. Renormalization conditions fix the
renormalization constants and thus give meaning to the
theory. Counterterms properly remove ultraviolet
divergences arising from loop diagrams. Infrared divergences from loop
diagrams are cancelled by real bremsstrahlung radiative
corrections. In the on-shell scheme, renormalization conditions are
chosen to have propagator poles at the physical particle masses with
residue $\!=\!1$~and to
recover the classical Thomson limit for the $ee\gamma$-vertex.
The common set of input parameters for the on-shell renormalization
scheme is
\beq
  e,~M_W,~M_Z,~M_H,~{\rm~and~the~fermion~masses}~m_f.
  \nonumber
\eeq
It is a vital property of the on-shell scheme that tree level
relations between masses, couplings and the electroweak mixing angle
$\theta_W$ remain intact also at loop level. Thus one may easily
switch between input parameters. In particular
\beq
  \SWS = 1 - \frac{M_W^2}{M_Z^2}
  \label{rensw}
\eeq
can be retained in the on-shell renormalization.
For the renormalized Fermi coupling constant \gmu~which is an
attractive input parameter, because it is experimentally known with
high precision, the following relation holds~\cite{jegerlect}:
\beq
  \sqrt{2} \, G_{\mu} \; = \; \frac{\pi \alpha}
                                {M_W^2\,\SWS\,(1-\Delta r)}~~.
\eeq
Thus, in the case of pure QED loop corrections to a process, even
\gmu~retains its tree level relation to the other parameters.
\para
As this thesis is concerned with QED corrections, the on-shell
renormalization for QED will be worked out in some detail
below. After choosing a set of input parameters, one starts from the
tree level QED Lagrangian
\ba
  \cl_0^{QED} & = & -\,\frac{1}{4}\,F_{0,\mu\nu} F_0^{\mu\nu} \; - \;
    {\bar \psi_0} \, \gamma^\mu \DMU \, \psi_0 \; - \;
    m_0 \, {\bar \psi_0} \psi_0 \; + \;
    \ri e_0 \:\! Q_{\!f} \, {\bar \psi_0} \, \gamma^\mu \, \psi_0 \,
    A_{0,\mu}
  \nl\nl
  F_0^{\mu\nu} & \equiv & \DMUUP A_0^{\nu} - \DNUUP A_0^{\mu}
\ea
with the photon field $A_0^\mu$ and introduces multiplicative
renormalization constants:
\ba
  \psi_0 & = & \sqrt{Z_\psi} \cdot \psi \hspace{2cm}
  A_0^{\mu} \;\; = \;\; \sqrt{Z_{\!A}} \cdot A^{\mu} \nl \nl
  m_0 & = & \frac{Z_m}{Z_\psi} \cdot m \hspace{2.28cm}
  e_0 \;\; = \;\; \frac{Z_e}{Z_\psi \sqrt{Z_A}} \cdot e~~~.
\ea
Therefore, expanding $Z_i \!=\! 1 \!+\! \delta Z_i$, the QED
Lagrangian is given by
\ba
  \cl^{QED} & = & -\,\frac{1}{4}\,F_{\mu\nu} F^{\mu\nu} \; - \;
    {\bar \psi} \, \gamma^\mu \DMU \, \psi \; - \;
    m \, {\bar \psi} \psi \; + \;
    \ri e \:\! Q_{\!f} \, {\bar \psi} \, \gamma^\mu \, \psi \, A_{\mu}
    \;\; + \;\; \delta \cl
  \label{renlagra}
\ea
with the counterterm Lagrangian
\ba
  \delta \cl & = & -\,\frac{\dza}{4} \!\cdot\! F_{\mu\nu} F^{\mu\nu}
    \; - \; \dzp \!\cdot\! {\bar \psi} \, \gamma^\mu \DMU \, \psi \;-\;
    \dzm \!\cdot\! m \, {\bar \psi} \psi \; + \;
    \dze \!\cdot\! \ri e \:\! Q_{\!f} \, {\bar \psi} \, \gamma^\mu \, \psi \,
    A_{\mu}~~. \nl
\ea
{}From the requirement of gauge invariance one concludes that the
counterterms $\dze$~and $\dzp$~are equal which is nothing but the QED
Ward identity in the language of renormalization
constants~\cite{mandl84}. In the next step, counterterm Feynman rules
are derived from $\delta \cl$. Counterterms are then added to physical
1-loop amplitudes and finally fixed by the on-shell renormalization
conditions. The QED counterterm Feynman rules are given on
page~\pageref{counterfeyn} after explicit expressions for the
counterterms have been derived.
\vspace{.8cm} \\
\underline{\bf The photon field counterterm $\dza$} \vspace{.1cm} \\
Renormalization of the photon propagator means to add an ultraviolet
divergent vacuum polarization diagram and a counterterm diagram as
given in figure~\ref{vacpol} to the photon propagator derived from the
renormalized Lagrangian $\cl$~in equation~(\ref{renlagra}). From the three
diagrams in figure~\ref{vacpol}, the 1-loop expression for the photon
propagator is derived as
\beq
  D_\gamma^{\mu\nu} \;\; = \;\;
    \frac{-\ri g^{\mu\nu} \left( 1-\dza \right)}{k^2-\ieps} \; + \;
    \frac{-\ri g^{\mu\alpha}}{k^2-\ieps} \, \ri \Pi_{\alpha\beta}
    \frac{-\ri g^{\beta\nu}}{k^2-\ieps}
\eeq
with the vacuum polarization given by~(see e.g.~\cite{dima80})
\ba
  \ri \Pi_{\alpha\beta} & = & - \, \left( e\:\! Q_{\!f}\right)^2
    \left. \, \mu^{(4-n)} \! \int\!\frac{d^np}{(2\pi)^n} \;
    \frac{{\rm Tr} \left[ \rule[0cm]{0cm}{.35cm} \gamma_\alpha
          \left( \dagg{p}+\ri m \right) \gamma_\beta
          \left( \dagg{p}-\dagg{k}+\ri m \right) \right]}
         {\left[ \rule[0cm]{0cm}{.35cm} p^2 + m^2 - \ieps \right]
          \left[ \rule[0cm]{0cm}{.35cm} (p-k)^2+m^2-\ieps \right] }
         ~\right|_{\mu=m_e} \nl
  & = & \frac{4 \ri \left( e\:\! Q_{\!f}\right)^2}{16\,\pi^2}
        \left[ \rule[0cm]{0cm}{.6cm}
        \frac{2}{3} \left( \rule[0cm]{0cm}{.35cm}
          k^2 g_{\alpha\beta} \!-\! k_\alpha k_\beta \right) \, {\rm P}
        \,+\, 2\,\left( \rule[0cm]{0cm}{.35cm}k^2 g_{\alpha\beta}
                        \!-\! k_\alpha k_\beta \right) \,
                        {\cal I}(k^2,m^2,m^2) \right] \nl
  & \equiv & \ri \, g_{\alpha\beta} \, \Pi^\gamma(k^2) \; + \;
    \ri \, k_\alpha k_\beta \, \Pi^{(2)}(k^2)~~,
  \label{sephot}
\ea
\vspace{-.3cm}
\ba
  {\cal I}(k^2,m^2,m^2)  & = & \int\limits_0^1 \! dx \; x (1-x)
  \!\cdot\!
  \myln \left[ \rule[0cm]{0cm}{.6cm} x (1-x) \frac{k^2}{\MES} +
                 \frac{m^2}{\MES} \right]~~.
  \label{vacpolar}
\ea
The dimensionally regularized ultraviolet divergence P is defined in
equation~(\ref{poleUV}). It follows from current conservation that, in
equation~(\ref{sephot}), the term proportional to the photon momentum
$k$~yields vanishing contributions~\cite{mandl84}. Therefore one finds
\beq
  D_\gamma^{\mu\nu} \;\; = \;\;
    \frac{-\ri g^{\mu\nu}}{k^2-\ieps} \left( \rule[0cm]{0cm}{.35cm}
    1 + \frac{\,\Pi^\gamma \:-\: k^2\,\dza\,}{k^2-\ieps} \right)~~.
  \label{renphotpr1}
\eeq
Introducing the renormalized vacuum polarization $\hat{\Pi}^\gamma =
\Pi^{\gamma} \:-\: k^2\,\dza$ which is a ``small'' quantity, one can
rewrite equation~(\ref{renphotpr1}):
\ba
  D_\gamma^{\mu\nu} & = &
    \frac{-\ri g^{\mu\nu}}{\,k^2-\ieps} \cdot
    \frac{1}{1-\frac{\hat{\Pi}^\gamma}{k^2-\ieps}} \;\;\; + \;\;\;
    {\rm higher~orders}
  \nl & = &
    \frac{-\ri g^{\mu\nu}}{\,k^2-\ieps-\hat{\Pi}^\gamma} \;\;\; + \;\;\;
    {\rm higher~orders}~~.
  \label{renphotpr2}
\ea
It is seen from equation~(\ref{renphotpr2}) that the photon propagator
has a pole strictly at $k^2=0$, if ~$\hat{\Pi}^\gamma\!=\!k^2
d\hat{\Pi}^{\gamma}/dk^2|_{k^2=0} +\! {\cal O}(k^4)$. Then, the
photon does not acquire a mass term through higher order effects and
hence there is no photon mass renormalization. The on-shell
renormalization condition requires the photon
propagator~(\ref{renphotpr1}) to have residue $\!=\!1$, i.e to retain its
tree level form in the limit $k^2=0$. Thus
{}~$\hat{\Pi}^\gamma/k^2|_{k^2=0} =\!0$
and therefore
\beq
  \dza \;\; = \;\; \left. \rule[0cm]{0cm}{.6cm}
    \frac{\,\Pi^\gamma(k^2)}{k^2} \right|_{k^2=0}~~,
\eeq
which is readily evaluated from equation~(\ref{vacpolar}), yielding
\beq
  \dza \;\; = \;\; \frac{\left( e\:\! Q_{\!f}\right)^2}{16\,\pi^2}
    \cdot \frac{4}{3} \left( \rule[0cm]{0cm}{.35cm}
    2 {\rm P} \,+\, \ln\frac{m^2}{m_e^2} \right)~~.
  \label{photct}
\eeq
For non-vanishing $k^2,~\hat{\Pi}^\gamma$~yields the charge
renormalization familiar from many
textbooks~\cite{itzykson,mandl84,halzen}.
\vspace{.6cm} \\
\begin{figure}[t]
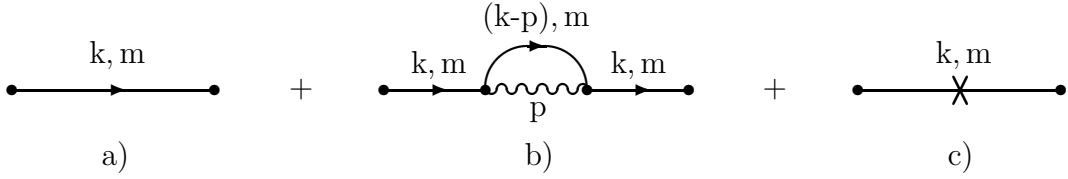

\vspace{1cm}
  \begin{Feynman}{150,12}{-37,-5.5}{0.9}
    \put(-40,0){\line(1,0){30}}
    \put(-24,0){\vector(1,0){1}}
    \put(-28.5,4){k,\,m}
    \put(-26.7,-11){a)}
    \put(1,-1){+}
    \put(15,0){\line(1,0){15}}
    \put(23.5,0){\vector(1,0){1}}
    \put(19,2.5){k,\,m}
    \put(37.5,0){\oval(15,13)[t]}
    \put(30,0){\photonrighthalf}
    \put(37.9,6.5){\vector(1,0){1}}
    \put(45,0){\line(1,0){15}}
    \put(53.5,0){\vector(1,0){1}}
    \put(48.5,2.5){k,\,m}
    \put(29.5,9.5){(k-p),\,m}
    \put(36.5,-3.8){p}
    \put(35.8,-11){b)}
    \put(71,-1){+}
    \put(85,0){\line(1,0){30}}
    \put(99.15,2){\line(1,-2){2}}
    \put(99.15,-2){\line(1,2){2}}
    \put(96.5,4){k,\,m}
    \put(98.3,-11){c)}
    \put(-40,0){\circle*{1.5}}
    \put(-10,0){\circle*{1.5}}
    \put(15,0){\circle*{1.5}}
    \put(30,0){\circle*{1.5}}
    \put(45,0){\circle*{1.5}}
    \put(60,0){\circle*{1.5}}
    \put(85,0){\circle*{1.5}}
    \put(115,0){\circle*{1.5}}
  \end{Feynman}
  \caption[Feynman diagrams for the fermion propagator renormalization]
    {\it Feynman diagrams for the fermion propagator renormalization.
      a) Renormalized fermion propagator, ~b) Fermion self energy, ~and
      c) Fermion propagator counterterm.}
  \label{selfen}
\end{figure}
\newpage
\noindent
\underline{\bf The fermion field and mass counterterms $\dzp$, $\dzm$,
  and $\dmf$.} \vspace{.1cm} \\
The Feynman diagrams for the renormalized fermion propagator are shown
in figure~\ref{selfen}. Decomposing the fermionic terms from the
counterterm Lagrangian,
\ba
  \dagg{k} \dzp \: - \: \ri m \, \dzm & \equiv &
  \left( \dagg{k} - \ri m \right) \dzp - \ri \, \dmf \nl
  \dmf & = & m \left( \dzm - \dzp \right)~~,
  \label{fercts}
\ea
the 1-loop expression for the fermion propagator corresponding to the
three diagrams is
\ba
  S^f & = &
    -\,\frac{1}{\,\dagg{k} \!-\! \ri m - \ieps\,}
      \left(1\!-\!\dzp \!+\!
      \frac{\ri\,\dmf}{\dagg{k} - \ri m} \right) \; - \;
      \frac{1}{\,\dagg{k}\!-\!\ri m\!-\!\ieps} \cdot \Sigma^f \cdot
      \frac{1}{\,\dagg{k}\!-\!\ri m\!-\!\ieps} \nl
\ea
with the fermion self energy given by
\ba
  \Sigma^f(k) & = & - \ri \left( e\:\! Q_{\!f}\right)^2 \,
    \left. \, \mu^{(4-n)} \! \int\!\frac{d^np}{(2\pi)^n} \;
    \frac{\gamma^\mu \left( \dagg{k} - \dagg{p} + \ri m \right) \gamma_\mu}
         {\left[ \rule[0cm]{0cm}{.35cm} (k-p)^2 + m^2 -\ieps \right]
          \left[ \rule[0cm]{0cm}{.35cm} p^2-\ieps \right]}
    ~\right|_{\mu=m_e} \nl
  & = & \Sigma^f(\ri m) \; + \;
    B^f \, \left( \rule[0cm]{0cm}{.35cm} \dagg{k} - \ri m \right)
    \;+\; {\cal O}\left[ \rule[0cm]{0cm}{.35cm}
                  \left( \dagg{k} - \ri m \right)^2 \right]
  \nl\nl
  \Sigma^f(\ri m) & = &
    - \, \frac{\left( e\:\! Q_{\!f}\right)^2}{16\,\pi^2} \; \ri m
    \left( \rule[0cm]{0cm}{.6cm} 6 {\rm P} - 4 +
           3\,\ln\frac{m^2}{m_e^2} \right) \nl
  B^f  & = & \hspace{.4cm}
    \frac{\left( e\:\! Q_{\!f}\right)^2}{16\,\pi^2} \,
    \left( \rule[0cm]{0cm}{.6cm} 2 {\rm P} + 4 {\rm P^{IR}} - 4 +
    3\,\ln\frac{m^2}{m_e^2} \right)~~.
  \label{ferseen}
\ea
The dimensionally regularized infrared singularity ${\rm P^{IR}}$~is
defined in equation~(\ref{poleIR}). Using the fermion self energy
decomposition from equation~(\ref{ferseen}), the renormalized fermion
propagator $S^f$~can be written as
\ba
  S^f & = & -\,\frac{1}{\,\dagg{k} \!-\! \ri m - \ieps\,}
   \left(1 - \dzp + B^f \:+\: {\cal O} \left(\dagg{k} - \ri m \right)
   \:+\:  \frac{\,\ri\,\dmf + \Sigma^f(\ri m)}{\dagg{k} \!-\! \ri m}
   \right)
  \label{fermprop}
  \\ \nl
   & = & -\,\frac{1}{\left( \dagg{k} \!-\! \ri m \right)
                     \left[ \rule[0cm]{0cm}{.35cm}
                     1+\dzp\!-\!B^f \,-\,
                     {\cal O}(\dagg{k} \!-\! \ri m) \right] -
                     \ri\,\dmf\!-\!\Sigma^f(\ri m)}
    \;\: + \;{\rm higher~orders}\,. \nl
  \label{resumprop}
\ea
Imposing the on-shell condition, i.e. requiring $S^f$~in
equation~(\ref{resumprop}) to have a pole at $\dagg{k}=\ri m$, the
result for the fermion mass counterterm reads
\beq
  \dmf \;\; = \;\; - \, \frac{\Sigma^f(\ri m)}{\ri} \;\; = \;\;
    m \; \frac{\left( e\:\! Q_{\!f}\right)^2}{16\,\pi^2}
    \left( \rule[0cm]{0cm}{.6cm}
    6 {\rm P} - 4 + 3\,\ln\frac{m^2}{m_e^2} \right)~~.
  \label{massren}
\eeq
In addition, as the pole of the 1-loop fermion propagator is required
to have residue $\!=\!1$, the second fermion propagator
renormalization condition reads
\beq
  \dzp \;\; = \;\; B^f \;\; = \;\;
    \frac{\left( e\:\! Q_{\!f}\right)^2}{16\,\pi^2} \,
    \left( \rule[0cm]{0cm}{.6cm} 2 {\rm P} + 4 {\rm P^{IR}}
           - 4 + 3\,\ln\frac{m^2}{m_e^2} \right)~~.
  \label{fermren}
\eeq
With the renormalization conditions~(\ref{massren})
and~(\ref{fermren}) it is ensured that external
on-shell fermion lines in Feynman diagrams do not receive any QED loop
corrections. The mass counterterm $\dzm$~is now easily derived by
introducing the results~(\ref{massren}) and~(\ref{fermren}) into
equation~(\ref{fercts}):
\ba
  \dzm & = & \dzp + \frac{\dmf}{m} \;\; = \;\;
    \frac{\left( e\:\! Q_{\!f}\right)^2}{16\,\pi^2} \,
    \left( \rule[0cm]{0cm}{.6cm} 8 {\rm P} + 4 {\rm P^{IR}} - 8 +
    6\,\ln\frac{m^2}{m_e^2} \right)~~. \\ \nl \nonumber
\ea
\underline{\bf The vertex counterterm $\dze$} \vspace{.1cm} \\
The renormalized 1-loop electromagnetic vertex consists of the three
Feynman diagrams depicted in figure~\ref{renvert}.
For on-shell external fermions it is given by
\beq
  - \, e\:\! Q_{\!f} \hat{\Gamma}^\lambda \;\; = \;\;
  - \, e\:\! Q_{\!f} \gamma^\lambda \,
    (1\!+\!\dze) \; - \; e\:\! Q_{\!f} \, \Gamma^\lambda
 \label{renvertex}
\eeq
with the vertex correction
\ba
  \Gamma^\lambda & = & - \ri \! \left( e\:\! Q_{\!f}\right)^2 \!
    \left. \mu^{(4-n)} \! \int \!\! \frac{d^np}{(2\pi)^n} \,
    \frac{\gamma^\mu \left( \dagg{k}_{\!2}\!-\!\dagg{p}\!+\!\ri m \right)
      \gamma^\lambda \left( \dagg{k}_{\!1}\!-\!\dagg{p}\!+\!\ri m \right)
      \gamma_\mu}
      {\left[ \rule[0cm]{0cm}{.35cm} (k_1\!-\!p)^2 \!+\! m^2 \!-\!
              \ieps \right] \!
       \left[ \rule[0cm]{0cm}{.35cm} (k_2\!-\!p)^2 \!+\! m^2 \!-\!
              \ieps \right] \!
       \left[ \rule[0cm]{0cm}{.35cm} p^2 \!-\! \ieps \right]}
    \:\right|_{\mu=m_e} \nl\nl
  & = & \frac{\left( e\:\! Q_{\!f}\right)^2}{16\,\pi^2}
    \left\{ \rule[0cm]{0cm}{.9cm} \;\;
    \gamma^\lambda \left[ \rule[0cm]{0cm}{.8cm} \!- 2 {\rm P}
    - 2 \left( q^2 \!+\! 2 m^2 \right){\cal J}(q^2,m^2,m^2) \, {\rm P^{IR}}
    - \ln\frac{m^2}{\MES} \, + \right. \right. \nl
  & & \hspace{2.84cm} \left. \left. \rule[0cm]{0cm}{.8cm}
    \left( \rule[0cm]{0cm}{.35cm} 3 q^2/2 \!+\! 4 m^2 \right)
      {\cal J}(q^2,m^2,m^2) -
    \left( q^2 \!+\! 2 m^2 \right){\cal K}(q^2,m^2,m^2)
    \right] \right. \nl
  & & \hspace{1.715cm} \left. \rule[0cm]{0cm}{.9cm}
    + \frac{\ri m}{2}
      \left( \gamma^\lambda \dagg{q} - \dagg{q} \gamma^\lambda \right)
      {\cal J}(q^2,m^2,m^2) \hspace{3cm} \right\} \nl\nl
  & \equiv & \gamma^\lambda \!\cdot\! F_1(q^2,m^2,m^2) \;\; + \;\;
    \left( \gamma^\lambda \dagg{q} - \dagg{q} \gamma^\lambda \right)
    \!\cdot\! F_2(q^2,m^2,m^2)
  \label{loopvert}
\ea
where
\ba
  {\cal J}(q^2,m^2,m^2) & = & \int\limits_0^1
    \frac{dx}{x (1-x) q^2 + m^2} \;\; \stackrel{q^2\!=0}{=} \;\;
    \frac{1}{m^2}~~, \nl
  {\cal K}(q^2,m^2,m^2) & = & \int\limits_0^1
    \frac{dx}{x (1-x) q^2 + m^2} \!\cdot\!
    \ln\frac{x (1-x) q^2 + m^2}{\MES} \;\; \stackrel{q^2\!=0}{=} \;\;
    \frac{1}{m^2} \!\cdot\! \ln\frac{m^2}{\MES}~~. \nl
\ea
%
%
%
\hspace{-.24cm} From equation~(\ref{loopvert}), the Dirac form factor
$F_1$~and the
Pauli form factor $F_2$~can easily be read off. The on-shell
renormalization condition requires the renormalized vertex
$\hat{\Gamma}^\lambda$~to recover $\gamma^\lambda$
in the Thomson limit $q \to 0$,
\beq
  \hat{\Gamma}^\lambda(q=0,m^2,m^2) \; = \; \gamma^\lambda~~.
  \label{vrencon}
\eeq
Equivalently one may require
\beq
  F_1(q^2=0,m^2,m^2) \;= \; - \dze
  \label{fermirencon}
\eeq
for the Dirac form factor. Introducing condition~(\ref{vrencon})
or~(\ref{fermirencon}) into equation~(\ref{renvertex}) and using the
result for $\Gamma^\lambda$~from equation~(\ref{loopvert}) with the
proper limits ${\cal J}(q^2=0,m^2,m^2)$~and ${\cal K}(q^2=0,m^2,m^2)$,
one obtains
\ba
  \dze\cdot\gamma^\lambda & = & - \, \Gamma^\lambda(0,m^2,m^2)
  \;\; = \;\; \frac{\left( e\:\! Q_{\!f}\right)^2}{16\,\pi^2}
  \, \left( \rule[0cm]{0cm}{.6cm} 2 {\rm P}
    + 4 {\rm P^{IR}} - 4 + 3 \ln\frac{m^2}{\MES} \right) \cdot
    \gamma^\lambda
  \label{chargect}
\ea
for the vertex counterterm. It is therefore seen that $\dzp$~and
$\dze$~are equal and thus indeed fulfill the QED Ward identity.
\para
It should be mentioned that the vertex counterterm derived above is
also valid for the QED renormalized $f\!{\bar f}\ZZ$~vertex. Briefly,
one could argue that the term $\left( g^V_{f\!{\bar f}\!Z} +
g^A_{f\!{\bar f}\!Z} \, \gamma_5 \right)$~from the replacement
\ba
  - \,e \:\! Q_{\!f}\, \gamma^\lambda \;\; \longrightarrow \;\;
    - \: g_Z \, g_2 \cdot \gamma^\lambda
    \left( g^V_{f\!{\bar f}\!Z} + g^A_{f\!{\bar f}\!Z} \, \gamma_5
    \right) \nonumber
\ea
is exactly pulled through to the end of the vertex correction,
\ba
  \Gamma^\lambda_{f\!{\bar f}Z} & = &
    -\, g_Z \, g_2\, \cdot \Gamma^\lambda \cdot
    \left( g^V_{f\!{\bar f}\!Z} + g^A_{f\!{\bar f}\!Z} \, \gamma_5
      \right) \;\; = \;\;
    \frac{-\,e}{\,4\,\SIW\,\COW\,} \cdot \Gamma^\lambda \cdot
    \left( g^V_{f\!{\bar f}\!Z} + g^A_{f\!{\bar f}\!Z} \, \gamma_5
      \right)~~. \nl
\ea
Therefore, as the weak mixing angle is not affected by the on-shell
QED renormalization,
\ba
  \hat{\Gamma}^\lambda_{f\!{\bar f}Z} & = &
    \frac{-\,e}{\,4\,\SIW\,\COW\,} \cdot \hat{\Gamma}^\lambda \cdot
    \left( g^V_{f\!{\bar f}\!Z} + g^A_{f\!{\bar f}\!Z} \, \gamma_5
      \right)~~.
\ea
Thus the renormalization condition~(\ref{vrencon}) and its
solution~(\ref{chargect}) yield the same multiplicative counterterm
for the $f\!{\bar f}\ZZ$~and the $f\!{\bar f}\gamma$~vertex.
\para
The same result may be obtained from equation (58) of
reference~\cite{hollect} or from equation (4.21) of
reference~\cite{dima82} by neglecting the counterterms that do not
belong to the on-shell QED renormalization.
\vspace{1.5cm}\\
\label{counterfeyn}
\noindent
\hspace*{-.31cm} \underline{\bf Counterterm Feynman rules} \vspace{.1cm} \\
With the above results, namely equations~(\ref{photct}),
(\ref{massren}), (\ref{fermren}), and~(\ref{chargect}), the
counterterms as needed in the 1-loop QED counterterm Feynman rules are
fixed. The QED counterterm Feynman rules are listed below and must be
used together with the set of rules given in section~\ref{feynrules}.
Subsequently the abbreviation c.t. will be used for ``counterterm''.
\vspace{.8cm} \\
{\underline {\bf Rule 9: Renormalized external lines}} \\
Do not add loop contributions nor counterterms to on-shell external
particle lines.
\vspace{.8cm} \\
{\underline {\bf Rule 10: QED Counterterms}} \\
\vspace{2.3cm}
\begin{Feynman}{120,50}{7,-30}{1.}
  \put(-10,40){\fermionrighthalf}
  \put(5,40){\fermionrighthalf}
  \put(-10,40){\circle*{1.5}}
  \put(20,40){\circle*{1.5}}
  \put(3,37.95){\line(1,1){4}}
  \put(3,41.95){\line(1,-1){4}}
  \put(3.8,34){$p$}
  \put(37,39){\rm fermion c.t.~:~~
              ${\ds \frac{-1}{\,\dagg{p} \!-\! \ri m \!-\! \ieps\,}
                    \!\cdot\! \left[ \rule[0cm]{0cm}{.35cm}
                           \left(\dagg{p} \!-\! \ri m\right) \dzp -
                           \ri\dmf \right] \!\cdot\!
                    \frac{-1}{\,\dagg{p} \!-\! \ri m \!-\! \ieps\,}}$}
  \put(-10,15){\photonright}
  \put(-10,15){\circle*{1.5}}
  \put(20,15){\circle*{1.5}}
  \put(-12,17.5){$\alpha$}
  \put(19,17.5){$\beta$}
  \put(4,9){$k$}
  \put(3.15,13.1){\line(1,1){4}}
  \put(3.15,17.1){\line(1,-1){4}}
  \put(37,14){\rm photon c.t. \hspace{.065cm}:~~
              ${\ds \frac{-\,\ri g^{\mu\alpha}}{\,k^2\!-\!\ieps\,}
                 \!\cdot\!
                 \left( -\,\ri g_{\alpha\beta} \, k^2 \, \dza \right)
                 \frac{-\,\ri g^{\beta\nu}}{\,k^2\!-\!\ieps\,}}$}
  \put(75,0) {${\ds \; = \; \frac{-\,\ri g^{\mu\nu}}{\,k^2\!-\!\ieps\,}
                 \!\cdot\!\left( \rule[0cm]{0cm}{.35cm}-\,\dza
                 \right)}$}
  \put(-2,-15){\fermionurhalf}
  \put(-2,-15){\fermionulhalf}
  \put(-2,-15){\photonrighthalf}
  \put(-3,-17){\line(1,2){2}}
  \put(-3,-13){\line(1,-2){2}}
  \put(-2.5,-12){$\alpha$}
  \put(-13,-8){${\bar f}$}
  \put(-13,-24){$f$}
  \put(15,-16.5){$\gamma,\,\ZZ$}
  \put(37,-16.5){\rm vertex c.t. \hspace{.21cm}:~~
                 ${\ds - \: g_B \, g_2 \cdot \gamma^\alpha \left(
                    g^V_{f\!{\bar f}\!B} + g^A_{f\!{\bar f}\!B} \,
                   \gamma_5 \right) \cdot \dze}$}
\end{Feynman}
{}From the above counterterm Feynman rules, several observations are
made:
\begin{itemize}
  \item In case of electron lines, the logarithms
    ~$\myln\left(m^2/\MES\right)$~vanish in all counterterms.
    Initially, these logarithms arose due to the choice of the
    electron mass as the scale for dimensional regularization.
  \item For massless fermions, the fermion propagator counterterm
    Feynman rule translates into a factor ~$-\dzp$ that must be
    multiplied to the tree level amplitude corresponding to the
    counterterm amplitude under consideration.
  \item Similarly, an amplitude with a photon propagator counterterm
    is obtained by multiplication of the corresponding tree level
    amplitude with $-\dza$.
  \item An amplitude with a QED vertex counterterm is obtained by
    multiplication of the corresponding tree level amplitude with
    $\dze$.
\end{itemize}
In conclusion, the counterterms for the QED initial state corrections
to process~(\ref{eezz4f}) may be multiplicatively taken into account
as it is done in equations~(\ref{ctpart}) and~(\ref{mcts}).
%
\chapter{Complex Logarithm and Polylogarithms}
\label{rellogs}
For the proper computation of many phase space and loop integrals, the
definitions of the complex logarithm and the derived polylogarithms
are needed. In this appendix, the necessary definitions are listed
together with some useful relations. An exhaustive treatment of
Polylogarithms is found in reference~\cite{lewin58}.
%
%
\section{The Logarithm Function}
\label{logfun}
Throughout this dissertation the logarithm is defined as the main
branch of the inverse of the complex exponential function:
\ba
  \myln z \; \equiv \; \myln \left( |z| \cdot e^{{\rm i\, arg}(z)} \right)
  \; = \;  \myln |z| + {\rm i\, arg}(z)~~,
  && z \seteps \numcomp \put(4,-1){\line(-1,3){3.4}}
  \;\: \numreal^- \nl
  && {\rm arg}(z)\;\, \epsilon\; \left]-\pi,\pi\right[
\ea
The logarithm of a product of two complex numbers $a$~and $b$~is given
by~\cite{thooft79}
\ba
  \myln \left( a\cdot b \right) & = & \ln a \; + \; \ln b \; + \;
    \eta\!\left( a,b \right) \\
  \eta\!\left( a,b \right) & = & 2\pi\ri \left[ \rule[0cm]{0cm}{.35cm}
    \: \Theta(-\IM\, a) \, \Theta(-\IM\, b) \, \Theta(\IM\, ab) \; - \;
    \Theta(\IM\, a) \, \Theta(\IM\, b) \, \Theta(-\IM\, ab)\,
    \right]
    \nonumber
\ea
with the Heavyside step function $\Theta(x)$. It can be concluded that
\ba
  \myln \left( a\cdot b \right) & = & \ln a \; + \; \ln b
  \hspace{2cm}
  {\rm if~~} \IM\, a {\rm ~~and~~} \IM\, b {\rm ~~have~different~sign.} \nl
  {\ds \myln \frac{a}{b} } & = & \ln a \; -\; \ln b
  \hspace{2cm}
  {\rm if~~} \IM\, a {\rm ~~and~~} \IM\, b {\rm ~~have~equal~sign.}
  \nonumber
\ea
{}From table~\ref{logsign}, where all possible sign combinations for
$\eta\!\left( a,b \right)$~are listed, it can be seen that for real
numbers $x$~and $y$ and infinitesimal $\varepsilon$
\ba
  \myln \left( xy - \ieps\right) & = & \myln \left( x - \ieps\right)
        \; + \; \myln \left( y - \ieps/x\right)~~.
\ea
%
%
\begin{table}[ht]
  \begin{center}
  \begin{tabular}{|c|c|c||c|} \hline
    & & & \\
    sgn$(\IM\, a)$ & sgn$(\IM\, b)$ & sgn$(\IM\, ab)$ &
    $\eta\!\left( a,b \right)$ \\
    & & & \\ \hline \hline
     +  &  +  & +      & \hspace{.1cm} 0 \\
     +  &  +  & $-$    & $-2\pi\ri$ \\
     +  & $-$ & $\pm$  & \hspace{.1cm} 0 \\
    $-$ &  +  & $\pm$  & \hspace{.1cm} 0 \\
    $-$ & $-$ & +      & $+2\pi\ri$ \\
    $-$ & $-$ & $-$    & \hspace{.1cm} 0 \\ \hline
  \end{tabular}
  \label{logsign}
  \caption[Imaginary parts for the multiplication of complex
    logarithms]
    {\it All possible values of $\eta\!\left( a,b \right)$.}
  \end{center}
\end{table}
%
%
\section{The Dilogarithm Function}
\label{dilog}
The dilogarithm function is defined by the integral
\ba
  \mysp\,(z) \; \equiv \; -\int\limits_0^1 {\ds
    \frac{\,\ln(1-tz)\,}{t} \, dt }~~,
  && z \; \epsilon \; \numcomp \backslash \!
     \left\{ x \, \epsilon \, \numreal : x > 1\right\}~~.
\ea
The following one parameter relations are valid for $z$ being any
complex number such that none of the involved arguments lie on the
corresponding logarithmic or dilogarithmic cuts.
\beq
  \mysp\,(0) \; = \; 0~, \hspace{1.5cm}
  \mysp\,(-1) \; = \; -\frac{\pi^2}{12}~, \hspace{1.5cm}
  \mysp\,(1) \; = \; \frac{\pi^2}{6}
\eeq
\beq
  \mysp\,(z) + \mysp\,(-z) \;\; = \;\;
    \frac{1}{2} \: \mysp\,(z^2)
\eeq
\beq
  \mysp\,(z) + \mysp\,(1/z) \;\; = \;\;
    - \frac{\pi^2}{6} \; - \; \frac{1}{2} \ln^2(-z)
\eeq
\beq
  \mysp\,(z) + \mysp\,(1-z) \;\; = \;\;
    \frac{\pi^2}{6} \; - \; \ln(z) \cdot \ln(1-z)
\eeq
\beq
  \mysp\,(z)  + \mysp\left( \frac{1}{1-1/z} \right) \;\; = \;\;
    -\frac{1}{2} \: \ln^2(1-z)
\eeq
\beq
  \mysp\,(1-z) + \mysp\,(1-1/z) \;\; = \;\;
    -\frac{1}{2} \: \ln^2(z)
\eeq
For an infinitesimal real number $\varepsilon$ and
$\,x\,\epsilon\,\numreal\,$ the dilogarithm satisfies
\beq
  \IM \left\{ \mysp\,(x+\ieps) \right\} \;\; = \;\;
    {\rm sgn}(\varepsilon) \cdot \Theta(x-1) \cdot \pi\,\myln|x|~~.
\eeq
Therefore, for $\,x\,\epsilon\,\numreal,\;x>1\,$  one can easily deduce
\ba
  \mysp\,(x+\ieps) + \mysp\,(x-\ieps) \;\; = \;\;
    {\rm sgn}(\varepsilon) \cdot 2\pi\ri \cdot \ln x~~. \nonumber
\ea
A general two parameter relation for complex-valued dilogarithms which
can be useful for special purposes and will not be repeated here is
given in reference~\cite{wimdiss}.
%
%
\section{The Trilogarithm Function}
\label{trilog}
Similar to the dilogarithm, the trilogarithm function is defined by an
integral:
\ba
  \mytri\,(y) \; \equiv \; \int\limits_0^y \frac{\mysp\,(z)}{z} \, dz
    \; = \; \int\limits_0^1 \frac{\mysp\,(yz)}{z} \, dz
    \; = \; \int\limits_0^1 \frac{\ln z \cdot \ln (1-yz)}{z} \, dz~~,
    \nl
    \hspace{6cm} y \; \epsilon \; \numcomp \backslash \!
    \left\{ x \, \epsilon \, \numreal : x > 1\right\}~~.
\ea
{}From this definition two integrals are easily derived:
\ba
  \int\limits_0^y \frac{\ln z \cdot \ln (1-z)}{z} \: dz ~& = &
    \mytri\,(y) \; - \; \ln y \cdot \mysp\,(y)~~,
    \hspace{1cm} y \; \epsilon \; \numcomp \backslash \numreal
    \; \cup \; \left[ 0,1 \right] \\
  \int\limits_0^y \frac{\ln z \cdot \ln (1+az)}{z} \: dz & = &
    \mytri\,(-ay) \; - \; \ln y \cdot \mysp\,(-ay)~~, \nl
    & &  y \; \epsilon \; \numcomp \backslash \numreal^-~,~~~
         ay\; \epsilon \; \numcomp \backslash \!
         \left\{ x \, \epsilon \, \numreal : x < -1\right\}~~.
\ea
Some useful one parameter relations are given below. Again, it is
understood that the arguments in each of all equations must not lie on
any of the corresponding above-defined cuts:
\beq
  \mytri\,(0) \; = \; 0
\eeq
\beq
  \mytri\,(1) \; = \; \zeta (3) \; = \; 1.20205690... ~~,
  \hspace{1.2cm} \mytri\,(-1) \; = \; -\frac{3}{4} \, \mytri\,(1)
\eeq
\beq
  \mytri\,(z) + \mytri\,(-z) \;\; = \;\; \frac{1}{4} \, \mytri\,(z^2)
\eeq
\beq
  \mytri\,(z) - \mytri\,(1/z) \;\; = \;\;
    - \frac{\pi^2}{6} \ln(-z) \; - \; \frac{1}{6} \myln^3(-z)
\eeq
Further, more complicated relations are found in chapter VI of
reference~\cite{lewin58}. Here, only some relations for real arguments
$x$ will be quoted in addition:
\beq
  \mytri\,(x) - \mytri\,(1/x) \;\; = \;\; \frac{\pi^2}{3} \ln x \; - \;
    \frac{1}{6} \myln^3 x \; - \; \frac{1}{2}\,\ri\pi \myln^2 x~~,
    \hspace{2cm} x\,>\,1
\eeq
\ba
  \mytri\,\left( \frac{-x}{1-x} \right) + \mytri\,(1-x) + \mytri\,(x)
  \;\; = \;\; \hspace{6.3cm} \nl
  \mytri\,(1) \: + \: \frac{\pi^2}{6} \ln(1-x) \: - \:
  \frac{1}{2}\,\ln x \!\cdot\! \ln^2(1-x) \: + \:
  \frac{1}{6} \myln^3 (1-x)~, \hspace{.5cm} 0<x<1
\ea
\ba
  \mytri\,\left( \frac{-x}{1-x} \right) +
  \mytri\,\left( \frac{1}{1-x} \right) + \mytri\,(x) \;\; = \;\;
  \hspace{6.02cm} \nl
  \mytri\,(1) \: - \: \frac{\pi^2}{6} \ln(1-x) \: - \:
  \frac{1}{2}\,\ln (-x) \!\cdot\! \ln^2(1-x) \: + \:
  \frac{1}{3} \myln^3 (1-x)~, \hspace{.5cm} x<0~.
\ea
%
\chapter{Description of the Phase Space}
\label{phasespa}
It is the intention of this appendix to familiarize the reader with
the phase space pa\-ram\-etri\-za\-tions used throughout this
thesis. For the Born case of process~(\ref{eezz4f}) and for the
virtual initial state corrections, the 2$\,\to\,$4 particle phase
space is needed, whereas the 2$\,\to\,$5 particle phase space
parametrization is used to describe bremsstrahlung.
%
\begin{figure}[hb]
  \vspace{12.5cm}
\caption[The 2$\to$4 body phase space]
{\it Graphical representation of a 2$\,\to\,$4 particle reaction. In
  general, the vectors  $\vec{p}_1,\vec{p}_2,\vec{p}_3$ and
  $\vec{p}_4$ do not lie in the plane of the picture.}
\label{ps24}
\end{figure}
%
\section{The 2$\,\to\,$4 Particle Phase Space}
\label{ps2to4}
A reaction of two particles to a four-particle final state has
eight kinematical degrees of freedom which can be parametrized in many
different ways. For semi-analytical calculations the following choice
of kinematical variables has proven to be
convenient~\cite{dl94,dimaww,muta86}: \vspace{.1cm} \\
\begin{tabular}{rcl}
  $\!\phi$        & : & Azimuth angle around the \epl~beam direction
                        $\vec{k}_2$. \\
  $\!\varth$      & : & Scattering angle of $\vec{v}_1$ with respect to
                        the \epl~direction $\vec{k}_2$ in the center
                        of \\
                  &   & mass frame. \\
  $\!\phi_{12}$   & : & Azimuth angle around the $\vec{v}_1$
                        direction.\\
  $\!\theta_{12}$ & : & Decay polar angle of $\vec{p}_1$ in the
                        $\vec{v}_1$ rest frame with the z axis along
                        $\vec{v}_1$.\\
  $\!\phi_{34}$   & : & Azimuth angle around the $\vec{v}_2$
                        direction.\\
  $\!\theta_{34}$ & : & Decay polar angle of $\vec{p}_3$ in the
                        $\vec{v}_2$ rest frame with the z axis along
                        $\vec{v}_2$.\\
  $\!\SONE$       & : & Invariant mass of the final state fermion pair
                        \fone\bfone~:~~$\SONE= -v_1^2=-(p_1+p_2)^2$. \\
  $\!\STWO$       & : & Invariant mass of the final state fermion pair
                        \ftwo\bftwo~:~~$\STWO= -v_2^2=-(p_3+p_4)^2$.
\end{tabular}
\vspace{.4cm} \\
Here, $k_1$ and $k_2$ are the momentum four-vectors of the initial
state electron and positron, $v_1$ and $v_2$ denote the four-momenta
of the intermediate bosons and $p_1,p_2,p_3$, and $p_4$ are the final
state four-momenta. The corresponding three-momenta are given by
superscript arrows. To illustrate the above kinematical
variables, a graphical representation of a 2$\,\to\,$4 particle
reaction is shown in figure~\ref{ps24}. The center of mass energy
squared is given by $s = - (k_1+k_2)^2 = - (p_1+p_2+p_3+p_4)^2$.
The decomposition of the $2\!\to\!n$~phase space into subsequently
decaying compounds is, for example, worked out in chapter 4.2 of
reference~\cite{was94}.
In terms of the above variables the four-particle Lorentz invariant
phase space has the form
\ba
  d\Gamma_4 & = & \frac{1}{(2\pi)^{12}} \cdot \frac{\,d^3p_1\,}{2p_1^0}\,
                \frac{\,d^3p_2\,}{2p_2^0}\,\frac{\,d^3p_3\,}{2p_3^0}\,
                \frac{\,d^3p_4\,}{2p_4^0}
                \times  \delta^{(4)}(k_1+k_2-\sum_{i=1}^4 p_i) \nl
          & = & \frac{1}{(2\pi)^{11}} \:
                \frac{\sqrt{\lambda(s,\SONE,\STWO)}}{8s} \:
                \frac{\sqrt{\lambda(\SONE,m_1^2,m_1^2)}}{8\SONE} \:
                \frac{\sqrt{\lambda(\STWO,m_2^2,m_2^2)}}{8\STWO} \times \nl
          &   & \hspace{4cm} d\SONE \; d\STWO \; d\!\mycos\varth \;
                d\phi_{12} \, d\!\mycos\theta_{12} \;
                d\phi_{34} \, d\!\mycos\theta_{34}~~,
  \label{gamma4}
\ea
with the usual definition of the $\lambda$ function,
\ba
\lambda(a,b,c) & = & a^2+b^2+c^2-2ab-2ac-2bc \nonumber
\ea
which is symmetric in all arguments and has the special cases
\ba
\lambda(a,b,b) & = & a^2 \cdot \left( 1 - 4b/a \right)~~, \nl
\lambda(a,b,0) & = & (a-b)^2 \hspace{1.48cm} , \nl
\lambda(a,0,0) & = & a^2 \hspace{2.5cm} . \nonumber
\ea
The limits of the phase space variables are: \vspace{.3cm} \\
$\begin{array}{cccccc}
 \hspace{.75cm}
 & 4\,m_1^2 & \leq & \SONE & \leq & \left( \sqrt{s} - 2\,m_2
                                    \right)^2 \\
 & 4\,m_2^2 & \leq & \STWO & \leq & \left( \sqrt{s} - \sqrt{\SONE}
                                    \right)^2 \\
 & -1 & \leq & \mycos\varth & \leq & +1       \\
 &~~0 & \leq & \phi_{\{12,34\}} & \leq & 2 \pi \\
 & -1 & \leq & \mycos\theta_{\{12,34\}} & \leq & \hspace{1.4cm} +1
 \hspace{1.3cm} .
\end{array}$
\vspace{.3cm} \\
In the presented semi-analytical \xsec calculation, the integrations
over the final state fermion decay angles $\phi_{12},~\theta_{12}$~and
$\phi_{34},~\theta_{34}$~are se\-pa\-ra\-ted from the other
integrations and are carried out analytically with the help of
invariant tensor integration (see appendix~\ref{tenint}). The
remaining integrations over $\varth,~\SONE,$~and $\STWO$~are
non-trivial, especially, if virtual QED corrections are included. The
angle $\varth$~is integrated analytically, $\SONE$~and $\STWO$~are
integrated numerically. To perform integrations over the above
variables, one must express the four-vectors of the four-particle
final state in terms of the phase space parameters. All scalar
pro\-ducts appearing in the \xsec calculation can then be expressed in
terms of these integration variables. In the center of mass frame with
a Cartesian coordinate system as drawn in figure~\ref{ps24}, the
initial state vectors $k_1$~and $k_2$ are given by
\vspace{.3cm} \\
$\begin{array}{lll}
  k_1 & = & \left( k^0,~k\,\mysin\varth,~0,~-k\,\mycos\varth \right) \\
  k_2 & = & \left( k^0,~-k\,\mysin\varth,~0,~k\,\mycos\varth \right)
\end{array}$
\hspace{1cm} with \hspace{1cm}
$\begin{array}{lll}
  k^0 & = & \frac{\sqrt{s}}{2} \vspace{.1cm} \\
  k   & = & \frac{\sqrt{s}}{2} \, \beta \vspace{.1cm} \\
  \beta & \equiv & \sqrt{1-\frac{4\MES}{s}}
\end{array}$
\vspace{-1.45cm}
\ba
  \label{kvect}
\ea
\vspace{.2cm} \\
and the electron mass $m_e$. Using the mass $m_1$~for \fone, and
$m_2$~for \ftwo~one obtains further
\vspace{.5cm} \\
$\begin{array}{lll}
  p_1 & \!\!\! = & \!\!\!\!
    \left( \gamma_{12}^0\!\:p_{12}^{B,0} +
                           \gamma_{12}\!\:p_{12}^B\mycos\theta_{12},\:
           p_{12}^B\!\:\mysin\theta_{12} \mycos\phi_{12},\:
           p_{12}^B\!\:\mysin\theta_{12} \mysin\phi_{12},\:
           \gamma_{12}^0\!\:p_{12}^B \mycos\theta_{12} +
                           \gamma_{12}\!\:p_{12}^{B,0}
            \right)\! \\ \\
  p_2 & \!\!\! = & \!\!\!\!
    \left( \gamma_{12}^0\!\:p_{12}^{B,0} -
                           \gamma_{12}\!\:p_{12}^B\mycos\theta_{12},\;
          -p_{12}^B\!\:\mysin\theta_{12} \mycos\phi_{12},\:
          -p_{12}^B\!\:\mysin\theta_{12} \mysin\phi_{12},\: \right.\\
    & & \hspace{9.94cm} \left.
          -\gamma_{12}^0\!\:p_{12}^B \mycos\theta_{12} +
                           \gamma_{12}\!\:p_{12}^{B,0}
            \right)\! \\ \\

  p_3 & \!\!\! = & \!\!\!\!
    \left( \gamma_{34}^0\!\:p_{34}^{B,0} -
                           \gamma_{34}\!\:p_{34}^B \mycos\theta_{34},\:
           p_{34}^B\!\:\mysin\theta_{34} \mycos\phi_{34},\:
           p_{34}^B\!\:\mysin\theta_{34} \mysin\phi_{34},\:
           \gamma_{34}^0\!\:p_{34}^B \mycos\theta_{34} -
                           \gamma_{34}\!\:p_{34}^{B,0}
            \right)\! \\ \\
  p_4 & \!\!\! = & \!\!\!\!
    \left( \gamma_{34}^0\!\:p_{34}^{B,0} +
                           \gamma_{34}\!\:p_{34}^B\mycos\theta_{34},\;
          -p_{34}^B\!\:\mysin\theta_{34} \mycos\phi_{34},\:
          -p_{34}^B\!\:\mysin\theta_{34} \mysin\phi_{34},\: \right.\\
    & & \hspace{9.94cm} \left.
          -\gamma_{34}^0\!\:p_{34}^B \mycos\theta_{34} -
                           \gamma_{34}\!\:p_{34}^{B,0}
            \right)\!
\end{array}$
\ba
  \label{pvect}
\ea
with
\vspace{.3cm} \\
\hspace*{1cm}
$\begin{array}{lll}
  p_{12}^{B,0}  & = & {\ds \frac{\sqrt{\SONE}}{2} } \vspace{.2cm} \\
  p_{12}^B      & = & {\ds \frac{\sqrt{\lambda(\SONE,m_1^2,m_1^2)}}
                                {2\sqrt{\SONE}} } \vspace{.2cm} \\
  \gamma_{12}^0 & = & {\ds \frac{s+\SONE-\STWO}{2\sqrt{s\SONE}} }
                                                  \vspace{.2cm} \\
  \gamma_{12}   & = & {\ds \frac{\sqrt{\lambda(s,\SONE,\STWO)}}
                                     {2\sqrt{s\SONE}} }
\end{array}$
\hspace{2cm}
$\begin{array}{lll}
  \;\; p_{34}^{B,0}  & = & {\ds \frac{\sqrt{\STWO}}{2} } \vspace{.2cm} \\
  \;\; p_{34}^B      & = & {\ds \frac{\sqrt{\lambda(\STWO,m_2^2,m_2^2)}}
                                {2\sqrt{\STWO}} } \vspace{.2cm} \\
  \;\; \gamma_{34}^0 & = & {\ds \frac{s-\SONE+\STWO}{2\sqrt{s\STWO}} }
                                                  \vspace{.2cm} \\
  \;\; \gamma_{34}   & = & {\ds \frac{\sqrt{\lambda(s,\SONE,\STWO)}}
                                     {2\sqrt{s\STWO}} }~~.
\end{array}$
\vspace{.5cm} \\
The superscript $B$ hints at the relation of the corresponding
quantities to the appropriate boson rest frame. The quantities
$p_{ij}^{B,0}, \; p_{ij}^B\!=\!\left|\vec{p}_{ij}^{\,B}\right|$, and
$\gamma_{ij}^0$~are e.g. found in chapter IV, equations~(3.4) and
(3.6) of reference~\cite{byckling73}.
$\gamma_{ij}\!=\!\left|\vec{\gamma}_{ij}\right|$~may be derived from
the equation ${\gamma_{ij}^0}^2 \!-\!\vec{\gamma}_{ij}^{\,2} \!=\! 1$.
For use in appendix~\ref{ps2to5} vector components in the above
equation~(\ref{pvect}) are rewritten in terms of the generic names
\beq
  p_i \; = \; \left( p^0_i(s),~p^x_i(s),~p^y_i(s),~p^z_i(s) \right)~~.
  \label{pinames}
\eeq
All possible scalar products built from the four-vectors $k_1, k_2,
p_1,p_2,p_3$, and $p_4$ can now be easily expressed in terms of the
phase space parameters and the particle masses. Scalar products
involving $v_1$~or $v_2$~are derived from $v_1=p_1+p_2$~and
$v_2=p_3+p_4$:
\ba
  v_1 & = & \left( \rule[0cm]{0cm}{.95cm}
              \frac{s+\SONE-\STWO}{2\sqrt{s}},~0,~0,~~
              \frac{\sqrt{\lambda(s,\SONE,\STWO)}}{2\sqrt{s}}
            \right)~~, \nl
  v_2 & = & \left( \rule[0cm]{0cm}{.95cm}
               \frac{s-\SONE+\STWO}{2\sqrt{s}},~0,~0,~
              -\frac{\sqrt{\lambda(s,\SONE,\STWO)}}{2\sqrt{s}}
            \right)~~.
\ea
Subsequently, the fermion masses
$m_1$~and $m_2$ are neglected. The initial state electron mass $m_e$
is neglected wherever reasonable which is called the
``ultrarelativistic approximation'' (URA). The following quantities
are often used throughout the calculation:
\ba
  \lambda \; \equiv \; \lambda(s,\SONE,\STWO)~~, \hspace{1cm}
  \sigma  \; \equiv \; \SONE+\STWO~~, \hspace{1cm}
  \delta  \; \equiv \; \SONE-\STWO~~, \nl \nl
  t_{min} \; \equiv \; \frac{1}{2} \left( s-\sigma-\SLAM \right)~~,
  \hspace{1.5cm}
  t_{max} \; \equiv \; \frac{1}{2} \left( s-\sigma+\SLAM \right)~~.
  \hspace{.4cm}
\ea
%
%
\vspace{1cm}
\section{The 2$\,\to\,$5 Particle Phase Space}
\label{ps2to5}
In the case of initial state photon bremsstrahlung a fifth
particle appears in the final state. The five-particle phase space has
eleven kinematical degrees of freedom which are chosen as
\vspace{.3cm} \\
\begin{tabular}{rcl}
  $\!\phi$        & : & Azimuth angle around the \epl~beam direction
                        $\vec{k}_2$. \\
  $\!\theta$      & : & Polar (scattering) angle of the photon with
                        respect to the \epl ~direction $\vec{k}_2$ in
                        \\
                  &   & the center of mass frame. \\
  $\!\phi_R$      & : & Azimuth angle around the photon direction
                        $\vec{p}$.\\
  $\!\theta_R$    & : & Polar angle of $\vec{v}_1$ in the
                        $(\vec{v}_1\!+\!\vec{v}_2)$ rest frame with the
                        z axis along $(\vec{v}_1\!+\!\vec{v}_2)$. \\
  $\!\phi_{12}$   & : & Azimuth angle around the $\vec{v}_1$
                        direction.\\
  $\!\theta_{12}$ & : & Decay polar angle of $\vec{p}_1$ in the
                        $\vec{v}_1$ rest frame with the z axis along
                        $\vec{v}_1$.\\
  $\!\phi_{34}$   & : & Azimuth angle around the $\vec{v}_2$
                        direction.\\
  $\!\theta_{34}$ & : & Decay polar angle of $\vec{p}_3$ in the
                        $\vec{v}_2$ rest frame with the z axis along
                        $\vec{v}_2$.\\
  $\!\SONE$       & : & Invariant mass of the final state fermion pair
                        \fone\bfone~:~~$\SONE= -v_1^2=-(p_1+p_2)^2$. \\
  $\!\STWO$       & : & Invariant mass of the final state fermion pair
                        \ftwo\bftwo~:~~$\STWO= -v_2^2=-(p_3+p_4)^2$. \\
  $\!s'$          & : & Reduced center of mass energy squared. This is
                        equivalent to the invariant \\
                  &   & mass of the final state four-fermion system: \\
                  &   & $s'= -(v_1+v_2)^2 = -(p_1+p_2+p_3+p_4)^2$.
\end{tabular}
\vspace{.4cm} \\
The symbol $p$ denotes the four-momentum of the photon radiated from
the initial state. The kinematics of a 2$\,\to\,$5 particle
reaction is graphically illustrated in figure~\ref{ps25}. The center
of mass energy squared is given by $s = -(k_1+k_2)^2 =
-(p+p_1+p_2+p_3+p_4)^2$. The angle $\theta_R$~corresponds to the angle
$\varth$~in the $2\!\to\!4$~phase space described in
appendix~\ref{ps2to4}, where the two boson rest frame and the center
of mass frame of the \ee~annihilation are identical. Due to the
radiated photon, the two boson rest frame and the \ee~center of mass
frame do not coincide for the $2\!\to\!5$~phase space.
%
%
\begin{figure}[t]
  \vspace{12.5cm}
\caption[The 2$\to$5 body phase space]
{\it Graphical representation of a 2$\,\to\,\!$5 particle reaction. In
  general, the vectors
  $\vec{v}_1,\vec{v}_2,\vec{p}_1,\vec{p}_2,\vec{p}_3,$~and
  $\vec{p}_4$ do not lie in the plane of the picture. In the figure,
  the angle $\theta$~is called $\theta_\gamma$ to emphasize that it
  belongs to the direction of the radiated photon.}
\label{ps25}
\vspace{.5cm}
\end{figure}
%
The five-particle Lorentz invariant phase space is parametrized by a
decomposition into subsequently decaying compounds,
\ba
  d\Gamma_5 & = & \frac{1}{(2\pi)^{14}} \: \frac{s-s'}{8s} \:
                  \frac{\sqrt{\lambda(s',\SONE,\STWO)}}{8s'} \:
                  \frac{\sqrt{\lambda(\SONE,m_1^2,m_1^2)}}{8\SONE} \:
                  \frac{\sqrt{\lambda(\STWO,m_2^2,m_2^2)}}{8\STWO}
                  \times \nl
            &   & \hspace{2cm} ds' \; d\SONE \; d\STWO \;
                  d\!\mycos\theta \; d\phi_{R} \, d\!\mycos\theta_{\!R} \;
                  d\phi_{12} \, d\!\mycos\theta_{12} \;
                  d\phi_{34} \, d\!\mycos\theta_{34}
\label{ph25el}
\vspace{.3cm}
\ea
with phase space limits \vspace{.3cm} \\
$\begin{array}{cccccc}
  \hspace{.75cm}
  & \left( 2\,m_1 + 2\,m_2 \right)^2 & \leq & s' & \leq & s \\
  & 4\,m_1^2 & \leq & \SONE & \leq & \left( \sqrt{s'} - 2\,m_2
                                     \right)^2 \\
  & 4\,m_2^2 & \leq & \STWO & \leq & \left( \sqrt{s'} - \sqrt{\SONE}
                                     \right)^2 \\
  & -1 & \leq & \mycos\theta & \leq & +1       \\
  &~~0 & \leq & \phi_{\{R,12,34\}} & \leq & 2 \pi \\
  & -1 & \leq & \mycos\theta_{\{R,12,34\}} & \leq & \hspace{1.4cm} +1
  \hspace{1.3cm} ,
\end{array}$
\vspace{-.85cm}
\ba
  \label{par25a}
\ea
or alternatively \vspace{.3cm} \\
$\begin{array}{cccccc}
  \hspace{.75cm}
  & 4\,m_1^2 & \leq & \SONE & \leq & \left( \sqrt{s} - 2\,m_2
                                     \right)^2 \\
  & 4\,m_2^2 & \leq & \STWO & \leq & \left( \sqrt{s} - \sqrt{\SONE}
                                     \right)^2 \\
  & \left( \sqrt{\SONE} + \sqrt{\STWO} \right)^2 & \leq & s' & \leq &
  s \\
  & -1 & \leq & \mycos\theta & \leq & +1       \\
  &~~0 & \leq & \phi_{\{R,12,34\}} & \leq & 2 \pi \\
  & -1 & \leq & \mycos\theta_{\{R,12,34\}} & \leq & \hspace{1.4cm} +1
  \hspace{1.3cm} .
\end{array}$
\vspace{-.85cm}
\ba
  \label{par25b}
\ea
In the semi-analytical bremsstrahlung \xsec calculation, after
integrating over $\phi_{12},~\theta_{12}$~and
$\phi_{34},~\theta_{34}$~with invariant tensor integration, the
non-trivial angular integrations over $\phi_R,~\theta_R,$~and $\theta$
are performed analytically. The invariant masses $\SONE,~\STWO,$~and
$s'$~are then integrated numerically, using the
parametrization~(\ref{par25b}).
Below, the four-vectors of the five-particle final state are expressed
in terms of the variables of the phase space~(\ref{ph25el}). Vectors
are given for the center of mass frame with the Cartesian coordinate
system of figure~\ref{ps25}. Similar to equation~(\ref{kvect}),
$k_1$~and $k_2$ are represented by
\vspace{.3cm} \\
$\begin{array}{lll}
  k_1 & = & \left( k^0,~k\,\mysin\theta,~0,~-k\,\mycos\theta \right)
  \\
  k_2 & = & \left( k^0,~-k\,\mysin\theta,~0,~k\,\mycos\theta \right)
\end{array}$
\hspace{1cm} with \hspace{1cm}
$\begin{array}{lll}
  k^0 & = & \frac{\sqrt{s}}{2} \vspace{.1cm} \\
  k   & = & \frac{\sqrt{s}}{2} \, \beta \vspace{.1cm} \\
  \beta & \equiv & \sqrt{1-\frac{4\MES}{s}}~~.
\end{array}$
\vspace{-1.45cm}
\ba
  \label{k5vect}
  \\
  \nonumber
\ea
Secondly, one finds
\beq
  p \; =\; (p^0,~0,~0,~p^0) \hspace{1.3cm} {\rm with} \hspace{1.3cm}
  p^0 \; =\; \frac{s-s'}{2\sqrt{s}}
  \label{photmom}
\eeq
for the four-momentum of the photon radiated along the $z$ axis. The
momenta of the final state fermions can be written down in terms of
the expressions originally introduced for the four-particle final
state in equation~(\ref{pinames}):
\ba
  p_i & = & \left(\!\!
  \begin{array}{c}
    \gamma^{(0)}\,p^0_i(s') - \gamma \left( \rule[0cm]{0cm}{.35cm}
      p^z_i(s')\mycos\theta_R - p^x_i(s')\mysin\theta_R \right) \\
    \left( \rule[0cm]{0cm}{.35cm}
      p^z_i(s')\mysin\theta_R + p^x_i(s')\mycos\theta_R \right)
      \mycos\phi_R  + p^y_i(s')\mysin\phi_R \\
    - \left( \rule[0cm]{0cm}{.35cm}
      p^z_i(s')\mysin\theta_R + p^x_i(s')\mycos\theta_R \right)
      \mysin\phi_R  + p^y_i(s')\mycos\phi_R \\
    \gamma^{(0)} \left( \rule[0cm]{0cm}{.35cm}
      p^z_i(s')\mycos\theta_R - p^x_i(s')\mysin\theta_R \right)
      - \gamma\,p^0_i(s')
   \end{array}
  \!\!\right)
  \label{5pvect}
\ea
with
\ba
  \gamma^{(0)} \; = \; \frac{s+s'}{2\sqrt{ss'}}~~, \hspace{2cm}
  \gamma   \; = \; \frac{s-s'}{2\sqrt{ss'}}~~.
  \label{kingamma}
\ea
In equation~(\ref{5pvect}) the components of the momentum four-vector have
been written in a column for reasons of
clarity. Equation~(\ref{5pvect}) is better understood, if one notes
that, in the $p_1+p_2+p_3+p_4=v_1+v_2$~rest frame, the fermion momenta
are given by
\beq
  p_i \; = \; \left( p^0_i(s'),~p^x_i(s'),~p^y_i(s'),~p^z_i(s') \right)~~,
  \label{pirestnames}
\eeq
where, compared to equation~(\ref{pinames}), only the center of mass
energy had to be adjusted due to photon radiation. After rotating and
boosting to the five-particle center of mass frame,
equation~(\ref{5pvect}) is obtained.
\para
The intermediate boson momenta $v_1$~and $v_2$~are given by
expressions similar to those in equation~(\ref{pvect}),
\ba
  v_1 & = & \left( \gamma^{(0)}\,p_{12}^{R,0} - \gamma \cp_R \mycos\theta_R,~
                   \cp_R \mysin\theta_R \mycos\phi_R,~
                  -\cp_R \mysin\theta_R \mysin\phi_R, \right. \nl
    & & \hspace{7.7cm} \left.
                   \gamma^{(0)} \cp_R \mycos\theta_R - \gamma \,p_{12}^{R,0}
            \right)~~, \nl
  v_2 & = & \left( \gamma^{(0)}\,p_{34}^{R,0} + \gamma \cp_R \mycos\theta_R,~
                  -\cp_R \mysin\theta_R \mycos\phi_R,~
                   \cp_R \mysin\theta_R \mysin\phi_R, \right. \nl
    & & \hspace{7.7cm} \left.
                  -\gamma^{(0)} \cp_R \mycos\theta_R - \gamma \,p_{34}^{R,0}
            \right)\,.
  \label{v5vect}
\ea
with
\ba
  \cp_R \: & = &
    \frac{\sqrt{\lambda(s',\SONE,\STWO)}\,}{2\sqrt{s'}} \nl \nl
  p_{12}^{R,0} & = & \frac{s'+\SONE-\STWO}{2\sqrt{s'}}~~, \hspace{3cm}
  p_{34}^{R,0} \; = \; \frac{s'-\SONE+\STWO}{2\sqrt{s'}}~~.
  \label{kinpijr0}
\ea
The four-momentum of the two-boson system recoiling against the
radiated photon is given by
\ba
  (v_1+v_2) & = &
    \left( \gamma^{(0)} \sqrt{s'},~0,~0,~-\gamma \sqrt{s'}\,\right) \; = \;
    \left( \frac{s+s'}{2\sqrt{s}},~0,~0,~\frac{s-s'}{2\sqrt{s}} \right)
\ea
Once more, some derived kinematical quantities are defined for use
throughout the calculation:
\ba
  \sprp & \equiv \; s + s'~~, \hspace{2cm} &
  \sprm \; \equiv \; s - s' \nl
  \lambda' & \equiv \; \lambda(s',\SONE,\STWO)~~, \hspace{1cm} &
  {\bar \lambda} \; \equiv \; \lambda(\sprm,-\SONE,-\STWO) \; =
  \; {\sprm}^2 + 2\sprm\sigma + \delta^2~~.
\ea
\\
Finally, a short remark on the integration of phase space in the above
$2\!\to\!4$~and $2\!\to\!5$ parametrizations is due: As all angular
phase space variables are integrated analytically, and as the remaining
invariant masses are integrated numerically without the use of Monte
Carlo techniques, high precision numerical results can be obtained.
%
\chapter{The Cross-Section: Matrix Elements, Details, and Techniques}
\label{xscalc}
In this appendix, matrix elements for process~(\ref{eezz4f}) will be
presented. This includes the Born, the initial state bremsstrahlung,
and the virtual initial state QED matrix elements. Together with the
matrix elements, the analytical integrations of the angular variables
are outlined. Results for the differential cross-sections
obtained from these integrations are presented. Unless stated
otherwise the ultrarelativistic approximation is used, that is the
electron mass is neglected wherever possible.
Correspondingly, final state masses are
neglected too. A short account is given of the soft photon resummation
technique which was applied to the \oal~results. Generalization of the
below results to process~(\ref{eeNC4f}) means to add photonic Feynman
diagrams with equal topology, as explained in chapter~\ref{xsZZGG}.
The necessary changes to couplings and propagators are most
efficiently implemented with equation~(\ref{ZZtoNC}).
\vspace{.5cm}
%
%
\section{Born Level Results}
\label{bornres}
At Born level, the two Feynman diagrams given in
figure~\ref{zzfeyn} represent the amplitude for the \zz~pair
production process~(\ref{eezz4f}). According to the Feynman
rules from appendix~\ref{feynrules}, using the four-particle phase
space parametrization as in appendix~\ref{ps2to4}, and neglecting the
masses of the initial state electrons and final state fermions, the
Born t-channel matrix element is
\beq
  \cm_t^B \; = \; \frac{g_{\beta \beta'}}{\,D_Z(\SONE)\,}\:
                  \frac{g_{\alpha \alpha'}}{\,D_Z(\STWO)\,}\;
                  \frac{1}{\,q_t^2+\MES-\ieps\,}\;
                  M_{12}^{\beta'}\,M_{34}^{\alpha'}\,B^{\alpha \beta}_t
  \label{tmat}
\eeq
with
\ba
  M_{ij}^\mu & = & {\bar u}(p_i) \cdot \gamma^\mu
    \left[  L_{f\!{\bar f}\!Z} \!\cdot\!\left( 1+\gamma_5 \right)
          + R_{f\!{\bar f}\!Z}\!\cdot\!\left( 1-\gamma_5 \right)
    \right] \cdot u(-p_j) \nl
  B^{\alpha \beta}_t & = & {\bar u}(-k_2) \cdot \gamma^\alpha
    \left[  L_{\EMI\!\EPL\!Z} \!\cdot\!\left( 1+\gamma_5 \right)
          + R_{\EMI\!\EPL\!Z} \!\cdot\!\left( 1-\gamma_5 \right)
    \right] \cdot \dagg{q}_{\!t} \; \times \nl
    & & \hspace{1.57cm} \gamma^\beta
    \left[  L_{\EMI\!\EPL\!Z} \!\cdot\!\left( 1+\gamma_5 \right)
          + R_{\EMI\!\EPL\!Z} \!\cdot\!\left( 1-\gamma_5 \right)
    \right] \cdot u(k_1) \nl
  & = & 2\,{\bar u}(-k_2) \cdot \gamma^\alpha
    \left[  {L_{\EMI\!\EPL\!Z}}^2 \!\cdot\!\left( 1+\gamma_5 \right)
          + {R_{\EMI\!\EPL\!Z}}^2 \!\cdot\!\left( 1-\gamma_5 \right)
    \right] \cdot \dagg{q}_{\!t} \, \gamma^\beta \cdot u(k_1)~~~.
  \label{bornmat}
\ea
Spin indices are suppressed. The electron couplings to the \zz~boson are
\ba
  L_{\EMI\!\EPL\!Z} & = & - \frac{e}{\,4\,\SIW\,\COW} \,
    \left( 1 - 2 \SWS \right) \nl
  R_{\EMI\!\EPL\!Z} & = & \hspace{.33cm}
    \frac{e}{\,2\,\SIW\,\COW} \; \SWS~~,
  \label{lrcoup}
\ea
fermion couplings are derived from equation~(\ref{LRcoup}), and the
t-channel electron momentum is given by
\ba
  q_t & = & k_1 - v_1 \; = \; -k_2 + v_2  \; = \;
    \frac{1}{2} \left( k_1 - k_2 - v_1 + v_2\right) \nl
      & = & k_1 - p_1 - p_2 \; = \; -k_2 + p_3 + p_4 \; = \;
    \frac{1}{2} \left( k_1 - k_2 - p_1 - p_2 + p_3 + p_4 \right)~~.
\ea
Using the representation of four-vectors in terms of phase space
parameters from appendix~\ref{ps2to4} and the ultrarelativistic
approximation (URA) one arrives at
\ba
  t & \equiv & q_t^2+\MES \; \stackrel{\rm URA}{=} \;
    \frac{1}{2} \left[ A + B \mycos\varth \right]~~, \nl
  & & A \;\: = \; s-\SONE-\STWO~~, \nl
  & & B \;\, = \; \SLAM ~~.
  \label{bornt}
\ea
The center of mass energy squared is given by $s$, the invariant
\zz~masses are $\SONE$~and $\STWO$, $\varth$~is the \zz~scattering
angle, and ~$\lambda = \lambda(s,\SONE,\STWO) = s^2 + \SONE^2 + \STWO^2
- 2s\SONE - 2s\STWO - 2\SONE\STWO$. See appendix~\ref{ps2to4} for
further details. Finite boson widths are introduced into the
\zz~propagators $D_Z(s)$ from equation~(\ref{bornmat}),
\beq
  D_Z(s_{ij}) \; \equiv \; s_{ij} - M_Z^2 + \ri \sqrt{s_{ij}} \,
  \Gamma_Z(s_{ij})~~,
  \label{propagator}
\eeq
where the \zz~width is given by
\beq
\Gamma_Z (s_{ij}) =
  \frac{G_{\mu}\, M_Z^2} {\,24\pi \sqrt{2}\,} \, \sqrt{s_{ij}} \,
  \sum_f \left( {g^V_{f\!{\bar f}\!Z}}^2 + {g^A_{f\!{\bar f}\!Z}}^2
         \right)\!\cdot\!N_c(f)~~.
  \label{gzoff}
\eeq
with the color factor $N_c(f)$. For the Born u-channel amplitude one
obtains
\ba
  \cm_u^B & = & \frac{g_{\alpha \alpha'}}{\,D_Z(\SONE)\,}\:
                \frac{g_{\beta \beta'}}{\,D_Z(\STWO)\,}\;
                \frac{1}{\,q_u^2+\MES-\ieps\,} \; M_{34}^{\beta'}\,
                M_{12}^{\alpha'}\,B^{\alpha \beta}_u~~, \nl \nl
  B^{\alpha \beta}_u & = &
    2\,{\bar u}(-k_2) \cdot \gamma^\alpha
    \left[  {L_{\EMI\!\EPL\!Z}}^2 \!\cdot\!\left( 1+\gamma_5 \right)
          + {R_{\EMI\!\EPL\!Z}}^2 \!\cdot\!\left( 1-\gamma_5 \right)
    \right] \cdot \dagg{q}_{\!\!u} \, \gamma^\beta \cdot u(k_1)~~, \nl
  q_u & = & k_1 - v_2  \; = \; -k_2 + v_1  \; = \;
    \frac{1}{2} \left( k_1 - k_2 + v_1 - v_2 \right) \nl
      & = & k_1 - p_3 - p_4 \; = \; -k_2 + p_1 + p_2 \; = \;
    \frac{1}{2} \left( k_1 - k_2 + p_1 + p_2 - p_3 - p_4 \right) \nl
  u  & = & q_u^2+\MES \; \stackrel{\rm URA}{=} \;
    \frac{1}{2} \left[ A - B \mycos\varth \right]~~.
  \label{umat}
\ea
It is appropriate to mention that the t-channel matrix element
$\cm_t^B$ and the u-channel matrix element $\cm_u^B$ are obtained from
each other by the interchanges
\beq
  v_1 \leftrightarrow v_2~~, \hspace{2cm}
  \cm_{12}^\mu \leftrightarrow \cm_{34}^\mu~~.
  \label{bornmatsym}
\eeq
After squaring the matrix element $\,\cm^B\!=\!\cm^B_t\!+\!\cm^B_u$,
application of Feynman rule 8 from appendix~\ref{feynrules}, use of
the phase space parametrization described in appendix~\ref{ps2to4},
and fivefold analytical integration over the angular degrees of
freedom of the four-particle phase space with the help of the
integrals listed in appendices~\ref{tenint} and~\ref{bornint} yields
the unpolarized cross-section
\ba
  \sigma^B(s) & = &
    \int\limits_{4m_1^2}^{(\sqrt{s} - 2m_2)^2} d\SONE \, \rho_Z(\SONE)
    \int\limits_{4m_2^2}^{(\sqrt{s} - \sqrt{\SONE})^2} d\STWO \, \rho_Z(\STWO)
    \; \times \nl & & \hspace{5cm}
    \sigma^B_4(s;\SONE,\STWO) \times 2 \cdot B\!R(1) \cdot B\!R(2)
  \label{sigzz}
\ea
with branching ratios $B\!R(1)$~and $B\!R(2)$ of the \zz~boson into
the fermion-antifermion pairs \fone\bfone~and \ftwo\bftwo~
respectively, and with
\ba
  \rho_Z(s_{ij}) & = & \frac{1}{\pi}
    \frac {\sqrt{s_{ij}} \, \Gamma_Z (s_{ij})}
          {|s_{ij} - M_Z^2 + i \sqrt{s_{ij}} \, \Gamma_Z (s_{ij}) |^2}~~~~
     -\!\!\!-\!\!\!-\!\!\!\!
      \longrightarrow_{\hspace{-1cm}_{\Gamma_Z \rightarrow 0}}
      ~~~~\delta (s_{ij} - M_Z^2)~~. \nonumber
  \label{rhoz}
\ea
A particularly convenient form for ~$\sigma^B_4(s;\SONE,\STWO)$ ~is
obtained by using
$L_e \equiv L_{\EMI\!\EPL\!Z}/(g_Z g_2)$ $= - ( 1 - 2 \SWS )$~~and~
$R_e \equiv R_{\EMI\!\EPL\!Z}/(g_Z g_2) = 2 \SWS $:
\ba
  \sigma^B_4(s;\SONE,\STWO) & = & \frac{\sqrt{\lambda}}{\,\pi\,s^2} \,
    {\ds \frac{\left(G_{\mu} M_Z^2 \right)^2}{8}}
      \left(L_e^4+R_e^4\right) {\cal G}_4^{t+u}(s;\SONE,\STWO)~~, \nl
  {\cal G}_4^{t+u}(s;\SONE,\STWO) & = &
    \frac{\,s^2+(\SONE+\STWO)^2\,}{s-\SONE-\STWO}\,\cl_B - 2~~, \nl
  {\cal L}_B & = & {\cal L}(s;\SONE,\STWO) \; = \;
    \frac{1}{\sqrt{\lambda}} \,
    \ln \frac{s-\SONE-\STWO+\sqrt{\lambda}}
                                     {s-\SONE-\STWO-\sqrt{\lambda}}~~.
\label{sigzz4}
\ea
${\cal G}_4^{t+u}(s;\SONE,\STWO)$ decomposes into three components
stemming from the t-channel the u-channel and their interference:
\ba
  {\cal G}_4^{t+u}(s;\SONE,\STWO) & = & {\cal G}_4^t + {\cal G}_4^u +
                                        {\cal G}_4^{tu} \nonumber
\ea
\ba
  {\cal G}_4^t & = & {\cal G}_4^u \;\; = \;\;
    \frac{1}{8\,\SONE\STWO} \left[ \rule[0cm]{0cm}{.6cm}
      \frac{\lambda}{6} + 2s(\SONE\!+\!\STWO) - 8\SONE\STWO +
      4\SONE\STWO(s\!-\!\SONE\!-\!\STWO){\cal L}_B \right] \nl
  {\cal G}_4^{tu} & = & \hspace{1.09cm} - \, \frac{1}{8\,\SONE\STWO}
    \left[ \rule[0cm]{0cm}{.7cm} \frac{\lambda}{3} +
      4s(\SONE\!+\!\STWO) +
      16\,s\,\SONE\STWO \left( \rule[0cm]{0cm}{.6cm}
      1 - \frac{s}{s\!-\!\SONE\!-\!\STWO} \right)
      {\cal L}_B \right]~~. \nl
  \label{borncon}
\ea
To obtain the analytical result in equation~(\ref{sigzz4}), use
was made of the representation of the processes Lorentz invariants in
terms of phase space variables as outlined in
appendix~\ref{ps2to4}. Throughout this thesis, algebraic manipulations
were carried out with the help of SCHOONSCHIP~\cite{schoonschip},
FORM~\cite{form}, and Mathematica~\cite{mathematica}.
\clearpage
%
%
\section{Virtual Initial State Corrections}
\label{virtres}
Applying on-shell renormalization, the Feynman diagrams for the
virtual initial state radiation to process~(\ref{eezz4f}) are shown in
figure~\ref{zzvirt}. The ``amputated'' diagrams presented there must
be connected to both the t- and the u-channel, resulting in a total
number of eight diagrams. For our calculation, it is sufficient
to use the t-channel virtual matrix element and compute its
interference with the t- and the u-channel Born matrix elements. The
interferences of the virtual u-channel with the Born matrix elements
are then obtained by symmetry arguments as outlined at the end of this
section. The virtual t-channel matrix element is obtained by
substituting $B^{\alpha \beta}_t\!/(q_t^2+\MES-\ieps) \to V^{\alpha
  \beta}_t$ in equation~(\ref{tmat}):
\beq
  \cm_t^V \; = \; \frac{g_{\beta \beta'}}{\,D_Z(\SONE)\,}\:
                  \frac{g_{\alpha \alpha'}}{\,D_Z(\STWO)\,}\;
                  M_{12}^{\beta'}\,M_{34}^{\alpha'}\,V^{\alpha
                  \beta}_t~~.
\eeq
Using $p$ as the loop photon momentum and applying dimensional
regularization~\cite{marciano75,thooft72}, one finds
\ba
  V^{\alpha\beta}_t & = & e^2 \cdot \left\{ \,
    \frac{\,V^{\alpha\beta}_{t,CT}\,}{16\pi^2} \;\; - \;\;
    \ri \left. \, \mu^{(4-n)} \! \int \hspace{-.1cm} \frac{d^n p}{(2\pi)^n}
    \left( V^{\alpha\beta}_{t,v1} + V^{\alpha\beta}_{t,v2} +
           V^{\alpha\beta}_{t,self}  + V^{\alpha\beta}_{t,box} \right)
    \, \right|_{\mu=m_e} \: \right\}~~\raisebox{-.2cm}[0cm][0cm]{.}
    \nl
\ea
The counterterm part $V^{\alpha\beta}_{t,CT}$ is given by
\ba
  V^{\alpha\beta}_{t,CT} & = &
    \frac{B^{\alpha \beta}_t}{q_t^2+\MES-\ieps}
    \left( \rule[0cm]{0cm}{.35cm} C_{v1} + C_{v2} + C_{self} \right) \nl
    & = & \frac{B^{\alpha \beta}_t}{q_t^2+\MES-\ieps}
    \left( \rule[0cm]{0cm}{.35cm} 2 {\rm P} + 4 {\rm P^{IR}} - 4
    \right)~~.
    \label{ctpart}
\ea
The counterterms $C_{v1}$~and $C_{v2}$~for the vertex corrections and
$C_{self}$~for the electron propagator self energy have been derived
in appendix~\ref{renorm}.
\ba
  C_{v1} \;\; & = & C_{v2}
         \;\; = \;\; {\textstyle \frac{16\pi^2}{e^2}} \, \dze
         \;\; = \hspace{.44cm} 2 {\rm P} + 4 {\rm P^{IR}} - 4 \nl
  C_{self}    & = &
         \hspace{1.25cm} - \, {\textstyle \frac{16\pi^2}{e^2}} \, \dzp
         \hspace{.16cm} = \; -2 {\rm P} - 4 {\rm P^{IR}} + 4
  \label{mcts}
\ea
The ultraviolet and infrared poles {\rm P} and ${\rm P^{IR}}$ are
given in equations~(\ref{poleUV}) and~(\ref{poleIR}) of
appendix~\ref{loopint1}.
The matrix elements $V^{\alpha\beta}_{t,v1}, V^{\alpha\beta}_{t,v2},$
$V^{\alpha\beta}_{t,self},$~and $V^{\alpha\beta}_{t,box}$ belong to the
positron and the electron vertex corrections, to the intermediate
electron self energy, and to the box correction as depicted in
figure~\ref{zzvirt}. In the ultrarelativistic approximation one finds
\ba
  V^{\alpha\beta}_{t,v1} & = & {\bar u}(-k_2) \cdot
    \frac{ \left( -2 k_2^\mu - \gamma^\mu \dagg{p} \right)
           \gamma^\alpha \left(\dagg{q}_{\!t} - \dagg{p} \right)
           \gamma_\mu \, \dagg{q}_{\!t} \, \gamma^\beta }
         { \left[ \rule[0cm]{0cm}{.35cm} p^2-\ieps \right]
           \left[ \rule[0cm]{0cm}{.35cm} (p - q_t)^2 +\MES-\ieps \right]
           \, (p^2 + 2 k_2 p) \,
           \left[ \rule[0cm]{0cm}{.35cm} q_t^2+\MES-\ieps \right]} \times
    \nl & & \hspace{4cm}
    2 \left[  {L_{\EMI\!\EPL\!Z}}^2 \!\cdot\!\left( 1+\gamma_5 \right)
            + {R_{\EMI\!\EPL\!Z}}^2 \!\cdot\!\left( 1-\gamma_5 \right)
      \right] \cdot u(k_1)~~~, \nl
  V^{\alpha\beta}_{t,v2} & = & {\bar u}(-k_2) \cdot
    \frac{ \gamma^\alpha \, \dagg{q}_{\!t} \, \gamma^\mu
           \left(\dagg{q}_{\!t} - \dagg{p} \right) \gamma^\beta
           \left( 2 k_{1,\mu} - \dagg{p} \gamma_\mu \right) }
         { \left[ \rule[0cm]{0cm}{.35cm} p^2-\ieps \right]
           \left[ \rule[0cm]{0cm}{.35cm} (p - q_t)^2 +\MES-\ieps \right]
           \, (p^2 - 2 k_1 p) \,
           \left[ \rule[0cm]{0cm}{.35cm} q_t^2+\MES-\ieps \right]} \times
    \nl & & \hspace{4cm}
    2 \left[  {L_{\EMI\!\EPL\!Z}}^2 \!\cdot\!\left( 1+\gamma_5 \right)
            + {R_{\EMI\!\EPL\!Z}}^2 \!\cdot\!\left( 1-\gamma_5 \right)
      \right] \cdot u(k_1)~~~, \nl \nl
  V^{\alpha\beta}_{t,self} & = & {\bar u}(-k_2) \cdot
    \frac{ \gamma^\alpha \, \dagg{q}_{\!t} \, \gamma^\mu
           \left(\dagg{q}_{\!t} - \dagg{p} \right) \gamma_\mu \,
           \dagg{q}_{\!t} \, \gamma^\beta }
         { \left[ \rule[0cm]{0cm}{.35cm} p^2-\ieps \right]
           \left[ \rule[0cm]{0cm}{.35cm} (p - q_t)^2 +\MES-\ieps \right]
           \left[ \rule[0cm]{0cm}{.35cm} q_t^2+\MES-\ieps \right]^2 }
    \times \nl & & \hspace{4cm}
    2 \left[  {L_{\EMI\!\EPL\!Z}}^2 \!\cdot\!\left( 1+\gamma_5 \right)
            + {R_{\EMI\!\EPL\!Z}}^2 \!\cdot\!\left( 1-\gamma_5 \right)
      \right] \cdot u(k_1)~~~, \nl \nl
  V^{\alpha\beta}_{t,box} & = & {\bar u}(-k_2) \cdot
    \frac{ \left( -2 k_2^\mu - \gamma^\mu \dagg{p} \right)
           \gamma^\alpha \left(\dagg{q}_{\!t} - \dagg{p} \right)
           \gamma^\beta
           \left( 2 k_{1,\mu} - \dagg{p} \gamma_\mu \right) }
         { \left[ \rule[0cm]{0cm}{.35cm} p^2-\ieps \right]
           (p^2 - 2 k_1 p) \, (p^2 + 2 k_2 p) \,
           \left[ \rule[0cm]{0cm}{.35cm} (p - q_t)^2 +\MES-\ieps \right] }
    \times \nl & & \hspace{4cm}
    2 \left[  {L_{\EMI\!\EPL\!Z}}^2 \!\cdot\!\left( 1+\gamma_5 \right)
            + {R_{\EMI\!\EPL\!Z}}^2 \!\cdot\!\left( 1-\gamma_5 \right)
      \right] \cdot u(k_1)~~~. \nonumber
\ea
Taking into account that the virtual u-channel matrix element
$\cm_u^V$ is obtained from $\cm_t^V$ with the help of the symmetry
relation~(\ref{bornmatsym}) one can now write down the virtual \oal~
corrections to process~(\ref{eezz4f}) as the interference between the
Born and the virtual matrix elements. Using the four-particle phase
space parametrization from equation~(\ref{gamma4}),
appendix~\ref{ps2to4}, one arrives at
\ba
  \sigma^V(s) & = & \frac{(2\pi)^4}{\,2\,s\,} \,
    \int \!\! d\Gamma_4 \;\;
    \frac{1}{4} \! \sum_{spins} 2 \, \RE \! \left[ \rule[0cm]{0cm}{.6cm}
    \! \left( \cm_t^V+\cm_u^V \right)
    \left( {\cm_t^B}^{*}+{\cm_u^B}^{*} \right) \right]\,.
  \label{virtxsec}
\ea
The evaluation of equation~(\ref{virtxsec}) is carried out in four
steps. First step: The final state fermion decay angles are integrated
with the invariant tensor integration formula~(\ref{dectenint}) from
appendix~\ref{tenint}. Second step: The result of the first step is
used as input for a computer algebra program written in
FORM~\cite{form}. Inside the FORM program, steps three and four are
carried out together with measures to simplify complicated
intermediate expressions. Among the tools used to achieve
simplifications, partial fraction decompositions played an important
r\^{o}le. Third step: The photon loop momentum $p$ is
integrated over, making use of the integrals listed in
appendix~\ref{loopint1}. In the result of the third step, ultraviolet
divergences from loop integrals and counterterms are explicitly
cancelled. Fourth step: The integration over the boson scattering
angle $\varth$~is performed with the help of the integrals given in
appendix~\ref{loopint2}.
As a result of the above four steps, one can express the cross-section
$\sigma^V$~by a convolution similar to equation~(\ref{sigzz})
\ba
  \sigma^V(s) & \equiv & \sigma^V_{uni} \; + \; \sigma^V_{nonuni} \nl \nl
    & = & \frac{\alpha}{\pi} \cdot \delta_{virtual}^{\rm \,IR} \cdot
    \sigma^B(s) \hspace{.5cm} + \nl \nl
  & & \frac{\alpha}{\pi} \!\cdot\!
    \frac{\,\left(G_{\mu} M_Z^2 \right)^2\,}{8\pi s}
    \left(L_e^4 \!+\! R_e^4\right)
    \times 2 \cdot B\!R(1) \cdot B\!R(2) \times \nl
  & & \hspace{1.8cm} \int\limits_{4m_1^2}^{(\sqrt{s} - 2m_2)^2}
    \hspace{-.7cm} d\SONE \, \rho_Z(\SONE)
    \int\limits_{4m_2^2}^{(\sqrt{s} - \sqrt{\SONE})^2} \hspace{-.7cm}
    d\STWO \, \rho_Z(\STWO) \; \sigma^V_{4,nonuni}(s;\SONE,\STWO) \nl
    \nl \nl
  \delta_{virtual}^{\rm \,IR} & = & \RE \left\{ \rule[0cm]{0cm}{.6cm}
    2\,{\rm P^{IR}} \left( 1 - l_0 \right) -
    \frac{l_0^2}{2} + \frac{3\,l_0}{2} + \frac{\pi^2}{6} - 2 \right\}
\label{xsvirt}
\ea
with $\alpha = e^2/(4\pi)$~and
\ba
 l_0 & = & \ln(s/\MES) - \ri\pi \;\; = \;\; l_\beta - \ri\pi~~.
\ea
$\delta_{virtual}^{\rm \,IR}$ is identical to the \oal~radiator obtained
for the virtual initial state QED corrections in fermion pair
production~\cite{kniehl88,berends88}. $\delta_{virtual}^{\rm \,IR}$
contains all leading logarithms of the virtual cross-section. The
infrared pole in $\delta_{virtual}^{\rm \,IR}$ is cancelled by the
soft part of the \oal~bremsstrahlung contribution. The
\oal~non-universal virtual cross-section does not contain leading
logarithms $l_\beta$, but is rather complicated. It contains four
parts corresponding to the four different matrix element products in
equation~(\ref{virtxsec}),
\vspace{.3cm}
\ba
  \sigma^V_{4,nonuni}(s;\SONE,\STWO) & = & \frac{1}{8s}
    \left( \rule[0cm]{0cm}{.35cm}
    \sigma^{V,tt}_{4,nonuni} \; + \; \sigma^{V,tu}_{4,nonuni} \; + \;
    \sigma^{V,ut}_{4,nonuni} \; + \; \sigma^{V,uu}_{4,nonuni} \right)
    \nonumber
\ea
with
\ba
  \lefteqn{ \sigma^{V,tt}_{4,nonuni} \;\; = \;\; \sigma^{V,uu}_{4,nonuni}
  \;\; = \;\; \hspace{.5cm} \frac{3\,\cl_{B3}}{2}
          \cdot \left[ \rule[0cm]{0cm}{.35cm}
          s\,l_+ + \delta\,l_- - s \left( s-\sigma \right)\,I_{12Q}
          \right] \; + \;} \nl
    & & \hspace{1cm} \, \cl_{B2} \cdot \left[ \rule[0cm]{0cm}{.35cm}
          (s-2\sigma)(s-\sigma)\,I_{12Q} +
          \frac{3\,\sigma\delta}{2\,s}\,l_- -
          (s-4\sigma)\,\frac{l_+}{2} + s - \sigma \right] \; + \; \nl
    & & \hspace{1cm} \: \cl_B \cdot \left[ \rule[0cm]{0cm}{.35cm}
          - \left( \lambda + \sigma^2 + s^2/2 \right)\,I_{12Q} -
          \frac{5\,\delta}{2}\,l_- - 4\sigma\,l_+ + 8\sigma - 9s
          \right]  \; + \; \nl
    & & \hspace{1cm} \:
          8\,\SLAM \left( \rule[0cm]{0cm}{.35cm} 2-l_+\right) \; + \;
          \left( \rule[0cm]{0cm}{.35cm} 3s-\sigma \right)\!\cdot\!
          \left( \rule[0cm]{0cm}{.35cm} 2 \cd_+ - \cl_-\,l_- \right)
  \nl \nl \nl
  \lefteqn{ \sigma^{V,tu}_{4,nonuni} \;\; = \;\; \sigma^{V,ut}_{4,nonuni}
  \;\; = \;\; \hspace{.5cm} \frac{3\,\cl_{B3}}{2}
          \cdot \left[ \rule[0cm]{0cm}{.35cm}
          s\,l_+ + \delta\,l_- - s (s-\sigma)\,I_{12Q} \right] \; + \;} \nl
    & & \hspace{1cm} \, \cl_{B2} \cdot \left[ \rule[0cm]{0cm}{.6cm}
          (s-\sigma)\,\left( \rule[0cm]{0cm}{.35cm}
            1 + 5\,s\,I_{12Q} + \frac{\delta}{2s}\,l_- \right) -
          \frac{9}{2} \left( \rule[0cm]{0cm}{.35cm} s\,l_+ +
            \delta\,l_- \right) \right] \; + \; \nl
    & & \hspace{1cm} \:\cl_B \cdot \left[ \rule[0cm]{0cm}{.7cm}
            \frac{s^2}{s-\sigma} \; \left\{ \rule[0cm]{0cm}{.6cm} \!
            -\frac{7}{3}\,\cl_B^2 - 16\,\cd_d + 4\,\cd_{d+} -
            4\,l_\sigma \left( \rule[0cm]{0cm}{.35cm}
              l_\sigma - 2\,l_+ +3/2 \right) +
            \right. \right. \nl & & \hspace{5.7cm}
            \left. \rule[0cm]{0cm}{.6cm} 4\,l_-\,l_{d-} -
            4 \left( \rule[0cm]{0cm}{.35cm} 1 - 3\pi^2/2 \right) -
            3\,l_+\left( \rule[0cm]{0cm}{.35cm} l_+ - 3 \right)
          \right\} \nl & & \hspace{2.19cm} + \;
            2\,s^2 \, \left\{ \rule[0cm]{0cm}{.6cm} \!
            \left( \rule[0cm]{0cm}{.35cm} l_\sigma - 3\,l_+/4 \right) \!\!
            \left( \rule[0cm]{0cm}{.35cm} 3\,h_{1+} + h_{2+} \right)
            + \frac{3}{4}\,l_-\,\left( \rule[0cm]{0cm}{.35cm}
              3\,h_{1-} + h_{2-} \right) + 2\,h_{1+} \right\} \nl
          & & \hspace{2.63cm} + \;
            s\, \left\{ \rule[0cm]{0cm}{.6cm} 2\,l_+ - 6\,l_\sigma -
            \frac{15}{2}\,s\,I_{12Q} - \frac{\SONE}{s-\STWO} -
            \frac{\STWO}{s-\SONE} - 3 \right\} \nl
          & & \left. \rule[0cm]{0cm}{.7cm} \hspace{2.5cm} + \;
            \sigma \left( \rule[0cm]{0cm}{.35cm} l_+ - 2\,l_\sigma \right)
            + \frac{\delta}{2}\,l_- \hspace{6.85cm} \right] \; +
            \hspace{-.4cm} \nl
    & & \hspace{1cm} \frac{8\,s^2}{s-\sigma}
          \cdot \left[ \rule[0cm]{0cm}{.7cm} \cf_{d-} - \cf_{t-} -
            \cf_\sigma + \left( \rule[0cm]{0cm}{.35cm} 3/4 - l_+/2 \right)
            \cd_\sigma + \frac{l_+}{2}\,\cd_{d-} \right] \; + \nl
    & & \hspace{1cm} s^2\,\left( \rule[0cm]{0cm}{.35cm}
          3\,h_{1+} + h_{2+} \right) \cdot \left[ \rule[0cm]{0cm}{.7cm}
            \cd_+ - 2\,\cd_\sigma - \frac{l_-}{2}\,\cl_- \right] \; + \nl
    & & \hspace{1cm} s\, \cdot \left[ \rule[0cm]{0cm}{.7cm}
          2\,\cd_+ + 6\,\cd_\sigma + l_- \left( \rule[0cm]{0cm}{.35cm}
          \SLAM\,h_{1-} - \cl_- \right) \right] \;\;\;\; + \;\;\;\;
          2\sigma\,\cd_\sigma~~.
  \label{nunivxs}
\ea
In equation~(\ref{nunivxs}) the following notations are used:
\ba
  d_{12} & = & \frac{s-\STWO}{\SONE} \hspace{5cm}
    {\rm (compare~appendix~\ref{loopint2})} \nl \nl
  d_{34} & = & \frac{s-\SONE}{\STWO} \hspace{5cm}
    {\rm (compare~appendix~\ref{loopint2})} \nl \nl
  h_{1\pm} & = & \frac{1}{s-\SONE} \; \pm \; \frac{1}{s-\STWO} \nl \nl
  h_{2\pm} & = & \frac{\STWO}{\,\left(s-\SONE\right)^{2}\,} \; \pm \;
                 \frac{\SONE}{\,\left(s-\STWO\right)^{2}\,} \nl \nl
  l_+ \; & = & \myln \frac{\SONE}{s} \; + \; \myln \frac{\STWO}{s} \nl \nl
  l_- \; & = & \myln \frac{\SONE}{s} \; - \; \myln \frac{\STWO}{s}
    \;\; = \;\; \myln \frac{\SONE}{\STWO} \nl \nl
  l_\sigma \; & = & \myln \frac{s-\sigma}{s} \nl \nl
  l_{d-} & = & \ln d_{12} \; - \; \ln d_{34} \nl \nl
  \cl_{B2} & = & \frac{\,s\,(s-\sigma)\,}{\lambda}
    \left[ \rule[0cm]{0cm}{.6cm}
           \cl_B \; - \; \frac{2\SLAM}{s-\sigma} \right] \nl \nl
  \cl_{B3} & = & \frac{\,s\,(s-\sigma)^3}{\lambda^2} \left[
    \rule[0cm]{0cm}{.6cm} \cl_B \; - \; \frac{2\SLAM}{s-\sigma} \;-\;
    \frac{2}{3} \left( \frac{\SLAM}{s-\sigma} \right)^{\!3}\, \right] \nl\nl
  \cl_- & = & \cl_{12} - \cl_{34} \nl \nl
  \cd_+ & = & \mysp \left( -\,\frac{\,t_{max}\,}{\SONE} \right) \:-\:
              \mysp \left( -\,\frac{\,t_{min}\,}{\SONE} \right) \:+\:
              \mysp \left( -\,\frac{\,t_{max}\,}{\STWO} \right) \:-\:
              \mysp \left( -\,\frac{\,t_{min}\,}{\STWO} \right) \nl\nl
  \cd_\sigma & = &  \mysp \left( \frac{t_{max}}{\,s-\sigma\,} \right)
            \; - \; \mysp \left( \frac{t_{min}}{\,s-\sigma\,} \right) \nl\nl
  \cd_d & = & \RE \left[ \rule[0cm]{0cm}{.35cm}
                \mysp \left(d_{12}\right) \; + \;
                \mysp \left(d_{34}\right) \right] \nl\nl
  \cd_{d+}\!\! & = & \mysp \left( \frac{t_{max}}{\,s-\STWO} \right) \:+\:
                  \mysp \left( \frac{t_{min}}{\,s-\STWO} \right) \:+\:
                  \mysp \left( \frac{t_{max}}{\,s-\SONE} \right) \:+\:
                  \mysp \left( \frac{t_{min}}{\,s-\SONE} \right)
                  \nl\nl
  \cd_{d-}\!\! & = & \mysp \left( \frac{t_{max}}{\,s-\STWO} \right) \:-\:
                  \mysp \left( \frac{t_{min}}{\,s-\STWO} \right) \:-\:
                  \mysp \left( \frac{t_{max}}{\,s-\SONE} \right) \:+\:
                  \mysp \left( \frac{t_{min}}{\,s-\SONE} \right)
                  \nl\nl
  \cf_\sigma & = & \mytri \left( \frac{t_{max}}{\,s-\sigma\,} \right)
           \; - \; \mytri \left( \frac{t_{min}}{\,s-\sigma\,} \right)
                   \nl\nl
  \cf_{d-} & = & \mytri \left( \frac{t_{max}}{\,s-\STWO} \right) \:-\:
                 \mytri \left( \frac{t_{min}}{\,s-\STWO} \right) \:+\:
                 \mytri \left( \frac{t_{max}}{\,s-\SONE} \right) \:-\:
                 \mytri \left( \frac{t_{min}}{\,s-\SONE} \right)
                 \nl\nl
  \cf_{t-} & = & \mytri \left( -\,\frac{t_{max}}{d_{12}\,t_{min}} \right)
           \:-\: \mytri \left( -\,\frac{t_{min}}{d_{12}\,t_{max}} \right)
                 \nl \nl & & \hspace{1cm}
           \:+\: \mytri \left( -\,\frac{t_{max}}{d_{34}\,t_{min}} \right)
           \:-\: \mytri \left( -\,\frac{t_{min}}{d_{34}\,t_{max}} \right)
  \label{nunivnota}
  \\ \nonumber
\ea
See appendix~\ref{loopint2} for the definitions of
$\cl_B, \cl_{12},$~and $\cl_{34}$. $I_{12Q}$~is found in integral 22.)
of appendix~\ref{loopint1}. It is noteworthy that the quantities
$\cl_{B2}$~and $\cl_{B3}$ have zero limit for $\lambda$ approaching
zero. This is easily verified by expanding $\cl_B$~in powers of
$\sqrt{\lambda}/(s-\sigma)$. It is seen from equation~(\ref{nunivxs}),
the \xsec~parts stemming from identical channels are much simpler than
those originating from different channels. More specifically,
{}~$\sigma^{V,tt}_{4,nonuni}$~~and ~$\sigma^{V,uu}_{4,nonuni}$~~do not
contain trilogarithms whereas ~$\sigma^{V,tu}_{4,nonuni}$~~and
{}~$\sigma^{V,ut}_{4,nonuni}$~~do.
\para
A word about the relations between the above twofold differential
cross-sections is in order. From inspection of the virtual Feynman
diagrams one finds
\ba
  \sigma^{V,tt}_{4,nonuni}(s,\SONE,\STWO) & = &
    \sigma^{V,uu}_{4,nonuni}(s,\STWO,\SONE) \nl
  \sigma^{V,tu}_{4,nonuni}(s,\SONE,\STWO) & = &
    \sigma^{V,ut}_{4,nonuni}(s,\STWO,\SONE)~~.
  \label{symm}
\ea
The equalities in equation~(\ref{nunivxs}) are due to the invariance
of ~$\sigma^{V,tt}_{4,nonuni}$~~and ~$\sigma^{V,tu}_{4,nonuni}$~~under
the interchange~$\SONE \leftrightarrow \STWO$. This is not surprising,
because final state masses are neglected in the matrix elements and
{}~$\sigma^{V,tt}_{4,nonuni}$~~and ~$\sigma^{V,tu}_{4,nonuni}$~~are
kinematical functions, resulting in kinematically equivalent bosons.
When adding the contributions given in equation~(\ref{nunivxs}) to
obtain $\sigma^V_{4,nonuni}$~, some minor simplifications of the
overall expression are possible, but the
representation~(\ref{nunivxs}) was chosen for reasons of transparency.
\newpage
\noindent
%
%
\section{Initial State Bremsstrahlung}
\label{bremres}
The amputated Feynman diagrams for initial state bremsstrahlung are
presented in figure~\ref{zzbrem}. They must be connected to both
t-channel and u-channel. Once more, the symmetry of the problem
simplifies the calculational task. It is sufficient to compute the
square of the t-channel bremsstrahlung matrix element and the
bremsstrahlung t-u interference. The u-channel contribution to the
cross-section is obtained from the t-channel contribution by
virtue of a symmetry operation similar to relation~(\ref{symm}).
The bremsstrahlung t-channel matrix element is derived from
equation~(\ref{tmat}) via the substitution
$B^{\alpha\beta}_t\!/(q_t^2+\MES-\ieps) \to e\,R^{\alpha\beta\mu}_t
\,\varepsilon^\lambda_\mu$ with polarization index $\lambda$.
Neglecting numerator electron masses, the Feynman rules in
appendix~\ref{feynrules} yield
\vspace{.3cm}
\beq
  \cm_t^{R,\lambda} \; = \; \frac{g_{\beta \beta'}}{\,D_Z(\SONE)\,}
                    \:\frac{g_{\alpha\alpha'}}{\,D_Z(\STWO)\,}\;
                    M_{12}^{\beta'}\,M_{34}^{\alpha'}\cdot e\,
                    R^{\:\!\alpha \beta \mu}_{\:\!t}\,
                    \varepsilon^\lambda_\mu
    \;\;\; \equiv \;\;\; \cm_t^{R,\mu}\,\varepsilon^\lambda_\mu
  \label{brtmat}
\eeq
\vspace{.15cm}
with
\vspace{.15cm}
\ba
  R^{\:\!\alpha \beta \mu}_{\:\!t} & = & {\bar u}(-k_2) \cdot
    \left\{ \rule[0cm]{0cm}{.6cm} \gamma^\alpha
      \frac{\,\dagg{v}_2 - \dagg{k}_2\,}{t_2} \, \gamma^\beta
      \frac{\,2k_1^\mu - \dagg{p} \gamma^\mu\,}{z_1} \;\; + \;\;
      \gamma^\alpha \frac{\,\dagg{v}_2 - \dagg{k}_2\,}{t_2} \,
      \gamma^\mu \frac{\,\dagg{k}_1 - \dagg{v}_1}{t_1} \, \gamma^\beta
    \right. \nl & & \hspace{2cm} \left. \rule[0cm]{0cm}{.6cm} + \;\;
      \frac{\,\gamma^\mu \dagg{p} - 2k_2^\mu\,}{z_2} \, \gamma^\alpha
      \frac{\,\dagg{k}_1 - \dagg{v}_1}{t_1} \, \gamma^\beta \right\}
    \times \nl & & \hspace{1.75cm}
      2 \left[
          {L_{\EMI\!\EPL\!Z}}^2 \!\cdot\!\left( 1+\gamma_5 \right)
        + {R_{\EMI\!\EPL\!Z}}^2 \!\cdot\!\left( 1-\gamma_5 \right)
      \right] \cdot u(k_1)~~.
\label{brtcurrent}
\ea
The momentum of the radiated photon is called $p$. Similarly, one
finds for the u-channel bremsstrahlung matrix element
\ba
  \cm_u^{B,\lambda} & = & \frac{g_{\alpha \alpha'}}{\,D_Z(\SONE)\,}
                \:\frac{g_{\beta \beta'}}{\,D_Z(\STWO)\,}\;
                M_{34}^{\beta'}\,M_{12}^{\alpha'} \cdot e \,
                R^{\:\!\alpha \beta \mu}_{\:\!u} \,
                \varepsilon^\lambda_\mu~~,
  \;\;\; \equiv \;\;\; \cm_u^{R,\mu}\,\varepsilon^\lambda_\mu
  \nl \nl
  R^{\:\!\alpha \beta \mu}_{\:\!u} & = & {\bar u}(-k_2) \cdot
    \left\{ \rule[0cm]{0cm}{.6cm} \gamma^\alpha
      \frac{\,\dagg{v}_1 - \dagg{k}_2\,}{u_2} \, \gamma^\beta
      \frac{\,2k_1^\mu - \dagg{p} \gamma^\mu\,}{z_1} \;\; + \;\;
      \gamma^\alpha \frac{\,\dagg{v}_1 - \dagg{k}_2\,}{u_2} \,
      \gamma^\mu \frac{\,\dagg{k}_1 - \dagg{v}_2\,}{u_1} \,
      \gamma^\beta
    \right. \nl & & \hspace{2cm} \left. \rule[0cm]{0cm}{.6cm} + \;\;
      \frac{\,\gamma^\mu \dagg{p} - 2k_2^\mu\,}{z_2} \, \gamma^\alpha
      \frac{\,\dagg{k}_1 - \dagg{v}_2\,}{u_1} \, \gamma^\beta \right\}
    \times \nl & & \hspace{1.75cm}
      2 \left[
          {L_{\EMI\!\EPL\!Z}}^2 \!\cdot\!\left( 1+\gamma_5 \right)
        + {R_{\EMI\!\EPL\!Z}}^2 \!\cdot\!\left( 1-\gamma_5 \right)
      \right] \cdot u(k_1)~~.
\label{brumat}
\ea
In the above matrix element the radiatively changed Mandelstam
variables were introduced. In terms of the five-particle phase space
parametrization from appendix~\ref{ps2to5} they are given by
\ba
  t_1 & = & \left( \rule[0cm]{0cm}{.35cm} k_1 - v_1 \right)^2 + \MES
      \;\; = \;\; a_1^t \; + \; b_1 \, \mycos\theta_R \; + \;
                  c \, \mysin\theta_R \mycos\phi_R \nl
  t_2 & = & \left( \rule[0cm]{0cm}{.35cm} v_2 - k_2 \right)^2 + \MES
      \;\; = \;\; a_2^t \; + \; b_2 \, \mycos\theta_R \; + \;
                  c \, \mysin\theta_R \mycos\phi_R \nl
  u_1 & = & \left( \rule[0cm]{0cm}{.35cm} k_1 - v_2 \right)^2 + \MES
      \;\; = \;\; a_1^u \; - \; b_1 \, \mycos\theta_R \; - \;
                  c \, \mysin\theta_R \mycos\phi_R \nl
  u_2 & = & \left( \rule[0cm]{0cm}{.35cm} v_1 - k_2 \right)^2 + \MES
      \;\; = \;\; a_2^u \; - \; b_2 \, \mycos\theta_R \; - \;
                  c \, \mysin\theta_R \mycos\phi_R \nl \nl
  z_1 & = & -2\,k_1 p \;\; = \;\;
            \frac{\sprm}{\,2\,} \left( \rule[0cm]{0cm}{.35cm}
            1+\beta\mycos\theta \right) \;\;\; \equiv \;\;\;
            \frac{\sprm}{\,2\,} \!\cdot\! {\bar z_1} \nl
  z_2 & = & -2\,k_2 p \;\; = \;\;
            \frac{\sprm}{\,2\,} \left( \rule[0cm]{0cm}{.35cm}
            1-\beta\mycos\theta \right) \;\;\; \equiv \;\;\;
            \frac{\sprm}{\,2\,} \!\cdot\! {\bar z_2}
  \label{tuz}
  \\ \nonumber
\ea
with the kinematical parameters
\vspace{.15cm}
\ba
  a_1^t & = & -\,\SONE \; + \; \frac{\sprp}{\,4\,s'\,} \, (s'+\delta)
              \; - \; \frac{\sprm}{\,4\,s'\,} \, (s'+\delta) \, \beta
                   \, \mycos\theta \nl
  a_2^t & = & -\,\STWO \; + \; \frac{\sprp}{\,4\,s'\,} \, (s'-\delta)
              \; + \; \frac{\sprm}{\,4\,s'\,} \, (s'-\delta) \, \beta
                   \, \mycos\theta \nl
  a_1^u & = & -\,\STWO \; + \; \frac{\sprp}{\,4\,s'\,} \, (s'-\delta)
              \; - \; \frac{\sprm}{\,4\,s'\,} \, (s'-\delta) \, \beta
                   \, \mycos\theta \nl
  a_2^u & = & -\,\SONE \; + \; \frac{\sprp}{\,4\,s'\,} \, (s'+\delta)
              \; +  \; \frac{\sprm}{\,4\,s'\,} \, (s'+\delta) \, \beta
                    \, \mycos\theta \nl \nl
  b_1 & = & \frac{\SLAMP}{\,4\,s'\,}
            \left( \sprp \, \beta \, \mycos\theta  - \sprm \right) \nl
  b_2 & = & \frac{\SLAMP}{\,4\,s'\,}
            \left( \sprp \, \beta \, \mycos\theta  + \sprm \right)
            \nl \nl
  c & = & - \, \frac{\SLAMP}{2} \, \sqrt{\frac{\,s\,}{s'}} \, \beta \,
          \mysin\theta \\ \nonumber
\ea
derived from the representations~(\ref{k5vect}) and (\ref{v5vect}) in
appendix~\ref{ps2to5}. The Lorentz invariants satisfy the relations
\ba
  t_1 + u_1 + z_1 & = & s - \sigma \nl
  t_2 + u_2 + z_2 & = & s - \sigma \nl
  z_1 + z_2 & = & \sprm~~~~.
\ea
The total bremsstrahlung \xsec is obtained by the integration
\ba
  \sigma^R(s) & = & \frac{(2\pi)^4}{\,2\,s\,} \,
    \int \!\! d\Gamma_5\;\;
    \frac{1}{4} \! \sum_{spins}
    \left( \cm_t^{R,\mu} + \cm_u^{R,\mu} \right)
    \left( {\cm_{t,\mu}^R}^* + {\cm_{u,\mu}^R}^* \right)
  \label{bremphint}
\ea
with the five-particle phase space $d\Gamma_5$ as parametrized in
equation~(\ref{par25b}), appendix~\ref{ps2to5}. After invariant
tensor integration over the boson decay degrees of freedom with
formula~(\ref{dectenint}) from appendix~\ref{tenint}, a
FORM~\cite{form} computer algebra program is used to further evaluate
equation~(\ref{bremphint}). Inside the FORM code, partial fraction
decomposition is used to simplify intermediate expressions, and the
following non-trivial scalar products are needed:
\ba
  k_1 \!\cdot\! v_2 & = & \left( t_1 + z_1 - s + \SONE \right)/2
    \;\;\; = \;\;\; - \left( u_1 + \STWO \right)/2 \nl
  k_2 \!\cdot\! v_1 & = & \left( t_2 + z_2 - s + \STWO \right)/2
    \;\;\; = \;\;\; - \left( u_2 + \SONE \right)/2 \nl
  k_1 \!\cdot\! v_1 & = & \left( u_1 + z_1 - s + \STWO \right)/2
    \,\;\; = \;\;\; - \left( t_1 + \SONE \right)/2 \nl
  k_2 \!\cdot\! v_2 & = & \left( u_2 + z_2 - s + \SONE \right)/2
    \,\;\; = \;\;\; - \left( t_2 + \STWO \right)/2 \nl
  p \!\cdot\! v_1   & = & \left( t_2 - t_1 - z_1 \right)/2
    \;\;\; = \;\;\; \left( u_1 - u_2 - z_2 \right)/2 \nl
  p \!\cdot\! v_2   & = & \left( t_1 - t_2 - z_2 \right)/2
    \;\;\; = \;\;\; \left( u_2 - u_1 - z_1 \right)/2~~.
\ea
The, after the tensor integration step, remaining three angles
$\theta_R$, $\phi_R$,~and finally $\theta$~(see appendix~\ref{ps2to5})
are integrated analytically with
the help of the integrals listed in appendices~\ref{bremint1}
and~\ref{bremint2}. The complexity of this threefold analytical
integration is due to the appearance of the angularly dependent
Mandelstam variables $t_1,~t_2,~u_1,~u_2,~z_1,$~and $z_2$~in the
denominators of the matrix elements~$\cm_t^{R,\lambda}$~and
$\cm_u^{R,\lambda}$, equations~(\ref{brtmat}) and~(\ref{brumat}). The
result for the total bremsstrahlung \xsec splits into a universal,
leading logarithm part with the Born \xsec factorizing and a
non-universal part,
\ba
  \sigma^R(s) & \equiv & \sigma^R_{uni} \; + \; \sigma^R_{nonuni}
    \nl \nl & = &
    \frac{\alpha}{\pi}
    \int\limits_{4m_1^2}^{(\sqrt{s} - 2m_2)^2} d\SONE \, \rho_Z(\SONE)
    \int\limits_{4m_2^2}^{(\sqrt{s} - \sqrt{\SONE})^2} d\STWO \,
    \rho_Z(\STWO)
    \int\limits_{(\sqrt{\SONE}+\sqrt{\STWO})^2}^s ds' \nl
  & & \hspace{.5cm} \left[ \rule[0cm]{0cm}{.8cm} \left(
    \myln \frac{s}{\MES} - 1 \right) \frac{\,s^2+{s'}^2}{s^2 \sprm}\;
    \sigma^B_4(s';\SONE,\STWO) \;\; + \right. \nl
  & & \left. \rule[0cm]{0cm}{.8cm} \hspace{.9cm}
    \frac{\left(G_{\mu} M_Z^2\right)^2}{8\pi s^2} \left(L_e^4+R_e^4\right)
    \sigma^R_{4,nonuni}(s,s';\SONE,\STWO) \; \right] \times
    2 \cdot B\!R(1) \cdot B\!R(2)~~. \nl
\label{bremxs}
\ea
The non-universal \xsec term $\sigma^R_{4,nonuni}$ is a sum of the
t-channel, the u-channel and the t-channel/u-channel interference
bremsstrahlung contributions,
\ba
  \sigma^R_{4,nonuni} (s,s';\SONE,\STWO) & = & \sigma^{R,t}_{4,nonuni}
  \; + \; \sigma^{R,tu}_{4,nonuni} \; + \; \sigma^{R,u}_{4,nonuni}
  \nonumber
\ea
with
\ba
  \lefteqn{ \hspace{-.5cm} \sigma^{R,t}_{4,nonuni} \;\; = \;\;
    \sigma^{R,u}_{4,nonuni} \;\; = \;\;
    \hspace{1cm}
    \SLAMP \left( \frac{1}{\sprm} - \frac{\sigma}{\,\LAMB\,} \right)
    \;\; +} \nl
   & & \hspace{-.5cm} \SLAMP \left( \rule[0cm]{0cm}{.65cm}
         \frac{s}{\,4 {s'}^2} + \frac{1}{\,4 s'} +
         \frac{\sigma}{\,2 \LAMB\,} +
         \frac{3\,\sigma\SONE\STWO}{\LAMB^2} \right)
         \left( L_{c3} + L_{c4} \right) \;\; + \;\;
       \frac{\sigma}{\,2\,} \left( \rule[0cm]{0cm}{.65cm}
         \frac{s}{{s'}^2} - \frac{1}{s} \right) L_{c5} \;\; - \;\; \nl
  & & \hspace{-.5cm}
      \frac{D^t_{12}}{\,4\SLAMB} \left[ \rule[0cm]{0cm}{.85cm} \;
        2\sigma + s + \sprm + \frac{\sigma}{\LAMB}
          \left( \rule[0cm]{0cm}{.65cm} s\,\left(\sprm + \sigma\right) -
                 \sigma^2 + \delta^2 \right) +
        12 \, \frac{\,s\,\SONE\,\STWO\,\sigma
          \left( \sprm+\sigma\right)\,}{\LAMB^2}
        \; \right] \;\; + \;\; \nl
  & & \hspace{-.5cm}
      \frac{s}{\,4 s'} \left( \rule[0cm]{0cm}{.65cm}
        \frac{\sigma}{s}+\frac{\sigma}{s'}-1 \right)
        \left( \rule[0cm]{0cm}{.5cm} D^t_1 \!+\! D^t_2 +\!
               D_{z1t2} \!+\! D_{z2t1} \right) \; - \; \frac{1}{\,4\,}
        \left( \rule[0cm]{0cm}{.5cm} D^t_1 \!+\! D^t_2 -\!
               D_{z1t2} \!-\! D_{z2t1} \right) \; + \hspace{-1cm} \nl
  & & \hspace{-.5cm}
      \frac{\delta}{\,4\sprm\,} \left( L_{c1} -  L_{c2} \right) \cdot
        \left[ \rule[0cm]{0cm}{.85cm} \; 2 - \frac{s'}{\,s\,} -
               \frac{s^2}{\,{s'}^2} +
               \frac{\,\sigma\left(3s-\sprm-\sigma\right)\,}{\LAMB} \; +
        \right. \nl
  & & \hspace{5.8cm} \left. \rule[0cm]{0cm}{.85cm}
               \frac{s}{\,\sprm\,} \left( \rule[0cm]{0cm}{.65cm}
                 1+\frac{\sigma^2}{\,\LAMB\,} \right) +
               12 \, \frac{\,s\,\SONE\,\STWO\,\sigma\,}{\LAMB^2}
        \right] \; + \hspace{-1cm} \nl
  & & \hspace{-.5cm}
      \frac{1}{\,4\,} \left( L_{c1} +  L_{c2} \right) \cdot
        \left[ \rule[0cm]{0cm}{.75cm} 1 + \frac{\sigma}{\,s\,} +
               \frac{\sigma}{\,s'\,} - \frac{\,s-\sigma\,}{\sprm} -
               \frac{\,2s\sigma\,}{{\sprm}^2} \, + \,
               \frac{\,2\sigma\,}{\LAMB}\left(\sprp-\sigma\right) \,+\,
               24\,\frac{\,s\,\SONE\,\STWO\,\sigma\,}{\LAMB^2} \; \right],
      \hspace{-.5cm} \nl \nl \nl \nl
  \lefteqn{ \hspace{-.5cm} \sigma^{R,tu}_{4,nonuni} \;\; = \;\;
    \frac{2\,\sprm \sigma}{s' \left(s'-\sigma\right)} \cdot L_{c5}
      \;\; + \;\; \frac{\sprp}{s'}
      \left( \rule[0cm]{0cm}{.35cm} D^t_1 + D^t_2 \right)} \nl \nl
  & & \hspace{-.5cm} - \;
      \frac{s^2+{s'}^2}{\,2 \, \sprm\left(s'-\sigma\right)\,}
        \left( \rule[0cm]{0cm}{.35cm} D_{z1t1} + D_{z2t2} \right)
      \;\; - \;\;
      \left( \rule[0cm]{0cm}{.7cm} \frac{s}{\,s'} \, + \,
             \frac{s^2+{s'}^2}{\,2\,\sprm\left(s'-\sigma\right)\,} \right)
      \left( \rule[0cm]{0cm}{.35cm} D_{z2t1} + D_{z1t2} \right) \nl \nl
  & & \hspace{-.5cm} - \;
      \frac{\,s^2+{s'}^2-4\sigma\left(s'-\sigma\right)\,}
           {2\,\left(s'-\sigma\right) \left(\sprp-2\sigma\right)}
        \cdot \left[ \rule[0cm]{0cm}{.45cm} L_{c8}
                     \left( \rule[0cm]{0cm}{.35cm} L_{c6} + L_{c7}
                     \right) \, + \, D^{tu}_{a1} + D^{tu}_{a2}
              \right] \nl \nl
  & & \hspace{-.5cm} - \;
      \frac{\,s-2\SONE\,}{\sqrt{s\SONE}} \, D_{t1u2} \;\; - \;\;
      \frac{\,s-2\STWO\,}{\sqrt{s\STWO}} \, D_{t2u1} \;\; + \;\;
      \frac{s^2+{s'}^2}{\,4\,\sprm \left(s'-\sigma \right)}
        \left( \rule[0cm]{0cm}{.35cm} D^z_{t1u2} + D^z_{t2u1} \right)
      \nl \nl
  & & \hspace{-.5cm} + \;
      \frac{\,\SLAMP \left[ \rule[0cm]{0cm}{.35cm}
                            s^2 + {s'}^2 - 4\sigma (s'-\sigma) \right]
            \left[ \rule[0cm]{0cm}{.35cm}
                   s (s'-\sigma) - \SONE (\sprp - 2\sigma) \right]}
           {2\,(\sprp - 2\sigma) (s'-\sigma)} \cdot D^a_{t1u2} \nl \nl
  & & \hspace{-.5cm} + \;
      \frac{\,\SLAMP \left[ \rule[0cm]{0cm}{.35cm}
                            s^2 + {s'}^2 - 4\sigma (s'-\sigma) \right]
            \left[ \rule[0cm]{0cm}{.35cm}
                   s (s'-\sigma) - \STWO (\sprp - 2\sigma) \right]}
           {2\,(\sprp - 2\sigma) (s'-\sigma)} \cdot D^a_{t2u1}~~~~~~.
  \label{brnunitu}
  \\ \nl \nonumber
\ea
The shorthand notations $L_{c1}, \,L_{c2}, \,L_{c3}, \,L_{c4}, \,L_{c5},
\,L_{c6}, \,L_{c7}, \,L_{c8}, \,D^t_1, D^t_2, \,D^t_{12}, \,D_{z1t1},
\,D_{z1t2},$ \\
$\,D_{z2t1}, \,D_{z2t2}, \,D^{tu}_{a1}, \,D^{tu}_{a2}, \,D_{t1u2},
\,D_{t2u1}, \,D^z_{t1u2}, \,D^z_{t2u1}, \,D^a_{t1u2},$~and
$D^a_{t2u1}$~are introduced in
appendix~\ref{bremint2}. For easier reference, the number of the
integral from appendix~\ref{bremint2} corresponding to each of these
notations is given in table~\ref{notafind}. Together with
$\sigma^V_{4,nonuni}$~from equation~(\ref{nunivxs}),
$\sigma^R_{4,nonuni}$~from equation~(\ref{brnunitu}) represents a main
result of this thesis.
%
\begin{table}[b]
  \begin{center}
    \begin{tabular}{|c||c|c|c|c|c|c|c|c|} \hline & & & & & & & & \\
      Notation & $L_{c1}$ & $L_{c2}$ & $L_{c3}$ & $L_{c4}$ & $L_{c5}$
      & $L_{c6}$ & $L_{c7}$ & $L_{c8}$ \\ \hline
      Number from & & & & & & & & \\
      appendix~\ref{bremint2} & 10.) & 13.) & 30.) & 31.) & 10.) &
      26.) & 27.) & 26.) \\ \hline
      \hline
      & & & & & & & & \\
      Notation & $D^t_1$ & $D^t_2$ & $D^t_{12}$ & $D_{z1t1}$ &
      $D_{z1t2}$ & $D_{z2t1}$ & $D_{z2t2}$ & $D^{tu}_{a1}$ \\ \hline
      Number from & & & & & & & & \\
      appendix~\ref{bremint2} & 18.) & 19.) & 46.) & 10.) & 11.) &
      12.) & 13.) & 26.) \\ \hline \hline
      & & & & & & & & \\
      Notation & $D^{tu}_{a2}$ & $D_{t1u2}$ & $D_{t2u1}$ &
      $D^z_{t1u2}$ & $D^z_{t2u1}$ & $D^a_{t1u2}$ & $D^a_{t2u1}$ & \\
      \hline
      Number from & & & & & & & & \\
      appendix~\ref{bremint2} & 27.) & 51.) & 52.) & 53.) & 54.) &
      57.) & 58.) & \\ \hline
    \end{tabular}
    \caption[Shorthand notation references for non-universal initial
      state bremsstrahlung contributions]{\it
      Shorthand notations from equation~(\ref{brnunitu}) together
      with the number of the integral from appendix~\ref{bremint2}
      where each notation is defined.}
    \label{notafind}
  \end{center}
\end{table}
%
%
\para
The radiator $(s^2+{s'}^2)/(s^2\sprm)$ obtained in
equation~(\ref{bremxs}) is known from \ee~annihilation into fermion
pairs~\cite{bonneau71}. It is seen from
equation~(\ref{photmom}) that the radiator's singularity in the factor
{}~$1/\sprm = 1/(s-s')\longrightarrow_{\hspace{-.65cm} _{s'\to s}} \;
\infty\:$ is related to the radiation of soft photons. This
so-called infrared singularity is identically cancelled by the
infrared singularity arising from the virtual corrections. The
cancellation of infrared singularities is worked out in
appendix~\ref{IRtreat}.
\para
It is important that the non-universal contribution
$\sigma^R_{nonuni}$ is not infrared divergent. This was explicitly
checked by taking the limit $s'\to s$ for all parts of
$\sigma^R_{4,nonuni}$
that contain factors $1/\sprm$. In addition it was checked that all
parts of $\sigma^R_{4,nonuni}$ have finite limits for $\SLAMP \to 0$.
As a concluding remark it is mentioned that, similar to the case of
virtual non-universal corrections, some simplifications are possible
when terms are collected from the individual contributions in
equation~(\ref{brnunitu}) to yield $\sigma^R_{4,nonuni}$. As these are
not major, the expression for $\sigma^R_{4,nonuni}$ is not rewritten.
%
%
\vspace{2cm}
\section{Treatment of Infrared Singularities}
\label{IRtreat}
To show how the infrared divergences from the virtual and the real
bremsstrahlung corrections cancel, a short derivation of this
cancellation is worked out below. One starts from the bremsstrahlung
matrix elements given in equations~(\ref{brtmat}) to~(\ref{brumat}).
The infrared divergence originates from the square of those parts of
the matrix elements which contain the bremsstrahlung photon momentum
$p$ in the denominator. In equations~(\ref{brtmat}) to~(\ref{brumat})
these are the parts containing $1/z_1$~and $1/z_2$~as is seen from
equation~(\ref{tuz}). Denoting the matrix element parts yielding the
divergence by $\cm^{\rm IR}$, using very small $p$ , i.e. neglecting
numerator terms proportional to $p$, and applying the Dirac equation
one gets
\beq
  \cm^{\rm IR} \;\; = \;\; \cm^B \cdot e \cdot
    \left( \rule[0cm]{0cm}{.8cm} \frac{k_1.\varepsilon}{k_1.p} \, - \,
           \frac{k_2.\varepsilon}{k_2.p} \right)
\eeq
with the Born matrix element $\cm^B = \cm^B_t + \cm^B_u$ taken from
equations~(\ref{tmat}) and~(\ref{umat}), and with the photon
polarization vector $\varepsilon$. Evaluation of the fully
differential bremsstrahlung \xsec from $\cm^{\rm IR}$~and summation
over photon polarizations yields the factorization
\ba
  d\sigma^{\rm IR} & = & d\sigma^B \cdot \frac{\alpha}{\pi} \cdot
    \delta^{\rm IR}
  \label{IRdiffxs}
  \\ \nl
  \delta^{\rm IR}  & = & \frac{1}{2\pi} \; \int \frac{d^3p}{2p^0}
    \left( \rule[0cm]{0cm}{.8cm} \frac{k_1}{k_1.p} \, - \,
           \frac{k_2}{k_2.p} \right)^2 \nl
    & = & \frac{1}{2\pi} \; \int \frac{d^3p}{2p^0} \left[
      \Theta(\Delta -p^0) + \Theta(p^0-\Delta) \right]
      \left( \rule[0cm]{0cm}{.8cm} \frac{k_1}{k_1.p} \, - \,
      \frac{k_2}{k_2.p} \right)^2 \nl
    & \equiv & \hspace{2.6cm} \delta^{\rm IR}_{soft} \hspace{.46cm} \;+\;
      \hspace{.43cm} \delta^{\rm IR}_{hard}
  \label{IRall}
\ea
with the fully differential Born \xsec $d\sigma^B$, the photon
energy $p^0$, the soft cutoff $\Delta\!>\!0$, the Heavyside step
function $\Theta(x)$, and the long known Low factor~\cite{low58}
\ba
  \cf^{\rm IR} = \left( \rule[0cm]{0cm}{.8cm} \frac{k_1}{k_1.p}
    \, - \, \frac{k_2}{k_2.p} \right)^2~~.
\ea
The infrared divergence is obviously contained
in $\delta^{\rm IR}_{soft}$. Following a procedure of dimensional
regularization similar to the one outlined in
reference~\cite{marciano75} and carried through in~\cite{bardin94} one
arrives at
\ba
  \delta^{\rm IR}_{soft} & = &
    2 \, \left[ {\rm P^{IR}} + \myln \omega \right]
    \left( \myln \frac{s}{\MES} \, - \, 1 \right) \; + \;
    \frac{1}{2} \, \myln^2 \frac{s}{\MES} \; - \; \frac{\pi^2}{3} \nl
  \;\; \omega & = & \frac{\,2 \Delta}{\sqrt{s}}~~.
  \label{IRsoft}
\ea
A derivation of~$\delta^{\rm IR}_{soft}$~is given at the end of this
section. The result may be verified against reference~\cite{berends88}
where regularization with a photon mass $\lambda$ is used by the
replacement
\ba
  {\rm P^{IR}} \;\; \rightarrow \;\; - \, \frac{1}{2} \,
    \myln \frac{\lambda^2}{\MES}~~.
\ea
The dependence of $\delta^{\rm IR}_{soft}$~on the cutoff $\omega$
is cancelled by a similar dependence in $\delta^{\rm IR}_{hard}$.
Integration of equation~(\ref{IRdiffxs}) over the Born phase space
yields the infrared divergent \xsec contribution with the factorized
Born \xsec $\sigma^B$
\ba
  \sigma^{\rm IR}_{soft} & = &
    \frac{\alpha}{\pi} \cdot \sigma^B(s) \cdot \delta_{soft}^{\rm IR}
    \;\;=\;\; \frac{\alpha}{\pi} \cdot \int\limits_{s'_{\!min}}^s
    \frac{ds'}{s} \;\, \delta\!\left(1-\frac{s'}{s}\right) \!\cdot\!
    \delta_{soft}^{\rm \,IR} \!\cdot\! \sigma^B(s')~~.
  \label{sigIRsoft}
\ea
Next, the contributions from \oal~virtual and bremsstrahlung
corrections must be added to cancel the infrared divergence
${\rm P^{IR}}$. As only universal cross-section parts are infrared
divergent, it is sufficient to consider those. Non-universal
contributions are therefore ignored for the remainder of this
section. First, the universal \xsec contributions from
equations~(\ref{xsvirt}) and~(\ref{bremxs}) are rewritten using the
definition ~$L \equiv \ln(s/\MES) -1$~:
\ba
  \sigma^V_{uni} & \equiv & \frac{\alpha}{\pi} \cdot
    \delta_{virtual}^{\rm \,IR} \cdot \sigma^B(s) \;\; = \;\;
    \frac{\alpha}{\pi} \cdot \int\limits_{s'_{\!min}}^s
    \frac{ds'}{s} \;\, \delta\!\left(1-\frac{s'}{s}\right) \!\cdot\!
    \delta_{virtual}^{\rm \,IR} \!\cdot\! \sigma^B(s')
  \label{IRvirt}
\ea
\ba
  \sigma^R_{uni} & = &
    \frac{\alpha}{\pi} \cdot
    \int\limits_{s'_{\!min}}^s \frac{ds'}{s} \;\,
    \frac{s^2+{s'}^2}{\,s\,(s-s')\,} \cdot L \cdot
    \sigma^B(s') \nl
    & = & \frac{\alpha}{\pi} \cdot \hspace{-.3cm}
      \int\limits_{s'_{\!min}}^{s\,(1-\omega)} \frac{ds'}{s} \;\,
      \frac{s^2+{s'}^2}{\,s\,(s-s')\,} \cdot L \cdot \sigma^B(s')
      \nl
    & & \hspace{2.5cm} \;\;+\;\;
      \frac{\alpha}{\pi} \cdot \hspace{-.3cm}
      \int\limits_{s\,(1-\omega)}^s \frac{ds'}{s} \;
      \left[ \frac{2\,s}{\,s-s'\,} - \frac{\,s+s'\,}{s} \right]
      \cdot L \cdot \sigma^B(s')~~.
    \label{IRbrem1}
\ea
One observes that addition or subtraction of non-divergent terms in the
second integrand of equation~(\ref{IRbrem1}) yields negligible effects for
sufficiently small infrared cutoff $\omega$. Therefore
\ba
  \sigma^R_{uni} & = & \frac{\alpha}{\pi} \cdot \hspace{-.3cm}
    \int\limits_{s'_{\!min}}^{s\,(1-\omega)} \frac{ds'}{s} \;\,
    \frac{s^2+{s'}^2}{\,s\,(s-s')\,} \cdot L \cdot \sigma^B(s')
    \hspace{.47cm} + \hspace{.47cm} \frac{\alpha}{\pi} \cdot
    \hspace{-.3cm} \int\limits_{s\,(1-\omega)}^s \frac{ds'}{s} \;
    \frac{2\,s}{\,s-s'\,}\cdot L \cdot \sigma^B(s')~~. \nl
  \label{IRbrem2}
\ea
It is evident that the infrared divergence of the bremsstrahlung
part is in the second term of equation~(\ref{IRbrem2}). The
infrared divergence is located at $s\!=\!s'$~and the corresponding
\xsec is given by
\ba
  \sigma^{\rm IR}_{soft} & = & \frac{\alpha}{\pi} \cdot
    \hspace{-.3cm} \int\limits_{s\,(1-\omega)}^s \frac{ds'}{s} \;
    \frac{2\,s}{\,s-s'\,}\cdot L \cdot \sigma^B(s')~~.
  \label{IRbrem3}
\ea
Replacing this expression by the expression given in
equation~(\ref{sigIRsoft}), one can rewrite equation~(\ref{IRbrem2}):
\ba
  \sigma^R_{uni} & = & \frac{\alpha}{\pi} \cdot \hspace{-.2cm}
    \int\limits_{s'_{\!min}}^s \frac{ds'}{s} \;\, \sigma^B(s') \cdot
    \left\{ \rule[-.2cm]{0cm}{1cm} \hspace{.5cm}
            \Theta\!\left(1-\frac{\,s'\,}{s}-\omega\right)\,
            \frac{s^2+{s'}^2}{\,s\,(s-s')\,} \cdot L \right. \nl
    & & \hspace{3.41cm} \left. \rule[-.2cm]{0cm}{1cm} \: + \:
            \delta\left(1-\frac{\,s'\,}{s}\right) \!\cdot\!
            \delta^{\rm IR}_{soft} \hspace{2.7cm} \right\}~~
            \raisebox{-.3cm}[0cm][0cm]{.}
  \label{IRbrem4} \\ \nonumber
\ea
The complete universal \oal~\xsec is obtained by adding the virtual
and the bremsstrahlung parts from equations~(\ref{IRvirt})
and~(\ref{IRbrem4}). The result, also known from the literature on
fermion pair production in
\ee~annihilation~\cite{berends88,bonneau71}, reads
\ba
  \sigma^{V+R}_{uni} & = & \sigma^{V}_{uni} \;+\; \sigma^{R}_{uni} \nl
    & = & \frac{\alpha}{\pi} \cdot \hspace{-.2cm}
    \int\limits_{s'_{\!min}}^s \frac{ds'}{s} \;\, \sigma^B(s') \cdot
    \left\{ \rule[-.2cm]{0cm}{1cm}
            \Theta\!\left(1-\frac{\,s'\,}{s}-\omega\right)
            H_e \! \left(\frac{\,s'\,}{s}\right) \: + \:
            \delta\left(1-\frac{\,s'\,}{s}\right) S_e(s) \right\} \nl
            \nl \nl
  H_e (s'/s) & = & \frac{s^2+{s'}^2}{\,s\,(s-s')\,} \cdot L
    \;\; = \;\; \frac{1+(s'/s)^2}{1-s'/s} \cdot L \nl \nl
  S_e (s) & = & \delta^{\rm IR}_{virtual} + \delta^{\rm IR}_{soft}
    \;\; = \;\; \left( \rule[-.2cm]{0cm}{.8cm} 2 \myln\omega +
    \frac{3}{2} \right) \!\cdot\! L  \: + \: \frac{\pi^2}{3}
    \: - \: \frac{1}{2}~~~
  \label{alxs}
\ea
with $\delta^{\rm IR}_{virtual}$~from equation~(\ref{xsvirt}) and
$\delta^{\rm IR}_{soft}$~from equation~(\ref{IRsoft}).
The dependence on the infrared cutoff $\omega$ cancels if the
integration is carried out. From the representation~(\ref{alxs}) of
the universal \oal~\xsec the cancellation of infrared divergences is
obvious. As stated earlier, the non-universal \oal~\xsec parts are
infrared finite. One can therefore present a singularity-free
full \oal~\xsec by adding non-universal \oal~virtual and
bremsstrahlung contributions from equations~(\ref{xsvirt})
and~(\ref{bremxs}) to the result~(\ref{alxs}).
\vspace{1cm} \\
{\underline {\bf Derivation of the soft bremsstrahlung infrared
  divergence $\delta^{\rm IR}_{soft}$}} \vspace{.45cm} \\
To derive the expression for $\delta^{\rm IR}_{soft}$ given in
equation~(\ref{IRsoft}) on starts from
equation~(\ref{IRall}). Introduction of a Feynman parameter $\alpha$
yields
\ba
  \delta^{\rm IR}_{soft} & = & \frac{1}{2\pi} \; \int\limits_0^1 d\alpha
    \int \frac{d^3p}{2p^0}\; \Theta(\Delta -p^0)
    \left\{ \rule[0cm]{0cm}{.8cm}
            \frac{s-2\MES}{\left(k_\alpha \!\cdot\! p \right)^2} \,-\,
            \frac{\MES}{\left(k_1 \!\cdot\! p \right)^2} \, - \,
            \frac{\MES}{\left(k_2 \!\cdot\! p \right)^2} \right\} \nl\nl
  \;\; k_\alpha  & = & k_1 \alpha + k_2 \left( 1-\alpha \right)~~.
  \label{IRsoft1}
\ea
Using the spatial angles $\theta_i,~~i=1,2,\alpha$~~between the photon
momentum $\vec{p}$~and the momenta $\vec{k}_i$, the above denominators are
expressed as squares of
\ba
  -k_i \!\cdot\! p & = & k_i^0 p^0
  \left( 1 - \beta_i \mycos \theta_i \right)
  \hspace{1.3cm} {\rm with~velocities} \hspace{1.3cm}
  \beta_i \;\; = \;\; \frac{\left| k_i \right|
    \raisebox{.14cm}[0cm][0cm]{\hspace{-.65cm} ${}^\rightarrow$} \,\,}
  {\,k_i^0\,}~~.
\ea
To dimensionally regularize equation~(\ref{IRsoft1}), the
transition from 3-dimensional to $(n-1)$-dimensional space has to be
performed.
As all three terms in the integrand of equation~(\ref{IRsoft1}) do not
depend on the polar angle $\theta$ of the photon emission, one has the
freedom to choose a coordinate system with $\theta\!=\!\theta_i$ for
each of the three integrand terms in eq.~(\ref{IRsoft1}). Physically
this freedom of choice is equivalent to the isotropy of soft photon
emission. With the mass scale $\mu$~ to maintain the dimension of the
expression, one gets
\ba
  \lefteqn{ \frac{(2\pi)^3}{2\pi} \; \int\limits_0^1 d\alpha
    \int \frac{d^3p}{(2\pi)^3} \frac{1}{2p^0}\; \theta(\Delta -p^0)
    \frac{1}{\left(k_i \!\cdot\! p\right)^2} } \nl
  & & \to \; \frac{2\!\cdot\! (2\pi)^2}
      {\left( \rule[0cm]{0cm}{.35cm} 2\sqrt{\pi}\right)^{\! n}
       \Gamma\!\left(\frac{n}{2}-1\right)} \;\,
    \int\limits_0^1 \!\! \frac{d\alpha}{\, \mu^{(n-4)}} \;
    \int\limits_0^\Delta \!\! dp^0 \left(p^0\right)^{(n-5)} \;
    \int\limits_0^\pi \! \frac{\mysin^{(n-3)} \theta_i \, d\theta_i}
      {\,\left(k_i^0\right)^2 \left( 1 - \beta_i \mycos \theta_i
      \right)^2\,} \; . \nl
\ea
Introducing the integrals
\ba
  \frac{1}{\,\mu^{(n-4)}} \int\limits_0^\Delta \!\! dp^0
    \left(p^0\right)^{(n-5)} & = &
    \frac{\left( \Delta/\mu \right)^{(n-4)}}{n-4} \;\; = \;\;
    \frac{1}{n-4} \cdot\! \left[\, \rule[-.2cm]{0cm}{.8cm} 1 \: + \:
      \left( n-4 \right) \myln \frac{\Delta}{\mu} \: + \:
      {\cal O}(n\!-\!4) \, \right] \hspace{-.5cm} \nl \nl
  \int\limits_0^\pi \! \frac{\mysin^{(n-3)} \theta_i \, d\theta_i}
    {\,\left( 1 - \beta_i \mycos \theta_i \right)^2\,} & = &
    \int\limits_{-1}^{+1}
    \frac{\left(1-x^2\right)^{\frac{n-4}{2}}\,dx}
         {\left(1-\beta_i x\right)^2} \nl & = &
    \int\limits_{-1}^{+1} \frac{dx}{\left(1-\beta_i x\right)^2}\,
    \left[\, \rule[-.2cm]{0cm}{.8cm} 1 \: + \: \frac{n-4}{2}
    \ln(1-x^2) \: + \: {\cal O}(n\!-\!4) \,\right]~~, \nl
\ea
neglecting terms of ${\cal O}(n-4)$, and taking into account that
$\beta_1 = \beta_2 = \beta = \sqrt{1\!-\!4\MES/s}\,$,~one finds
\ba
  \delta^{\rm IR}_{soft} & = & \frac{1}{2} \left[ \rule[-.1cm]{0cm}{.35cm}
    {\rm P^{IR}} + \myln \frac{\Delta}{\mu} \right] \cdot
    \int\limits_0^1 \!\! d\alpha \int\limits_{-1}^{+1} \!\!dx \;
    \cf(\alpha,x) \; + \;
    \frac{1}{4} \int\limits_0^1 \!\! d\alpha
    \int\limits_{-1}^{+1} \!\! dx \; \ln(1-x^2) \, \cf(\alpha,x)
  \nl \nl & &
  \cf(\alpha,x) \;\; = \;\;
    \frac{s-2\MES}{\left(k_\alpha^0\right)^2
                   \left(1-\beta_\alpha x \right)^2} \; - \;
    \frac{4\MES}{s\,\left(1-\beta x \right)^2} \;-\;
    \frac{4\MES}{s\,\left(1-\beta x \right)^2}~~.
  \label{IRmedres}
\ea
Using the two integrals
\ba
  \int\limits_{-1}^{+1} \!\!dx \; \frac{1}{\,\left(1-\beta_i x\right)^2\,}
    & = & \frac{2}{\,1-\beta_i^2\,} \nl\nl
  \int\limits_{-1}^{+1} \!\!dx \;
    \frac{\ln(1-x^2)}{\,\left(1-\beta_i x\right)^2\,} & = &
    \frac{2}{\,1-\beta_i^2\,} \left[ \; \rule[-.2cm]{0cm}{.8cm} 2\ln 2
    \,-\, \frac{1}{\beta_i} \, \myln \frac{1+\beta_i}{1-\beta_i} \right]
\ea
results in
\ba
  \frac{1}{2} \int\limits_0^1 \!\! d\alpha \int\limits_{-1}^{+1}
    \!\!dx \;\cf(\alpha,x) \hspace{1.92cm} & = &
    \left( s\!-\!2\MES \right) \!\cdot\! \int\limits_0^1 \!
    \frac{d\alpha}{\,-k_\alpha^2\,} \;\;\;\; - \;\;\;\; 2 \nl\nl
  \frac{1}{4} \int\limits_0^1 \!\! d\alpha \int\limits_{-1}^{+1}
    \!\!dx \; \ln(1-x^2)\,\cf(\alpha,x) & = &
    \myln 2 \cdot \left[\, \rule[-.2cm]{0cm}{.8cm}
      \left( s\!-\!2\MES \right) \!\cdot\! \int\limits_0^1 \!
      \frac{d\alpha}{\,-k_\alpha^2\,} \;\;\;\; - \;\;\;\; 2 \, \right]
    \nl & + & \left( \rule[-.2cm]{0cm}{.6cm}
    \frac{s\!-\!2\MES}{2} \right) \!\cdot\!\!
    \int\limits_0^1 \! \frac{d\alpha}{\,-\beta_\alpha k_\alpha^2\,}\,
    \myln \frac{1\!-\!\beta_\alpha}{1\!+\!\beta_\alpha} \;\;\;\, - \;\;\;\,
    \frac{1}{\beta} \, \myln \frac{1\!-\!\beta}{1\!+\!\beta} \nl
  \label{IRints}
\ea
with the Feynman parameter integrals
\ba
  \int\limits_0^1 \! \frac{d\alpha}{\,-k_\alpha^2\,} & = &
    \int\limits_0^1 \! \frac{d\alpha}
    {\left(s\!-\!4\MES\right) \alpha \left(1\!-\!\alpha\right) + \MES}
    \;\;=\;\; \frac{2}{s\beta} \; \myln\frac{1\!+\!\beta}{1\!-\!\beta}
  \nonumber
\ea
\ba
  \int\limits_0^1 \! \frac{d\alpha}{\,-\beta_\alpha k_\alpha^2\,}\,
  \myln \frac{1\!-\!\beta_\alpha}{1\!+\!\beta_\alpha} & = &
  \int\limits_0^1 \! \frac{d\alpha}
    {\left(s\!-\!4\MES\right) \alpha \left(1\!-\!\alpha\right) + \MES}
    \!\cdot\! \frac{1}{\beta \sqrt{\left(1\!-\!2 \alpha\right)^2}}
    \cdot \myln \frac{1\!-\!\beta \sqrt{\left(1\!-\!2 \alpha\right)^2}}
                     {1\!+\!\beta \sqrt{\left(1\!-\!2 \alpha\right)^2}}
    \nl & = & \frac{4}{s\beta} \cdot\! \int\limits_0^1 \!
    \frac{dx}{x \left(1\!-\!\beta x\right) \left(1\!+\!\beta x\right)}
    \cdot \myln \frac{1\!-\!\beta x}{1\!+\!\beta x} \nl & = &
    \frac{2}{s\beta} \cdot \left[ \rule[-.2cm]{0cm}{.9cm} \;
    2 \mysp\left(-\beta\right) \;-\; 2 \mysp\left(\beta\right) \;+\;
    \mysp\left(\frac{1\!+\!\beta}{2}\right) \;-\;
    \mysp\left(\frac{1\!-\!\beta}{2}\right)
    \right. \nl & & \hspace{1.02cm} \left. \rule[-.2cm]{0cm}{.9cm} +\;
    \frac{1}{2} \myln^2 \frac{1\!+\!\beta}{2} \;-\;
    \frac{1}{2} \myln^2 \frac{1\!-\!\beta}{2} \; \right]~~.
  \label{IRfeynint}
\ea
Introducing the results~(\ref{IRints}) and~(\ref{IRfeynint}) into
equation~(\ref{IRmedres}), using
\beq
  \myln \frac{1\!+\!\beta}{1\!-\!\beta} \;\;\; \stackrel{URA}{=} \;\;\;
  \myln \frac{s}{\MES} \;\; = \;\; l_\beta~~,
\eeq
taking the result to the ultrarelativistic limit, and evaluating it at
the scale $\mu\!=\!\MES$ yields exactly the result given in
equation~(\ref{IRsoft}),
\ba
  \delta^{\rm IR}_{soft} & = & 2 \!\cdot\! \left[ \rule[-.1cm]{0cm}{.35cm}
    {\rm P^{IR}} + \myln \frac{2\Delta}{\mu} \right] \!\cdot\!
    \left( \myln \frac{s}{\MES} \, - \, 1 \right) \; - \;
    \frac{\pi^2}{3} \; - \;
    \frac{1}{2} \, \myln^2 \frac{s}{\MES} \; + \; \myln\frac{s}{\MES}
  \nl \nl & = &
    2 \, \left[ {\rm P^{IR}} + \myln \omega \right]
    \left( \myln \frac{s}{\MES} \, - \, 1 \right) \; + \;
    \frac{1}{2} \, \myln^2 \frac{s}{\MES} \; - \; \frac{\pi^2}{3}~~.
\ea
%
%
\section{Soft Photon Exponentiation}
\label{expo}
The factor $\frac{1}{1-s'/s}$~in the universal part
$H_e$~(see equation~(\ref{alxs})) yields large contributions at the phase
space edge close to $s\!=\!s'$. Therefore resummation of higher order
contributions is suggested to obtain reasonable precision. One can
follow the derivation of the resummed \xsec given
in reference~\cite{berends88}, because its authors treat precisely the
collinear leading logarithms that are found in the universal \xsec
parts of the presented calculation. To connect to the formulae of
ref.~\cite{berends88}, one starts from equation~(\ref{alxs}) and
introduces the notations
\ba
  v & \equiv & 1 - \frac{s'}{s} \nl
  \beta_e & \equiv & \frac{\,2\alpha\,}{\pi} \cdot L \nl
  {\bar S_1} & \equiv & \frac{\,\alpha\,}{\pi}
    \left[ \rule[0cm]{-.1cm}{.35cm} \;
    S_e (s) - 2 \myln \omega \!\cdot\! L \, \right]
    \;\;\; = \;\;\; \frac{\,\alpha\,}{\pi} \left[
    \rule[0cm]{-.1cm}{.35cm} \frac{\pi^2}{3} - \frac{1}{2} \right]
           \; + \; \frac{3}{4} \, \beta_e \nl
  {\bar H_1} & \equiv & \frac{\,\alpha\,}{\pi} \left[
    \rule[0cm]{-.1cm}{.65cm} \; H_e(s'/s) - \frac{2}{v} \!\cdot\! L \,
    \right]
    \;\;\; = \;\;\; - \frac{1}{2} \left[ \rule[0cm]{-.1cm}{.35cm} \;
    2-v \right] \!\cdot\! \beta_e~~.
  \label{radal}
\ea
{}From equation~(\ref{alxs}) one obtains the cross-section for
process~(\ref{eezz4f}) with universal initial state QED corrections
\ba
  \sigma^{ISR} & = & \int\limits_{0}^{v_{max}} dv \;\, \sigma^B(s') \;
    \left\{ \rule[0cm]{0cm}{.8cm} \hspace{.3cm} \delta(v) \left[
    \rule[0cm]{0cm}{.7cm} 1 \; + \; \beta_e \myln \omega
    \; + \; {\bar S_1} \; + \; \frac{\beta_e^2}{2} \, \myln^2 \omega
    \; + \; \beta_e \myln \omega \: {\bar S_1} \; + \; {\bar S_2}
    \right] \right. \hspace{-.5cm} \nl
  & & \hspace{3cm} + \; \Theta(v-\omega) \left[
    \rule[0cm]{0cm}{.7cm} \beta_e \, \frac{1}{v} \; + \; {\bar H_1}
    \; + \; \beta_e^2 \, \frac{\ln v}{v} \; + \;
    \beta_e \, \frac{1}{v} \, {\bar S_1} \; + \; {\bar H_2} \right] \nl
  & & \left. \rule[0cm]{0cm}{.8cm} \hspace{2.86cm} + \,
    {\cal O}(\alpha^3) \hspace{.5cm} \right\}~~,
  \label{ISRal2}
\ea
with the second order initial state radiation taken from
reference~\cite{berends88} where expressions for the second order
virtual+soft part ${\bar S_2}$ and the second order hard part
${\bar H_2}$ can be found. Before carrying on, it is worth mentioning
that the infrared divergent second order soft initial state radiation
part $\delta^{\rm IR}_{soft,soft}$ is given by
\beq
  \delta^{\rm IR}_{soft,soft} \;\; = \;\; \frac{1}{2} \:
    {\delta^{\rm IR}_{soft}}^2 \; + \; c_2
\eeq
with infrared finite $c_2$. This remarkable property
indicates the possibility to obtain higher order initial state
soft-collinear divergences from an exponentiation of the lowest-order
correction $\delta^{\rm IR}_{soft}$~ apart from a finite correction
factor $c_n$~\cite{berends88}.
\para
Equation~(\ref{ISRal2}) indicates that the cross-section will be
enhanced by terms ~$\left[\ln^{\!n-1} v\right]\!/v$ which become
especially important near the phase space boundary $v \to 0$,
i.e. the enhancement is due to soft photon emission. It will be
present in all orders $n$ of perturbation theory, and therefore an
appropriate resummation technique is needed. As the logarithms
originate from infrared singularities they can be
resummed~\cite{berends88}, and, according to the Bloch-Nordsieck
theorem~\cite{bloch37} and Yennie, Frautschi, and
Suura~\cite{yennie61}, they indeed exponentiate to all orders of
perturbation theory. Therefore equation~(\ref{ISRal2}) represents the
truncated power series expansion of the universal \xsec to
process~(\ref{eezz4f}) with soft photon exponentiation which reads
\ba
  \sigma^{ISR}_{exp} & = & \int\limits_{0}^{v_{max}} dv \;\,
    \sigma^B(s') \; \left\{ \rule[0cm]{0cm}{.35cm}
    \delta(v) \cdot {\bar S} \cdot e^{\beta_e \myln \omega}
    \; + \; \Theta(v-\omega) \cdot H \right\}
  \label{expoall1}
\ea
with the regular soft+virtual part ${\bar S}$ and the hard part $H$,
\ba
  {\bar S} & = & \sum_{n=0}^{\infty} {\bar S_n} \hspace{2cm}
    {\rm with} \hspace{2cm} {\bar S_0} \; = \; 1~~~{\rm
      for~the~Born~contribution}~~, \nl\nl
  H & = & \sum_{n=1}^{\infty} \;
    \left[ \rule[0cm]{0cm}{.7cm} \, \frac{{\beta_e}^n}{(n-1)!} \!\cdot\!
    \frac{\ln^{\!n-1} v}{v} \!\cdot\! {\bar S} \; + \; {\bar H_n} \,
    \right]~~.
  \label{barsh}
\ea
All ${\bar H_n}$~are integrable in $v$, and the infrared divergent
term in $H$~is completely determined by the requirement that
the~$\myln \omega$~terms appearing in the $v$ integration of
$H$~cancel all $\myln \omega$~terms in the virtual+soft part
${\bar S} \!\cdot\! e^{\beta_e \myln \omega}$. To rid
equation~(\ref{expoall1}) of the unphysical parameter
$\omega$~one introduces
${\bar H} \equiv \sum_{n=1}^{\infty} {\bar H_n}$~, uses the equality
\ba
  \sum_{n=1}^{\infty} \; \frac{{\beta_e}^n}{(n-1)!} \!\cdot\!
    \frac{\ln^{\!n-1} v}{v} & = & \beta_e v^{\beta_e - 1}
\ea
and rewrites
\ba
  \sigma^{ISR}_{exp} & = & \int\limits_{0}^{v_{max}} dv \;\,
  \sigma^B(s') \; \left\{ \rule[0cm]{0cm}{.5cm} \left[
  \rule[0cm]{0cm}{.35cm}
  \delta(v) \!\cdot\! e^{\beta_e \myln \omega} \: + \:
  \Theta(v\!-\!\omega) \!\cdot\! \beta_e v^{\beta_e - 1} \right]
  \!\cdot\! {\bar S}
  \;\; + \;\; \Theta(v\!-\!\omega) \!\cdot\! {\bar H} \right\}~.
  \nl
  \label{expodiss}
\ea
For the three above integral terms one finds
\beq
  \int\limits_{0}^{v_{max}} dv \;\, \sigma^B(s') \!\cdot\! \delta(v)
    \!\cdot\! e^{\beta_e \myln \omega} \!\cdot\! {\bar S}
    \;\; = \;\;
    \sigma^B(s) \!\cdot\! {\bar S} \!\cdot\! \omega^{\beta_e}~,
  \label{expodiss1}
\eeq
\ba
  \lefteqn{ \int\limits_{0}^{v_{max}} dv \;\, \sigma^B(s') \!\cdot\!
    \Theta(v\!-\!\omega) \!\cdot\! \beta_e v^{\beta_e - 1}
    \!\cdot\! {\bar S}} \nl
  & & \hspace{1.5cm} = \; \int\limits_{0}^{v_{max}} dv \;\,
    \sigma^B(s') \!\cdot\! \beta_e v^{\beta_e - 1} \!\cdot\! {\bar S}
    \;\; - \;\; \int\limits_{0}^{\omega} dv \;\, \sigma^B(s')
    \!\cdot\! \beta_e v^{\beta_e - 1} \!\cdot\! {\bar S} \nl
  & & \hspace{1.5cm} = \; \int\limits_{0}^{v_{max}} dv \;\,
    \sigma^B(s') \!\cdot\! \beta_e v^{\beta_e - 1} \!\cdot\! {\bar S}
    \;\; - \;\; \sigma^B(s) \!\cdot\! {\bar S} \!\cdot\!
    \omega^{\beta_e}~,
  \label{expodiss2}
\ea
and, because ${\bar H}$ is integrable and $\omega$ is
sufficiently small,
\beq
  \int\limits_{0}^{v_{max}} dv \;\, \sigma^B(s') \!\cdot\!
    \Theta(v\!-\!\omega) \!\cdot\! {\bar H} \;\; = \;\;
    \int\limits_{0}^{v_{max}} dv \;\, \sigma^B(s') \!\cdot\! {\bar H}~~~.
  \label{expodiss3}
\eeq
Adding the results of equations~(\ref{expodiss1}) to~(\ref{expodiss3})
one notices that the contributions from~(\ref{expodiss1}) and from the
second term of~(\ref{expodiss2}) cancel each other. Thus a new and
physical form of the initial state radiation corrected cross-section
with soft photon exponentiation is obtained from
equation~(\ref{expodiss}):
\ba
  \sigma^{ISR}_{exp} \; = \; \int\limits_{s'_{\!min}}^s
  {\ds \frac{ds'}{s}} \;\, \sigma^B(s') \left\{ \rule[-.1cm]{0cm}{.4cm}
    \beta_e v^{\beta_e - 1} \!\cdot\! {\bar S} + {\bar H} \right\}~~.
  \label{expoxsec}
\ea
A \xsec with ${\cal O}(\alpha^n)$ soft photon exponentiation is now
obtained by replacing the terms ${\bar S}$~and ${\bar H}$
in~(\ref{expoxsec}) with corresponding truncated sums from
equation~(\ref{barsh}). For an \oal~ISR corrected \xsec with soft
photon exponentiation the series'~(\ref{barsh}) are truncated after
the first order term which means
\ba
  {\bar S} & = & 1 + {\bar S_1} + {\cal O}(\alpha^2)~~, \nl
  {\bar H} & = & \hspace{.64cm} {\bar H_1} + {\cal O}(\alpha^2)~~.
  \nonumber
\ea
Returning to process~(\ref{eezz4f}) and aiming at the \oal~ISR \xsec
with soft photon exponentiation, it is necessary to include
non-universal \xsec parts. Since the virtual and the bremsstrahlung
non-universal parts as derived in appendices~\ref{virtres}
and~\ref{bremres} are both infrared finite and do not contain any
singularity, they may be readily added to the regular first order
expressions ${\bar S_1}$~and ${\bar H_1}$ without changing the
argument leading to equation~(\ref{expoxsec}). Therefore accounting
for \oal~non-universal \xsec parts means to replace:
\ba
  \sigma^B(s') \!\cdot\! \left( 1 + {\bar S_1} \right) & \rightarrow &
    \sigma^B(s') \!\cdot\! \left( 1 + {\bar S_1} \right) \; + \;
    \sigma^V_{nonuni} \hspace{.9cm} \equiv \;\; {\sf S}_1 \nl
  \sigma^B(s') \!\cdot\! {\bar H_1} \hspace{1.04cm} & \rightarrow &
    \sigma^B(s') \!\cdot\! {\bar H_1} \hspace{1.15cm}  + \;
    s\,\frac{d\sigma^R_{nonuni}}{ds'} \;\;\; \equiv \;\; {\sf H}_1
  \label{nuniexp}
\ea
with $\sigma^V_{nonuni}$ from equation~(\ref{xsvirt}) and
$\sigma^R_{nonuni}$ taken from equation~(\ref{bremxs}).
Thus one obtains for the \oal~ISR corrected \xsec with soft photon
exponentiation for process~(\ref{eezz4f})
\ba
  \sigma^{ISR,\OAL}_{exp} \; = \; \int\limits_{s'_{\!min}}^s
  {\ds \frac{ds'}{s}} \;\, \left\{ \rule[-.1cm]{0cm}{.4cm}
    \beta_e v^{\beta_e - 1} \!\cdot\! {\sf S}_1 \; + \; {\sf H}_1
  \right\}~~.
  \label{expooal}
\ea
%
%
\chapter{Evaluation of Loop Integrals}
\label{loops}
Calculations of radiative corrections within the Electroweak Standard
Model contain loop integrals as basic building blocks. For the
one-loop case, loop integrals have been intensively studied by several
authors~\cite{wimdiss,thooft79,thooft73,passarino79}, and the
literature also contains pedagogical
treatments~\cite{holliklect,jegerlect}. It is noteworthy that scalar
one-loop diagrams with five or more external lines can be reduced to
the scalar 1-, 2-, 3-, and 4-point functions~\cite{thooft79}. As
tensor $m$-point functions are reducible to linear combinations of
scalar $k$-point functions with $k\leq m$, it is sufficient to know
all possible scalar 1-, 2-, 3-, and 4-point functions together with
the appropriate tensor reduction rules. As it is not the intention of
this appendix to repeat the results obtained and comprehensively
presented earlier (see e.g. references~\cite{wimdiss,passarino79}), it
will be restricted to the formulae relevant for this thesis. This
means a restriction to the 1-, 2-, 3-point functions, because the only
4-point function that had to be dealt with could be easily evaluated
by direct integration. It should be noted that, in special cases, the
more general results obtained below may be subject to simplifications
by virtue of the relations presented in appendix~\ref{rellogs}. Such
simplifications can often be more easily seen, if the specific loop
integral is directly evaluated. In this case, however, the general
result is still very useful for cross-checks.
Subsequently, the following quantities will be investigated:
\ba
  A_0(m_1^2) & \equiv & \; \mu^{-(n-4)}
    \int\!\!\frac{d^{n}q}{(2\pi)^n} \: \frac{1}{D_1} \nl
  B_{(0,\mu,\mu\nu)}(p_1,m_1^2,m_2^2) & \equiv & \mu^{-(n-4)}
    \int\!\!\frac{d^{n}q}{(2\pi)^n} \:
    \frac{\{1,\,q_\mu,\,q_\mu q_\nu\}}{D_1\,D_2} \nl
  C_{(0,\mu,\mu\nu,\mu\nu\!\rho)}(p_1,p_2,m_1^2,m_2^2,m_3^2) & \equiv &
    \mu^{-(n-4)} \int\!\!\frac{d^{n}q}{(2\pi)^n} \:
    \frac{\{1,\,q_\mu,\,q_\mu q_\nu,\,q_\mu q_\nu q_\rho\}}
         {D_1\,D_2\,D_3}~~.
\label{loopints}
\ea
Dimensional regularization is applied to render all expressions
well-defined. $\mu$ is an arbitrary mass scale. The following
definitions are used:
\ba
  \hspace*{4cm}
  D_1 & = & q^2 + m_1^2 - \ieps \nl
  D_2 & = & (q+p_1)^2 + m_2^2 - \ieps \nl
  D_3 & = & (q+p_1+p_2)^2 + m_3^2 - \ieps~~.
  \label{loopden}
\ea
The $n$-dimensional Dirac algebra needed for the evaluation of loop
expressions can be found in~\cite{jegerlect,leibbrandt75} or in standard
textbooks such as~\cite{itzykson}. Care is required if $\gamma_5$
matrices are involved. Note that the expressions in
equation~(\ref{loopints}) are invariant under shifts of the loop
momentum $q$.
Before proceeding further a lemma will be presented that is going to
be very useful in section~\ref{mptfun}.
\begin{lemma} \hspace{.2cm}
  Let a be real. The numbers b,c and $y_0$ may be complex. Let further
  $-\varepsilon$ be an infinitesimal real number carrying the sign of
  $I\!=\!\IM(ay^2\!+\!by\!+\!c)$, where $I$ is supposed to not change
  sign for $y\:\!\epsilon\:\![0,1]$. Let the infinitesimal real number
  $-\delta$ carry the sign of $\IM(ay_0^2\!+\!by_0\!+\!c)$. Let $y_1$
  and $y_2$ be the roots of $ay^2\!+\!by\!+\!c = 0$. Then the
  identity
  \ba
    \lefteqn{\int\limits_0^1 \frac{dy}{y-y_0}
      \left[ \rule[0cm]{0cm}{.35cm}
      \ln(ay^2+by+c) - \ln(ay_0^2+by_0+c) \right]\;=} \nl
    & & \hspace{1cm} \myln \frac{y_0-1}{y_0}
      \cdot \left[ \rule[0cm]{0cm}{.35cm}
      \eta(-y_1,-y_2) \; - \; \eta(y_0-y_1,y_0-y_2) \; - \;
      \eta(a-\ieps,a+\ri\delta) \right] \nl
    & & \hspace{1cm} + \; {\rm R}(y_0,y_1) \; + \; {\rm R}(y_0,y_2)
    \label{binint}
  \ea
holds, if the two parameter function R is defined by
  \ba
    {\rm R}(y_0,y_i) & = & \int\limits_0^1 \frac{dy}{y-y_0}
      \left[ \rule[0cm]{0cm}{.35cm} \ln(y-y_i) - \ln(y_0-y_i) \right]
      \nl & = & \hspace{.42cm}
      \mysp \left( \frac{y_0}{y_0-y_i} \right) \; - \;
      \mysp \left( \frac{y_0-1}{y_0-y_i} \right)
      \nl & & + \;
      \eta\!\left(-y_i,\frac{1}{y_0-y_i}\right)\!\cdot\!
        \myln \frac{y_0}{y_0-y_i} \; - \;
      \eta\!\left(1\!-\!y_i,\frac{1}{y_0-y_i}\right)\!\cdot\!
        \myln \frac{y_0-1}{y_0-y_i}~.
  \ea
  \label{intlem}
\end{lemma}
The proof of Lemma~\ref{intlem} can be found in appendix B of
reference~\cite{thooft79}. See appendix~\ref{logfun} for the
definition of $\eta\!\left(a,b\right)$.
%
\vspace{1.5cm}
\section{Scalar $m$-Point Functions}
\label{mptfun}
The evaluation of scalar $m$-point functions in $n$ dimensions is
based on the representation of $n$-dimensional space in spherical
coordinates as outlined in the appendix of reference~\cite{marciano75}.
For $m \geq 2$, the second important ingredient is Feynman
parametrization. One arrives at expressions which become rather
handy for $n=4-\varepsilon,~|\varepsilon|\ll 1$. Integrals converge for
$\varepsilon > 0$ in the case of an ultraviolet (UV) divergence and for
$\varepsilon < 0$ in the case of infrared divergences.
\para
The scalar 1-point function is given by
\ba
  A_0(m_1^2) & = & \iospi\: m_1^2 \cdot
    \left[ \Delta - 1 + \ln \frac{m_1^2-\ieps}{\mu^2} \right]~, \nl
  \Delta & = & \frac{2}{n-4} + \gamma_E - \ln (4\pi)~~,
\ea
with Euler's constant $\gamma_E = 0.577216...$~.
\newpage
\noindent
The scalar 2-point function is given by~\cite{thooft79}
\ba
  B_0(p_1,m_1^2,m_2^2) & = & - \, \iospi
    \left\{ \rule[0cm]{0cm}{.7cm} \indent
            \Delta \; + \; \ln\frac{-p_1^2 - \ieps}{\mu^2} \right. \nl
    & & \hspace{1.7cm} \left. \; + \; \sum_{i=1}^2
          \left[ \ln(1-x_i) - x_i \!\cdot\!\ln \frac{x_i-1}{x_i} - 1 \right]
    \rule[0cm]{0cm}{.7cm} \right\}~~, \\ \nonumber
\ea
where $x_1$~and $x_2$~are the roots of~~
$-p_1^2 x^2 \, + \, (p_1^2 + m_2^2 - m_1^2) x + m_1^2 - \ieps \; = \;
0$~. If $p_1^2=0$, there is only one root, and $\ln(-p_1^2 - \ieps)$
should be replaced by $\ln(m_2^2 - m_1^2 - \ieps)$. If in addition
$m_1^2 = m_2^2 = m^2$ one finds
\ba
  B_0(0,m^2,m^2) & = & - \, \iospi \left\{ \rule[0cm]{0cm}{.7cm}
    \Delta \; + \; \ln \frac{m^2-\ieps}{\mu^2} \right\}~~.
\ea
The pathological case $p_1^2 = m_1^2 = m_2^2 = 0$ is infrared
divergent and would require appropriate special treatment.
\para
The scalar 3-point function has a more complicated structure, but is
not ultraviolet divergent. It can be presented as
\ba
  C_0(p_1,p_2,m_1^2,m_2^2,m_3^2) & = &
    \iospi \int\limits_0^1 dx \int\limits_0^x dy
    \left[ ax^2 + by^2 + cxy + dx + ey +f \right]^{-1} \nl
  & \equiv & \iospi \; {\tilde C_0}
\ea
with
\ba
  \indent
  a = -p_2^2, ~~~~ b = -p_1^2, ~~~~
  c = -2p_1p_2 = p_1^2 + p_2^2 - (p_1+p_2)^2, \hspace{.86cm} \nl
  \indent
  d = m_2^2-m_3^2+p_2^2, ~~~~
  e = m_1^2-m_2^2+p_1^2+2p_1p_2, ~~~~ f = m_3^2-i\epsilon~~.
  \nonumber
\ea
Defining $\alpha$ as one of the roots of ~$b \alpha^2 + c \alpha + a =
0$~, the first integration yields:
\ba
  {\tilde C_0} & = & \hspace{.3cm}
    \int\limits_0^1 \! dy \,
      \frac{\, \myln \left[ \rule[0cm]{0cm}{.4cm}
                          by^2 + (c+e)\:\!y + a + d + f \right] \; - \;
               \myln \left[ \rule[0cm]{0cm}{.4cm}
                          by_1^2 + (c+e)\:\!y_1 + a + d + f \right] \, }
           {(c+2\alpha b)\:\!y + d + e\alpha + 2a + c \alpha} \nl
  & & \!\!\! - \;
    \int\limits_0^1 \! dy \,
      \frac{\, (1-\alpha)}
           {(c+2\alpha b)(1-\alpha)\:\!y + d + e\alpha} \times \nl
  & & \hspace{1.15cm}
      \left\{ \rule[0cm]{0cm}{.5cm}
              \myln \left[ \rule[0cm]{0cm}{.4cm}
                  (a\!+\!b\!+\!c)\:\!y^2 + (d\!+\!e)\:\!y + f \right] \;-\;
              \myln \left[ \rule[0cm]{0cm}{.4cm}
                  (a\!+\!b\!+\!c)\:\!y_2^2 + (d\!+\!e)\:\!y_2 + f \right]
            \right\} \nl \nl
  & & \!\!\! + \;
    \int\limits_0^1 \! dy \,
      \frac{\, \alpha \cdot \left\{ \rule[0cm]{0cm}{.5cm}
            \myln \left[ \rule[0cm]{0cm}{.4cm} ay^2 + dy + f \right] \;-\;
            \myln \left[ \rule[0cm]{0cm}{.4cm} ay_3^2 + dy_3 + f \right]
            \right\}\,} {(c+2\alpha b)\:\!\alpha y - (d + e\alpha)}
\label{3ptf}
\ea
with\vspace{.2cm} \\
$ \left.
  \begin{array}{crcl}
    \hspace{2cm} & y_1 & = & y_0 + \alpha \vspace{.5cm} \\
                 & y_2 & = & {\ds \frac{y_0}{1-\alpha} } \vspace{.2cm} \\
                 & y_3 & = & - \, {\ds \frac{y_0}{\alpha} } \\
  \end{array} \rule[0cm]{0cm}{1.6cm}
  \right\} \hspace{.5cm} y_0 \; = \; - \, {\ds
                                  \frac{d + e\alpha}{c+2\alpha b} }~~.
$ \vspace{.2cm} \\
Taking into account that for each term of equation~(\ref{3ptf}) the
non-logarithmic factor can be written as
$(c+2\alpha b)\!\cdot\![y-y_i]$ and using the identity~(\ref{binint})
presented in Lemma~\ref{intlem} at the end of the previous section, it
is straightforward to further integrate equation~(\ref{3ptf}). This
solution is valid for all physical cases, i.e. for physical
four-momenta $p_1$ and $p_2$ and physical masses~\cite{thooft79}.
Physical four-momenta are real, masses may be complex but must have a
positive real part.
\para
For real roots $\alpha$~and stable particles, the 3-point
function can be cast into a rather simple form:
\ba
  \lefteqn{C_0(p_1,p_2,m_1^2,m_2^2,m_3^2) \;\;\;
    \stackrel{\ds \alpha,m_i \,\epsilon\, \numreal}{\ds =} } \nl
  & & \frac{\ri}{16\pi^2(c+2 \alpha b)} \sum_{l=1}^3 \sum_{j=1}^2
      (-1)^{l+1} \left[ \rule[0cm]{0cm}{.7cm}
      \mysp \left( \frac{x_l}{x_l - y_{lj}} \right) \; - \;
      \mysp \left( \frac{x_l-1}{x_l - y_{lj}} \right) \right]
\ea
with
\ba
  x_1 = -\,\frac{d+2a+(e+c)\alpha}{c+2 \alpha b}~, ~~&~~&~~
  y_{1j} = \frac{-c-e \pm \sqrt{(c+e)^2 - 4b(a+d+f)}}{2b}~,\nl
  x_2 = -\,\frac{d+e\alpha}{(1-\alpha)(c+2 \alpha b)}
    \hspace{.3cm}, ~~&~~&~~
  y_{2j} = \frac{-d-e \pm \sqrt{(d+e)^2 - 4f(a+b+c)}}{2(a+b+c)}~,\nl
  x_3 = \hspace{.4cm} \frac{d+e\alpha}{\alpha(c+2 \alpha b)}
    \hspace{1.34cm}, ~~&~~&~~
  y_{3j} = \frac{-d \pm \sqrt{d^2 - 4af}}{2a}~. \nonumber
\ea
\para
Among the special cases of the scalar 3-point function, one is rather
interesting, namely the infrared divergent case which is encountered, if
one propagator mass vanishes and the momenta $p_1$ and $p_1+p_2$ are
on-shell. Without loss of generality one can choose
\beq
 m_1 = 0~, ~~~~~ p_1^2 = -m_2^2~, ~~~~~ (p_1+p_2)^2 = -m_3^2~.
\eeq
The infrared divergent scalar 3-point function is then given by
\ba
  \lefteqn{C^{\rm IR}_0(p_1,p_2,0,m_2^2,m_3^2) \; =} \nl
  & & \mu^{-(n-4)} \int\!\!\frac{d^{n}q}{(2\pi)^n} \:
    \frac{1}{(q^2-\ieps)(q^2+2qp_1-\ieps)(q^2+2q(p_1+p_2)-\ieps)}~.
\label{IR3p}
\ea
{}From power counting in equation~(\ref{IR3p}) it is seen that~
$C^{\rm IR}_0$~ is infrared divergent for $n=4$, but regularized by
$n>4$. Expansion around $n=4$ yields
\ba
  C^{\rm IR}_0(p_1,p_2,0,m_2^2,m_3^2) & = & \iospi \cdot \frac{1}{2}
    \int\limits_0^1 \frac{\,dy\,}{-p_y^2} \left[ \rule[0cm]{0cm}{.75cm}
    \Delta^{\rm IR} \, + \, \ln \left( \frac{-p_y^2}{\mu^2} \right) \right]
\label{IRinter}
\ea
with~ $-p_y^2 = -p_2^2 y^2 + (p_2^2-m_2^2+m_3^2)\:\!y + m_2^2 - \ieps$~
and~ $\Delta^{\rm IR} = \!\Delta$, the superscript IR indicating its
infrared origin. The case $p_2^2\!=\!0$ yields a simplified integral and
will not be treated here. In addition, the cases $m_2\!=\!0$~and
$m_3\!=\!0$~are excluded from further consideration to ensure that the
roots $y_1$ and $y_2$ of $-p_y^2 = 0$ do not coincide with the
integration edges. Defining
\ba
  -p_{y_1}^2 & \equiv & -p_2^2 y_1^2 + (p_2^2-m_2^2+m_3^2)\:\!y_1 +
                        m_2^2 - \ieps \nl
  -p_{y_2}^2 & \equiv &  -p_2^2 y_2^2 + (p_2^2-m_2^2+m_3^2)\:\!y_2 +
                        m_2^2 - \ieps \nonumber
\ea
one can rewrite
\ba
  \lefteqn{C^{\rm IR}_0(p_1,p_2,0,m_2^2,m_3^2) \; = } \nl
  & & \iospi \! \cdot \! \frac{1}{2} \! \cdot \!
    \frac{1}{\,-p_2^2\!\cdot\! (y_1-y_2)\,}
    \left\{ \rule[0cm]{0cm}{.85cm} \;
      \int\limits_0^1 \frac{dy}{y-y_1} \left[
      \Delta^{\rm IR} + \ln \frac{-p_{y_1}^2}{\mu^2} +
      \ln (-p_y^2) - \ln(-p_{y_1}^2) \right] \right. \nl
  & & \left. \rule[0cm]{0cm}{.85cm} \hspace{4.5cm} - \;
      \int\limits_0^1 \frac{dy}{y-y_2} \left[
      \Delta^{\rm IR} + \ln \frac{-p_{y_2}^2}{\mu^2} +
      \ln (-p_y^2) - \ln(-p_{y_2}^2) \right] \; \right\} \nl
\ea
Using
\beq
  \int\limits_0^1 \frac{dy}{y-y_i} = \myln \frac{y_i-1}{y_i}
\eeq
and identity~(\ref{binint}) from Lemma~\ref{intlem} further
integration is straightforward. Despite the generality of this result
direct integration of the {\sc r.h.s.} of equation~(\ref{IRinter}) is
simpler in many practical cases.
\vspace{2.5cm}
\section{Reduction of Tensor Integrals}
\label{tenred}
Because of Lorentz covariance, tensor loop integrals can be decomposed
into linear combinations of all possible tensors with identical rank
that can be built from the involved external four-vectors and the
metric tensor.
\ba
  B_\mu \;\;   & = & p_{1 \mu}\, B_1 \label{tendec1} \\
  B_{\mu\nu}\; & = & p_{1 \mu}p_{1 \nu}\, B_{21} \; + \; \mten B_{22}
  \label{tendec2} \\ \nl
  C_\mu \;\;   & = & p_{1 \mu}\, C_{11} \; + \; p_{2 \mu}\, C_{12}
  \label{tendec3} \\
  C_{\mu\nu}\; & = & p_{1 \mu}p_{1 \nu}\, C_{21} \; + \;
                     p_{2 \mu}p_{2 \nu}\, C_{22} \; + \;
                     \{p_1 p_2\}_{\mu\nu}\, C_{23} \; + \;
                     \mten C_{24} \label{tendec4} \\
  C_{\mu\nu\!\rho} & = & p_{1 \mu}p_{1 \nu}p_{1 \rho}\, C_{31} \; + \;
                       p_{2 \mu}p_{2 \nu}p_{2 \rho}\, C_{32} \; + \;
                       \{p_2 p_1 p_1\}_{\mu\nu\!\rho}\, C_{33} \nl
                 & & + \; \{p_1 p_2 p_2\}_{\mu\nu\!\rho}\, C_{34} \; + \;
                       \{p_1\,g\}_{\mu\nu\!\rho}\, C_{35} \; + \;
                       \{p_2\,g\}_{\mu\nu\!\rho}\, C_{36}
                       \label{tendec5}
\ea
The linear factors $B_{i(j)}$ and $C_{ij}$ are linear combinations of
scalar loop integrals and depend on the Lorentz invariants that can be
formed from the momenta $p_i$ of the involved external four-vectors
and on the propagator masses squared~$m_i^2$. For the sake of
simplicity, no arguments are exposed in equations~(\ref{tendec1})
to~(\ref{tendec5}). If nothing else is given, arguments are always
assumed to be as exhibited in equation~(\ref{loopints}). The
convenient shorthand prescriptions
\ba
  \{p_1 p_2\}_{\mu\nu} & \equiv &
    p_{1 \mu} \, p_{2 \nu} \; + \; p_{2 \mu} \, p_{1 \nu} \nl
  \{p_i p_j p_j\}_{\mu\nu\!\rho} & \equiv &
    p_{i \mu} \, p_{j \nu} \, p_{j \rho} \; + \;
    p_{j \mu} \, p_{i \nu} \, p_{j \rho} \; + \;
    p_{j \mu} \, p_{j \nu} \, p_{i \rho} \nl
  \{p_i g\}_{\mu\nu\!\rho} & \equiv & p_{i \mu} \, g_{\nu\rho} \; + \;
    p_{i \nu} \, g_{\mu\rho} \; + \; p_{i \rho} \, g_{\mu\nu}
\ea
were taken from reference~\cite{passarino79}, the other notations are
rather standard (see e.g.~\cite{wimdiss}).
Now the computation of the linear factors $B_{i(j)}$ and $C_{ij}$ is
described. To evaluate $B_1$, both sides of~(\ref{tendec1}) are
multiplied by $p_1^\mu$. This is followed by quadratic
supplementation of the integrand's numerator on the \lhs~Then
division by $p_1^2$ yields the well-known result
\beq
  B_1 \;\; = \;\; \frac{1}{2\,p_1^2} \left[ \rule[0cm]{0cm}{.4cm}
                  f_1 B_0 + A_0(m_1^2) - A_0(m_2^2) \right]
\eeq
with ~$f_1 = m_1^2-m_2^2-p_1^2$. Similarly, multiplication of
both sides of~(\ref{tendec2}) with $g^{\mu\nu}$~and $p_1^\mu p_1^\nu$
respectively yields a linear system of equations,
\beq \left(
  \begin{array}{cc}
     p_1^2 & n \\
     p_1^2 & 1
  \end{array} \right) \left(
  \begin{array}{c}
    \!\! B_{21} \!\! \\ \!\! B_{22} \!\!
  \end{array} \right) \;\; = \;\; \left(
  \begin{array}{c}
     A_0(m_2^2) - m_1^2\,B_0 \\
     \! \frac{1}{2} A_0(m_2^2) + \frac{1}{2}\,f_1 B_1 \!
  \end{array} \right)~~,
\eeq
which has the solution
\ba
  B_{22} & = & \frac{1}{n-1} \left[ \rule[0cm]{0cm}{.58cm}
    - \frac{1}{2}\,f_1 B_1 - m_1^2 B_0 + \frac{1}{2} A_0(m_2^2) \right]
  \nl \nl
  B_{21} & = & \frac{1}{p_1^2} \left[ \rule[0cm]{0cm}{.58cm} \;
    \frac{1}{2}\,f_1 B_1 + \frac{1}{2} A_0(m_2^2) - B_{22} \right]~~.
\ea
The above already reveals the algorithm for the reduction of
tensor $n$-point functions. {\it Multiplication of both sides of the
  tensor integral's linear decomposition (see equations~(\ref{tendec1})
  to~(\ref{tendec5})) by the integral's tensor coefficients and
  subsequent evaluation of the resulting integrals with the help of
  quadratic supplementation of the integrand yields a linear system of
  equations for the linear factors $B_{i(j)}$ and $C_{ij}$ in terms of
  scalar $m$-point functions.}
This resulting system originally consists of $k$ equations for the $k$
linear factors to be evaluated. A more elegant way with reduced
dimension of the system of equations to be evaluated is given in
references~\cite{wimdiss,passarino79}. Subsequently, we shall follow
their approach and present the solutions for the remaining linear
factors $C_{ij}$. First define
\ba
  B_{(0,\mu,\mu\nu)}(k,l) & \equiv & \;
    \mu^{-(n-4)}\int\!\!\frac{d^{n}q}{(2\pi)^n} \left.
    \: \frac{\{1,\,q_\mu,\,q_\mu q_\nu\}}{D_k\,D_l}
    \right|_{Q_k = 0}
\ea
with~$Q_1 \! = \! 0$~and~$Q_2 \! = \! p_1$. The denominators for this
definition are found in equation~(\ref{loopden}). The arguments
$(k,l)$,~$k\!<\!l$~will be correspondingly used for the
linear factors~$B_{i(j)}(k,l)$ which can be derived by
virtue of equations~(\ref{tendec1}) and~(\ref{tendec2}). Using ~$f_2
\equiv \!m_2^2 - m_3^2 + p_1^2 - (p_1+p_2)^2$~ and considering
\begin{center}
$ X \; = \;
  \left( \begin{array}{cc}
    \! p_1^2 & p_1 p_2 \! \\
    \! p_1 p_2 & p_2^2 \!
  \end{array} \right)
$
\end{center}
an invertible matrix (i.e. assuming~ $(p_1 p_2)^2 - p_1^2 p_2^2 \neq
0$), the linear factors $C_{1j}$ for the \rhs~of equation~(\ref{tendec3})
are given by
\ba
  \left( \begin{array}{c}
    \!\! C_{11} \!\! \\ \!\! C_{12} \!\!
  \end{array} \right)
  & = & X^{-1}
  \left( \begin{array}{c}
    \!\! R_1 \!\! \\ \!\! R_2 \!\!
  \end{array} \right) \\
  R_1 & \equiv & p_1^\mu C_\mu \;\; = \;\;
    \frac{1}{2} \left[ \rule[0cm]{0cm}{.4cm} \,
    f_1 C_0 + B_0(1,3) - B_0(2,3) \right] \nl
  R_2 & \equiv & p_2^\mu C_\mu \;\; = \;\;
    \frac{1}{2} \left[  \rule[0cm]{0cm}{.4cm} \,
    f_2 C_0 + B_0(1,2) - B_0(1,3) \right]~~. \nonumber
\ea
The linear factors $C_{2j}$ on the \rhs~of equation~(\ref{tendec4}) are
given by similar matrix relations, namely
\ba
    \left( \begin{array}{c}
    \!\! C_{21} \!\! \\ \!\! C_{23} \!\!
  \end{array} \right)
  & = & X^{-1}
  \left( \begin{array}{c}
    \!\! R_3 - C_{24} \!\! \\ \!\! R_4 \!\!
  \end{array} \right)~, \hspace{1.2cm}
  \left( \begin{array}{c}
    \!\! C_{23} \!\! \\ \!\! C_{22} \!\!
  \end{array} \right)
  \;\; = \;\; X^{-1}
  \left( \begin{array}{c}
    \!\! R_5 \!\! \\ \!\! R_6 - C_{24} \!\!
  \end{array} \right)~.
\ea
The UV divergent linear factor
\ba
  C_{24} & = & \frac{1}{n-2} \left[  \rule[0cm]{0cm}{.58cm}
    -m_1^2 C_0 - \frac{1}{2} \left( \rule[0cm]{0cm}{.35cm}
    f_1 C_{11} + f_2 C_{12} - B_0(2,3) \right) \right] \nonumber
\ea
is obtained from the original~ $5\times5$~system of linear equations
and
\ba
  R_3 & \equiv & \frac{1}{p_1 p_2} \left[ \rule[0cm]{0cm}{.4cm} \,
    p_1^\mu p_2^\nu \, C_{\mu\nu} - p_2^2 \, R_5 \right]
    \hspace{1.26cm} = \;\;
     \frac{1}{2} \left[ \rule[0cm]{0cm}{.4cm} \,
     f_1 C_{11} + B_1(1,3) + B_0(2,3) \right]~~, \nl
  R_4 & \equiv &
    \frac{(p_1 p_2) \cdot p_2^\mu p_2^\nu -
          p_2^2 \cdot p_1^\mu p_2^\nu}
         {(p_1 p_2)^2 - p_1^2 p_2^2} \: C_{\mu\nu} \hspace{.53cm} = \;\;
     \frac{1}{2} \left[ \rule[0cm]{0cm}{.4cm} \,
     f_2 C_{11} + B_1(1,2) - B_1(1,3) \right]~~, \nl
  R_5 & \equiv &
    \frac{(p_1 p_2) \cdot p_1^\mu p_1^\nu -
          p_1^2 \cdot p_1^\mu p_2^\nu}
         {(p_1 p_2)^2 - p_1^2 p_2^2} \: C_{\mu\nu} \hspace{.52cm} = \;\;
     \frac{1}{2} \left[ \rule[0cm]{0cm}{.4cm} \,
     f_1 C_{12} + B_1(1,3) - B_1(2,3) \right]~~, \nl
  R_6 & \equiv & \frac{1}{p_1 p_2} \left[ \rule[0cm]{0cm}{.4cm} \,
    p_1^\mu p_2^\nu \, C_{\mu\nu} - p_1^2 \, R_4 \right]
    \hspace{1.26cm} = \;\;
     \frac{1}{2} \left[ \rule[0cm]{0cm}{.4cm} \,
     f_2 C_{12} - B_1(1,3) \right]~. \nonumber
\ea
Finally, the matrix relations for the linear factors $C_{3j}$ are
given by
\ba
  \left( \begin{array}{c}
    \!\! C_{35} \!\! \\ \!\! C_{36} \!\!
  \end{array} \right)
  & = & X^{-1}
  \left( \begin{array}{c}
    \!\! R_{10} \!\! \\ \!\!  R_{11} \!\!
  \end{array} \right)~,
  \nl \nl
  \left( \begin{array}{c}
    \!\! C_{31} \!\! \\ \!\! C_{33} \!\!
  \end{array} \right)
  & = & X^{-1}
  \left( \begin{array}{c}
    \!\! R_{12} - 2\,C_{35} \!\! \\ \!\!  R_{13} \!\!
  \end{array} \right)~, \hspace{.9cm}
  \left( \begin{array}{c}
    \!\! C_{34} \!\! \\ \!\! C_{32} \!\!
  \end{array} \right)
  \;\; = \;\; X^{-1}
  \left( \begin{array}{c}
    \!\! R_{14} \!\! \\ \!\!  R_{15} - 2\,C_{36} \!\!
  \end{array} \right).
\ea
In addition there is the redundant matrix relation
\ba
  \left( \begin{array}{c}
    \!\! C_{33} \!\! \\ \!\! C_{34} \!\!
  \end{array} \right)
  & = & X^{-1}
  \left( \begin{array}{c}
    \!\! R_{16} - C_{36} \!\! \\ \!\!  R_{17} - C_{35} \!\!
  \end{array} \right)~.
\ea
The scalar factors needed are
\ba
  R_{10} & \equiv & \frac{1}{2} \left[ \rule[0cm]{0cm}{.4cm} \,
    f_1 C_{24} + B_{22}(1,3) - B_{22}(2,3) \right]~~, \nl
  R_{11} & \equiv & \frac{1}{2} \left[ \rule[0cm]{0cm}{.4cm} \,
    f_2 C_{24} + B_{22}(1,2) - B_{22}(1,3) \right]~~, \nl
  R_{12} & \equiv & \frac{1}{2} \left[ \rule[0cm]{0cm}{.4cm} \,
    f_1 C_{21} + B_{21}(1,3) - B_0(2,3) \right]~~, \nl
  R_{13} & \equiv & \frac{1}{2} \left[ \rule[0cm]{0cm}{.4cm} \,
    f_2 C_{21} + B_{21}(1,2) - B_{21}(1,3) \right]~~, \nl
  R_{14} & \equiv & \frac{1}{2} \left[ \rule[0cm]{0cm}{.4cm} \,
    f_1 C_{22} + B_{21}(1,3) - B_{21}(2,3) \right]~~, \nl
  R_{15} & \equiv & \frac{1}{2} \left[ \rule[0cm]{0cm}{.4cm} \,
    f_2 C_{22} - B_{21}(1,3) \right]~~, \nl
  R_{16} & \equiv & \frac{1}{2} \left[ \rule[0cm]{0cm}{.4cm} \,
    f_1 C_{23} + B_{21}(1,3) + B_1(2,3) \right]~~, \nl
  R_{17} & \equiv & \frac{1}{2} \left[ \rule[0cm]{0cm}{.4cm} \,
    f_2 C_{23} - B_{21}(1,3) \right]~~. \nonumber
\ea
This completes our outline of the reduction of tensor to scalar loop
integrals and the appendix on loop integrals as a whole.
%
\chapter{Table of Analytical Integrals}
\label{tabofint}
In this appendix the analytical solutions of the integrals needed for
the semi-analytical \xsec calculation of process~(\ref{eezz4f}) are
presented. These integrals are necessary to obtain the twofold
differential cross-section $d^2\sigma/ d\SONE d\STWO$ in the Born and
virtual ISR cases and the threefold differential \xsec
$d^3\sigma/ds' d\SONE d\STWO$ for initial state bremsstrahlung.
This list is restricted to integrals needed for the t-channel
contribution and the t-channel/u-channel interference, because the
contributions from the u-channel and, in the virtual ISR case, from
the u-channel/t-channel interference are obtained by symmetry
arguments as detailed in appendix~\ref{xscalc}.
Integration of the azimuth around the beam axis is
trivial. The integrals presented below have been numerically checked.
\vspace{.5cm}
%
\section{Fermion Decay Angle Tensor Integrals}
\label{tenint}
Integrations over final state fermion decay angles are carried out
with the help of invariant tensor integration. Using the decay matrix
element $M_{ij}^\mu$~from equation~(\ref{bornmat}) one obtains
\ba
  \sum_{Spins~i,j} \, \int d\Gamma_{ij} \: M_{ij}^\mu M_{ij}^\nu
  & \equiv &
  \frac{\sqrt{\lambda(s_{ij},m_{ij}^2,m_{ij}^2)}}{8 \, s_{ij}} \:
    \sum_{Spins~i,j} \: \int_{-1}^{+1} d\!\mycos\theta_{ij} \:
    \int_0^{2\pi} d\phi_{ij} \: M_{ij}^\mu M_{ij}^\nu \nl
  & = & \frac{\sqrt{\lambda(s_{ij},m_{ij}^2,m_{ij}^2)}}{8 \, s_{ij}}
    \; \frac{2\pi}{3}
    \left( {g^V_{f\!{\bar f}\!B}}^2 + {g^A_{f\!{\bar f}\!B}}^2 \right)
    \times \nl
  & & \hspace{2.4cm}
    \left\{ \rule[0cm]{0cm}{.35cm} s_{ij} \, g^{\mu\nu} +
      \left(p_i^\mu + p_j^\mu\right) \left(p_i^\nu + p_j^\nu\right)
    \right\}
  \label{dectenint}
\ea
with $m_{12}\!\equiv\!m_1$~and $m_{34}\!\equiv\!m_2$. The vector and
axial vector boson-fermion couplings
$g^V_{f\!{\bar f}\!B}$~and $g^A_{f\!{\bar f}\!B}$~are  found in
appendix~\ref{feynrules}. The result of the above decay phase space
integral factorizes in the cross-section. Therefore one can use the
above formula~(\ref{dectenint}) to integrate the final state fermion
decay angles for both the $2\!\to\!4$~and the initial
state bremsstrahlung $2\!\to\!5$~particle phase space. The factor
$\left({g^V_{f\!{\bar f}\!B}}^2 + {g^A_{f\!{\bar f}\!B}}^2 \right)$
enters the decay width $\Gamma_Z (s_{ij})$, equation~(\ref{gzoff1}),
and thus the Breit-Wigner density $\rho_Z(s_{ij})$,
equation~(\ref{rhos}).
\newpage
\noindent
%
\section{Born Level Phase Space Integrals}
\label{bornint}
For the calculation of the Born level \xsec of
process~(\ref{eezz4f}) only a few integrals over the boson scattering
angle $\varth$ are needed in addition to those over the final state
fermion decay angles. They are listed below, making use of the phase
space parametrization given in appendix~\ref{ps2to4}, the
ultrarelativistic approximation (URA), and the notation
\beq
  \hspace{4.7cm} \left[A\right]_t \equiv \frac{1}{2}
  \int\limits_{-1}^{+1} d\mycos\varth~A ~~.
\eeq
\vspace{.3cm}
$\begin{array}{llll}
  1.) & \left[1\right]_t & = & 1 \medskip
\end{array}$
\newline
$\begin{array}{llll}
  2.) & \left[\mycos\varth\right]_t & = & 0 \medskip
  4.) & {\ds \left[ \frac{1}{\,t\,} \right]_t } & = &
        {\ds \frac{1}{\SLAM} \, \myln \frac{s\!-\!\sigma\!+\!\SLAM}
             {s\!-\!\sigma\!-\!\SLAM} } \medskip
\end{array}$
$\begin{array}{llll}
\hspace{2cm}
  3.) & \left[\mycos^2\varth\right]_t & = & 1/3 \medskip
\hspace{2cm}
  5.) & {\ds \left[ \frac{1}{\,u\,} \right]_t } & = &
        {\ds \frac{1}{\SLAM} \, \myln \frac{s\!-\!\sigma\!+\!\SLAM}
             {s\!-\!\sigma\!-\!\SLAM} } \medskip
\end{array}$
\newline
$\begin{array}{llll}
  6.) & {\ds \left[ \frac{1}{\,t^2\,} \right]_t } & = &
        {\ds \frac{1}{\SONE \STWO} } \medskip
\end{array}$
\vspace{1cm}
%
\section{Bremsstrahlung Integrals -- First Series}
\label{bremint1}
For the calculation of the bremsstrahlung process~(\ref{eezz4f}),
the phase space was parametrized as exposed in
equation~(\ref{par25b}), appendix~\ref{ps2to5}. After integration
over the final state fermion decay angles, the first series of
brems\-strah\-lung integrals is over the scattering azimuth and polar
angles $\phi_R$ and $\theta_R$ of the boson three-vector $v_1$ in the
two-boson rest frame. Using the double integral notation
\begin{equation}
  \left[A\right]_R \equiv \frac{1}{4\pi}\int\limits_{0}^{2\pi} d\phi_R
                   \int\limits_{-1}^{+1} d\mycos\theta_R~A
\end{equation}
and the ultrarelativistic approximation, i.e. neglecting
the electron mass wherever possible, the bremsstrahlung integrals for
the first series are listed below:
\vspace{.3cm} \\
$\begin{array}{llll}
 1.) & \left[1\right]_R & = & 1 \hugeskip
 3.) & \left[\mycos\theta_R\right]_R & = & 0 \hugeskip
 5.) & \left[\mycos^2\theta_R\right]_R & = & 1/3 \hugeskip
\end{array}$
$\begin{array}{llll}
\hspace{3.5cm} 2.) & \left[\mycos\phi_R \cdot f(\mycos\theta_R)\right]_R
                    & = & 0 \hugeskip
\hspace{3.5cm} 4.) & \left[\mycos^2\phi_R\right]_R & = & 1/2 \hugeskip
\hspace{3.5cm} 6.) & \left[\mycos^2\phi_R \cdot
                           \mycos^2\theta_R\right]_R & = & 1/6
                     \hugeskip
\end{array}$
\newline
$\begin{array}{llll}
 7.) & {\ds \left[\frac{1}{t_1}\right]_R } & = &
       {\ds \frac{2s'}{\SLAMP (\sprp - \sprm\mycos\theta)} \: \cdot }
         \myln \left(
            \frac{\sprp(s'+\SLAMP) - 2s'\sigma + \sprm\delta -
                (s'+\delta+\SLAMP)\sprm \mycos\theta}
                {\sprp(s'-\SLAMP) - 2s'\sigma + \sprm\delta -
                (s'+\delta-\SLAMP)\sprm \mycos\theta}
            \right)^{\footnotemark[2]} \smskip
         & & \equiv &
         {\ds \frac{2s'}{\SLAMP \, S_{d1}} \cdot l_{t1} }
       \bigskip \\
 8.) & {\ds \left[\frac{1}{t_2}\right]_R } & = &
       {\ds \frac{2s'}{\SLAMP(\sprp + \sprm\mycos\theta)} \: \cdot }
         \myln \left(
            \frac{\sprp(s'+\SLAMP) - 2s'\sigma - \sprm\delta +
                (s'-\delta+\SLAMP)\sprm \mycos\theta}
                {\sprp(s'-\SLAMP) - 2s'\sigma - \sprm\delta +
                (s'-\delta-\SLAMP)\sprm \mycos\theta}
            \right) \smskip
         & & \equiv &
         {\ds \frac{2s'}{\SLAMP \, S_{d2}} \cdot l_{t2} }
        \bigskip \\
 9.) & {\ds \left[\frac{1}{u_1}\right]_R } & = &
       {\ds \frac{2s'}{\SLAMP(\sprp - \sprm\mycos\theta)} \: \cdot }
         \myln \left(
            \frac{\sprp(s'+\SLAMP) - 2s'\sigma - \sprm\delta -
                (s'-\delta+\SLAMP)\sprm \mycos\theta}
                {\sprp(s'-\SLAMP) - 2s'\sigma - \sprm\delta -
                (s'-\delta-\SLAMP)\sprm \mycos\theta}
            \right) \smskip
         & & = & {\ds {\cal S\!}
           \left(\left[\frac{1}{t_2}\right]_R\right)
           \raisebox{.35cm}[0cm][0cm]{$\!\!\footnotemark[3]\,\,$}
           \; \equiv \; \frac{2s'}{\SLAMP \, S_{d1}} \cdot l_{u1}  }
        \bigskip \\
 10.) & {\ds \left[\frac{1}{u_2}\right]_R } & = &
        {\ds \frac{2s'}{\SLAMP(\sprp + \sprm\mycos\theta)} \: \cdot }
         \myln \left(
            \frac{\sprp(s'+\SLAMP) - 2s'\sigma + \sprm\delta +
                (s'+\delta+\SLAMP)\sprm \mycos\theta}
                {\sprp(s'-\SLAMP) - 2s'\sigma + \sprm\delta +
                (s'+\delta-\SLAMP)\sprm \mycos\theta}
            \right) \smskip
         & & = & {\ds {\cal S\!}\left(\left[\frac{1}{t_1}\right]_R\right)
           \; \equiv \; \frac{2s'}{\SLAMP \, S_{d2}} \cdot l_{u2}  }
        \bigskip \\
 11.) & {\ds \left[ \frac{\cos\theta_R}{t_1}\right]_R } & = &
        {\ds \frac{16\;s'^2 b_1}{\lambda^{'} \cdot {S_{d1}}^2} \cdot
            \left( 1 - a^t_1 \left[ \frac{1}{t_1}\right]_R \right) }
        \bigskip \\
 12.) & {\ds \left[ \frac{\cos\theta_R}{t_2}\right]_R } & = &
        {\ds \frac{16\;s'^2 b_2}{\lambda^{'} \cdot {S_{d2}}^2} \cdot
            \left( 1 - a^t_2 \left[ \frac{1}{t_2}\right]_R \right) }
        \bigskip \\
\end{array}$
\footnotetext[2]{As used throughout the thesis, $s$~($s'$) is the
  (reduced) center of mass energy squared, and $\SONE$, $\STWO$ are
  the invariant masses of final state fermion pairs. The definitions
  of the derived quantities $t, u, t_1, t_2, u_1, u_2, z_1, z_2,
  {\bar z_1}, {\bar z_2}, \sprm, \sprp, \lambda, \lambda^{'},
  {\bar \lambda}, \sigma, \delta, a_1^t, a_2^t, a_1^u, a_2^u, b_1,
  b_2,$~and $c$~are given in appendices~\ref{phasespa}
  and~\ref{xscalc} where details of the applied phase space
  parametrizations and the cross-section computation are worked out.}
\footnotetext[3]{The symmetry operations $\cs$~and $\ct$~are
  defined by: \hspace*{.5cm} $\cs\left[\rule[0cm]{0cm}{.35cm}
    f(\mycos\theta)\right] \;\; \equiv \;\; f(-\mycos\theta)~~,$
  \vspace{.2cm} \\
    \hspace*{8.625cm} $\ct\left[\rule[0cm]{0cm}{.35cm}
  g(\SONE,\STWO)\right] \;\; \equiv \;\; g(\STWO,\SONE)~~.$ }
\newpage
$\begin{array}{llll}
 13.) & {\ds \left[ \frac{1}{t_1^2}\right]_R } & = &
        {\ds \frac{4s'}{\SONE \cdot
         \left(s - \delta + \SLAMP - \sprm \cdot\mycos\theta \right)
         \left(s - \delta - \SLAMP - \sprm \cdot\mycos\theta \right) } }
      \smskip
      & & \equiv & {\ds \frac{4s'}{\SONE \cdot d_1^+ \, d_1^- } }
        \bigskip \\
 14.) & {\ds \left[ \frac{1}{t_2^2}\right]_R } & = &
        {\ds \frac{4s'}{\STWO \cdot
         \left(s + \delta + \SLAMP + \sprm \cdot\mycos\theta \right)
         \left(s + \delta - \SLAMP + \sprm \cdot\mycos\theta \right) } }
      \smskip
      & & \equiv & {\ds \frac{4s'}{\STWO \cdot d_2^+ \, d_2^- } }
        \bigskip \\
 15.) & {\ds \left[ \frac{1}{t_1t_2}\right]_R } & = &
        {\ds \frac{1}{2 S_{t_1t_2}} \cdot
         \myln \left( \frac{A^t_{1-} \cdot A^t_{2+}}
                        {A^t_{1+} \cdot A^t_{2-}} \right)
         \;\; \equiv \;\; \frac{1}{2 S_{t_1t_2}} \cdot l_{t12} }
       \bigskip
     & \indent S_{t_1t_2} & = & {\ds \frac{\SLAMP}{8s'} \: \cdot
         \left( \sqrt{\bar{\lambda}} + \delta + \sprm\mycos\theta
         \right)
         \left( \sqrt{\bar{\lambda}} - \delta - \sprm\mycos\theta
         \right) } \miniskip
     & & \equiv & {\ds \frac{\SLAMP}{8s'} \: \cdot
                       S^t_{12+} \, S^t_{12-} }
        \smskip
     & \indent A^t_{1-} & = & a_{1-} \; + \; b_{1-} \cdot \mycos\theta
       \smskip
     & \indent A^t_{1+} & = & a_{1+} \; + \; b_{1+} \cdot \mycos\theta
       \smskip
     & \indent A^t_{2-} & = & a_{2-} \; + \; b_{2-} \cdot \mycos\theta
       \smskip
     & \indent A^t_{2+} & = & a_{2+} \; + \; b_{2+} \cdot \mycos\theta
       \bigskip
     & \indent\; a_{1-} & = & {\ds s^2\sigma - s\delta^2 +
         \frac{s}{2} \sprm \left( s' + \delta - 2\sigma\right ) -
         \SLAMP \left( s\delta - \frac{s}{2}\sprm \right) }
       \smskip
     & \indent\; b_{1-} & = &
         {\ds - \left( \SLAMP + s' + \delta \right) \frac{s}{2} \sprm }
       \smskip
     & \indent\; a_{1+} & = & {\ds s^2\sigma - s\delta^2 +
         \frac{s}{2} \sprm \left( s' + \delta - 2\sigma\right ) +
         \SLAMP \left( s\delta - \frac{s}{2}\sprm \right) }
       \smskip
     & \indent\; b_{1+} & = &
         {\ds + \left( \SLAMP - s' - \delta \right) \frac{s}{2} \sprm }
       \smskip
     & \indent\; a_{2-} & = & {\ds s^2\sigma - s\delta^2 +
           \frac{s}{2} \sprm \left( s' - \delta - 2\sigma\right ) -
           \SLAMP \left( s\delta + \frac{s}{2}\sprm \right) }
       \smskip
     & \indent\; b_{2-} & = &
         {\ds - \left( \SLAMP - s' + \delta \right) \frac{s}{2} \sprm }
       \smskip
     & \indent\; a_{2+} & = & {\ds s^2\sigma - s\delta^2 +
           \frac{s}{2} \sprm \left( s' - \delta - 2\sigma\right ) +
           \SLAMP \left( s\delta + \frac{s}{2}\sprm \right) }
       \smskip
     & \indent\; b_{2+} & = &
         {\ds + \left( \SLAMP + s' - \delta \right) \frac{s}{2} \sprm }
       \bigskip \\
\end{array}$
\newpage
$\begin{array}{llll}
 16.) & {\ds \left[ \frac{1}{u_1u_2}\right]_R } & = &
        {\ds {\cal S\!} \left(\left[\frac{1}{t_1t_2}\right]_R\right)
          \; = \; \frac{1}{2 S_{u_1u_2}} \cdot
          \myln \left( \frac{A^u_{1-} \cdot A^u_{2+}}
                        {A^u_{1+} \cdot A^u_{2-}} \right)
         \;\; \equiv \;\; \frac{1}{2 S_{u_1u_2}} \cdot l_{u12} }
       \bigskip
     & \indent S_{u_1u_2} & = & {\ds \frac{\SLAMP}{8s'} \: \cdot
         \left( \sqrt{\bar{\lambda}} + \delta - \sprm\mycos\theta
         \right)
         \left( \sqrt{\bar{\lambda}} - \delta + \sprm\mycos\theta
         \right) } \miniskip
     & & \equiv & {\ds \frac{\SLAMP}{8s'} \: \cdot
                       S^u_{12+} \, S^u_{12-} }
       \smskip

     & \indent A^u_{1-} & = & a_{1-} \; - \; b_{1-} \cdot \mycos\theta
       \smskip
     & \indent A^u_{1+} & = & a_{1+} \; - \; b_{1+} \cdot \mycos\theta
       \smskip
     & \indent A^u_{2-} & = & a_{2-} \; - \; b_{2-} \cdot \mycos\theta
       \smskip
     & \indent A^u_{2+} & = & a_{2+} \; - \; b_{2+} \cdot \mycos\theta
       \bigskip \\
 17.) & {\ds \left[ \frac{1}{t_1u_1}\right]_R } & = &
        {\ds \frac{1}{a^{ut}_1} \left(
            \left[\frac{1}{t_1}\right]_R + \left[\frac{1}{u_1}\right]_R
                             \right) }
        \bigskip
      & \indent a^{ut}_1 & = &
          {\ds \frac{1}{2} \left( \sprp - \sprm\mycos\theta \right)
            - \sigma
          \;\; \equiv \;\; \frac{a}{2}
            \left( 1 - b \! \cdot \! \mycos\theta \right) }
        \bigskip \\
 18.) & {\ds \left[ \frac{1}{t_2u_2}\right]_R } & = &
        {\ds \frac{1}{a^{ut}_2} \left(
            \left[\frac{1}{t_2}\right]_R + \left[\frac{1}{u_2}\right]_R
                             \right) }
        \bigskip
      & \indent a^{ut}_2 & = &
          {\ds \frac{1}{2} \left( \sprp + \sprm\mycos\theta \right)
           - \sigma
          \;\; \equiv \;\; \frac{a}{2}
            \left( 1 + b \! \cdot \! \mycos\theta \right) }
        \bigskip \\
 19.) & {\ds \left[ \frac{1}{t_1u_2}\right]_R } & = &
        {\ds \frac{1}{2 \sqrt{C_{12}}} \cdot
          \left[ \rule[-.2cm]{0cm}{1cm} \;
          \myln \left( \frac{A^-_1 \cdot B^-_2}{A^+_1 \cdot B^+_2}
                \right)
          + 2 \myln \left( \frac{a_{s1} + b_d}{a_{s1} - b_d} \right)
                                       \; \right]
        \; \equiv \; \frac{1}{2 \sqrt{C_{12}}} \cdot l_{t_1u_2} }
        \bigskip \\
      & \indent\;\;\; A_1^- & = &
          \sqrt{C_{12}} \cdot b_d \cdot \left( a^t_1 + b_1 \right) +
          C_{12} + B^t_1 \cdot \left( a_{s1} - b_d \right)/2
        \smskip
      & \indent\;\;\; A_1^+ & = &
          \sqrt{C_{12}} \cdot b_d \cdot \left( a^t_1 - b_1 \right) +
          C_{12} + B^t_1 \cdot \left( a_{s1} + b_d \right)/2
        \smskip
      & \indent\;\;\; B_2^- & = &
          \sqrt{C_{12}} \cdot b_d \cdot \left( a^u_2 - b_2 \right) +
          C_{12} + B^u_2 \cdot \left( a_{s1} - b_d \right)/2
        \smskip
      & \indent\;\;\; B_2^+ & = &
          \sqrt{C_{12}} \cdot b_d \cdot \left( a^u_2 + b_2 \right) +
          C_{12} + B^u_2 \cdot \left( a_{s1} + b_d \right)/2
        \smskip \\
      & \indent\;\;\; C_{12} & = &
          {\ds \left( \frac{\SLAMP}{2s'} \right)^2 \cdot
          \left( A_{12} \cdot \mycos^2\theta + B_{12} \right) }
        \smskip
      & \indent\;\;\; A_{12} & = &
          - s'^{\;2}_- \cdot \SONE \cdot \left( s - \SONE \right)
        \smskip
      & \indent\;\;\; B_{12} & = &
          s'^{\;2}_- \cdot s \cdot \left( s - \SONE \right) -
          2 s \sprm \left( s - \sigma \right) \left( s - \SONE \right)
          + s^2 \left( s - \sigma \right)^2
\end{array}$
\newpage
$\begin{array}{llll}
      & \indent \indent B^t_1 & = &
          -2 \left[ \rule[0cm]{0cm}{.4cm}
               b_1 \left( a^t_1 b_2 + a^u_2 b_1 \right) + a_{s1}c^2
             \right]
        \smskip
      & \indent \indent B^u_2 & = &
          -2 \left[ \rule[0cm]{0cm}{.4cm}
               b_2 \left( a^t_1 b_2 + a^u_2 b_1 \right) + a_{s1}c^2
             \right]
        \smskip
      & \indent \indent a_{s1} & = & {\ds a^t_1 + a^u_2 \; = \;
          \frac{\sprp}{2s'} \left( s' + \delta \right) - 2 \SONE }
        \smskip
      & \indent \indent b_d & = & b_2 - b_1 \; = \;
          {\ds \frac{\SLAMP}{2s'} \sprm }
        \bigskip \\
 20.) & {\ds \left[ \frac{1}{t_2u_1}\right]_R } & = &
        {\ds {\cal T\!} \left( \left[\frac{1}{t_1u_2}\right]_R \right)}
        \smskip
      & & = &
        {\ds \frac{1}{2 \sqrt{C_{21}}} \cdot
        \left[ \rule[-.2cm]{0cm}{1cm} \;
          \myln \left( \frac{A^-_2 \cdot B^-_1}{A^+_2 \cdot B^+_1}
                \right)
          + 2 \myln \left( \frac{a_{s2} + b_d}{a_{s2} - b_d} \right)
                                    \; \right]
        \; \equiv \; \frac{1}{2 \sqrt{C_{21}}} \cdot l_{t_2u_1} }
        \smskip \\
      & \indent \indent A_2^- & = &
          \sqrt{C_{21}} \cdot b_d \cdot \left( a^t_2 - b_2 \right) +
          C_{21} + B^t_2 \cdot \left( a_{s2} - b_d \right)/2
        \smskip
      & \indent \indent A_2^+ & = &
          \sqrt{C_{21}} \cdot b_d \cdot \left( a^t_2 + b_2 \right) +
          C_{21} + B^t_2 \cdot \left( a_{s2} + b_d \right)/2
        \smskip
      & \indent \indent B_1^- & = &
          \sqrt{C_{21}} \cdot b_d \cdot \left( a^u_1 + b_1 \right) +
          C_{21} + B^u_1 \cdot \left( a_{s2} - b_d \right)/2
        \smskip
      & \indent \indent B_1^+ & = &
          \sqrt{C_{21}} \cdot b_d \cdot \left( a^u_1 - b_1 \right) +
          C_{21} + B^u_1 \cdot \left( a_{s2} + b_d \right)/2
        \smskip
      & \indent \indent C_{21} & = &
          {\ds \left( \frac{\SLAMP}{2s'} \right)^2 \cdot
          \left( A_{34} \cdot \mycos^2\theta + B_{34} \right) }
        \smskip
      & \indent \indent A_{34} & = &
          - s'^{\;2}_- \cdot \STWO \cdot \left( s - \STWO \right)
        \smskip
      & \indent \indent B_{34} & = &
          s'^{\;2}_- \cdot s \cdot \left( s - \STWO \right) -
          2 s \sprm \left( s - \sigma \right) \left( s - \STWO \right)
          + s^2 \left( s - \sigma \right)^2
        \smskip
      & \indent \indent B^t_2 & = &
          -2 \left[ \rule[0cm]{0cm}{.4cm}
               b_2 \left( a^u_1 b_2 + a^t_2 b_1 \right) + a_{s2}c^2
             \right]
        \smskip
      & \indent \indent B^u_1 & = &
          -2 \left[ \rule[0cm]{0cm}{.4cm}
               b_1 \left( a^u_1 b_2 + a^t_2 b_1 \right) + a_{s2}c^2
             \right]
        \smskip
      & \indent \indent a_{s2} & = & a^t_2 + a^u_1 \; = \;
          {\ds \frac{\sprp}{2s'} \left( s' - \delta \right) - 2 \STWO }
        \bigskip \\
 21.) & {\ds \left[ \frac{\cos\theta_R}{t_1^2}\right]_R } & = &
        {\ds \frac{\,16\:s'^2 \, b_1\,}{\lambda^{'} {S_{d1}}^2} \cdot
             \left( \left[ \frac{1}{t_1}\right]_R
                   - a^t_1 \left[ \frac{1}{t_1^2}\right]_R \right) }
        \bigskip \\
 22.) & {\ds \left[ \frac{\cos\theta_R}{t_2^2}\right]_R } & = &
        {\ds \frac{\,16\:s'^2\,b_2\,}{\lambda^{'} {S_{d2}}^2} \cdot
             \left( \left[ \frac{1}{t_2}\right]_R
                   - a^t_2 \left[ \frac{1}{t_2^2}\right]_R \right) }
        \bigskip \\
\end{array}$
\newpage
$\begin{array}{llll}
 23.) & {\ds \left[ \frac{1}{t_1^2t_2}\right]_R } & = &
        {\ds \frac{1}{{S_{t_1t_2}}^2}
          \left( -a^t_d \; + \;  a^t_1
                 \left( a^t_1\,a^t_d - b_1\,b_d \right)
                 \left[ \frac{1}{t_1^2}\right]_R \right. }
        \smskip & & & \indent \indent \;\: {\ds \left. +
                 \left( b_2 \, (a^t_1 b_2 - a^t_2 b_1) - a^t_d \, c^2
                 \right)
                 \left[ \frac{1}{t_1t_2}\right]_R
             \right) }
        \medskip
      & \indent a^t_d & = & a^t_2 -a^t_1 \; = \;
        {\ds - \frac{\sprm}{2\,s'}
             \left(\delta - s' \, \mycos\theta \right) }
        \bigskip \\
 24.)  & {\ds \left[ \frac{1}{t_1t_2^2}\right]_R } & = &
         {\ds \frac{1}{{S_{t_1t_2}}^2}
           \left( a^t_d \; - \; a^t_2
                  \left( a^t_2\,a^t_d - b_2\,b_d \right)
                  \left[ \frac{1}{t_2^2}\right]_R \right. }
         \smskip & & & \indent \indent \;\: {\ds \left. -
                  \left( b_1 \, (a^t_1 b_2 - a^t_2 b_1) - a^t_d \, c^2
                  \right)
                  \left[ \frac{1}{t_1t_2}\right]_R
              \right) }
        \bigskip \\
 25.)  & {\ds \left[ \frac{1}{t_1t_2u_1}\right]_R } & = &
         {\ds \frac{1}{a^{ut}_1} \left(
           \left[ \frac{1}{t_1t_2} \right]_R +
           \left[ \frac{1}{t_2u_1} \right]_R
                                 \right) }
        \bigskip \\
 26.)  & {\ds \left[ \frac{1}{t_1t_2u_2}\right]_R } & = &
         {\ds \frac{1}{a^{ut}_2} \left(
           \left[ \frac{1}{t_1t_2} \right]_R +
           \left[ \frac{1}{t_1u_2} \right]_R
                                 \right) }
        \bigskip \\
 27.)  & {\ds \left[ \frac{1}{t_1u_1u_2}\right]_R } & = &
         {\ds \frac{1}{a^{ut}_1} \left(
           \left[ \frac{1}{u_1u_2} \right]_R +
           \left[ \frac{1}{t_1u_2} \right]_R
                                 \right) }
        \bigskip \\
 28.)  & {\ds \left[ \frac{1}{t_2u_1u_2}\right]_R } & = &
         {\ds \frac{1}{a^{ut}_2} \left(
           \left[ \frac{1}{u_1u_2} \right]_R +
           \left[ \frac{1}{t_2u_1} \right]_R
                                 \right) }
	        \bigskip \\
 29.)  & {\ds \left[ \frac{1}{t_1t_2u_1u_2}\right]_R } & = &
         {\ds \frac{1}{a^{ut}_1} \cdot \frac{1}{a^{ut}_2}
         \left(
           \left[ \frac{1}{t_1t_2} \right]_R +
           \left[ \frac{1}{u_1u_2} \right]_R +
           \left[ \frac{1}{t_1u_2} \right]_R +
           \left[ \frac{1}{t_2u_1} \right]_R
         \right)~~. }
        \bigskip \\
\end{array}$
\newpage
%
\section{Bremsstrahlung Integrals -- Second Series}
\label{bremint2}
Below, the integrals needed for the last analytical integration in the
brems\-strah\-lung case are given. This integration is over the
photon scattering angle $\theta$. Using the electron mass $m_e$, the
ultrarelativistic approximation\footnote[2]{To obtain the correct
  result, $m_e$~must be retained in all integrals containing factors
  $1/z_i$. In addition, keeping $m_e$~is sometimes suggested for the
  sake of stability of the numerical invariant mass
  integrations that follow the $\theta$~integration described in this
  section.}, and the notation
\begin{equation}
  \left[A\right]_\theta \equiv
    \frac{1}{2} \int\limits_{-1}^{+1} d\mycos\theta~A~~~,
\end{equation}
the following integrals are required for the ``second series'' of
brems\-strah\-lung integrals:
\vspace{.3cm}
\begin{center}
$\begin{array}{llllrlll}
 1.) & {\ds \; \left[ 1 \right]_\theta} & = & 1
     & & & \smskip \\
 2.) & {\ds \left[ \frac{1}{\bar z_1} \right]_\theta} & = &
       {\ds \frac{1}{2} \; \myln \left( \frac{s}{m_e^{\;2}} \right)
         \; \equiv \; \frac{1}{2} \; l_\beta } &
       \hspace{3cm}
 3.) & {\ds \left[ \frac{1}{\bar z_2} \right]_\theta} & = &
       {\ds \frac{1}{2} \; l_\beta}
     \smskip \\
 4.) & {\ds \left[ \frac{m_e^{\;2}}{\bar z_1^{\;2}} \right]_\theta}
       & = & {\ds \frac{s}{4} } &
       \hspace{3cm}
 5.) & {\ds \left[ \frac{m_e^{\;2}}{\bar z_2^{\;2}} \right]_\theta}
       & = & {\ds \frac{s}{4} }
     \smskip \\
 6.) & {\ds \left[ \frac{1}{S_{d1}} \right]_\theta } & = &
       {\ds \frac{1}{2 \sprm} \: \myln\left( \frac{s}{s'} \right) } &
       \hspace{3cm}
 7.) & {\ds \left[ \frac{1}{S_{d2}} \right]_\theta } & = &
       {\ds \frac{1}{2 \sprm} \: \myln\left( \frac{s}{s'} \right) }
     \smskip \\
 8.) & {\ds \left[ \frac{1}{{S_{d1}}^2}  \right]_\theta } & = &
       {\ds \frac{1}{4\,ss'} } &
       \hspace{3cm}
 9.) & {\ds \left[ \frac{1}{{S_{d2}}^2}  \right]_\theta } & = &
       {\ds \frac{1}{4\,ss'} }
     \smskip \\
\end{array}$
\end{center}
$\begin{array}{llll}
 10.) & {\ds \left[ \frac{1}{\bar z_1}  \, l_{t1} \right]_\theta} & = &
        {\ds \frac{1}{2} \left( \frac{}{} l_\beta \cdot L_{t1}
             \; - \; D_{z1t1} \right) }
        \smskip
      & \indent L_{t1} & = & L_{c5} - L_{c1} \smskip
      & \indent L_{c1} & = &
        {\ds \myln \left( \frac{\sprm(s'+\SLAMP) + s'\sigma - s\delta}
                               {\sprm(s'-\SLAMP) + s'\sigma - s\delta}
                   \right) }
        \smskip
      & \indent L_{c5} & = &
        {\ds \myln \left(
                     \frac{s' - \sigma + \SLAMP}{s' - \sigma - \SLAMP}
                   \right) }
        \smskip
      & \indent D_{z1t1} & = & {\ds
            \mysp \left( \frac{\sprm(s + \sprm - \delta - \SLAMP)}
                              {2s(\sprm - \delta) + s'(\sigma+\delta)}
                    \right)
          - \mysp \left( \frac{\sprm(s + \sprm - \delta + \SLAMP)}
                              {2s(\sprm - \delta) + s'(\sigma+\delta)}
                  \right) }
      \smskip \\
\end{array}$
\newpage
$\begin{array}{llll}
 11.) & {\ds \left[ \frac{1}{\bar z_1}  \, l_{t2} \right]_\theta} & = &
        {\ds \frac{1}{2} \left( \frac{}{} l_\beta \cdot L_{c5}
             \; - \; D_{z1t2} \right) }
        \bigskip
      & \indent  D_{z1t2} & = & {\ds
            \mysp \left( \frac{-\sprm(s' + \delta - \SLAMP)}
                                {s'(\sigma+\delta)}
                    \right)
          - \mysp \left( \frac{-\sprm(s' + \delta + \SLAMP)}
                                {s'(\sigma+\delta)}
                    \right) }
      \bigskip \\
 12.) & {\ds \left[ \frac{1}{\bar z_2}  \, l_{t1} \right]_\theta} & = &
        {\ds \frac{1}{2} \left( \frac{}{} l_\beta \cdot L_{c5}
             \; - \; D_{z2t1} \right) }
        \bigskip
      & \indent  D_{z2t1} & = & {\ds
           \mysp \left( \frac{-\sprm(s' - \delta - \SLAMP)}
                               {s'(\sigma-\delta)}
                    \right)
         - \mysp \left( \frac{-\sprm(s' - \delta + \SLAMP)}
                               {s'(\sigma-\delta)}
                    \right) }
      \bigskip \\
 13.) & {\ds \left[ \frac{1}{\bar z_2}  \, l_{t2} \right]_\theta} & = &
        {\ds \frac{1}{2} \left( \frac{}{} l_\beta \cdot L_{t2}
             \; - \; D_{z2t2} \right) }
        \bigskip
      & \indent L_{t2} & = & L_{c5} - L_{c2} \smskip
      & \indent L_{c2} & = &
        {\ds \myln \left( \frac{\sprm(s'+\SLAMP) + s'\sigma + s\delta}
                               {\sprm(s'-\SLAMP) + s'\sigma + s\delta}
                   \right) }
        \smskip
      & \indent  D_{z2t2} & = & {\ds
           \mysp \left( \frac{\sprm(s + \sprm + \delta - \SLAMP)}
                               {2s(\sprm + \delta) + s'(\sigma-\delta)}
                    \right)
         - \mysp \left( \frac{\sprm(s + \sprm + \delta + \SLAMP)}
                                {2s(\sprm + \delta) +
                                  s'(\sigma-\delta)}
                    \right) }
      \bigskip \\
 14.)  & {\ds \left[ \frac{1}{\bar z_1}  \, l_{u1} \right]_\theta} & = &
         {\ds \left[ \,{\cal S\!} \left( \frac{1}{\bar z_2}  \, l_{t2}
                     \right) \right]_\theta \; = \;
           \left[ \frac{1}{\bar z_2}  \, l_{t2} \right]_\theta }
      \bigskip \\
 15.)  & {\ds \left[ \frac{1}{\bar z_1}  \, l_{u2} \right]_\theta} & = &
         {\ds \left[ \,{\cal S\!} \left( \frac{1}{\bar z_2}  \, l_{t1}
                     \right) \right]_\theta \; = \;
           \left[ \frac{1}{\bar z_2}  \, l_{t1} \right]_\theta }
      \bigskip \\
 16.)   & {\ds \left[ \frac{1}{\bar z_2}  \, l_{u1} \right]_\theta} & = &
         {\ds \left[ \,{\cal S\!} \left( \frac{1}{\bar z_1}  \, l_{t2}
                     \right) \right]_\theta \; = \;
           \left[ \frac{1}{\bar z_1}  \, l_{t2} \right]_\theta }
      \bigskip \\
 17.)   & {\ds \left[ \frac{1}{\bar z_2}  \, l_{u2} \right]_\theta} & = &
         {\ds \left[ \,{\cal S\!} \left( \frac{1}{\bar z_1}  \, l_{t1}
                     \right) \right]_\theta \; = \;
           \left[ \frac{1}{\bar z_1}  \, l_{t1} \right]_\theta }
      \bigskip \\
\end{array}$
\newpage
$\begin{array}{llll}
 18.) & {\ds \left[\frac{1}{S_{d1}} \, l_{t1}\right]_\theta} & = &
        {\ds \frac{1}{2 \sprm} \; \RE \! \left[ \rule[-.2cm]{0cm}{1cm} \;
             -\; \mysp \left( \frac{s\,(s'+\delta+\SLAMP)}
                                   {s'(\sigma + \delta)} \right)
             \;+\; \mysp \left( \frac{s'+\delta+\SLAMP}
                                   {\sigma + \delta} \right)
                           \right. }
          \smskip & & & \indent \indent \;\;\: {\ds \left.
             \hspace{-.05cm}
             +\; \mysp \left( \frac{s\,(s'+\delta-\SLAMP)}
                                   {s'(\sigma + \delta)} \right)
             \;-\; \mysp \left( \frac{s'+\delta-\SLAMP}
                                   {\sigma + \delta} \right)
             \rule[-.2cm]{0cm}{1cm} \; \right]^{\;\footnotemark[2]} }
      \bigskip
        & & \equiv & {\ds \frac{1}{2\, \sprm} \cdot D^t_1 }
      \bigskip \\
 19.) & {\ds \left[\frac{1}{S_{d2}} \, l_{t2}\right]_\theta} & = &
        {\ds \left[ \,{\cal ST\!} \left(\frac{1}{S_{d1}}
          \, l_{t1}\right) \right]_\theta \; = \;
          {\cal T\!} \left( \left[\frac{1}{S_{d1}} \,
          l_{t1}\right]_\theta \right)} \smskip
        & & = &
        {\ds \frac{1}{2 \sprm} \; \RE \! \left[ \rule[-.2cm]{0cm}{1cm} \;
             -\; \mysp \left( \frac{s\,(s'-\delta+\SLAMP)}
                                   {s'(\sigma - \delta)} \right)
             \;+\; \mysp \left( \frac{s'-\delta+\SLAMP}
                                     {\sigma - \delta} \right)
                           \right. }
          \smskip & & & \indent \indent \;\;\: {\ds \left.
             \hspace{-.05cm}
             +\; \mysp \left( \frac{s\,(s'-\delta-\SLAMP)}
                                   {s'(\sigma - \delta)} \right)
             \;-\; \mysp \left( \frac{s'-\delta-\SLAMP}
                                     {\sigma - \delta} \right)
             \rule[-.2cm]{0cm}{1cm} \; \right] }
      \bigskip
        & & \equiv & {\ds \frac{1}{2\, \sprm} \cdot D^t_2 }
      \bigskip \\
 20.) & {\ds \left[\frac{1}{S_{d1}} \, l_{u1}\right]_\theta} & = &
        {\ds \left[ \,{\cal S\!} \left( \frac{1}{S_{d2}} \, l_{t2}
                    \right) \right]_\theta \; = \;
             \left[ \frac{1}{S_{d2}} \, l_{t2} \right]_\theta }
      \bigskip \\
 21.) & {\ds \left[\frac{1}{S_{d2}} \, l_{u2}\right]_\theta} & = &
        {\ds \left[ \,{\cal S\!} \left( \frac{1}{S_{d1}} \, l_{t1}
                    \right) \right]_\theta \; = \;
             \left[ \frac{1}{S_{d1}} \, l_{t1} \right]_\theta }
      \bigskip \\
 22.) & {\ds \left[\frac{1}{\bar z_1} \, l_{t12} \right]_\theta} & = &
        {\ds \frac{1}{2} \left( \frac{}{} l_\beta \cdot L_{c1}
                               + D_{z1t1} + D_{z1t2} \right) }
      \bigskip \\
 23.) & {\ds \left[\frac{1}{\bar z_2} \, l_{t12} \right]_\theta} & = &
        {\ds \frac{1}{2} \left( \frac{}{} l_\beta \cdot L_{c2}
                               + D_{z2t1} + D_{z2t2} \right) }
\end{array}$
\footnotetext[2]{
A comment on the treatment of arguments of logarithms and
dilogarithms is in order here. The logarithm is undefined for
negative real values, the Dilogarithm has a cut for real
numbers larger than 1 as is explained in appendix~\ref{rellogs}.
This means that, in principle, infinitesimal imaginary parts from the
propagator denominators have to be carried along in the whole
calculation. This, however, is an unnecessary nuisance, because all
integrands are real and regular. If infinitesimal imaginary parts were
needed, because logarithms or Dilogarithms lay on cuts in final
results, they were attributed to the invariant masses
$\SONE$~and $\STWO$. As integrands are real and regular, this is a
correct treatment, because it is then irrelevant for the
integrand how an infinitesimal imaginary part is entered. This
technique will be used in subsequent integrals without further notice.}
\newpage
$\begin{array}{llll}
 24.) & {\ds \left[\frac{1}{\bar z_1} \, l_{u12} \right]_\theta} & = &
        {\ds \left[ \,{\cal S\!} \left( \frac{1}{\bar z_2} \, l_{t12}
                    \right) \right]_\theta \; = \;
             \left[ \frac{1}{\bar z_2} \, l_{t12}\right]_\theta}
      \bigskip \\
 25.) & {\ds \left[\frac{1}{\bar z_2} \, l_{u12} \right]_\theta} & = &
        {\ds \left[ \,{\cal S\!} \left( \frac{1}{\bar z_1} \, l_{t12}
                    \right) \right]_\theta \; = \;
             \left[ \frac{1}{\bar z_1} \, l_{t12}\right]_\theta }
      \bigskip \\
 26.) & {\ds \left[\frac{1}{a^{ut}_1} l_{t12} \right]_\theta} & = &
        {\ds \frac{1}{\sprm} \cdot \RE \!
          \left[ \rule[-.2cm]{0cm}{.35cm} \,
                 L_{c8} \cdot L_{c6} \; + \; D^{tu}_{a1} \, \right] }
       \medskip & \indent \indent
         L_{c8} & = & {\ds \myln \left( \frac{s-\sigma}{s'-\sigma}
                                 \right) }
       \medskip & \indent \indent
         L_{c6} & = & {\ds \myln
            \frac{(a_{1-}\,b + b_{1-})(a_{2+}\,b + b_{2+})}
                 {(a_{1+}\,b + b_{1+})(a_{2-}\,b + b_{2-})} }
         \hspace{.75cm} \left( \rule[0cm]{0cm}{.35cm}
                      {\rm see~15.)~\&~17.)~in~app.~\ref{bremint1}}
                    \right)
       \medskip & \indent \indent
         D^{tu}_{a1} & = & {\ds
          - \; \mysp \left( \frac{b_{1-}(1+b)}{a_{1-}\,b + b_{1-}} \right)
       \; + \; \mysp \left( \frac{b_{1-}(1-b)}{a_{1-}\,b + b_{1-}} \right) }
          \smskip & & & {\ds
          - \; \mysp \left( \frac{b_{2+}(1+b)}{a_{2+}\,b + b_{2+}} \right)
       \; + \; \mysp \left( \frac{b_{2+}(1-b)}{a_{2+}\,b + b_{2+}} \right) }
          \smskip & & & {\ds
          + \; \mysp \left( \frac{b_{1+}(1+b)}{a_{1+}\,b + b_{1+}} \right)
       \; - \; \mysp \left( \frac{b_{1+}(1-b)}{a_{1+}\,b + b_{1+}} \right) }
          \smskip & & & {\ds
          + \; \mysp \left( \frac{b_{2-}(1+b)}{a_{2-}\,b + b_{2-}} \right)
       \; - \; \mysp \left( \frac{b_{2-}(1-b)}{a_{2-}\,b + b_{2-}} \right) }
       \bigskip \\
 27.) & {\ds \left[\frac{1}{a^{ut}_2} l_{t12} \right]_\theta } & = &
        {\ds \frac{1}{\sprm} \cdot \RE \!
          \left[ \rule[-.2cm]{0cm}{.35cm} \,
                 L_{c8} \cdot L_{c7} \; + \; D^{tu}_{a2} \, \right] }
       \medskip & \indent \indent
         L_{c7} & = & {\ds \myln
            \frac{(a_{1-}\,b - b_{1-})(a_{2+}\,b - b_{2+})}
                 {(a_{1+}\,b - b_{1+})(a_{2-}\,b - b_{2-})} }
         \hspace{.75cm} \left( \rule[0cm]{0cm}{.35cm}
                      {\rm see~15.)~\&~17.)~in~app.~\ref{bremint1}} \right)
       \medskip & \indent \indent
         D^{tu}_{a2} & = & {\ds
          - \; \mysp \left( \frac{-b_{1-}(1+b)}{a_{1-}\,b - b_{1-}} \right)
       \; + \; \mysp \left( \frac{-b_{1-}(1-b)}{a_{1-}\,b - b_{1-}} \right) }
          \smskip & & & {\ds
          - \; \mysp \left( \frac{-b_{2+}(1+b)}{a_{2+}\,b - b_{2+}} \right)
       \; + \; \mysp \left( \frac{-b_{2+}(1-b)}{a_{2+}\,b - b_{2+}} \right) }
          \smskip & & & {\ds
          + \; \mysp \left( \frac{-b_{1+}(1+b)}{a_{1+}\,b - b_{1+}} \right)
       \; - \; \mysp \left( \frac{-b_{1+}(1-b)}{a_{1+}\,b - b_{1+}} \right) }
          \smskip & & & {\ds
          + \; \mysp \left( \frac{-b_{2-}(1+b)}{a_{2-}\,b - b_{2-}} \right)
       \; - \; \mysp \left( \frac{-b_{2-}(1-b)}{a_{2-}\,b - b_{2-}} \right) }
\end{array}$
\newpage
$\begin{array}{llll}
 28.) & {\ds \left[\frac{1}{a^{ut}_1} l_{u12} \right]_\theta } & = &
        {\ds \left[ \,{\cal S\!} \left( \frac{1}{a^{ut}_2} l_{t12}
                    \right) \right]_\theta \; = \;
          \left[\frac{1}{a^{ut}_2} l_{t12} \right]_\theta }
       \vspace{.7cm} \\
 29.) & {\ds \left[\frac{1}{a^{ut}_2} l_{u12} \right]_\theta } & = &
        {\ds \left[ \,{\cal S\!} \left( \frac{1}{a^{ut}_1} l_{t12}
                    \right) \right]_\theta \; = \;
          \left[\frac{1}{a^{ut}_1} l_{t12} \right]_\theta }
       \bigskip \\
 30.) & {\ds \left[\frac{1}{{S_{d1}}^2} \, l_{t1}\right]_\theta} & = &
        {\ds \frac{1}{4\,\sprm s' \SONE} \left(
          \SLAMP \, \myln \frac{s'}{s} \; + \;
          \frac{\SLAMP}{2} \, L_{c3} \; - \;
          \frac{s' - \sigma}{2} \, L_{c5} \right. }
          \smskip & & & \indent \indent \indent \; {\ds \left. + \;
          \frac{s(s'+\delta) - s'(\sigma+\delta)}{2s} \, L_{t1}
          \right) }
        \smskip
      & \indent \indent L_{c3} & = &
        {\ds \myln \left( 1 + \frac{\sprm (s-\delta)}{s'\STWO} \right)
          }
      \bigskip \\
 31.) & {\ds \left[\frac{1}{{S_{d2}}^2} \, l_{t2}\right]_\theta} & = &
        {\ds \frac{1}{4\,\sprm s' \STWO} \left(
          \SLAMP \, \myln \frac{s'}{s} \; + \;
          \frac{\SLAMP}{2} \, L_{c4} \; - \;
          \frac{s' - \sigma}{2} \, L_{c5} \right. }
          \smskip & & & \indent \indent \indent \; {\ds \left. + \;
          \frac{s(s'-\delta) - s'(\sigma-\delta)}{2s} \, L_{t2}
          \right) }
        \smskip
      & \indent \indent L_{c4} & = &
        {\ds \myln \left( 1 + \frac{\sprm (s+\delta)}{s'\SONE} \right)
          }
      \bigskip \\
 32.) & {\ds \left[\frac{1}{{S_{d1}}^3} \, l_{t1}\right]_\theta} & = &
        {\ds \frac{1}{16 \, \sprm} \left[ \rule[-.2cm]{0cm}{1cm} \:
             - \; \frac{\sprm \SLAMP}{s {s'}^2 \SONE}
          \; + \; \frac{(s' + \delta) \SLAMP}{2\,{s'}^2 \SONE^2} \,
          L_{c3} \right. }
          \smskip & & & \hspace{1.47cm} {\ds + \;\:
          \frac{\sigma^2 - \delta^2 + 2\sigma\delta +
                2s'(\sigma-\delta) - 2{s'}^2}
               {4\,{s'}^2 \SONE^2} \, L_{c5} }
          \smskip & & & \hspace{1.42cm} {\ds \left. + \;\:
          \frac{(s' + \delta) \SLAMP}{{s'}^2 \SONE^2}
          \myln\frac{s'}{s} \; +
          \left( \frac{{s'}^2 - s'(\sigma \!-\! \delta) + \delta^2}
                      {2\,{s'}^2 \SONE^2} - \frac{1}{s^2} \right) \! L_{t1}
          \, \rule[-.2cm]{0cm}{1cm} \right] } \hspace{-.5cm}
      \bigskip \\
 33.) & {\ds \left[\frac{1}{{S_{d2}}^3} \, l_{t2}\right]_\theta} & = &
        {\ds \frac{1}{16 \, \sprm} \left[ \rule[-.2cm]{0cm}{1cm} \:
             - \; \frac{\sprm \SLAMP}{s {s'}^2 \STWO}
          \; + \; \frac{(s' - \delta) \SLAMP}{2\,{s'}^2 \STWO^2} \,
          L_{c4}
          \right. }
          \smskip & & & \hspace{1.47cm} {\ds + \;\:
          \frac{\sigma^2 - \delta^2 - 2\sigma\delta +
                2s'(\sigma+\delta) - 2{s'}^2}
               {4\,{s'}^2 \STWO^2} \, L_{c5} }
          \smskip & & & \hspace{1.42cm} {\ds \left. + \;\:
          \frac{(s' - \delta) \SLAMP}{{s'}^2 \STWO^2}
          \myln\frac{s'}{s} \; +
          \left( \frac{{s'}^2 - s'(\sigma \!+\! \delta) + \delta^2}
                      {2\,{s'}^2 \STWO^2} - \frac{1}{s^2} \right) \!
                      L_{t2}
          \, \rule[-.2cm]{0cm}{1cm} \right] } \hspace{-.5cm}
        \medskip \\
\end{array}$
\newpage
$\begin{array}{llll}
 34.) & {\ds \left[\frac{m_e^{\;2}}{\bar z_1^{\;2}} \, l_{t2}
             \right]_\theta } & = &
        {\ds \frac{s}{4} \cdot L_{c5} }
      \bigskip \\
 35.) & {\ds \left[\frac{m_e^{\;2}}{\bar z_2^{\;2}} \, l_{t1}
             \right]_\theta } & = &
        {\ds \frac{s}{4} \cdot L_{c5} }
      \bigskip \\
 36.) & {\ds \left[\frac{m_e^{\;2}}{\bar z_1^{\;2}} \, l_{u2}
             \right]_\theta } & = &
        {\ds \left[ \,{\cal S\!} \left( \frac{m_e^{\;2}}{\bar z_2^{\;2}}  \,
          l_{t1} \right) \right]_\theta \; = \;
          \frac{s}{4} \cdot L_{c5} }
      \bigskip \\
 37.) & {\ds \left[\frac{m_e^{\;2}}{\bar z_2^{\;2}} \, l_{u1}
             \right]_\theta } & = &
        {\ds \left[ \,{\cal S\!} \left( \frac{m_e^{\;2}}{\bar z_1^{\;2}}  \,
          l_{t2} \right) \right]_\theta }
        \; = \;
        {\ds \frac{s}{4} \cdot L_{c5} }
      \bigskip \\
 38.) & {\ds \left[\frac{1}{d_1^+ \, d_1^-} \right]_\theta} & = &
        {\ds \frac{1}{4\,\sprm\SLAMP} \cdot L_{c1} }
        \bigskip \\
 39.) & {\ds \left[\frac{1}{d_2^+ \, d_2^-} \right]_\theta} & = &
        {\ds \frac{1}{4\,\sprm\SLAMP} \cdot L_{c2} }
        \bigskip \\
 40.) & {\ds \left[\frac{\mycos\theta}{d_1^+ \, d_1^-} \right]_\theta} & = &
        {\ds \frac{1}{4\,{\sprm}^2} \left( \,
          \frac{s-\delta}{\SLAMP} \, L_{c1} \; - \; L_{c3} \, \right) }
        \bigskip \\
 41.) & {\ds \left[\frac{\mycos\theta}{d_2^+ \, d_2^-} \right]_\theta} & = &
        {\ds - \, \frac{1}{4\,{\sprm}^2} \left( \,
          \frac{s+\delta}{\SLAMP} \, L_{c2} \; - \; L_{c4} \, \right) }
        \bigskip \\
 42.) & {\ds \left[\frac{1}{S^t_{12+} \, S^t_{12-} } \right]_\theta} & = &
        {\ds \frac{1}{4\,\sprm\SLAMB}  \cdot L_{\bar \lambda} }
        \bigskip
      & \indent \indent \indent L_{\bar \lambda} & = & {\ds
          \myln \left(
            \frac{\sprm + \sigma + \SLAMB}{\sprm + \sigma - \SLAMB}
          \right) }
        \bigskip \\
 43.) & {\ds \left[\frac{\mycos\theta}{S^t_{12+} \, S^t_{12-} }
             \right]_\theta} & = &
        {\ds \frac{1}{4\,{\sprm}^2} \left( \,
          \myln \frac{\SONE}{\STWO}
          \; - \; \frac{\delta}{\SLAMB}\,L_{\bar \lambda} \, \right) }
        \bigskip \\
\end{array}$
\newpage
$\begin{array}{llll}
 44.) & {\ds \left[\frac{1}{{S^t_{12+}}^2\,{S^t_{12-}}^2}
             \right]_\theta} & = &
        {\ds \frac{1}{8\,\sprm {\bar \lambda}} \left( \,
          \frac{1}{\SLAMB}\,L_{\bar \lambda} \; + \;
          \frac{\sprm \sigma + \delta^2}{2\,\sprm \SONE \STWO} \, \right) }
        \bigskip \\
 45.) & {\ds \left[\frac{\mycos\theta}{{S^t_{12+}}^2\,{S^t_{12-}}^2}
             \right]_\theta} & = &
        {\ds \frac{\delta}{8\,{\sprm}^2\LAMB} \left(
          \: - \; \frac{1}{\SLAMB}\,L_{\bar \lambda} \; + \;
          \frac{\sprm + \sigma}{2\,\SONE \STWO} \, \right) }
        \bigskip \\
 46.) & {\ds \left[\frac{l_{t12}}{S^t_{12+} \, S^t_{12-}}
             \right]_\theta } & = &
        {\ds \frac{1}{4\, \sprm \SLAMB} \; \RE \!
        \left[ \rule[-.2cm]{0cm}{1cm}
           - \; \mysp \left( + \frac{c_{++} a_{--} e_+}{d} \right)
        \; + \; \mysp \left( + \frac{c_{-+} a_{--} e_+}{d} \right) \right. }
        \smskip & & & \hspace{2.09cm} {\ds
        \; + \; \mysp \left( - \frac{c_{+-} a_{--} e_+}{d} \right)
        \; - \; \mysp \left( - \frac{c_{--} a_{--} e_+}{d} \right) }
        \smskip & & & \hspace{2.08cm} {\ds
        \; - \; \mysp \left( - \frac{c_{++} a_{-+} e_+}{d} \right)
        \; + \; \mysp \left( - \frac{c_{-+} a_{-+} e_+}{d} \right) }
        \smskip & & & \hspace{2.09cm} {\ds
        \; + \; \mysp \left( + \frac{c_{+-} a_{-+} e_+}{d} \right)
        \; - \; \mysp \left( + \frac{c_{--} a_{-+} e_+}{d} \right) }
        \smskip & & & \hspace{2.08cm} {\ds
        \; - \; \mysp \left( + \frac{c_{+-} a_{++} e_-}{d} \right)
        \; + \; \mysp \left( + \frac{c_{--} a_{++} e_-}{d} \right) }
        \smskip & & & \hspace{2.09cm} {\ds
        \; + \; \mysp \left( - \frac{c_{++} a_{++} e_-}{d} \right)
        \; - \; \mysp \left( - \frac{c_{-+} a_{++} e_-}{d} \right) }
        \smskip & & & \hspace{2.08cm} {\ds
        \; - \; \mysp \left( - \frac{c_{+-} a_{+-} e_-}{d} \right)
        \; + \; \mysp \left( - \frac{c_{--} a_{+-} e_-}{d} \right) }
        \smskip & & & \hspace{2.045cm} {\ds \left.
        \; + \; \mysp \left( + \frac{c_{++} a_{+-} e_-}{d} \right)
        \; - \; \mysp \left( + \frac{c_{-+} a_{+-} e_-}{d} \right)
        \rule[-.2cm]{0cm}{1cm} \right] }
        \bigskip
          & & \equiv & {\ds \frac{1}{4\, \sprm \SLAMB} \cdot D^t_{12} }
        \bigskip
      & \indent \indent \;\;\;\;\;
          a_{\pm\pm} & = & s \pm \SLAMP \pm \SLAMB  \smskip
      & \indent \indent \;\;\;\;\;
          c_{\pm\pm} & = & \delta \pm \sprm \pm \SLAMB  \smskip
      & \indent \indent \;\;\;\;\;\;
          e_{\pm} & = & s' - \sigma \pm \SLAMP   \smskip
      & \indent \indent \;\;\;\;\;\;\:
          d & = & 8\,s\,\SONE\,\STWO
\end{array}$
\newpage
$\begin{array}{llll}
 47.) & {\ds \left[\frac{l_{t12}}{{S^t_{12+}}^2 \, {S^t_{12-}}^2}
             \right]_\theta } & = &
        {\ds \frac{1}{2\,{\bar \lambda}}
          \left[\frac{l_{t12}}{S^t_{12+} \, S^t_{12-}} \right]_\theta }
        \smskip & & &
        {\ds + \; \frac{1}{8\,\sprm \LAMB} \left[ \indent
          \frac{\SLAMP\SLAMB}{2\,s\SONE\STWO} \, L_{\bar \lambda} \; + \;
          \frac{\sprm - \delta}{2\,\sprm \SONE} \, L_{c1} \; + \;
          \frac{\sprm + \delta}{2\,\sprm \STWO} \, L_{c2} \right. }
        \smskip & & & \hspace{1.8cm} {\ds
          \; + \; \frac{{\bar \lambda} - s(\sprm+\sigma)}{4\,s\SONE\STWO}
          \left( L_{c1} + L_{c2} \right) }
        \smskip & & & \hspace{1.75cm} {\ds \left.
          \; - \; \frac{\SLAMP(\sprm+\sigma)}{4\,s\SONE\STWO}
          \left( L_{c3} + L_{c4} \right)
        \hspace{.5cm} \right] }
        \hugeskip \\
 48.) & {\ds \left[\frac{\mycos\theta \; \cdot \; l_{t12}}
                        {{S^t_{12+}}^2 \, {S^t_{12-}}^2}
             \right]_\theta } & = &
        {\ds - \; \frac{\delta}{\sprm}
          \left[\frac{l_{t12}}{{S^t_{12+}}^2 \, {S^t_{12-}}^2}
          \right]_\theta }
        \smskip & & &
        {\ds + \; \frac{1}{8\,{\sprm}^2} \left[
          \hspace{.54cm}
          \frac{\SLAMP}{2\,s'\SONE\STWO} \myln \frac{\SONE}{\STWO}
          \; - \;
          \frac{1}{2\,\sprm \SONE} \, L_{c1} \; + \;
          \frac{1}{2\,\sprm \STWO} \, L_{c2} \right. }
        \smskip & & & \hspace{1.58cm} {\ds \left.
          \; - \; \frac{s'-\sigma}{4\,s\SONE\STWO}
          \left( L_{c1} \! - \! L_{c2} \right)
          \; - \; \frac{\SLAMP}{4\,s\SONE\STWO}
          \left( L_{c3} \! - \! L_{c4} \right)
        \: \right] }
        \bigskip \\
 49.) & {\ds \left[\frac{l_{t12}}{{S^t_{12+}}^3 \, {S^t_{12-}}^3}
             \right]_\theta } & = &
        {\ds \frac{3}{4 \LAMB}
          \left[\frac{l_{t12}}{{S^t_{12+}}^2 \, {S^t_{12-}}^2}
          \right]_\theta
          \;\;\;\;  + \;\;\;\; \frac{1}{32\,\sprm \SLAMB^{\,3}} \times }
        \smskip & & & {\ds \left[ \hspace{.56cm}
          \frac{\SLAMB\,\SLAMP\,\left( \sprm \sigma + 2\delta^2 \right)}
               {4 \, s \sprm \SONE^2 \STWO^2} \; + \;
          \frac{\SLAMB\,\delta}{4 \,\SONE^2 \STWO^2} \, L_{c12}
          \; + \; \frac{\SLAMB}{4 \,\sprm \SONE^2} \, L_{c1} \right. }
        \smskip & & & \hspace{.35cm} {\ds
             + \; \frac{\SLAMB}{4 \,\sprm \STWO^2} \, L_{c2} \; - \;
             \frac{\SLAMP \left( 2s\SONE\STWO + \LAMB (s'-\sigma)\right)}
                  {4\,s^2 \SONE^2 \STWO^2} \, L_{\LAMB}}
        \smskip & & & \hspace{.35cm} {\ds
             - \; \frac{\SLAMB \left( 6\,s\SONE\STWO -
                        s^2(\sprm+\sigma) + \LAMB(\sprp-\sigma) \right) }
                       {8\,s^2 \SONE^2 \STWO^2}
                  \left( L_{c1} + L_{c2} \right) }
        \smskip & & & \hspace{.35cm} {\ds \left.
             + \; \frac{\SLAMB\,\SLAMP\,\left( s(\sprm+\sigma) -
                        2\,\SONE\STWO - \LAMB \right) }
                       {8\,s^2 \SONE^2 \STWO^2}
                  \left( L_{c3} + L_{c4} \right)
        \hspace{.92cm} \right] }
        \medskip
      & \indent \indent \indent L_{c12} & = & {\ds
          \frac{1}{\sprm} \cdot \left( \frac{\SONE^2}{\sprm}\,L_{c2}
            - \frac{\STWO^2}{\sprm}\,L_{c1}
            - \frac{\delta\,\SLAMP}{s} \right) }
\end{array}$
\newpage
$\begin{array}{llll}
 50.) & {\ds \left[\frac{\mycos\theta \; \cdot \; l_{t12}}
                        {{S^t_{12+}}^3 \, {S^t_{12-}}^3}
             \right]_\theta } & = &
        {\ds \hspace{.44cm} \frac{1}{\,4\,\LAMB\,}
          \left[\frac{\mycos\theta \; \cdot \; l_{t12}}
                     {{S^t_{12+}}^2 \, {S^t_{12-}}^2}
          \right]_\theta \; + \;
          \frac{\delta}{4\,\sprm\LAMB}
          \left[\frac{l_{t12}}{{S^t_{12+}}^2 \, {S^t_{12-}}^2}
          \right]_\theta }
        \vspace{.25cm} \\ & & & {\ds
          - \; \frac{\delta}{\sprm}
          \left[\frac{l_{t12}}{{S^t_{12+}}^3 \, {S^t_{12-}}^3}
          \right]_\theta }
        \smskip & & & \hspace{-.11cm} {\ds \; + \;
          \frac{1}{32\,{\sprm}^2 \LAMB}
          \left[ \hspace{.6cm}
          \frac{\SLAMP\,\LAMB\,\delta}{2\,s\,\sprm\,\SONE^2\STWO^2}
          \; + \;
          \frac{\LAMB}{4\,\SONE^2 \STWO^2} \, L_{c12} \right. }
        \vspace{.25cm} \\ & & & \hspace{2.13cm} {\ds \; + \;
          \frac{1}{2\,\sprm\,\SONE}\,L_{c1} \; - \;
          \frac{1}{2\,\sprm\,\STWO}\,L_{c2} }
        \vspace{.25cm} \\ & & & \hspace{2.12cm} {\ds \; - \;
          \frac{\SLAMP \left( (s'\!-\!\sigma)\,\LAMB +
                2\,s\,\SONE\STWO \right)}{8\,s^2 \SONE^2 \STWO^2}}
          \times \\ & & & \hspace{6.5cm} {\ds
          \left( 2\myln \frac{\SONE}{\STWO} - L_{c3} + L_{c4} \right) }
        \hspace{-1cm} \smskip & & & \hspace{2.1cm} {\ds \left. \; + \;
          \frac{\LAMB\LAMP + 2\,\SONE\STWO \left( s\,(s'\!-\!\sigma)
                + \LAMB \right)}
               {8\,s^2 \SONE^2 \STWO^2}
          \left( L_{c1} - L_{c2} \right)
        \hspace{.3cm} \right] } \hspace{-.75cm}
        \vspace{.7cm} \\
 51.) & {\ds \left[ \frac{l_{t_1u_2}}{\,2\,\sqrt{C_{12}}\,} \right]_\theta }
      & \hspace{-.5cm} = &
        {\ds \frac{\ri \cdot c_{12}}{x_0} \cdot
          \left[ \rule[-.2cm]{0cm}{1cm} \hspace{.5cm}
          \mysp \left( \frac{\beta_{12} + \ri\:\!x_0}{\tau_1} \right)
          \; + \;
          \mysp \left( \frac{\beta_{12} + \ri\:\!x_0}{\tau_2} \right)
          \right. }
        \vspace{.1cm} \\ & & & \hspace{1.63cm} {\ds + \;
          \mysp \left( \frac{\beta_{12} + \ri\:\!x_0}{\tau_1^*} \right)
          \; + \;
          \mysp \left( \frac{\beta_{12} + \ri\:\!x_0}{\tau_2^*} \right) }
        \vspace{.25cm} \\ & & & \hspace{1.625cm} {\ds - \;
          \mysp \left(-\frac{\beta_{12} + \ri\:\!x_0}{\tau_1} \right)
          \; - \;
          \mysp \left(-\frac{\beta_{12} + \ri\:\!x_0}{\tau_2} \right) }
        \vspace{-.2cm} \\ & & & \hspace{1.58cm} {\ds \left. - \;
          \mysp \left(-\frac{\beta_{12} + \ri\:\!x_0}{\tau_1^*}
        \right)
          \; - \;
          \mysp \left(-\frac{\beta_{12} + \ri\:\!x_0}{\tau_2^*}
        \right)
        \hspace{.2cm} \rule[-.2cm]{0cm}{1cm} \right]^{\;\footnotemark[2]} }
        \\
      & & \hspace{-.5cm} \equiv & c_{12} \cdot D_{t1u2}
        \vspace{.5cm} \\
      & \indent \indent c_{12} & \hspace{-.5cm} = &
        {\ds \frac{s'}{\sprm\sqrt{\LAMP\,s\,\SONE}} } \vspace{.18cm} \\
      & \indent \indent x_0 & \hspace{-.5cm} = &
        {\ds \sqrt{\frac{s-\SONE}{s}} } \vspace{.18cm} \\
      & \indent \indent \beta_{12} & \hspace{-.5cm} = &
        {\ds \sqrt{\frac{\SONE-4\ME^2}{s}} }
\end{array}$
\vspace{.26cm}
\newline
$\begin{array}{llll}
      \hspace{1.5cm} & \indent \indent \tau_1 & = &
        {\ds \frac{+ q_{12} - \ri\:\!p_{12}}{b_{12} + \ri\:\!r_{12}} }
        \smskip
      \hspace{1.5cm} & \indent \indent \tau_2 & = &
        {\ds \frac{- q_{12} - \ri\:\!p_{12}}{b_{12} + \ri\:\!r_{12}} }
        \smskip
\end{array}$
$\begin{array}{llll}
      & \indent \indent q_{12} & = &
        {\ds \left(ss' - s\STWO -s'\SONE\right)
             \sqrt{\frac{\SONE}{s}} } \vspace{.1cm} \vspace{-.05cm} \\
      & \indent \indent p_{12} & = & \sprm\,\SONE \cdot x_0
      \vspace{.12cm} \\
      & \indent \indent r_{12} & = &
        \sqrt{\LAMP\,s\,\SONE} \cdot x_0 \vspace{.12cm} \\
      & \indent \indent b_{12} & = & \SONE \left(s-\delta\right)
\end{array}$
\footnotetext[2]{Integrals 51.) to 60.) were not directly calculated
  from the result of the $\Omega_R$ integration. Instead, a Feynman
  parametrization was used to ``linearize'' $1/t_1u_2$
  and $1/t_2u_1$: $1/t_1u_2 = \int_0^1 d\alpha/\left[t_1\alpha +
  u_2(1-\alpha)\right]^2$. Then, $\Omega_R$, $\theta$, and finally the
  Feynman parameter $\alpha$~were integrated.}
\newpage
$\begin{array}{llll}
 52.) & {\ds \left[ \frac{l_{t_2u_1}}{\,2\,\sqrt{C_{21}}\,}
         \right]_\theta } \;\; & = &
        {\ds {\cal T\!} \left( \left[
          \frac{l_{t_1u_2}}{\,2\,\sqrt{C_{12}}\,}
        \right]_\theta \right)^{\;\footnotemark[2]} } \;\; \equiv \;\;
        c_{34} \cdot D_{t2u1}
       \vspace{1cm} \\
 53.) & {\ds \left[ \frac{l_{t_1u_2}}{\,{\bar z_1} \cdot 2\,\sqrt{C_{12}}\,}
             \right]_\theta } & = & \hspace{.2cm}
        {\ds \frac{s'}{4\,\SLAMP\, \left(ss' - s\STWO -s'\SONE\right)}
          \times \left\{ \rule[-.3cm]{0cm}{1.4cm} \;\;
          2\, l_{\beta} \cdot \left( 2 L_{c5} - L_{c1} \right) \right. }
\end{array}$
\ba
  \hspace*{1.5cm}
  &  + &  \myln^2 \left(\frac{2\,x_1}{x_1 - 1} \right)
  \; - \; \myln^2 \left(\frac{2\,x_1}{x_1 + 1} \right)
  \; + \; \mysp \left(-\frac{x_1 + 1}{x_1 - 1} \right)
  \; - \; \mysp \left(-\frac{x_1 - 1}{x_1 + 1} \right) \nl \nl
  &  - &  \myln^2 \left(\frac{2\,x_2}{x_2 - 1} \right)
  \; + \; \myln^2 \left(\frac{2\,x_2}{x_2 + 1} \right)
  \; - \; \mysp \left(-\frac{x_2 + 1}{x_2 - 1} \right)
  \; + \; \mysp \left(-\frac{x_2 - 1}{x_2 + 1} \right) \nl \nl
  &  - &  \myln \frac{x_1-x_2}{x_1+x_2} \cdot \left(
          \myln \frac{x_2 - 1}{x_2 + 1} \; - \;
          \myln \frac{x_1 - 1}{x_1 + 1} \right) \nl \nl
  &  + &  \myln \frac{x_1 - 1}{x_1-x_2} \cdot
          \myln \left| \frac{x_2 - 1}{x_1-x_2} \right|
  \; - \; \myln \frac{x_1 + 1}{x_1-x_2} \cdot
          \myln \left| \frac{x_2 + 1}{x_1-x_2} \right| \nl \nl
  &  + &  \myln \frac{x_1 + 1}{x_1+x_2} \cdot
          \myln \left| \frac{x_2 - 1}{x_1+x_2} \right|
  \; - \; \myln \frac{x_1 - 1}{x_1+x_2} \cdot
          \myln \left| \frac{x_2 + 1}{x_1+x_2} \right| \nl \nl
  &  + &  2 \: \mysp \left( - \frac{x_2 - 1}{x_1-x_2} \right)
  \; - \; 2 \: \mysp \left( - \frac{x_2 + 1}{x_1-x_2} \right) \nl \nl
  &  + & \frac{\sprm\,\sqrt{\SONE\rule[0cm]{0cm}{.3cm}}}{\sqrt{s\LAMP}}
         \cdot \RE \left[ \rule[-.3cm]{0cm}{1.3cm} \hspace{.45cm}
         \sum_{i=1}^2 \;
         \frac{(-1)^{i+1}\cdot(x_i-a_{12})}
              {\sqrt{\beta^2\,x_i^2-x_0^2}}
         \; \times \right. \nl \nl
  & &    \sum_{k=1}^8 (-1)^{k+1}
         \!\left( \rule[-.3cm]{0cm}{1.2cm} \hspace{.42cm}
         \myln \frac{\,t^{(i)}_{k+}\,}{t^0_k} \! \cdot \!\!
         \left( \rule[-.1cm]{0cm}{.6cm}
         \myln \left[u^{(i)}_+(+\beta)\right] -
         \myln \left[u^{(i)}_+(-\beta)\right] -
         2 \pi \ri \!\cdot\! \Theta(x_i) \right) \right. \nl
  & & \hspace{2.37cm} - \myln \frac{\,t^{(i)}_{k-}\,}{t^0_k} \! \cdot \!\!
         \left( \rule[-.1cm]{0cm}{.6cm}
         \myln \left[u^{(i)}_-(+\beta)\right] -
         \myln \left[u^{(i)}_-(-\beta)\right] -
         2 \pi \ri \!\cdot\! \Theta(-x_i) \right) \nl
  & & \hspace{2.37cm} - \,
         \mysp \left[ -\frac{u^{(i)}_+(+\beta)}{t^{(i)}_{k+}} \right] \; + \;
         \mysp \left[ -\frac{u^{(i)}_+(-\beta)}{t^{(i)}_{k+}} \right] \nl
  & & \hspace{2.25cm} \left. \left. \left. + \,
         \mysp \left[ -\frac{u^{(i)}_-(+\beta)}{t^{(i)}_{k-}} \right] \; - \;
         \mysp \left[ -\frac{u^{(i)}_-(-\beta)}{t^{(i)}_{k-}} \right]
         \hspace{.3cm} \rule[-.3cm]{0cm}{1.2cm} \right)
         \;\;\;\; \rule[-.3cm]{0cm}{1.3cm} \right]
         \indent\; \rule[-.3cm]{0cm}{1.4cm} \right\} \nl \nl \nl
  & \equiv & \frac{s'}{4\,\SLAMP\, \left(ss' - s\STWO -s'\SONE\right)}
         \cdot
         \left\{ \rule[-.3cm]{0cm}{.35cm}
                 2\, l_{\beta} \cdot \left( 2 L_{c5} - L_{c1} \right)
                 \; + \; D^z_{t1u2} \right\}  \nonumber
\ea
\footnotetext[2]{Here and in the remainder of this section the
  expressions obtained by the symmetry operation ${\cal T}$~ will not be
  rewritten. The reader can easily do that by making the reciprocal
  replacement $\SONE \leftrightarrow \STWO$.}
\newpage
$\begin{array}{lcclccl}
  & \indent a_{12}  & = & {\ds \frac{s-\delta}{\sprm} } \miniskip
  & \indent      t^0_1 & = & \tau_1   \hspace{4.5cm} &
    \hspace{1cm} t^0_2 & = & - \tau_1 \\
  & \indent      t^0_3 & = & \tau_2   \hspace{4.5cm} &
    \hspace{1cm} t^0_4 & = & - \tau_2 \\
  & \indent      t^0_5 & = & \tau_2^* \hspace{4.5cm} &
    \hspace{1cm} t^0_6 & = & - \tau_2^* \\
  & \indent      t^0_7 & = & \tau_1^* \hspace{4.5cm} &
    \hspace{1cm} t^0_8 & = & - \tau_1^*
  \medskip
  & \indent t^{(i)}_{k\pm}  & = & t^0_k \; + \; t^{(i)}_{\pm}
  & \indent u^{(i)}_{\pm}(x)  & = & t(x) \; - \; t^{(i)}_{\pm}
    \smskip
  & \indent t^{(i)}_{\pm} & = & {\ds
      \frac{\ri\,\!x_0 \pm \sqrt{\beta^2 x_i^2 - {x_0}^2}}{x_i} }
  & \hspace{1.1cm}    t(x) & = & {\ds
      \frac{\ri\,\!x_0 + \sqrt{x^2-{x_0}^2}}{x} } \medskip
\end{array}$
\newline
$\begin{array}{lccl}
  & \hspace{1.34cm} x_{1/2} \hspace{1pt} & = & {\ds
      \frac{\sprm\,\SONE(s-\delta) \pm \SLAMP\,(ss' - s\STWO - s'\SONE)}
           {{\sprm}^2\SONE - s \LAMP} }
    \smskip
  & \hspace{1.3cm} \beta  & = &
    {\ds \sqrt{1 - \frac{4\,\ME^2}{s}} }
\end{array}$
\bigskip
\bigskip
\bigskip
$\begin{array}{llll}
 54.) & {\ds \left[ \frac{l_{t_2u_1}}
          {\,{\bar z_1} \cdot 2\,\sqrt{C_{21}}\,} \right]_\theta } & = &
        {\cal T\!} \left( \rule[-.2cm]{0cm}{1cm}
          {\ds \left[ \frac{l_{t_1u_2}}
            {\,{\bar z_1} \cdot 2\,\sqrt{C_{12}}\,}
               \right]_\theta }
          \rule[-.2cm]{0cm}{1cm} \right) \vspace{.3cm} \\
      & & \equiv &
        {\ds \frac{s'}
                  {4\,\SLAMP\, \left(ss' - s\SONE - s'\STWO\right)} }
          \cdot
          \left\{ \rule[-.3cm]{0cm}{.35cm}
                  2\, l_{\beta} \cdot \left( 2 L_{c5} - L_{c2} \right)
                  \; + \; D^z_{t2u1} \right\}
      \bigskip \\
 55.) & {\ds \left[ \frac{l_{t_1u_2}}{\,{\bar z_2} \cdot 2\,\sqrt{C_{12}}\,}
             \right]_\theta } & = &
          {\ds \left[ \frac{l_{t_1u_2}}{\,{\bar z_1} \cdot 2\,\sqrt{C_{12}}\,}
               \right]_\theta }
      \bigskip \\
 56.) & {\ds \left[ \frac{l_{t_2u_1}}{\,{\bar z_2} \cdot 2\,\sqrt{C_{21}}\,}
             \right]_\theta } & = &
        {\cal T\!} \left( \rule[-.2cm]{0cm}{1cm}
          {\ds \left[ \frac{l_{t_1u_2}}{\,{\bar z_1} \cdot 2\,\sqrt{C_{12}}\,}
               \right]_\theta }
          \rule[-.2cm]{0cm}{1cm} \right)
\end{array}$
\newpage
$\begin{array}{llll}
 57.) & {\ds \left[ \frac{l_{t_1u_2}}{\,a^{ut}_1 \cdot 2\,\sqrt{C_{12}}\,}
             \right]_\theta } & = &
        {\ds \frac{s'}{\sprm} \cdot \RE
          \left( \rule[-.1cm]{0cm}{.5cm} X_0 + X_{1} + X_{2} \right) }
      \nl \nl & & \equiv & {\ds \frac{s'}{\sprm} \cdot D^a_{t1u2} }
      \medskip
\end{array}$
\begin{list}{}{\leftmargin=1.9cm}
  \item For the calculation of the integrals $X_i$ the following
    quantities are needed: \vspace{-.2cm}
\end{list}
\ba
  \hspace*{2cm}
  A_3 & = & \SONE\,(\sprp -2\sigma)^2 \; - \; \beta^2 s \LAMP
   \hspace{5.4cm} \nl
  B_3 & = & \SONE\,(\sprp -2\sigma)\,(s-\delta) \nl
  \Delta_3 & = & \LAMP \cdot \left[ \rule[-.1cm]{0cm}{.7cm}
    \left\{ \rule[-.1cm]{0cm}{.5cm}
      ss' - s\STWO -s'\SONE -2\SONE\,(s-\sigma) \right\}^2 \right. \nl
    & & \left. \hspace{1cm} \; - \;
    4\ME^2 \left\{ \SONE\,(s-\delta)^2 + \LAMP\,(s-\SONE) \right\}
    \rule[-.1cm]{0cm}{.7cm} \right]
    \nl \nl
  x_{3/4}  & = & {\ds \beta \cdot \frac{-B_3 \pm \sqrt{\Delta_3}}{A_3} }
    \nonumber
\ea
\begin{list}{}{\leftmargin=1.9cm}
  \item In principle, depending on the values of $x_3$ and $x_4$, four
    cases have to be distin\-guished: \vspace{.2cm}
\end{list}
$\begin{array}{lrcccccc}
 \hspace*{1.75cm} & (i) & x_{3/4} \, \epsilon \, \numreal ; & \;\;
                         |x_4| & \geq & |x_3| & > & \beta \\
                 & (ii) & x_{3/4} \, \epsilon \, \numreal ; & \;\;
                         |x_4| & > & \beta & \geq & |x_3| \\
                & (iii) & x_{3/4} \, \epsilon \, \numreal ; & \;\;
                         \beta & \geq & |x_4| & \geq & |x_3| \\
                 & (iv) & \Delta_3 \leq 0 ; & \;\;
                         x_3 & = & x_4^* & \epsilon & \numcomp
 \vspace{.2cm}
\end{array}$
\begin{list}{}{\leftmargin=1.9cm}
\item However, since the integrand is regular, it is sufficient to
  present the solution for case $(i)$. The solutions for the other
  three cases are then obtained by analytical continuation. In
  case $(iv)$, which is relevant for only a very small fraction of the
  phase space, the problem of a numerically correct treatment arises.
  This problem was solved by computing the integral's value
  for a very nearby phase space point so that the expression of case
  $(i)$~could be used. Having in mind that the integral is regular, it
  is clear that thus only a negligible error is introduced.
  The expressions $X_0,~X_1$, and $X_2$ for case $(i)$~are presented
  below.
\end{list}
\ba
  \hspace*{2cm}
  X_0 & = & \frac{2}{\,\sqrt{\Delta_3}\,} \cdot
            L_{c8} \cdot \left[ \rule[-.3cm]{0cm}{1.1cm}
            \myln \left( \frac{\beta-x_3}{\beta+x_3} \right) \; - \;
            \myln \left( \frac{\beta-x_4}{\beta+x_4} \right) \right]
            \hspace{1.9cm} \nonumber
\ea
\newpage
\ba
  \hspace*{.75cm}
  X_1 & = & \frac{\,\sqrt{\SONE\rule[0cm]{0cm}{.3cm}} \!\cdot\!
                  \left( \sprp-\sigma \right)\,}
                 {2\,\sqrt{s\,\LAMP\,\Delta_3}} \; \cdot \;
            \sum_{i=3}^4 \frac{(-1)^{i+1}\cdot(x_i-a_{34})}
                              {\sqrt{x_i^2-x_0^2}}   \times \nl
  & & \indent \! \sum_{k=1}^8 (-1)^{k+1}
         \!\left( \rule[-.3cm]{0cm}{1.2cm} \hspace{.42cm}
         \myln \frac{\,t^{(i)}_{k+}\,}{t^0_k} \! \cdot \!\!
         \left( \rule[-.1cm]{0cm}{.6cm}
         \myln \left[u^{(i)}_+(+\beta)\right] -
         \myln \left[u^{(i)}_+(-\beta)\right] -
         2 \pi \ri \!\cdot\! \Theta(x_i) \right) \right. \nl
  & & \hspace{3.06cm} \! - \myln \frac{\,t^{(i)}_{k-}\,}{t^0_k} \!
         \cdot \!\! \left( \rule[-.1cm]{0cm}{.6cm}
         \myln \left[u^{(i)}_-(+\beta)\right] -
         \myln \left[u^{(i)}_-(-\beta)\right] -
         2 \pi \ri \!\cdot\! \Theta(-x_i) \right) \! \nl
  & & \hspace{3.04cm} \! - \,
         \mysp \left[ -\frac{u^{(i)}_+(+\beta)}{t^{(i)}_{k+}} \right]
         \; + \;
         \mysp \left[ -\frac{u^{(i)}_+(-\beta)}{t^{(i)}_{k+}} \right]
         \nl
  & & \hspace{3cm} \! \left. + \,
         \mysp \left[ -\frac{u^{(i)}_-(+\beta)}{t^{(i)}_{k-}} \right]
         \; - \;
         \mysp \left[ -\frac{u^{(i)}_-(-\beta)}{t^{(i)}_{k-}} \right]
         \hspace{.3cm} \rule[-.3cm]{0cm}{1.2cm} \right) \nonumber
\ea
$\begin{array}{lrclrcl}
  & \hspace{2cm} a_{34} & = & {\ds - \frac{s-\delta}{\sprp - 2\sigma} }
  \medskip
  & \indent t^{(i)}_{\pm\,k}  & = & t^0_k \; + \; t^{(i)}_{\pm}
  & \indent u^{(i)}_{\pm}(x)  & = & t(x) \; - \; t^{(i)}_{\pm}
    \smskip
  & \indent t^{(i)}_{\pm} & = & {\ds
      \frac{\ri\,\!x_0 \pm \sqrt{x_i^2 - {x_0}^2}}{x_i} }
  & \hspace{2cm}    t(x) & = & {\ds
      \frac{\ri\,\!x_0 + \sqrt{x^2-{x_0}^2}}{x} } \bigskip \bigskip
\end{array}$
\ba
  \hspace*{.75cm}
  X_2 & = & -\frac{1}{\,2\,\sqrt{\Delta_3}\,} \; \times \; \nl
      & & \sum_{i=1}^2 \; \sum_{j=1}^2 \; (-1)^{j+1} \,
            \left( \rule[-.3cm]{0cm}{1.2cm} \hspace{.58cm}
            \myln \frac{x_i-x_j-\ieps}{x_i+x_j-\ieps} \cdot
            \myln \frac{x_j-\beta}{x_j+\beta} \right. \nl \nl
 & & \hspace{3.36cm} - \; \mysp \left[
            -\frac{x_j-\beta}{x_i-x_j-\ieps} \right] \; + \;
            \mysp \left[
            -\frac{x_j+\beta}{x_i-x_j-\ieps} \right] \nl \nl
 & & \hspace{3.33cm} \left. + \; \mysp \left[
            \frac{x_j-\beta}{x_i+x_j-\ieps} \right] \; - \;
            \mysp \left[
            \frac{x_j+\beta}{x_i+x_j-\ieps} \right]
            \rule[-.3cm]{0cm}{1.2cm} \hspace{.8cm} \right) \; \nonumber
\ea
\newpage
$\begin{array}{llll}
 58.) & {\ds \left[ \frac{l_{t_2u_1}}{\,a^{ut}_1 \cdot 2\,\sqrt{C_{21}}\,}
             \right]_\theta } & = &
        {\cal T\!} \left( \rule[-.2cm]{0cm}{1cm}
          {\ds \left[ \frac{l_{t_1u_2}}{{\,a^{ut}_1} \cdot 2\,
            \sqrt{C_{12}}\,} \right]_\theta } \right)
          {\ds \;\;\; \equiv \;\;\; \frac{s'}{\sprm} \cdot D^a_{t2u1} }
      \bigskip \\
 59.) & {\ds \left[ \frac{l_{t_1u_2}}{\,a^{ut}_2 \cdot 2\,\sqrt{C_{12}}\,}
             \right]_\theta } & = &
        {\ds \left[ \frac{l_{t_1u_2}}{\,a^{ut}_1 \cdot 2\,\sqrt{C_{12}}\,}
             \right]_\theta }
      \bigskip \\
 60.) & {\ds \left[ \frac{l_{t_2u_1}}{\,a^{ut}_2 \cdot 2\,\sqrt{C_{21}}\,}
             \right]_\theta } & = &
        {\cal T\!} \left( \rule[-.2cm]{0cm}{1cm}
          {\ds \left[ \frac{l_{t_1u_2}}{\,a^{ut}_1 \cdot 2\,\sqrt{C_{12}}\,}
               \right]_\theta } \right)
 \vspace{5cm}
\end{array}$
%
\section{Loop Integrals}
\label{loopint1}
\indent
For the computation of the virtual initial state QED corrections to
the studied process~(\ref{eezz4f}), apart from the integration over
final state fermion decay angles, two sets of integrals are needed.
The second set is given in section~\ref{loopint2}. The first set is
presented in this section and consists of the integration over the
loop momentum $p$.
\para
It is sufficient to
present the integrals for the virtual t-channel graphs, because the
interferences of the virtual u-channel graphs with the Born level t-
and u-channel graphs equal the interferences of the virtual
t-channel graphs with the Born level u-channel and t-channel
graphs. This is easily verified by symmetry arguments. First, some
definitions are introduced:
\ba
  \hspace*{.75cm} \nl
  \frac{1}{\Pi_0} & = & \frac{1}{p^2-\ieps} \nl
  \frac{1}{\Pi_1} & = & \frac{1}{(p-k_1)^2+\ME^2-\ieps} \; = \;
                        \frac{1}{p^2-2pk_1-\ieps} \nl
  \frac{1}{\Pi_2} & = & \frac{1}{(p+k_2)^2+\ME^2-\ieps} \; = \;
                        \frac{1}{p^2+2pk_2-\ieps} \nl
  \frac{1}{\Pi_q} & = & \frac{1}{(p-q_t)^2+\ME^2-\ieps} ~~~~.
  \\ \nonumber
\ea
The momenta $k_1$, $k_2$, and $q_t$
are carried by the incoming electron, the incoming positron and the
intermediate initial state electron. For brevity, $q\!\equiv\!q_t$~with
$q^2\!=\!t$~(compare equation~(\ref{bornt})) will be used subsequently.
Dimensional regularization is used to consistently treat divergent
loop integrals. The following notations are employed:
\ba
  \left[A\right]_n & \equiv &
  \left. \, \mu^{(4-n)} \! \int \frac{d^np}{(2\pi)^n} \; A
  ~\right|_{\mu=m_e}
  \hspace{2.1cm}
  {\rm with} \hspace{.5cm} \left| n-4 \right| \; \ll \; 1~,
  \hspace{1.6cm} \\ \nl
  {\rm P} & \equiv & \frac{1}{n-4} + \frac{1}{2} \, \gamma_E +
               \myln \frac{\ME}{\mu} - \myln \left(2\sqrt{\pi}\right)
               \; \rule[-.45cm]{.02cm}{1.15cm}_{\;\mu=\ME}
  \hspace{1cm} n \; = \; 4 - \varepsilon~,
  \label{poleUV} \\ \nl
  {\rm P^{IR}} & \equiv & \frac{1}{n-4} + \frac{1}{2}\, \gamma_E +
               \myln \frac{\ME}{\mu} - \myln \left(2\sqrt{\pi}\right)
               \; \rule[-.45cm]{.02cm}{1.15cm}_{\;\mu=\ME}
  \hspace{1cm} n \; = \; 4 + \varepsilon~,
  \label{poleIR}
\ea
and
\ba
  \lz & \equiv & \myln \left( -\frac{s}{\ME^2-\ieps} \right)
      \;\; = \;\;
      \myln \frac{s}{\ME^2} - \ri\pi \; = \; l_\beta - \ri\pi
      \nl
  l_1 & \equiv & \myln \left( -\frac{\SONE}{\ME^2-\ieps} \right)
      \;\; = \;\;
      \myln \frac{\SONE}{\ME^2} - \ri\pi \nl
  l_2 & \equiv & \myln \left( -\frac{\STWO}{\ME^2-\ieps} \right)
      \;\; = \;\;
      \myln \frac{\STWO}{\ME^2} - \ri\pi \nl
  \rule[-.1cm]{0cm}{.8cm}
  l_t & \equiv & \myln \left( \frac{q^2 + \ME^2}{\ME^2} \right)
      \;\; \stackrel{\rm URA}{=} \;\; \myln\frac{q^2}{\ME^2}
      \hspace{1cm} .
  \label{logdefs}
  \\ \nonumber
\ea
Euler's constant has the value $\gamma_E = 0.577216...$~. The
difference between the ultraviolet divergence $\,{\rm P}\,$~and the
infrared divergence $\,{\rm P^{IR}}\,$~is that $\,{\rm P}\,$~is
regularized by $\:n=4-\varepsilon\:$ and $\,{\rm P^{IR}}\,$~by
$\:n=4+\varepsilon\:$~with $\:0 < \varepsilon \ll 1\:$. In the URA one
obtains: \vspace{1cm} \\
$\begin{array}{llll}
 1.) & {\ds \left[ \frac{1}{\Pi_0\Pi_1} \right]_n } & = &
       {\ds \iospi \left( \rule[0cm]{0cm}{.4cm}
         -2{\rm P} + 2 \right) } \medskip
 3.) & {\ds \left[ \frac{p^{\,\mu}}{\Pi_0\Pi_1} \right]_n } & = &
       {\ds \frac{\,\ri \cdot k_1^{\,\mu}}{16\,\pi^2}
            \left( \rule[-.1cm]{0cm}{.7cm} \!\!
                   -{\rm P} + \frac{1}{2}\right) } \bigskip \smallskip
\end{array}$
$\begin{array}{llll}
 \hspace{1.55cm}
 2.) & {\ds \left[ \frac{1}{\Pi_0\Pi_2} \right]_n } & = &
       {\ds \iospi \left( \rule[0cm]{0cm}{.4cm}
         -2{\rm P} + 2 \right) } \medskip
 \hspace{1.55cm}
 4.) & {\ds \left[ \frac{p^{\,\mu}}{\Pi_0\Pi_2} \right]_n } & = &
       {\ds - \,\frac{\,\ri \cdot k_2^{\,\mu}}{16\,\pi^2}
            \left( \rule[-.1cm]{0cm}{.7cm} \!\!
                   -{\rm P} + \frac{1}{2}\right) } \bigskip \smallskip
\end{array}$
\newline
$\begin{array}{llll}
 5.) & {\ds \left[ \frac{1}{\Pi_0\Pi_q} \right]_n } & = &
       {\ds \iospi \left( \rule[0cm]{0cm}{.4cm}
         - 2{\rm P} - l_t + 2 \right) }
       \bigskip
 6.) & {\ds \left[ \frac{p^{\,\mu}}{\Pi_0\Pi_q} \right]_n } & = &
       {\ds \frac{\,\ri \cdot q^{\,\mu}}{16\,\pi^2}
         \left( \rule[-.1cm]{0cm}{.7cm}
         - {\rm P} - \frac{l_t}{2} + 1 \right) }
       \bigskip
 7.) & {\ds \left[ \frac{1}{\Pi_1\Pi_2} \right]_n } & = &
       {\ds \iospi \left( \rule[0cm]{0cm}{.4cm}
         - 2{\rm P} - \lz + 2 \right) }
       \bigskip
 8.) & {\ds \left[ \frac{p^{\,\mu}}{\Pi_1\Pi_2} \right]_n } & = &
       {\ds \iospi \left( k_1 - k_2 \right)^\mu
         \left( \rule[-.1cm]{0cm}{.7cm}
           - {\rm P} - \frac{\lz}{2} + 1 \right) }
\end{array}$
\clearpage
$\begin{array}{llll}
 9.) & {\ds \left[ \frac{1}{\Pi_1\Pi_q} \right]_n } & = &
       {\ds \iospi \left( \rule[0cm]{0cm}{.4cm}
         - 2{\rm P} - l_1 + 2 \right) }
       \bigskip
 10.) & {\ds \left[ \frac{1}{\Pi_2\Pi_q} \right]_n } & = &
        {\ds \iospi \left( \rule[0cm]{0cm}{.4cm}
          - 2{\rm P} - l_2 + 2 \right) }
       \bigskip
 11.) & {\ds \left[ \frac{p^{\,\mu}}{\Pi_1\Pi_q} \right]_n } & = &
        {\ds \iospi \left( q^{\,\mu} + k_1^{\,\mu} \right)
          \left( \rule[0cm]{0cm}{.4cm}
          -{\rm P} - \frac{l_1}{2} +1 \right) }
       \bigskip
 12.) & {\ds \left[ \frac{p^{\,\mu}}{\Pi_2\Pi_q} \right]_n } & = &
        {\ds \iospi \left( q^{\,\mu} - k_2^{\,\mu} \right)
          \left( \rule[0cm]{0cm}{.4cm}
          -{\rm P} - \frac{l_2}{2} +1 \right) }
       \bigskip
 13.) & {\ds \left[ \frac{1}{\Pi_0\Pi_1\Pi_2} \right]_n } & = &
       {\ds \iospi \left[ \rule[-.1cm]{0cm}{.8cm}
         - \frac{2\,\lz}{s} \cdot {\rm P^{IR}} \; + \;
         \frac{1}{s} \left( \frac{\,\pi^2}{6} - \frac{\lz^2}{2} \right)
         \right] }
       \bigskip
 14.) & {\ds \left[ \frac{p^{\,\mu}}{\Pi_0\Pi_1\Pi_2} \right]_n } & = &
       {\ds - \iospi \left( k_1 - k_2 \right)^\mu \cdot \frac{\lz}{s} }
       \bigskip
 15.) & {\ds \left[ \frac{p^{\,\mu} p^{\,\nu}}
                         {\Pi_0\Pi_1\Pi_2} \right]_n } & = &
        {\ds \iospi \left[ \rule[-.1cm]{0cm}{.8cm} \hspace{.64cm}
          g^{\mu\nu} \cdot \left( -\frac{1}{2} {\rm P}
          - \frac{\lz}{4} + \frac{3}{4} \right) \right. } \\
      & & & \hspace{1.25cm} {\ds \left. \rule[-.1cm]{0cm}{.8cm} + \;
         \left( k_1^{\,\mu} k_1^{\,\nu} + k_2^{\,\mu} k_2^{\,\nu} \right)
         \cdot \frac{1-\lz}{2s} \; + \;
         \left( k_1^{\,\mu} k_2^{\,\nu} + k_2^{\,\mu} k_1^{\,\nu} \right)
         \cdot \frac{1}{2s}
         \; \right] }
       \bigskip
 16.) & {\ds \left[ \frac{1}{\Pi_0\Pi_1\Pi_q} \right]_n } & = &
        {\ds \iospi \cdot \frac{1}{t+\SONE} \cdot \left[ \,
          \rule[-.1cm]{0cm}{.8cm}
          \frac{1}{2} \left( \rule[0cm]{0cm}{.4cm} l_t - l_1 \right) \!
                      \left( \rule[0cm]{0cm}{.4cm} 3l_1 - l_t \right)
          - 2 \, \mysp \left( \rule[-.1cm]{0cm}{.7cm}
          \frac{t \! + \! \SONE \! - \! \ieps}{\SONE} \right) \, \right] }
       \bigskip
 17.) & {\ds \left[ \frac{1}{\Pi_0\Pi_2\Pi_q} \right]_n } & = &
        {\ds \iospi \cdot \frac{1}{t+\STWO} \cdot \left[ \,
          \rule[-.1cm]{0cm}{.8cm}
          \frac{1}{2} \left( \rule[0cm]{0cm}{.4cm} l_t - l_2 \right) \!
                      \left( \rule[0cm]{0cm}{.4cm} 3l_2 - l_t \right)
          - 2 \, \mysp \left( \rule[-.1cm]{0cm}{.7cm}
          \frac{t \! + \! \STWO \! - \! \ieps}{\STWO} \right) \, \right] }
       \bigskip
 18.) & {\ds \left[ \frac{p^{\,\mu}}{\Pi_0\Pi_1\Pi_q} \right]_n } & = &
        {\ds \iospi \cdot \frac{1}{t+\SONE} \cdot \left\{
          \rule[-.3cm]{0cm}{1.2cm} \;
          q^{\,\mu} \left( \rule[0cm]{0cm}{.4cm} l_t - l_1 \right)
          \right. } \\
      & & & {\ds \hspace{1.48cm} \left. + \;
          k_1^{\,\mu} \left( \rule[-.1cm]{0cm}{.8cm} - l_1 \; + \; t \!
          \cdot \!\!
          \left[ \frac{1}{\Pi_0\Pi_1\Pi_q} \right]_n
          - \; \frac{2t}{t+\SONE}
          \left( \rule[0cm]{0cm}{.4cm} l_t - l_1 \right) \right)
          \rule[-.3cm]{0cm}{1.2cm} \right\} }
       \bigskip
 19.) & {\ds \left[ \frac{p^{\,\mu}}{\Pi_0\Pi_2\Pi_q} \right]_n } & = &
        {\ds \iospi \cdot \frac{1}{t+\STWO} \cdot \left\{
          \rule[-.3cm]{0cm}{1.2cm} \;
          q^{\,\mu} \left( \rule[0cm]{0cm}{.4cm} l_t - l_2 \right)
          \right. } \\
      & & & {\ds \hspace{1.48cm} \left. - \;
          k_2^{\,\mu} \left( \rule[-.1cm]{0cm}{.8cm} - l_2 \; + \; t
          \!
          \cdot \!\!
          \left[ \frac{1}{\Pi_0\Pi_2\Pi_q} \right]_n
          - \; \frac{2t}{t+\STWO}
          \left( \rule[0cm]{0cm}{.4cm} l_t - l_2 \right) \right)
          \rule[-.3cm]{0cm}{1.2cm} \right\} }
       \bigskip
\end{array}$
\clearpage
$\begin{array}{llll}
 20.) & {\ds \left[ \frac{p^{\,\mu}p^{\,\nu}}{\Pi_0\Pi_1\Pi_q}
         \right]_n } & = &
        {\ds \iospi \cdot \left\{ \rule[-.3cm]{0cm}{1.2cm}
          \hspace{.5cm} g^{\mu\nu} \!\cdot\!
          \left( \!\! \rule[-.2cm]{0cm}{1cm} -
          \frac{{\rm P}}{2} + \frac{3}{4} \; - \;
          \frac{1}{4\left(t+\SONE\right)} \left( \rule[0cm]{0cm}{.4cm}
          t \cdot l_t + \SONE \cdot l_1 \right) \right) \right. } \\
      & & & {\ds \hspace{1.64cm} + \;
          \frac{k_1^{\,\mu} k_1^{\,\nu}}{t+\SONE} \cdot
          \left\{ \rule[-.2cm]{0cm}{1cm} \hspace{.34cm}
          \left( \frac{1}{2} + \frac{t}{t+\SONE} \right)
          \left( \rule[0cm]{0cm}{.4cm} 1 - l_1 \right) \right. } \\
      & & & {\ds \hspace{3.95cm} - \;
          \frac{3\, t^2}{\left( t+\SONE \right)^2}
          \left( \rule[0cm]{0cm}{.4cm} l_t - l_1 \right) } \\
      & & & {\ds \left. \hspace{3.92cm} + \;
          \frac{t^2}{t+\SONE} \cdot
          \left[ \frac{1}{\Pi_0\Pi_1\Pi_q} \right]_n
          \rule[-.2cm]{0cm}{1cm} \hspace{.5cm} \right\} } \\
      & & & {\ds \hspace{1.64cm} + \;
          \frac{\,k_1^{\,\mu} q^{\,\nu} + k_1^{\,\nu} q^{\,\mu}}
               {2 \left(t+\SONE\right)} \cdot
          \left\{ \rule[-.2cm]{0cm}{1cm} \frac{t}{t+\SONE}
          \left(\rule[0cm]{0cm}{.4cm} l_t - l_1 \right) \; - \; 1
          \right\} } \\
      & & & {\ds \left. \hspace{1.59cm} + \;
          \frac{\,q^{\,\mu} q^{\,\nu}}{2 \left(t+\SONE\right)} \cdot
          \left(\rule[0cm]{0cm}{.4cm} l_t - l_1 \right)
          \rule[-.3cm]{0cm}{1.2cm} \hspace{4.6cm}\right\} }
       \bigskip \medskip
 21.) & {\ds \left[ \frac{p^{\,\mu}p^{\,\nu}}{\Pi_0\Pi_2\Pi_q}
         \right]_n } & = &
        {\ds \iospi \cdot \left\{ \rule[-.3cm]{0cm}{1.2cm}
          \hspace{.5cm} g^{\mu\nu} \!\cdot\!
          \left( \!\! \rule[-.2cm]{0cm}{1cm}
          -\frac{{\rm P}}{2} + \frac{3}{4} \; - \;
          \frac{1}{4\left(t+\STWO\right)} \left( \rule[0cm]{0cm}{.4cm}
          t \cdot l_t + \STWO \cdot l_2 \right) \right) \right. } \\
      & & & {\ds \hspace{1.64cm} + \;
          \frac{k_2^{\,\mu} k_2^{\,\nu}}{t+\STWO} \cdot
          \left\{ \rule[-.2cm]{0cm}{1cm} \hspace{.34cm}
          \left( \frac{1}{2} + \frac{t}{t+\STWO} \right)
          \left( \rule[0cm]{0cm}{.4cm} 1 - l_2 \right) \right. } \\
      & & & {\ds \hspace{3.95cm} - \;
          \frac{3\, t^2}{\left( t+\STWO \right)^2}
          \left( \rule[0cm]{0cm}{.4cm} l_t - l_2 \right) } \\
      & & & {\ds \left. \hspace{3.92cm} + \;
          \frac{t^2}{t+\STWO} \cdot
          \left[ \frac{1}{\Pi_0\Pi_2\Pi_q} \right]_n
          \rule[-.2cm]{0cm}{1cm} \hspace{.5cm} \right\} } \\
      & & & {\ds \hspace{1.64cm} - \;
          \frac{\,k_2^{\,\mu} q^{\,\nu} + k_2^{\,\nu} q^{\,\mu}}
               {2 \left(t+\STWO\right)} \cdot
          \left\{ \rule[-.2cm]{0cm}{1cm} \frac{t}{t+\STWO}
          \left(\rule[0cm]{0cm}{.4cm} l_t - l_2 \right) \; - \; 1
          \right\} } \\
      & & & {\ds \left. \hspace{1.59cm} + \;
          \frac{\,q^{\,\mu} q^{\,\nu}}{2 \left(t+\STWO\right)} \cdot
          \left(\rule[0cm]{0cm}{.4cm} l_t - l_2 \right)
          \rule[-.3cm]{0cm}{1.2cm} \hspace{4.6cm}\right\} }
       \bigskip \medskip
\end{array}$
\clearpage
$\begin{array}{llll}
 22.) & {\ds \left[ \frac{1}{\Pi_1\Pi_2\Pi_q} \right]_n } & = &
        {\ds \frac{\rm i}{16\,\pi^2 \!\cdot\!\SLAM} \!\cdot\!
          \left\{ \rule[-.2cm]{0cm}{1cm}
          2 \,\mysp \left(x_2\right) \: - \:
          2 \,\mysp \left(x_1\right) \: + \:
          \myln\,(x_1x_2) \!\cdot\! \myln\left(
          \rule[-.05cm]{0cm}{.7cm} \frac{1-x_2}{1-x_1}\right) \right\} }
        \medskip & & \equiv & {\ds\rm \iospi \!\cdot\! I_{12q} }
        \bigskip
      & \indent \indent x_{1/2} & = &
        {\ds \frac{s-\delta \pm \SLAM}{2\,s} } \hspace{1.3cm}
        \Rightarrow
        \hspace{1.3cm} x_1\!\cdot\!x_2 \; = \; {\ds \frac{\STWO}{s} }
        \smskip
      & & & \hspace{5.3cm} {\ds \frac{1-x_2}{1-x_1} \; = \;
        \frac{s+\delta+\SLAM}{s+\delta-\SLAM} }
      \bigskip \medskip
 23.) & {\ds \left[ \frac{p^{\,\mu}}{\Pi_1\Pi_2\Pi_q} \right]_n } & = &
        {\ds \frac{\rm i}{16\,\pi^2 \!\cdot\! \lambda} \; \times } \\
      & & & {\ds \left\{ \rule[-.3cm]{0cm}{1.2cm} \hspace{.49cm}
          q^{\,\mu} \left(\rule[-.2cm]{0cm}{1cm}
          s \left(\rule[-.05cm]{0cm}{.4cm} s-\sigma\right) {\rm I_{12q}}
          \; - \; \left( s+\delta \right) \myln \frac{\SONE}{s}
          \; - \; \left( s-\delta \right) \myln \frac{\STWO}{s}
          \right) \right. } \\
      & & & {\ds \;\;\; + \; k_1^{\,\mu} \left(\rule[-.2cm]{0cm}{1cm}
          -\STWO\left(\rule[-.05cm]{0cm}{.4cm} s+\delta\right)
          {\rm I_{12q}} \; + \; \left( s-\sigma \right) \myln
          \frac{\SONE}{s} \; + \; 2\,\STWO \myln \frac{\STWO}{s}
          \right) } \\
      & & & {\ds \left. \hspace{.31cm} + \; k_2^{\,\mu}
          \left(\rule[-.2cm]{0cm}{1cm}
          \SONE\left(\rule[-.05cm]{0cm}{.4cm} s-\delta\right)
          {\rm I_{12q}} \; - \; 2\,\SONE \myln \frac{\SONE}{s} \; - \;
          \left( s-\sigma \right) \myln \frac{\STWO}{s} \right)
          \rule[-.3cm]{0cm}{1.2cm} \hspace{.32cm} \right\} }
      \bigskip \medskip
\end{array}$
\clearpage
$\begin{array}{llll}
 24.) & {\ds \left[ \frac{p^{\,\mu} p^{\,\nu}}
                         {\Pi_1\Pi_2\Pi_q} \right]_n } & = &
        {\ds \iospi \!\cdot\!\left\{ \rule[-.1cm]{0cm}{.6cm}
          \hspace{.65cm} g^{\mu\nu} \!\cdot\! F_{21} \; + \;
          q^{\,\mu} q^{\,\nu} \!\cdot\! F_{22} \; + \;
          k_1^{\,\mu} k_1^{\,\nu} \!\cdot\! F_{23} \; + \;
          k_2^{\,\mu} k_2^{\,\nu} \!\cdot\! F_{24} \right. } \\
      & & & \hspace{1.35cm} {\ds \; + \;
          \left(\rule[-.05cm]{0cm}{.4cm}
          k_1^{\,\mu} k_2^{\,\nu} + k_1^{\,\nu} k_2^{\,\mu} \right)
          \!\cdot\! F_{25} \; + \;
          \left(\rule[-.05cm]{0cm}{.4cm}
          k_1^{\,\mu} q^{\,\nu} + k_1^{\,\nu} q^{\,\mu} \right)
          \!\cdot\! F_{26} } \\
      & & & \hspace{1.31cm} {\ds \left. \; + \;
          \left(\rule[-.05cm]{0cm}{.4cm}
          k_2^{\,\mu} q^{\,\nu} + k_2^{\,\nu} q^{\,\mu} \right)
          \!\cdot\! F_{27} \hspace{.3cm} \rule[-.1cm]{0cm}{.6cm}
          \right\} }
        \bigskip
      & \indent \indent \; F_{21} & = &
          {\ds \frac{1}{4}
            \left\{ \rule[-.3cm]{0cm}{1.2cm} \; \frac{1}{\lambda}
            \left[ \rule[-.2cm]{0cm}{1cm}
            \SONE\!\left(\rule[-.05cm]{0cm}{.4cm} s \!-\! \delta \right)
            \myln \frac{\SONE}{s} \; + \;
            \STWO\!\left(\rule[-.05cm]{0cm}{.4cm} s \!+\! \delta \right)
            \myln \frac{\STWO}{s} \; - \;
            2 s\SONE\STWO \, {\rm I_{12q}} \right]
            \right. } \\
      & & & \hspace{.7cm} {\ds \left. \rule[-.3cm]{0cm}{1.2cm}
          - 2 {\rm P} + 3 - l_0 \rule[-.3cm]{0cm}{1.2cm} \hspace{.3cm}
          \right\} }
        \medskip
      & \indent \indent \; F_{22} & = &
          {\ds \frac{1}{\lambda}
            \left\{ \rule[-.3cm]{0cm}{1.2cm} -s \; - \:
            \frac{1}{2} \left[ \rule[-.2cm]{0cm}{1cm} 3s \, + \,
            \delta  \, + \, \frac{6s\SONE}{\lambda}
            \left(\rule[-.05cm]{0cm}{.4cm} s \!-\! \delta \right)
            \right] \!\cdot\! \myln \frac{\SONE}{s} \right. } \\
      & & & \hspace{1.52cm} {\ds \; + \;\,
            \frac{1}{2} \left[ \rule[-.2cm]{0cm}{1cm} 3s \, - \,
            \delta  \, + \, \frac{6s\STWO}{\lambda}
            \left(\rule[-.05cm]{0cm}{.4cm} s \!+\! \delta \right)
            \right] \!\cdot\! \myln \frac{\STWO}{s} } \\
      & & & \hspace{1.6cm} {\ds \left. + \;
            s^2 \left[ \rule[-.2cm]{0cm}{1cm}
            1 \, + \, \frac{6\,\SONE\STWO}{\lambda} \right] \!\cdot\!
            {\rm I_{12q}} \hspace{2.5cm}
            \rule[-.3cm]{0cm}{1.2cm} \right\} }
        \medskip
      & \indent \indent \; F_{23} & = &
          {\ds \frac{1}{\lambda}
            \left\{ \rule[-.3cm]{0cm}{1.2cm} -\STWO \; + \:
            \frac{1}{2} \left[ \rule[-.2cm]{0cm}{1cm}
            s \, - \,  \SONE \, - \, 3\STWO \, - \,
            \frac{6\,\SONE\STWO}{\lambda}
            \left(\rule[-.05cm]{0cm}{.4cm} s \!-\! \delta \right)
            \right] \!\cdot\! \myln \frac{\SONE}{s} \right. } \\
      & & & \hspace{.76cm} {\ds \left. - \;
            \frac{3\,\STWO^2}{\lambda}
            \left(\rule[-.05cm]{0cm}{.4cm} s \!+\! \delta \right)
             \myln \frac{\STWO}{s} \;\; + \;\;
            \STWO^2 \left[ \rule[-.2cm]{0cm}{1cm}
            1 \, + \, \frac{6\,s\SONE}{\lambda} \right] \!\cdot\!
            {\rm I_{12q}} \hspace{1.3cm}
            \rule[-.3cm]{0cm}{1.2cm} \right\} }
        \medskip
      & \indent \indent \; F_{24} & = &
          {\ds \frac{1}{\lambda}
            \left\{ \rule[-.3cm]{0cm}{1.2cm} -\SONE \; + \:
            \frac{1}{2} \left[ \rule[-.2cm]{0cm}{1cm}
            s \, - \,  \STWO \, - \, 3\SONE \, - \,
            \frac{6\,\STWO\SONE}{\lambda}
            \left(\rule[-.05cm]{0cm}{.4cm} s \!+\! \delta \right)
            \right] \!\cdot\! \myln \frac{\STWO}{s} \right. } \\
      & & & \hspace{.76cm} {\ds \left. - \;
            \frac{3\,\SONE^2}{\lambda}
            \left(\rule[-.05cm]{0cm}{.4cm} s \!-\! \delta \right)
             \myln \frac{\SONE}{s} \;\; + \;\;
            \SONE^2 \left[ \rule[-.2cm]{0cm}{1cm}
            1 \, + \, \frac{6\,s\STWO}{\lambda} \right] \!\cdot\!
            {\rm I_{12q}} \hspace{1.3cm}
            \rule[-.3cm]{0cm}{1.2cm} \right\} }
        \medskip
      & \indent \indent \; F_{25} & = &
          {\ds \frac{1}{2\lambda}
            \left\{ \rule[-.3cm]{0cm}{1.2cm} \; s - \sigma \; + \;
            \SONE \left[ \rule[-.2cm]{0cm}{1cm}
            1 \, + \, \frac{6\,\STWO}{\lambda}
            \left(\rule[-.05cm]{0cm}{.4cm} s \!+\! \delta \right)
            \right] \!\cdot\! \myln \frac{\SONE}{s} \right. } \\
      & & & \hspace{2.1cm} {\ds \; + \hspace{.195cm}
            \STWO \left[ \rule[-.2cm]{0cm}{1cm}
            1 \, + \, \frac{6\,\SONE}{\lambda}
            \left(\rule[-.05cm]{0cm}{.4cm} s \!-\! \delta \right)
            \right] \!\cdot\! \myln \frac{\STWO}{s} } \\
      & & & \hspace{1.465cm} {\ds \left. + \; 2\,\SONE\STWO
            \left[ \rule[-.2cm]{0cm}{1cm}
            1 \, - \, \frac{3\,s}{\lambda}
            \left(\rule[-.05cm]{0cm}{.4cm} s \!-\! \sigma \right)
            \right] \!\cdot\! {\rm I_{12q}} \rule[-.3cm]{0cm}{1.2cm}
            \rule[-.3cm]{0cm}{1.2cm} \hspace{1cm} \right\} }
\end{array}$
\clearpage
$\begin{array}{llll}
      & \indent \indent \; F_{26} & = &
          {\ds \frac{1}{\lambda}
            \left\{ \rule[-.3cm]{0cm}{1.2cm} \;\; \frac{s-\delta}{2}
            \; + \; \left[ \rule[-.2cm]{0cm}{1cm}
            \frac{s+\STWO}{2} \, + \, \frac{6\,s\SONE\STWO}{\lambda}
            \right] \!\cdot\! \myln \frac{\SONE}{s} \right. } \\
      & & & \hspace{2.13cm} {\ds  - \hspace{.145cm}
            \left[ \rule[-.2cm]{0cm}{1cm}
            \frac{\STWO}{2} \, - \, \frac{3\,s\STWO}{\lambda}
            \left(\rule[-.05cm]{0cm}{.4cm} s \!-\! \sigma \right)
            \right] \!\cdot\! \myln \frac{\STWO}{s} } \\
      & & & \hspace{1.415cm} {\ds \left. - \; s\STWO
            \left[ \rule[-.2cm]{0cm}{1cm}
            1 \, + \, \frac{3\,\SONE}{\lambda}
            \left(\rule[-.05cm]{0cm}{.4cm} s \!-\! \delta \right)
            \right] \!\cdot\! {\rm I_{12q}}
            \rule[-.3cm]{0cm}{1.2cm} \hspace{1cm} \right\} }
        \medskip
      & \indent \indent \; F_{27} & = &
          {\ds - \frac{1}{\lambda}
            \left\{ \rule[-.3cm]{0cm}{1.2cm} \;\; \frac{s+\delta}{2}
            \; + \; \left[ \rule[-.2cm]{0cm}{1cm}
            \frac{s+\SONE}{2} \, + \, \frac{6\,s\SONE\STWO}{\lambda}
            \right] \!\cdot\! \myln \frac{\STWO}{s} \right. } \\
      & & & \hspace{2.45cm} {\ds  - \hspace{.145cm}
            \left[ \rule[-.2cm]{0cm}{1cm}
            \frac{\SONE}{2} \, - \, \frac{3\,s\SONE}{\lambda}
            \left(\rule[-.05cm]{0cm}{.4cm} s \!-\! \sigma \right)
            \right] \!\cdot\! \myln \frac{\SONE}{s} } \\
      & & & \hspace{1.73cm} {\ds \left. - \; s\SONE
            \left[ \rule[-.2cm]{0cm}{1cm}
            1 \, + \, \frac{3\,\STWO}{\lambda}
            \left(\rule[-.05cm]{0cm}{.4cm} s \!+\! \delta \right)
            \right] \!\cdot\! {\rm I_{12q}}
            \rule[-.3cm]{0cm}{1.2cm} \hspace{1cm} \right\} }
        \bigskip \bigskip
\end{array}$
\begin{list}{}{\leftmargin=0cm}
  \item As mentioned in appendix~\ref{virtres}, a FORM computer
    algebra program was used to handle the many terms in the virtual
    corrections' calculation. Inspection of the FORM code at an
    intermediate level showed that, for the calculation of virtual
    QED corrections to process~(\ref{eezz4f})~(see
    appendix~\ref{virtres}), the four-point function is only
    needed in one specific Lorentz contraction. Precisely this Lorentz
    contraction is given below.
\end{list}
$\begin{array}{llll}
 25.) & {\ds \left[ \frac{2\,p\!\cdot\!q}
                         {\Pi_0\Pi_1\Pi_2\Pi_q} \right]_n } & = &
        {\ds \iospi \cdot \left\{ \rule[-.3cm]{0cm}{1.2cm}
          {\rm I_{12q}} \; + \;
          \frac{1}{s} \left[ \rule[-.2cm]{0cm}{1cm}
          l_0 \left(\rule[-.05cm]{0cm}{.4cm} l_1 + l_2 - 2\,l_t \right)
          \; + \; \frac{1}{2} \myln^2 \frac{\SONE}{\STWO} \right]
          \right\} }
\end{array}$
%
\newpage
\section{Virtual Corrections' Phase Space Integrals}
\label{loopint2}
After the integration over the loop momentum, there is one non-trivial
integration left when the virtual initial state corrections to
process~(\ref{eezz4f}) are to be
evaluated. This is the integration over the boson scattering angle
$\varth$ in the center of mass system. It should be recalled that it
is sufficient to present the integrals for the t-channel case and the
t-channel/u-channel interference for reasons of symmetry. Making use
of the URA and the notations
\ba
  \hspace*{2.5cm}
  \left[A\right]_V \;\; \equiv \;\;\:
    \frac{\SLAM}{2} \int\limits_{-1}^{+1} & &
      \hspace{-.9cm} d\mycos\varth~A \;\;\; = \;\;
    \int\limits_{t_{min}}^{t_{max}} dt~A \\ \nl
 t_{12} & = & t+\SONE \nl
 t_{34} & = & t+\STWO \nl\nl
 d_{12} \; = \; \frac{s-\STWO}{\SONE} & &
 t_{max}\; = \; \frac{s-\sigma+\SLAM}{2} \nl
 d_{34} \; = \; \frac{s-\SONE}{\STWO} & &
 t_{min}\; = \; \frac{s-\sigma-\SLAM}{2} \nonumber
\ea
the following list of integrals is obtained. \vspace{.3cm} \\
$\begin{array}{llll}
 \hspace{.205cm}
 1.) & \, \left[\, 1 \, \right]_V & = & \SLAM \medskip
\end{array}$
\newline
$\begin{array}{llll}
 \hspace{.205cm}
 2.) & \, \left[\, t \, \right]_V & = &
       {\ds \frac{\SLAM}{2}\cdot \left(s-\sigma\right) } \medskip
\end{array}$
$\begin{array}{rlll}
 \hspace{3.03cm}
  3.) & \,\! \left[\, t^2 \,\right]_V & = & {\ds \frac{\SLAM}{3} \!\cdot\!
       \left\{ \left(s \! - \!\sigma\right)^2 - \SONE\STWO \right\} }
     \medskip
\end{array}$
\newline
$\begin{array}{llll}
 \hspace{.205cm}
 4.) & {\ds \left[ \frac{1}{\,t\,} \right]_V } & = &
       {\ds \myln \frac{s\!-\!\sigma\!+\!\SLAM}{s\!-\!\sigma\!-\!\SLAM}
       \; \equiv \; {\cal L}_B } \medskip
\end{array}$
$\begin{array}{rlll}
 \hspace{1.53cm}
 5.) & {\ds \left[ \frac{1}{\,u\,} \right]_V } & = & {\cal L}_B
     \medskip
\end{array}$
\newline
$\begin{array}{llll}
 \hspace{.205cm}
 6.) & {\ds \left[ \frac{1}{t_{12}} \right]_V } & = &
       {\ds \myln \frac{s\!+\!\delta\!+\!\SLAM}{s\!+\!\delta\!-\!\SLAM}
       \; \equiv \; {\cal L}_{12} } \medskip
\end{array}$
$\begin{array}{rlll}
 \hspace{1.35cm}
 7.) & {\ds \left[ \frac{1}{t_{34}} \right]_V } & = &
       {\ds \myln \frac{s\!-\!\delta\!+\!\SLAM}{s\!-\!\delta\!-\!\SLAM}
       \; \equiv \; {\cal L}_{34} }
     \medskip
\end{array}$
\newline
$\begin{array}{llll}
 \hspace{.205cm}
 8.) & {\ds \left[ \frac{1}{t^2} \right]_V } & = &
       {\ds \frac{\SLAM}{\SONE\STWO} }
     \medskip
\end{array}$
$\begin{array}{rlll}
 \hspace{3.96cm}
 9.) & {\ds \left[ \frac{1}{t \cdot u} \right]_V } & = &
       {\ds \frac{2}{s-\sigma} \cdot {\cal L}_B }
     \medskip
\end{array}$
\newline
$\begin{array}{llll}
 10.) & {\ds \left[ \frac{1}{{t_{12}}^2} \right]_V } & = &
       {\ds \frac{\SLAM}{s\SONE} }
     \bigskip
\end{array}$
$\begin{array}{rlll}
 \hspace{3.75cm}
 11.) & {\ds \left[ \frac{1}{{t_{34}}^2} \right]_V } & = &
       {\ds \frac{\SLAM}{s\STWO} }
     \bigskip
\end{array}$
\vspace{.3cm} \newline
$\begin{array}{llll}
 12.) & {\ds \; \left[\, l_t \,\right]_V } & = &
        {\ds \frac{s-\sigma}{2} \!\cdot\! {\cal L}_B \; + \;
          \frac{\SLAM}{2}\!\cdot\! {\cal L}_S \; - \; \SLAM }
        \medskip
      & \indent {\cal L}_S & = & {\ds \left( \rule[-.15cm]{0cm}{.8cm}
          2 \,l_\beta \, + \, \myln \frac{\SONE}{s} \, + \,
          \myln \frac{\STWO}{s} \right) }
\end{array}$
\newpage
$\begin{array}{llll}
 13.) & {\ds \left[ \frac{\,l_t\,}{t} \right]_V } & = &
        {\ds \frac{1}{2} \, {\cal L}_B \!\cdot\! {\cal L}_S }
        \bigskip
 14.) & {\ds \left[ \frac{\,l_t\,}{u} \right]_V } & = &
        {\ds \myln \frac{s-\sigma}{\ME^2} \!\cdot\! {\cal L}_B \; - \;
          \mysp \left( \frac{\,t_{max}\,}{s-\sigma} \right) \; + \;
          \mysp \left( \frac{\,t_{min}\,}{s-\sigma} \right) }
      \bigskip
 15.) & {\ds \left[ \frac{l_t}{t_{12}} \right]_V } & = &
        {\ds \frac{{\cal L}_{12} \!\cdot\! {\cal L}_S}{2} \; - \;
             \frac{{\cal L}_B}{2} \!\cdot\! \myln \frac{\SONE}{s}
          \; + \; \mysp \left( -\,\frac{\,t_{max}\,}{\SONE} \right)
          \; - \; \mysp \left( -\,\frac{\,t_{min}\,}{\SONE} \right) }
      \bigskip
 16.) & {\ds \left[ \frac{l_t}{t_{34}} \right]_V } & = &
        {\ds \frac{{\cal L}_{34} \!\cdot\! {\cal L}_S}{2} \; - \;
             \frac{{\cal L}_B}{2} \!\cdot\! \myln \frac{\STWO}{s}
          \; + \; \mysp \left( -\,\frac{\,t_{max}\,}{\STWO} \right)
          \; - \; \mysp \left( -\,\frac{\,t_{min}\,}{\STWO} \right) }
      \bigskip
 17.) & {\ds \left[ \frac{\,l_t\,}{t^2} \right]_V } & = &
        {\ds \frac{1}{2\,\SONE\STWO} \left( \rule[-.05cm]{0cm}{.6cm}
          \SLAM \!\cdot\! \left( {\cal L}_S + 2 \right) \; - \;
          \left( s-\sigma \right) \!\cdot\! {\cal L}_B \right) }
      \bigskip
 18.) & {\ds \left[ \frac{l_t}{{t_{12}}^2} \right]_V } & = &
        {\ds \frac{1}{\SONE} \left( \rule[-.25cm]{0cm}{1cm}
          \frac{s-\delta}{2s} \!\cdot\! {\cal L}_B \; - \;
          {\cal L}_{12} \; + \; \frac{\SLAM}{2s} \!\cdot\! {\cal L}_S
          \right) }
      \bigskip
 19.) & {\ds \left[ \frac{l_t}{{t_{34}}^2} \right]_V } & = &
        {\ds \frac{1}{\STWO} \left( \rule[-.25cm]{0cm}{1cm}
          \frac{s+\delta}{2s} \!\cdot\! {\cal L}_B \; - \;
          {\cal L}_{34} \; + \; \frac{\SLAM}{2s} \!\cdot\! {\cal L}_S
          \right) }
      \bigskip
 20.) & {\ds \hspace{.015cm} \left[\, {l_t}^2 \, \right]_V } & = &
        {\ds \frac{s-\sigma}{2}\!\cdot\!
          \left( \rule[-.05cm]{0cm}{.4cm}{\cal L}_S - 2 \right)
          \!\cdot\! {\cal L}_B \; + \; \SLAM \cdot\!
          \left( \rule[-.15cm]{0cm}{.8cm} \frac{\,{{\cal L}_B}^2}{4}
          +\frac{\,{{\cal L}_S}^2}{4} - {\cal L}_S +2 \right) }
        \bigskip
 21.) & {\ds \left[ \frac{{l_t}^2}{t} \right]_V } & = &
        {\ds {\cal L}_B \!\cdot\! \left( \rule[-.25cm]{0cm}{1cm}
          \frac{\,{{\cal L}_B}^2}{12} \: + \:
          \frac{\,{{\cal L}_S }^2}{4} \right) }
      \bigskip
 22.) & {\ds \left[ \frac{{l_t}^2}{t^2} \right]_V } & = &
        {\ds \frac{1}{\SONE\STWO} \left[ \rule[-.25cm]{0cm}{1cm}
          \SLAM \cdot\! \left(
          \frac{{{\cal L}_B}^2}{4} + \frac{{{\cal L}_S}^2}{4} \; + \;
          {\cal L}_S \; + \; 2 \right) \; - \;
          \frac{s-\sigma}{2} \!\cdot\! \left( \rule[-.05cm]{0cm}{.4cm}
          {\cal L}_S + 2 \right) \!\cdot\! {\cal L}_B \right] }
      \hspace{-.5cm} \bigskip
 23.) & {\ds \left[ \frac{{l_t}^2}{u} \right]_V } & = &
        {\ds - \;\; \myln^2 \left( \frac{t_{max}}{\ME^2} \right) \cdot
               \myln \left( \frac{t_{min}}{s-\sigma} \right) \;\; + \;\;
          \myln^2 \left(\frac{t_{min}}{\ME^2} \right) \cdot
          \myln \left( \frac{t_{max}}{s-\sigma} \right) }
        \medskip & & & {\ds - \;\;
          2 \, \myln \left( \frac{t_{max}}{\ME^2} \right) \cdot
               \mysp \left( \frac{t_{max}}{s-\sigma} \right) \;\; + \;\;
          2 \, \myln \left( \frac{t_{min}}{\ME^2} \right) \cdot
               \mysp \left( \frac{t_{min}}{s-\sigma} \right) }
        \medskip & & & {\ds + \;\;
          2 \, \mytri \left( \frac{t_{max}}{s-\sigma} \right) \;\; - \;\;
          2 \, \mytri \left( \frac{t_{min}}{s-\sigma} \right) }
\end{array}$
\newpage
$\begin{array}{llll}
 24.) & {\ds \left[ \rule[-.15cm]{0cm}{.8cm}
          \mysp \left( \frac{t_{12}-\ieps}{\SONE} \right) \right]_V }
        & = & \hspace{.53cm}
        {\ds \frac{s-\sigma}{2} \cdot\! {\cal L}_B \; + \;
          \frac{\SLAM}{2} \cdot\! \left( \rule[-.05cm]{0cm}{.4cm}
          {\cal L}_S - 2\,l_1 -2 \right) }
        \medskip & & & {\ds + \;
          \left( \rule[-.05cm]{0cm}{.4cm} \SONE + t_{max} \right)
          \!\cdot\! \mysp \left(
          1 + \frac{t_{max}-\ieps}{\SONE} \right) }
        \medskip & & & {\ds - \;
          \left( \rule[-.05cm]{0cm}{.4cm} \SONE + t_{min} \right)
          \!\cdot\! \mysp \left(
          1 + \frac{t_{min}-\ieps}{\SONE} \right) }
      \medskip \medskip
 25.) & {\ds \left[ \rule[-.15cm]{0cm}{.8cm}
          \mysp \left( \frac{t_{34}-\ieps}{\STWO} \right) \right]_V }
        & = & \hspace{.53cm}
        {\ds \frac{s-\sigma}{2} \cdot\! {\cal L}_B \; + \;
          \frac{\SLAM}{2} \cdot\! \left( \rule[-.05cm]{0cm}{.4cm}
          {\cal L}_S - 2\,l_2 -2 \right) }
        \medskip & & & {\ds + \;
          \left( \rule[-.05cm]{0cm}{.4cm} \STWO + t_{max} \right)
          \!\cdot\! \mysp \left(
          1 + \frac{t_{max}-\ieps}{\STWO} \right) }
        \medskip & & & {\ds - \;
          \left( \rule[-.05cm]{0cm}{.4cm} \STWO + t_{min} \right)
          \!\cdot\! \mysp \left(
          1 + \frac{t_{min}-\ieps}{\STWO} \right) }
      \medskip \medskip
 26.) & {\ds \left[ \rule[-.15cm]{0cm}{.8cm} \frac{1}{\,t\,} \cdot\!
          \mysp \left( \frac{t_{12}-\ieps}{\SONE} \right) \right]_V }
        & = & \hspace{.55cm}
        {\ds \frac{\,\pi^2}{6}\cdot\! {\cal L}_B \; - \;
          2\,\mytri \left( -\frac{\,t_{max}\,}{\SONE}  \right) \; + \;
          2\,\mytri \left( -\frac{\,t_{min}\,}{\SONE}  \right) }
        \medskip & & & {\ds + \;
          \left( \rule[-.15cm]{0cm}{.8cm}
          \myln \frac{\,t_{max}\,}{\ME^2} - l_1 \right) \!\cdot\!
          \mysp \left( -\frac{\,t_{max}\,}{\SONE} \right) }
        \medskip & & & {\ds - \;
          \left( \rule[-.15cm]{0cm}{.8cm}
          \myln \frac{\,t_{min}\,}{\ME^2} - l_1 \right) \!\cdot\!
          \mysp \left( -\frac{\,t_{min}\,}{\SONE} \right) }
      \medskip \medskip
 27.) & {\ds \left[ \rule[-.15cm]{0cm}{.8cm} \frac{1}{\,t\,} \cdot\!
          \mysp \left( \frac{t_{34}-\ieps}{\STWO} \right) \right]_V }
        & = & \hspace{.55cm}
        {\ds \frac{\,\pi^2}{6}\cdot\! {\cal L}_B \; - \;
          2\,\mytri \left( -\frac{\,t_{max}\,}{\STWO}  \right) \; + \;
          2\,\mytri \left( -\frac{\,t_{min}\,}{\STWO}  \right) }
        \medskip & & & {\ds + \;
          \left( \rule[-.15cm]{0cm}{.8cm}
          \myln \frac{\,t_{max}\,}{\ME^2} - l_2 \right) \!\cdot\!
          \mysp \left( -\frac{\,t_{max}\,}{\STWO} \right) }
        \medskip & & & {\ds - \;
          \left( \rule[-.15cm]{0cm}{.8cm}
          \myln \frac{\,t_{min}\,}{\ME^2} - l_2 \right) \!\cdot\!
          \mysp \left( -\frac{\,t_{min}\,}{\STWO} \right) }
\end{array}$
\newpage
$\begin{array}{ll}
 \hspace*{-.61cm}
 28.) & {\ds \RE \left(\rule[-.25cm]{0cm}{1cm}
          \left[ \rule[-.15cm]{0cm}{.8cm} \frac{1}{\,u\,} \cdot\!
          \mysp \left( \frac{t_{12}-\ieps}{\SONE} \right) \right]_V
          \right) } \;\; =
        \medskip & \hspace{1.8cm}
        {\ds {\cal L}_B \!\cdot\! \left[ \rule[-.15cm]{0cm}{.8cm}
          \hspace{.5cm} \frac{{{\cal L}_B}^2}{6} \; - \;
          \frac{1}{2} \!\cdot\! \myln \left(d_{12}\right) \!\cdot\!
          \myln \frac{\SONE}{\STWO} \; - \;
          \myln^2 \left(d_{12}\right) \; - \;
          2\, \mysp \left( \frac{1}{d_{12}} \right) \; + \;
          \frac{\,2\pi^2}{3} \rule[-.15cm]{0cm}{.8cm} \;\; \right] }
        \medskip & \hspace{1.25cm} {\ds - \;\;\!
          \myln \frac{t_{max}}{\SONE} \!\cdot\!
          \mysp \left( \frac{t_{min}}{\,s-\STWO\,} \right)
          \;\; + \;\; \myln \frac{t_{min}}{\SONE} \!\cdot\!
          \mysp \left( \frac{t_{max}}{\,s-\STWO\,} \right) }
        \medskip & \hspace{1.25cm} {\ds + \;\;
          \mytri \left( -\frac{\,t_{max}\,}{\SONE} \right) \;\; - \;\;
          \mytri \left( -\frac{\,t_{min}\,}{\SONE} \right) \;\; - \;\;
          \mytri \left( \frac{t_{max}}{\,s-\STWO\,} \right)
          \;\; + \;\;
          \mytri \left( \frac{t_{min}}{\,s-\STWO\,} \right) }
        \medskip & \hspace{1.25cm} {\ds + \;\;
          \mytri \left( -\frac{t_{max}}{\,d_{12}\,t_{min}}\, \right)
          \;\; - \;\;
          \mytri \left( -\frac{t_{min}}{\,d_{12}\,t_{max}}\, \right) }
      \medskip \medskip
 \hspace*{-.61cm}
 29.) & {\ds \RE \left(\rule[-.25cm]{0cm}{1cm}
          \left[ \rule[-.15cm]{0cm}{.8cm} \frac{1}{\,u\,} \cdot\!
          \mysp \left( \frac{t_{34}-\ieps}{\STWO} \right) \right]_V
          \right) } \;\; =
        \medskip & \hspace{1.8cm}
        {\ds {\cal L}_B \!\cdot\! \left[ \rule[-.15cm]{0cm}{.8cm}
          \hspace{.5cm} \frac{{{\cal L}_B}^2}{6} \; + \;
          \frac{1}{2} \!\cdot\! \myln \left(d_{34}\right) \!\cdot\!
          \myln \frac{\SONE}{\STWO} \; - \;
          \myln^2 \left(d_{34}\right) \; - \;
          2\, \mysp \left( \frac{1}{d_{34}} \right) \; + \;
          \frac{\,2\pi^2}{3} \rule[-.15cm]{0cm}{.8cm} \;\; \right] }
        \medskip & \hspace{1.25cm} {\ds - \;\;\!
          \myln \frac{t_{max}}{\STWO} \!\cdot\!
          \mysp \left( \frac{t_{min}}{\,s-\SONE\,} \right)
          \;\; + \;\; \myln \frac{t_{min}}{\STWO} \!\cdot\!
          \mysp \left( \frac{t_{max}}{\,s-\SONE\,} \right) }
        \medskip & \hspace{1.25cm} {\ds + \;\;
          \mytri \left( -\frac{\,t_{max}\,}{\STWO} \right) \;\; - \;\;
          \mytri \left( -\frac{\,t_{min}\,}{\STWO} \right) \;\; - \;\;
          \mytri \left( \frac{t_{max}}{\,s-\SONE\,} \right)
          \;\; + \;\;
          \mytri \left( \frac{t_{min}}{\,s-\SONE\,} \right) }
        \medskip & \hspace{1.25cm} {\ds + \;\;
          \mytri \left( -\frac{t_{max}}{\,d_{34}\,t_{min}}\, \right)
          \;\; - \;\;
          \mytri \left( -\frac{t_{min}}{\,d_{34}\,t_{max}}\, \right) }
      \medskip \medskip
 \hspace*{-.61cm}
 30.) & {\ds \left[ \rule[-.15cm]{0cm}{.8cm} \frac{1}{\,t^2} \cdot\!
          \mysp \left( \frac{t_{12}-\ieps}{\SONE} \right) \right]_V }
        \;\; =
        \medskip & \hspace{1.25cm} {\ds - \;
          \frac{1}{\,t_{max}\,} \!\cdot\!
          \mysp \left( 1 + \frac{t_{max}-\ieps}{\SONE} \right)
          \;\; + \;\; \frac{1}{\,t_{min}\,} \!\cdot\!
          \mysp \left( 1 + \frac{t_{min}-\ieps}{\SONE} \right) }
        \medskip & \hspace{1.25cm} {\ds + \;
          \frac{1}{\SONE} \!\cdot\! \left[ \rule[-.15cm]{0cm}{.8cm} \;
          \mysp \left( 1 + \frac{\SONE}{t_{max}-\ieps} \right) \;\; - \;\;
          \mysp \left( 1 + \frac{\SONE}{t_{min}-\ieps} \right) \;
          \right] }
\end{array}$
\newpage
$\begin{array}{ll}
 \hspace*{-.61cm}
 31.) & {\ds \left[ \rule[-.15cm]{0cm}{.8cm} \frac{1}{\,t^2} \cdot\!
          \mysp \left( \frac{t_{34}-\ieps}{\STWO} \right) \right]_V }
        \;\; =
        \medskip & \hspace{1.25cm} {\ds - \;
          \frac{1}{\,t_{max}\,} \!\cdot\!
          \mysp \left( 1 + \frac{t_{max}-\ieps}{\STWO} \right)
          \;\; + \;\; \frac{1}{\,t_{min}\,} \!\cdot\!
          \mysp \left( 1 + \frac{t_{min}-\ieps}{\STWO} \right) }
        \medskip & \hspace{1.25cm} {\ds + \;
          \frac{1}{\STWO} \!\cdot\! \left[ \rule[-.15cm]{0cm}{.8cm} \;
          \mysp \left( 1 + \frac{\STWO}{t_{max}-\ieps} \right) \;\; - \;\;
          \mysp \left( 1 + \frac{\STWO}{t_{min}-\ieps} \right) \;
          \right] }
\end{array}$
%
%
%
\addcontentsline{toc}{chapter}{\protect\numberline{}{References}}

%
%
\chapter*{Acknowledgments}
\addcontentsline{toc}{chapter}{\protect\numberline{ }{Acknowledgments}}
I am very grateful to my thesis advisor Dr. T. Riemann whose guidance
and teaching have been essential for the preparation of this thesis.
I acknowledge many discussions with Dr.~D.~Bardin from whom I have
learnt many things. It is a pleasure to thank the DESY directorate,
and in particular to Prof. Dr. P.~S"oding, for giving me the
opportunity to not only prepare this thesis at IfH Zeuthen, but also
to attend conferences and a summer school which has considerably
widened my horizon. I am very much indebted to
Prof. Dr. M.~M"uller-Preu{\ss}ker for kindly acting as my official
thesis supervisor.
\para
I thank A.~Leike for supplying me with his {\sc Fortran} code {\tt
4fAN} (see reference~\cite{alNC}) which helped me to verify my Born
results.
\para
In addition I acknowledge scientific discussions with W.~Beenakker,
J. Bl"umlein, J. Botts, P.~Christova, A.~Denner, A.~Leike, T.~Mohaupt,
U.~M"uller, and S.~Riemann. I am indebted to the DESY-IfH Zeuthen
computer center staff, especially to W.~Friebel, A.~K"ohler,
W.~Niepraschk, and C.~Rethfeld. I thank the Technical Support group at
DESY-IfH for their permanent availability and M.~Gohr for preparing
the phase space illustrations in appendix~\ref{phasespa}.
\para
Careful reading of the manuscript by Dr.~T. Riemann and J. Wegmann
as well as comments on the presentation of the topic by Dr.~D.~Bardin and
Prof. Dr. M.~M"uller-Preu{\ss}ker are acknowledged.
\para
Thanks to Arnd, Christian, Denny, Dima, Jana, Jim, Lida, Penka,
Sabine, Thomas, Tord, Uwe, and all the others for the wonderful time.
%
%
\chapter*{Curriculum Vitae}
\addcontentsline{toc}{chapter}{\protect\numberline{}{Curriculum Vitae}}
\begin{tabular}{ll}
  Name & Dietrich Lehner \medskip
  2.11.1967  & geboren in M"unchen als viertes Kind von
               Prof.~Dr.~rer.~nat. G"unther \\
             & Lehner (geb. 1931) und dessen Ehefrau Dr.~rer.~nat.
               Lore~Lehner \\
             & (1932-1984), geborene Ro{\ss}ner. \medskip
  1973--1975 & Besuch der Toni-Pf"ulf-Grundschule in M"unchen. \medskip
  1975       & Umzug der Familie nach B"oblingen. \medskip
  1975--1977 & Besuch der Eduard-M"orike-Grundschule in
               B"oblingen. \medskip
  1977--1986 & Besuch des Otto-Hahn-Gymnasiums in B"oblingen. \medskip
  1983       & Dreimonatiger Aufenthalt an den ``University of Toronto
               Schools'', \\
             & Toronto, Ontario, Kanada als Austauschsch"uler. \medskip
  10.6.1986  & Abitur, Durchschnittsnote 1.3 . \medskip
  Nov. 1986  & Aufnahme des Physikstudiums an der
               Ludwig-Maximilians-Universit"at \\
             & M"unchen. \medskip
  3.11.1988  & Vordiplom, Gesamturteil ``sehr gut''. \medskip
  1989       & Sommerstudent am Europ"aischen Laboratorium f"ur
               Teilchenphysik \\
             & CERN, Genf. \medskip
  1989--1992 & Stipendiat der Friedrich-Naumann-Stiftung,
               Berlin. \medskip
  1990       & Laborpraktikum am CERN. \medskip
  April 1991 & Beginn der Diplomarbeit ``Messung der
               $e^{+}e^{-}$--Annihilation in die \\
             & 4--leptonischen Endzust\"ande $e^+ e^- e^+ e^-$,
               $e^+ e^- \mu^+ \mu^-$ und $\mu^+ \mu^- \mu^+ \mu^-$ \\
             & an der \zz--Resonanz'' bei Priv. Doz. Dr. W. Blum am
               Max-Planck- \\
             & Institut f"ur Physik in M"unchen. \medskip
  2.12.1992  & Diplom, Gesamturteil ``sehr gut''. \medskip
  Seit 1.2.1993 & Doktorand bei Dr. T. Riemann am
               DESY-IfH Zeuthen.
\end{tabular}
%
%
\chapter*{List of Publications and Talks}
\addcontentsline{toc}{chapter}{\protect\numberline{}{List of
    Publications and Talks}}
%
\section*{Publications}
\addcontentsline{toc}{section}{\protect\numberline{}{Publications}}
\begin{itemize}
  \item[\protect{[1]}] D. Lehner: \\
    Messung der $e^{+}e^{-}$--Annihilation in die 4--leptonischen
    Endzust\"ande $e^{+}e^{-}e^{+}e^{-}$, \\
    $e^{+}e^{-}{\mu^{+}\mu^{-}}$
    und ${\mu^{+}\mu^{-}\mu^{+}\mu^{-}}$ an der $Z^0$--Resonanz. \\
    MPI Munich preprint MPI--PhE/93--02; \vspace{.1cm} \\
    Measurement of the 4--lepton final states $e^{+}e^{-}e^{+}e^{-}$,
    $e^{+}e^{-}{\mu^{+}\mu^{-}}$~and ${\mu^{+}\mu^{-}\mu^{+}\mu^{-}}$
    in the 1990 and 1991 Data. \\
    ALEPH collaboration internal note  ``ALEPH 93-019, PHYSIC 93-012''.
  \item[\protect{[2]}]
    D. Bardin, M. Bilenky, D. Lehner, A. Olshevski, T. Riemann: \\
    Semi-Analytical Approach to Four-Fermion Production in
    $e^{+}e^{-}$~Annihilation. \\
    Preprints CERN-TH.7295/94, DESY 94-093, Proceedings of the {\it
    Zeuthen Workshop on Elementary Particle
    Theory ``Physics at LEP200 and Beyond'', Teupitz, Germany, April
    $10^{th}-15^{th}$ 1994}, \\
    Nucl. Phys. {\bf B} (Proc. Suppl.) 37B (1994) 148--157.
  \item[\protect{[3]}] D. Lehner: \\
    Initial State Radiation to Off-Shell $Z^0$ Pair Production in
    $e^{+}e^{-}$~Annihilation.  \\
    Mod. Phys. Lett. {\bf A9} (1994) 2937.
  \item[\protect{[4]}]
    The ALEPH collaboration (D. Buskulic, ... , D. Lehner et al.): \\
    Study of the four-fermion final state at the $Z^0$ resonance. \\
    Z. Phys. {\bf C66} (1995) 3.
  \item[\protect{[5]}] D. Bardin, D. Lehner, T. Riemann: \\
    Complete Initial State Radiation to Off-Shell $Z^0$~Pair
    Production in $e^{+}e^{-}$~An\-ni\-hi\-la\-tion. \\
    Preprint DESY 94--216, to appear in the proceedings of the~~{\it
    $IX^{th}$ International Workshop on ``High Energy Physics and
    Quantum Field Theory'', Zvenigorod, Mos\-cow Region, Russia,
    September $15^{th}-22^{nd}$ 1994}.
\end {itemize}
\newpage
\noindent
\section*{Talks}
\addcontentsline{toc}{section}{\protect\numberline{}{Talks}}
\begin{enumerate}
\item {\it Measurement of the Final States $e^{+}e^{-}e^{+}e^{-}$,~
   $e^+e^-\mu^+\mu^-$~ and $\mu^+\mu^-\mu^+\mu^-$~in the ALEPH 1990
   and 1991 Data} \vspace{-.25cm}
   \begin{itemize}
     \item[$\to$] DESY--IfH Zeuthen, September $6^{th}$ 1992,
     \item[$\to$] Meeting of the ALEPH Searches group, CERN, March
                  $1^{st}$ 1993,
     \item[$\to$] 1993 German Physical Society Spring Meeting, Mainz,
                  Germany, March 1993. \vspace{.01cm}
   \end{itemize}
\item {\it Initial State Radiation to Off-Shell $Z^0$ Pair Production in
     $e^{+}e^{-}$~Annihilation} \vspace{-.25cm}
   \begin{itemize}
     \item[$\to$] 1994 German Physical Society Spring Meeting,
                  Dortmund, Germany, March 1994,
     \item[$\to$] Humboldt--University, Berlin, Germany, May $16^{th}$
                  1994,
     \item[$\to$] $XVII^{th}$ Kazimierz Meeting on Elementary Particle
                   Physics, Kazimierz, Po\-land, May 1994.  \vspace{.51cm}
   \end{itemize}
\item {\it Complete Initial State Radiation to Off-Shell $Z^0$ Pair
    Production in $e^{+}e^{-}$~Annihilation} \vspace{-.25cm}
   \begin{itemize}
     \item[$\to$] $IX^{th}$ International Workshop on ``High Energy
                  Physics and Quantum Field Theory'', Zvenigorod,
                  Moscow Region, Russia, September 1994,
     \item[$\to$] DESY Theory Workshop, Hamburg, Germany, September
                  1994,
     \item[$\to$] 1995 German Physical Society Spring Meeting,
                  Karlsruhe, Germany, March 1995.  \vspace{.51cm}
   \end{itemize}
\item {\it Semi-Analytical Approach to Standard Model Four-Fermion
      Production at LEP2 and Beyond} \vspace{-.25cm}
   \begin{itemize}
     \item[$\to$] First Meeting of the LEP2 Working Group ``Standard
                  Model Proces\-ses'', \\  CERN, November $25^{th}$ 1994,
     \item[$\to$] 1995 German Physical Society Spring Meeting,
                  Karlsruhe, Germany, March 1995,
     \item[$\to$] Meeting of the ``L3-LEP2 Working Group'', CERN,
                  March $31^{st}$ 1995,
     \item[$\to$] Third Meeting of the LEP2 Working Group ``Standard
                  Model Processes'', CERN, April $5^{th}$
                  1995 (Update),
     \item[$\to$] Seminar at the ``Centre de Physique des Particules
                  de Marseille'', Marseille, France, April $10^{th}$
                  1995.
   \end{itemize}
\end{enumerate}
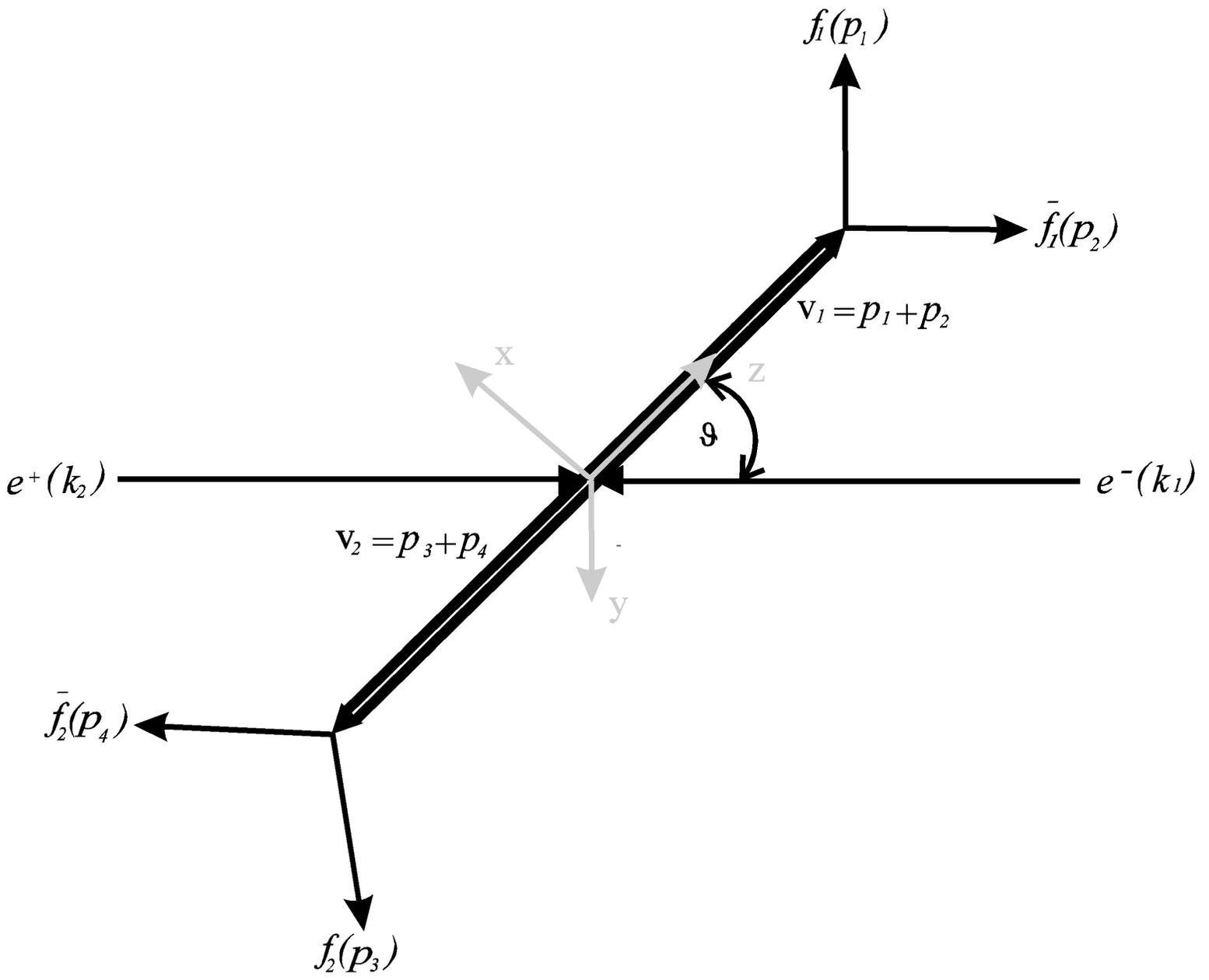
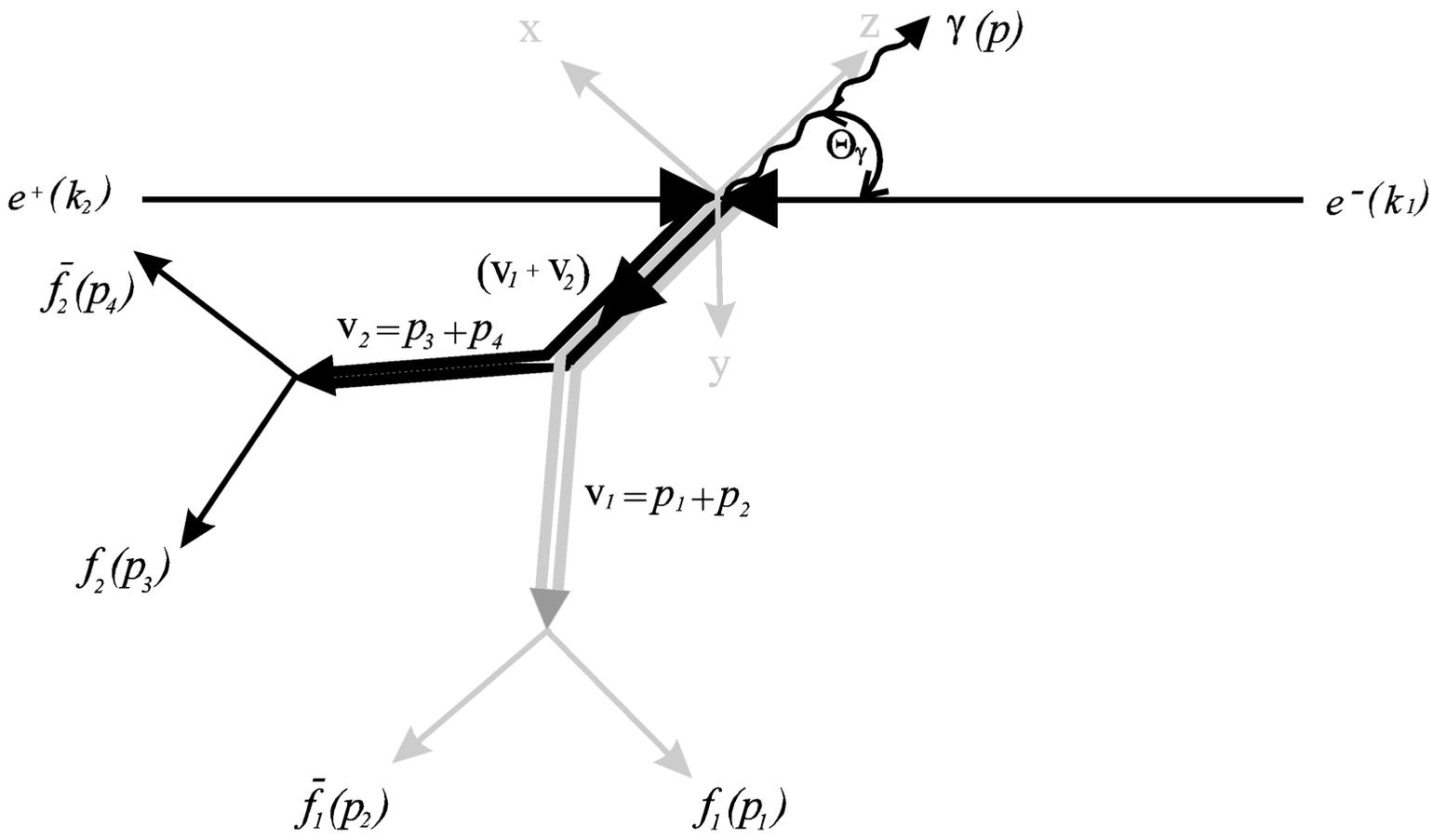
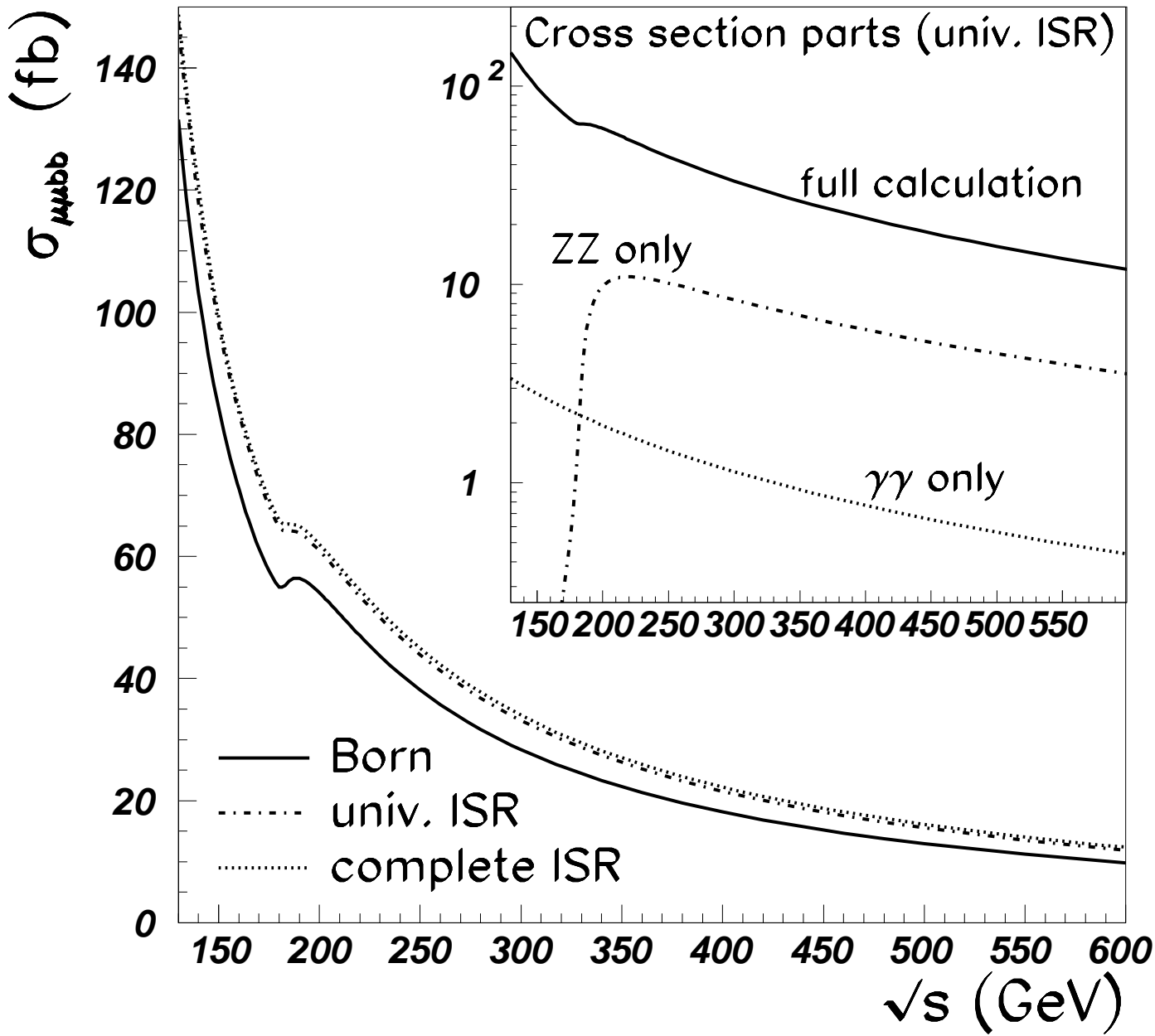
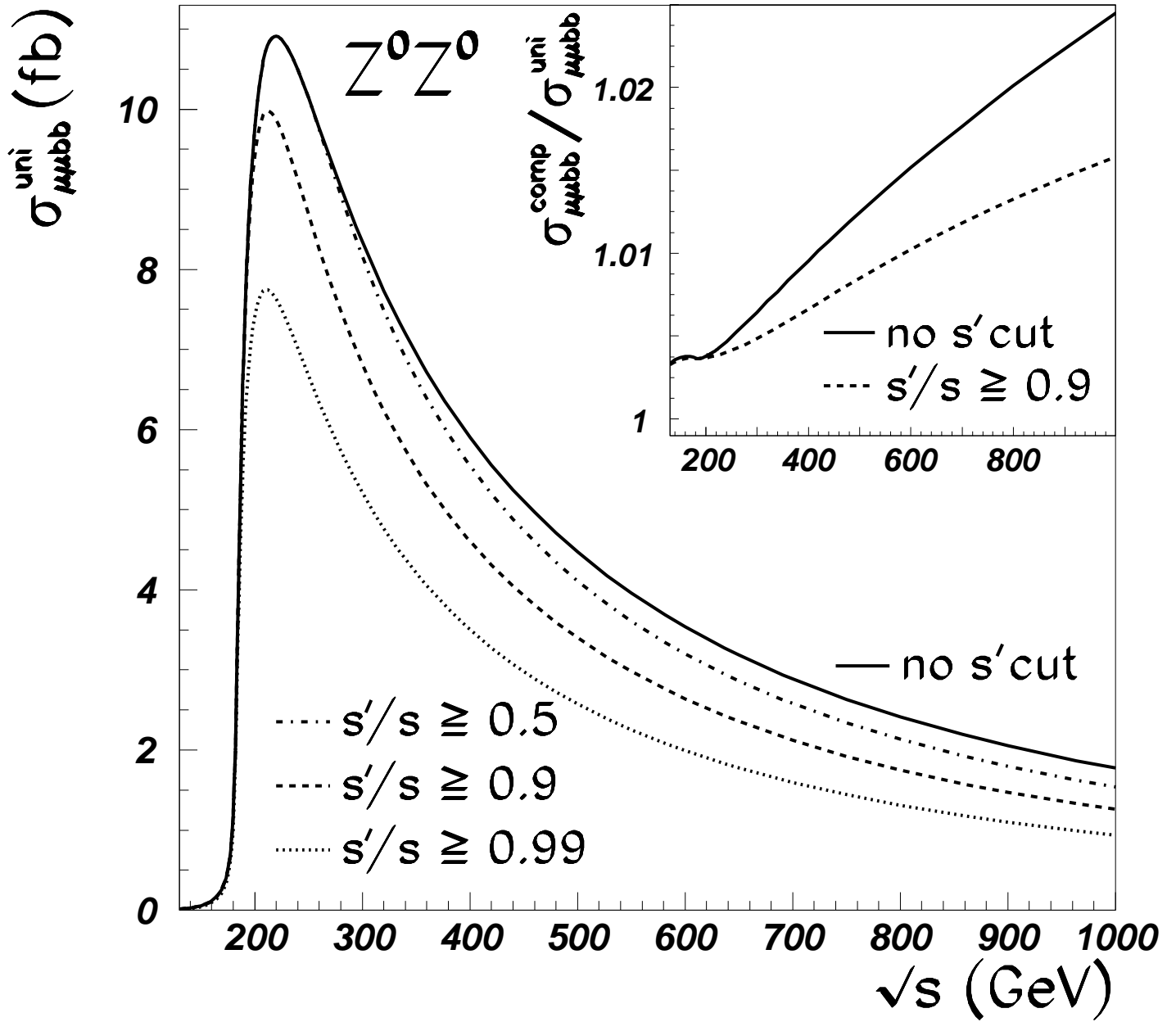
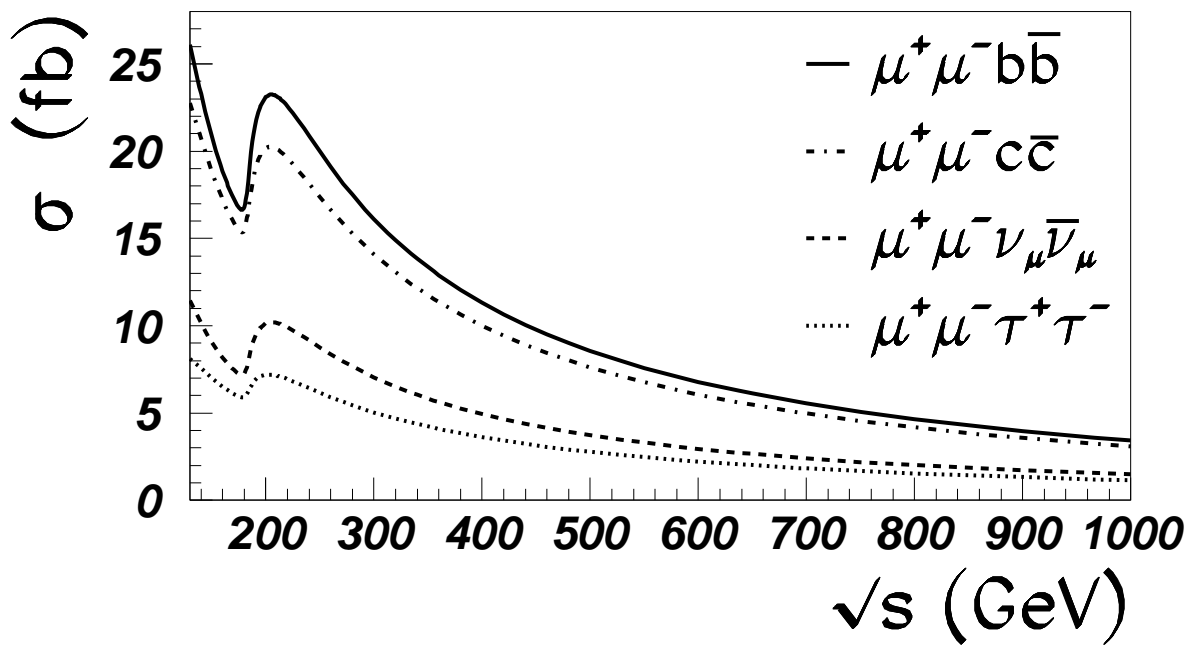
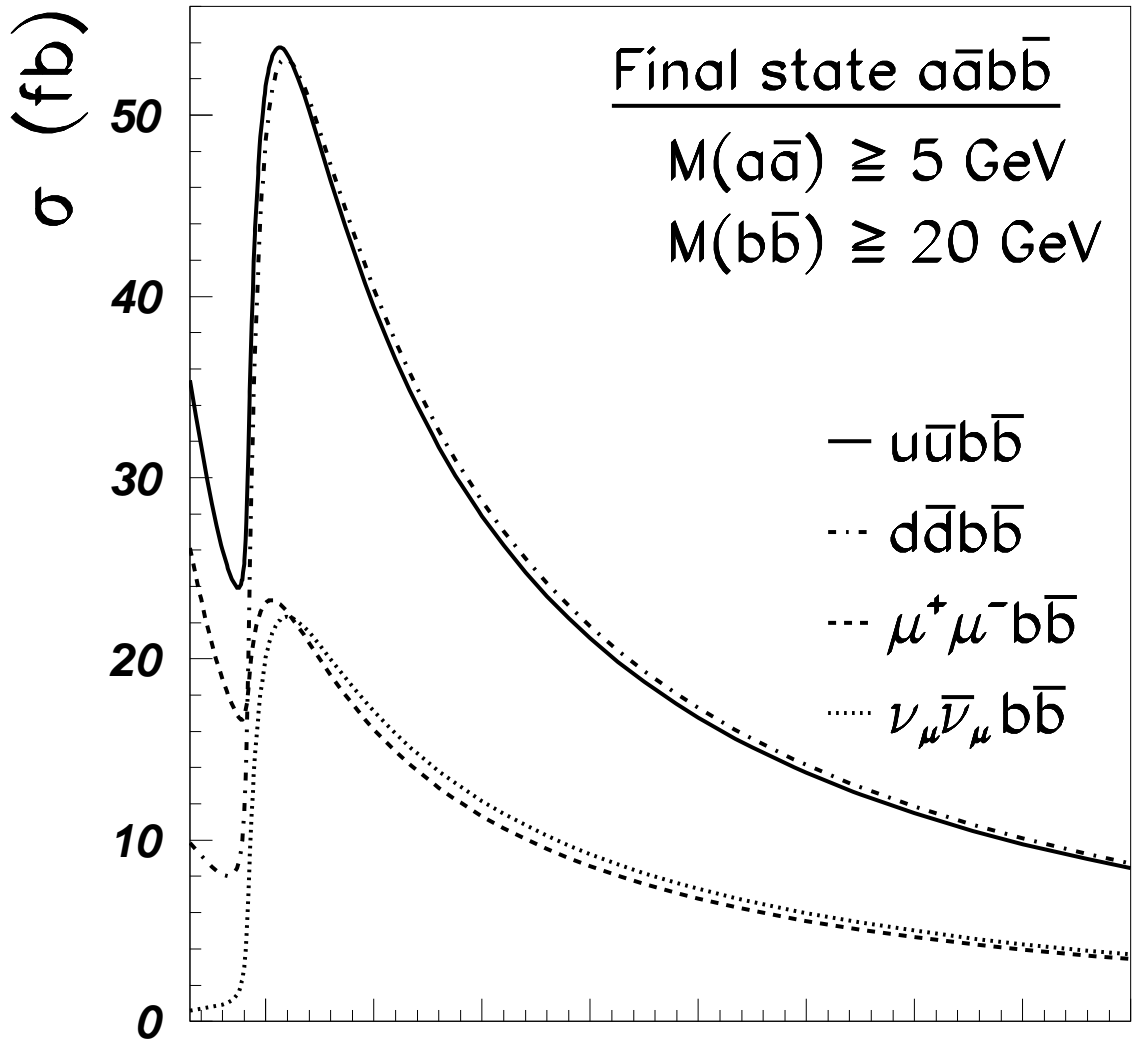
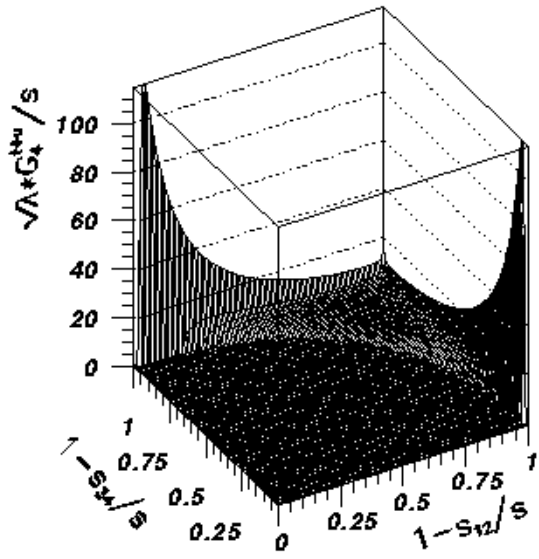
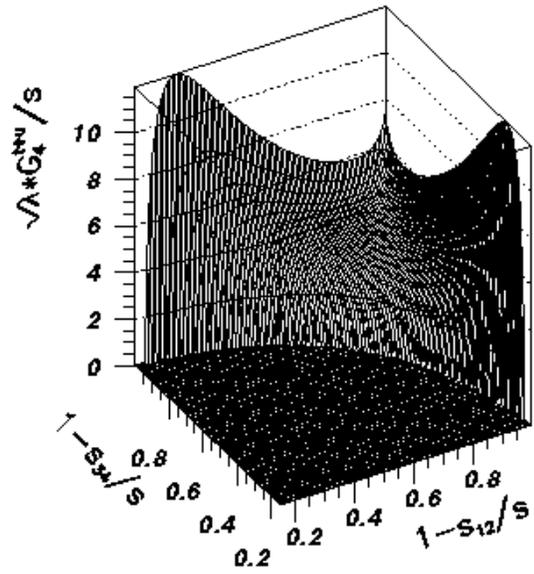
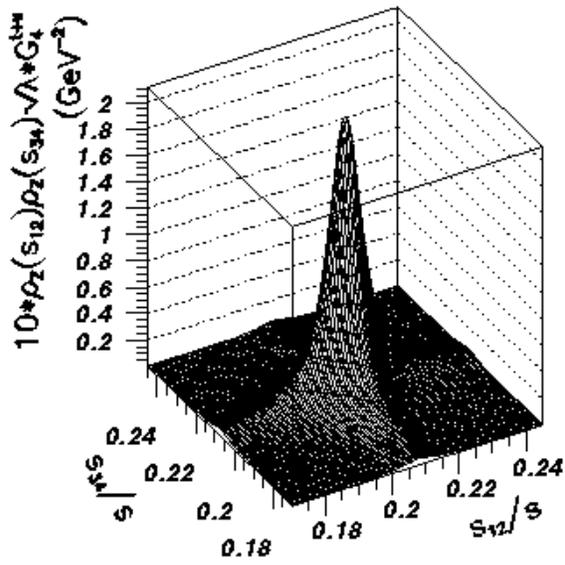
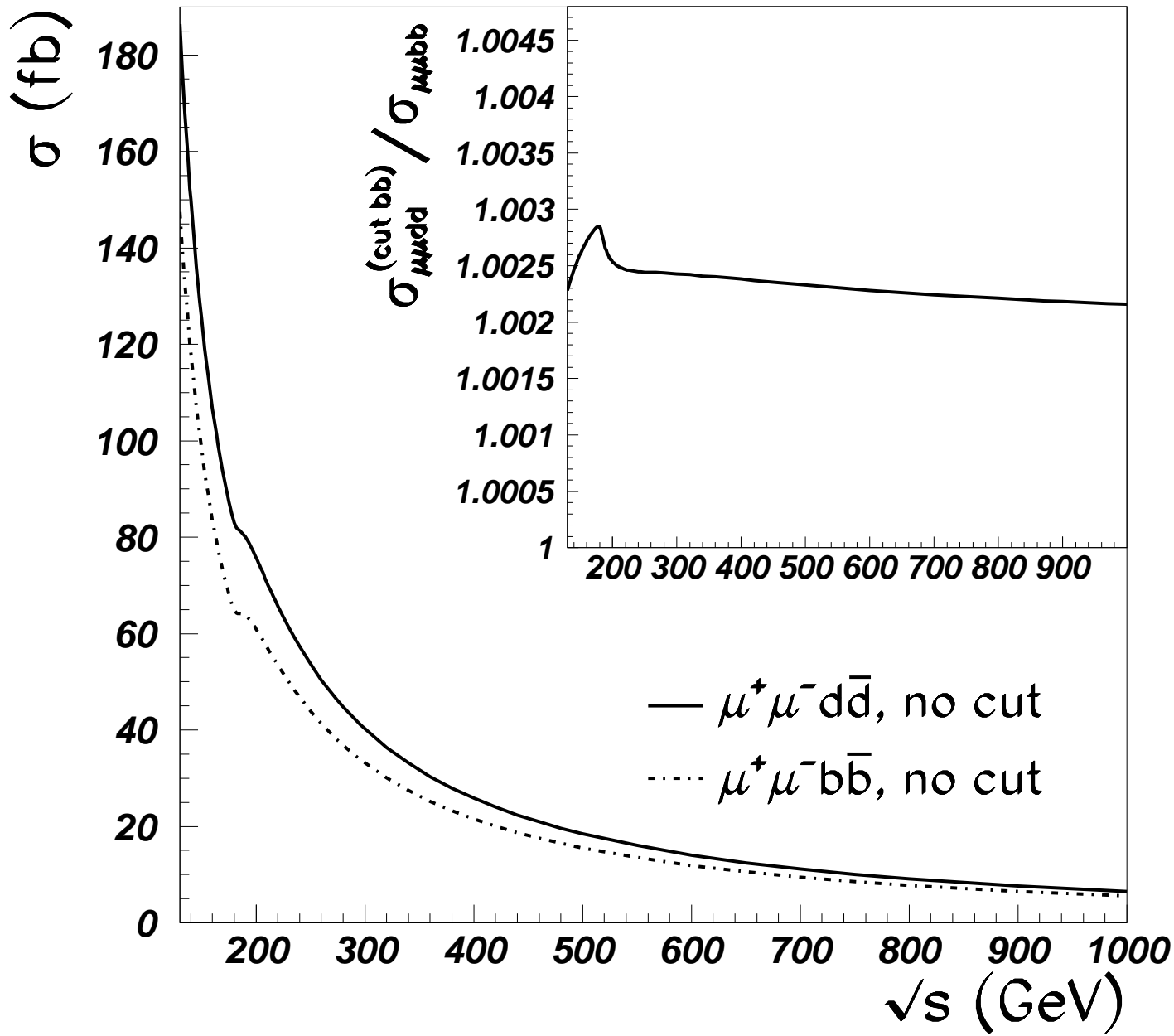
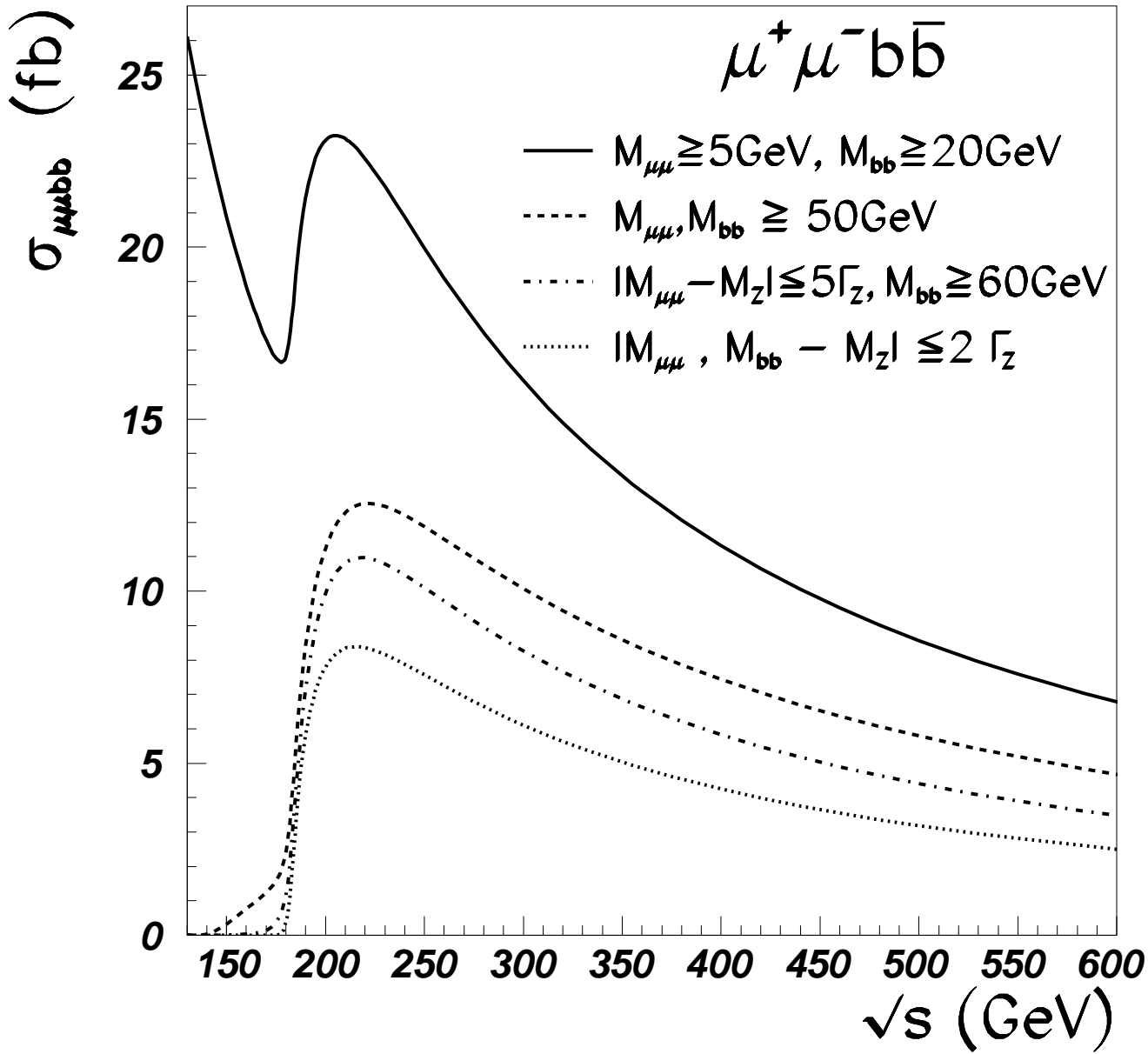
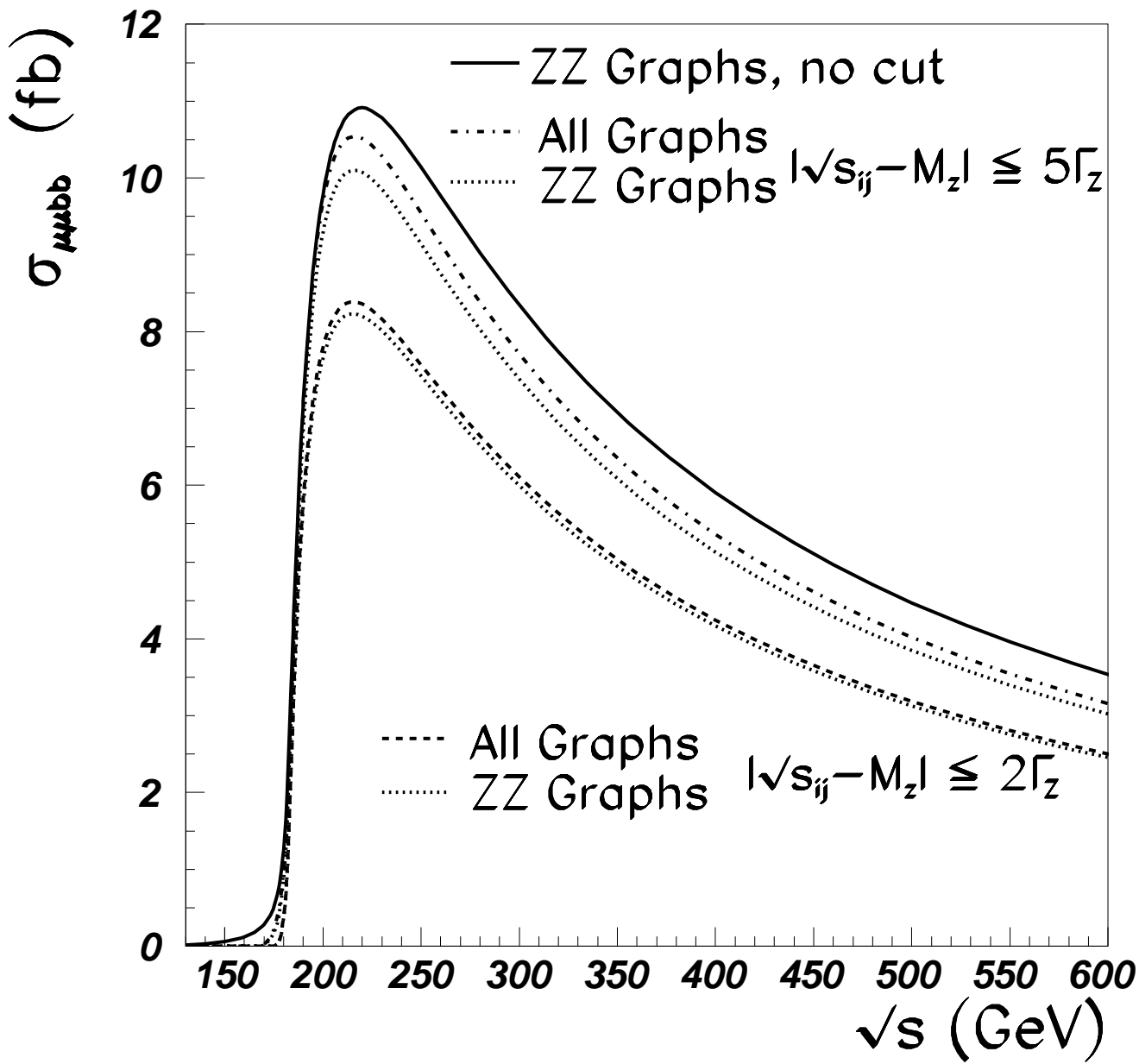
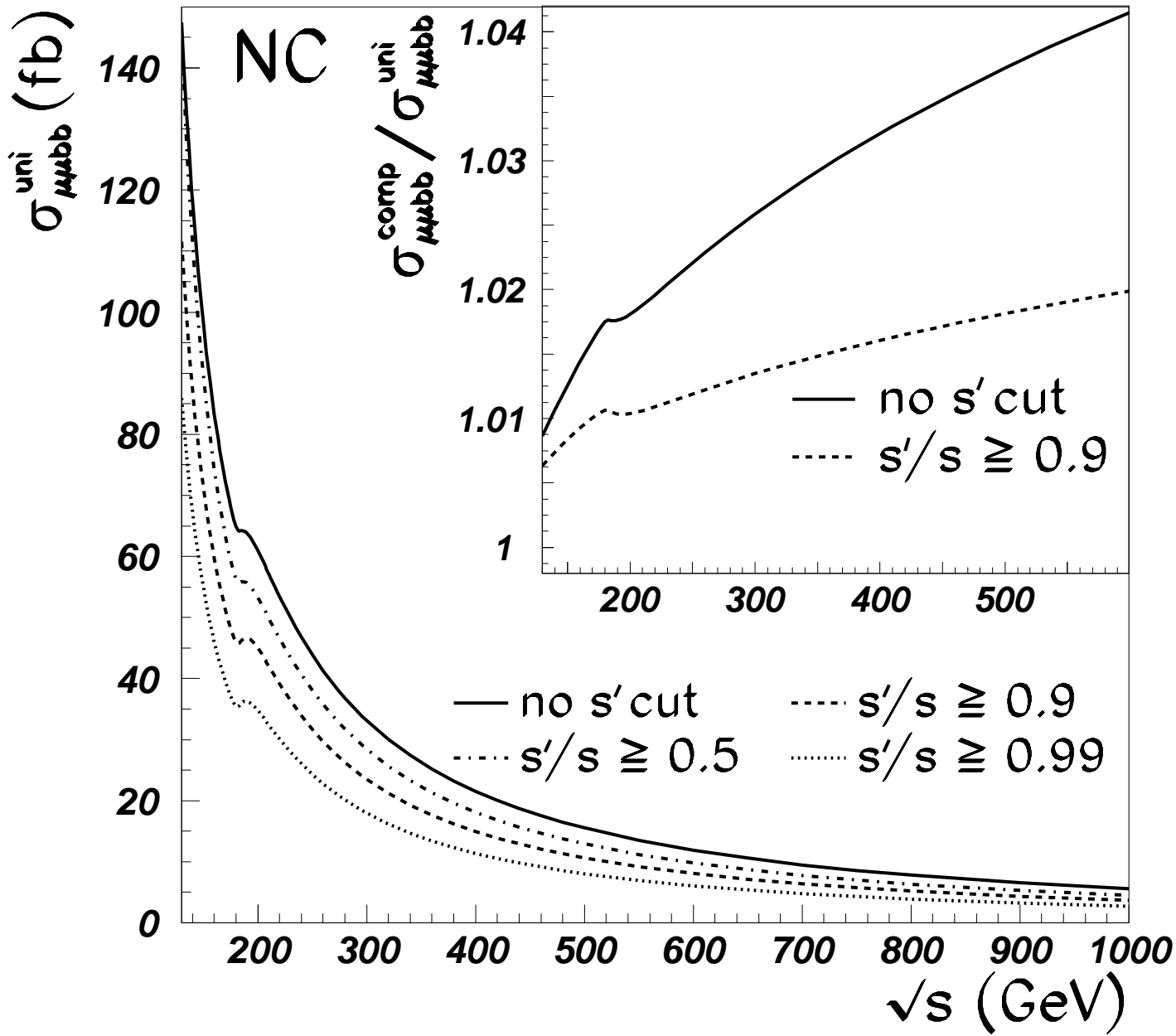
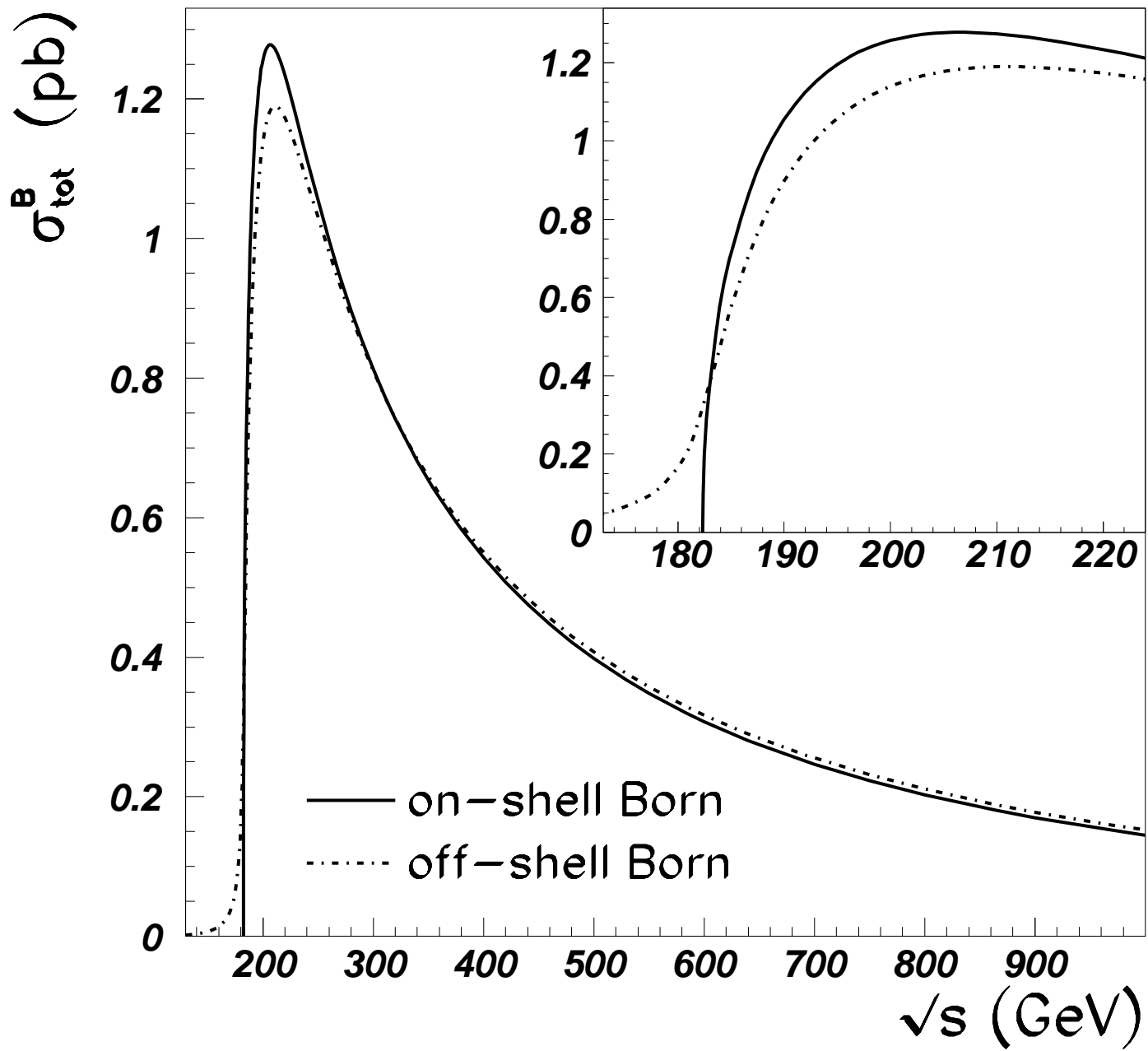
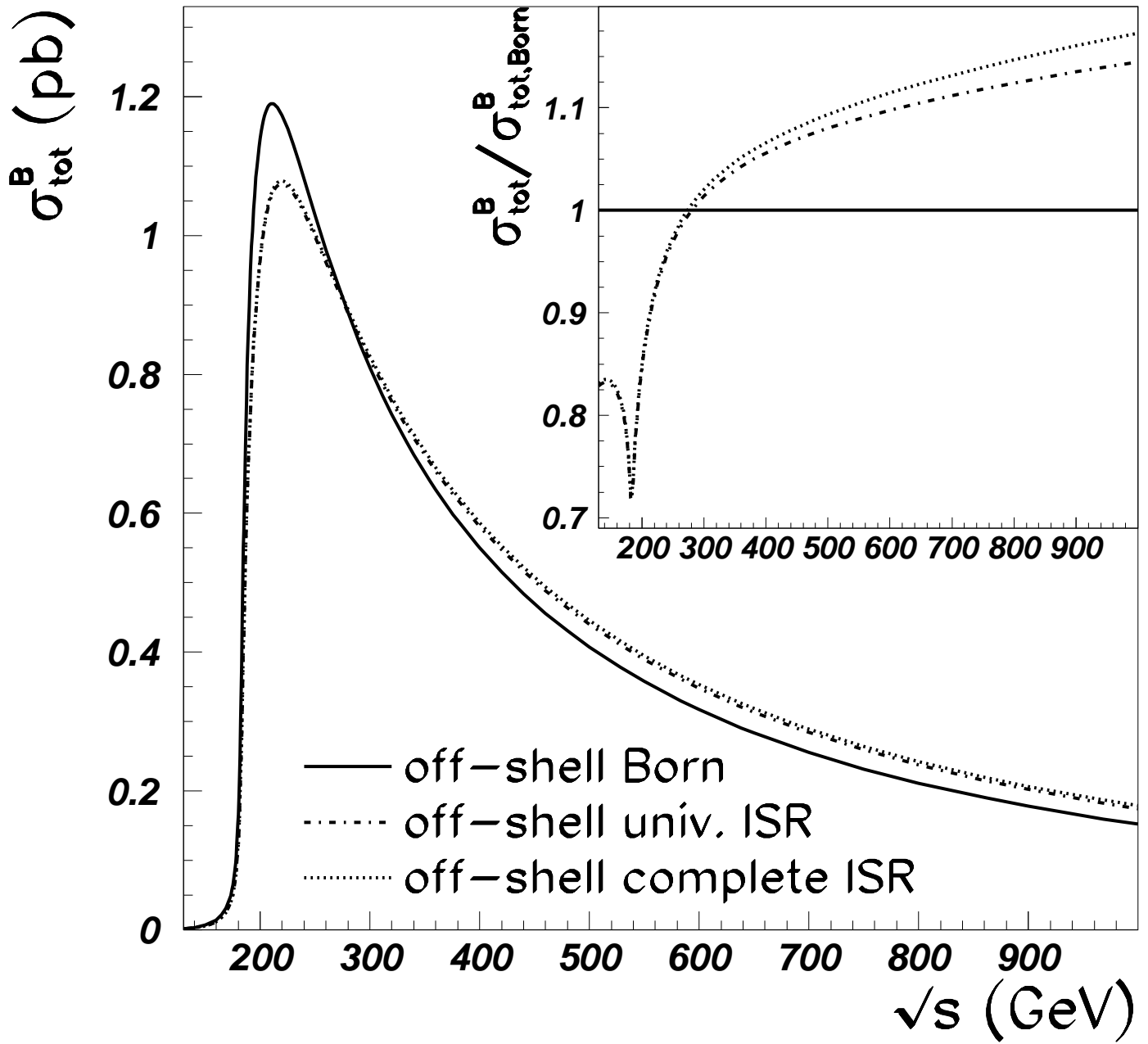
\end{document}